\documentclass[pdflatex,sn-mathphys-num]{sn-jnl}

\usepackage{graphicx}%
\usepackage{multirow}%
\usepackage{amsmath,amssymb,amsfonts}%
\usepackage{amsthm}%
\usepackage{mathrsfs}%
\usepackage[title]{appendix}%
\usepackage{xcolor}%
\usepackage{textcomp}%
\usepackage{manyfoot}%
\usepackage{booktabs}%
\usepackage{algorithm}%
\usepackage{algorithmicx}%
\usepackage{algpseudocode}%
\usepackage{listings}%
\usepackage{subcaption}
\usepackage{siunitx}
\usepackage[table,xcdraw]{xcolor}
\usepackage[version=4,arrows=pgf-filled]{mhchem}
\usepackage{mathtools}
\usepackage{nowidow}
\usepackage[normalem]{ulem}
\usepackage[percent]{overpic}
\usepackage{xcolor,colortbl}
\usepackage[font=small,justification=justified]{caption}

\newcommand{\ith}{$I_\mathrm{C}$}
\newcommand{\itr}{$I_\mathrm{C}^\mathrm{ref}$}
\newcommand{\id}{$I_\mathrm{D}$}
\newcommand{\vd}{$V_\mathrm{D}$}
\newcommand{\vg}{$V_\mathrm{G}$}

\newcommand{\idvg}{$I_\mathrm{D}(V_\mathrm{G})$}
\newcommand{\idvd}{$I_\mathrm{D}(V_\mathrm{D})$}

\newcommand{\dvth}{$\Delta V_\mathrm{th}$}

\newcommand{\er}{$E_\mathrm{R}$}
\newcommand{\et}{$E_\mathrm{T}$}
\newcommand{\xt}{$X_\mathrm{T}$}

\newcommand{\bisbeta}{$\beta$-\ce{Bi2SeO5}}
\newcommand{\bisalpha}{$\alpha$-\ce{Bi2SeO5}}
\newcommand{\bissc}{\ce{Bi2O2Se}}
\newcommand{\bisox}{\ce{Bi2SeO5}}

\newcommand{\fig}[1]     {Fig.~\ref{#1}}
\newcommand{\tab}[1]    {Table~\ref{#1}}

\newcommand{\figs}[2]    {Figs.~\ref{#1}$-$\ref{#2}}
\newcommand{\plan}   {TG-FET}
\newcommand{\fin}   {Fin-FET}
\newcommand{\sgaa}   {Semi-GAA-FET}
\newcommand{\gaa}   {GAA-FET}

\newcommand{\figheader}{
&&&\\&&&\\&&&\\
\includegraphics[width=0.035\linewidth]{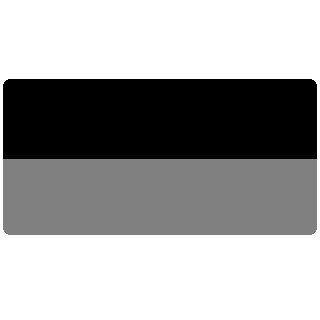} \begin{minipage}[p]{0.07\textwidth}\vspace{-4mm}\textsf{\small{\plan{}}}\end{minipage}&
\hspace{1mm}\hspace{-1mm}\includegraphics[width=0.035\linewidth]{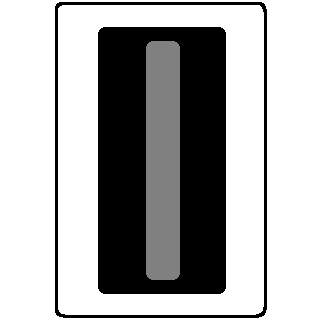} \begin{minipage}[p]{0.07\textwidth}\vspace{-4mm}\hspace{-1mm}\textsf{\small{\fin{}}}\end{minipage}&
\includegraphics[width=0.035\linewidth]{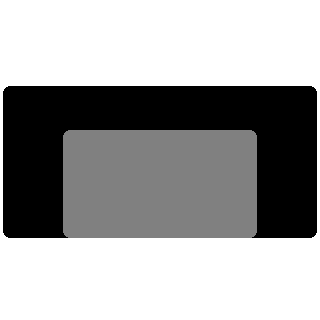} \begin{minipage}[p]{0.13\textwidth}\vspace{-4mm}\textsf{\small{\sgaa{}}}\end{minipage}&
\includegraphics[width=0.035\linewidth]{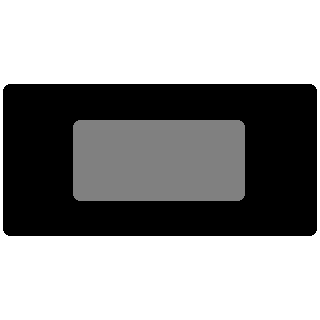} \begin{minipage}[p]{0.085\textwidth}\vspace{-4mm}\textsf{\small{\gaa{}}}\end{minipage}\\
}

\definecolor{plan_color}{HTML}{E6F4F4}
\definecolor{fin_color}{HTML}{FBEAEA}
\definecolor{sgaa_color}{HTML}{FBF9E5}
\definecolor{gaa_color}{rgb}{0.94,0.94,1}

\newcolumntype{p}{>{\columncolor{plan_color}}c}
\newcolumntype{f}{>{\columncolor{fin_color}}c}
\newcolumntype{s}{>{\columncolor{sgaa_color}}c}
\newcolumntype{g}{>{\columncolor{gaa_color}}c}

\begin{document}

\title[Article Title]{Towards ultra-scaled nanoelectronics using the zipper material system \bissc{}/\bisox{}}

\author[1]{\fnm{Mohammad Rasool} \sur{Davoudi}}
\author[1]{\fnm{Mina} \sur{Bahrami}}
\author[1]{\fnm{Axel} \sur{Verdianu}}
\author[1]{\fnm{Pedram} \sur{Khakbaz}}
\author[1]{\fnm{Dominic} \sur{Waldhoer}}
\author[1]{\fnm{Mahdi} \sur{Pourfath}}
\author[1]{\fnm{Alexander} \sur{Karl}}
\author[1]{\fnm{Christoph} \sur{Wilhelmer}}
\author[1,2]{\fnm{Yury} \sur{Illarionov}}
\author[1]{\fnm{Seyed Mehdi} \sur{Sattari-Esfahlan}}
\author[3]{\fnm{Yichi} \sur{Zhang}}
\author[3]{\fnm{Junchuan} \sur{Tang}}
\author[4]{\fnm{Aftab} \sur{Nazir}}
\author[3]{\fnm{Ye} \sur{Li}}
\author[3]{\fnm{Xiaoying} \sur{Gao}}
\author[3]{\fnm{Congwei} \sur{Tan}}
\author[4]{\fnm{Yu} \sur{Zheng}}
\author[4]{\fnm{Changze} \sur{Liu}}
\author[1]{\fnm{Michael} \sur{Waltl}}
\author[3]{\fnm{Hailin} \sur{Peng}}
\author[1]{\fnm{Theresia} \sur{Knobloch}}
\author*[1]{\fnm{Tibor} \sur{Grasser}}\email{grasser@iue.tuwien.ac.at}

\affil[1]{\orgdiv{Institute for Microelectronics}, \orgname{Technical University Vienna}, \orgaddress{\country{Austria}}}
\affil[2]{\orgname{Southern University of Science and Technology}, \orgaddress{\state{Shenzhen}, \country{China}}}
\affil[3]{\orgname{Peking University}, \orgaddress{\state{Beijing}, \country{China}}}
\affil[4]{\orgname{Huawei Technologies Research and Development Belgium N.V.}, \orgaddress{\country{Belgium}}}

\abstract{
Two-dimensional (2D) materials could overcome the scaling bottleneck of nanoelectronics by enabling atomically thin channels, superior electrostatic control, and reduced short-channel effects. However, progress is limited by the lack of semiconductor--insulator interfaces being simultaneously scalable, stable, and reliable. Conventional 2D interfaces are often low quality or require transferring dissimilar materials, limiting reproducibility and scalability. We show that the zipper heterostructure formed by the high-mobility 2D semiconductor \bissc{} and its native high-$\kappa$ oxide \bisox{} addresses these challenges,
providing an atomically sharp, chemically matched interface with excellent electrostatics and promising scaling potential.
We further present the first comprehensive multiscale assessment of \bissc{}/\bisox{} transistors targeting thermal stability, reliability, and scalability by linking atomic-scale structure and defects to device-level behavior across \textit{four} transistor generations (top-gated, fin, and two gate-all-around architectures).
Benchmarking against IRDS-2035 targets indicates that \bissc{}/\bisox{} devices could deliver high drive current with low gate leakage under aggressive scaling, potentially surpassing the targets in the upper-bound region of the sensitivity analysis.
Finally, we identify oxygen-related oxide defects as the dominant origin of hysteresis consistent with a beneficial role of encapsulation and oxygen-rich annealing. Together, our findings support the potential of this zipper material system as a technologically-credible and manufacturing-relevant platform for future nanoelectronics.
}


\maketitle
\newpage

\section{Introduction}\label{sec:intro}
The relentless drive for faster, more energy-efficient, and more densely integrated electronics relies on the continued miniaturization of transistors~\cite{roadmap_BC,roadmap_MM}. However, scaling silicon-based transistors into the nanometer regime leads to challenges such as mobility degradation and severe short-channel effects~\cite{multigateTransistors}, prompting a search for alternatives to conventional silicon technology~\cite{desai2016mos2}. Two-dimensional (2D) semiconductors have emerged as promising candidates: their atomically thin channels mitigate short-channel effects~\cite{lemmeNature,ChauIEDM2019}, they retain good mobility at extreme thicknesses~\cite{Introducing2DFET,10319515}, and they enable channel lengths below \SI{12}{nm}~\cite{Introducing2DFET,lemmeNature,desai2016mos2}. In addition, to reach their potential, 2D field-effect transistors (FETs) require high-$\kappa$ insulators to suppress short-channel effects~\cite{LiuNature2021}, while maintaining low leakage~\cite{ChauMicroEng,Knobloch2021} and low defect densities to ensure long-term reliability~\cite{illarionov2020insulators}. This is particularly critical in 2D FETs since large hysteresis in the transfer characteristics and sizable threshold voltage drifts leading to bias temperature instabilities (BTI) are frequently observed~\cite{Illarionov2017c,knobloch2022finding,StathisMicRel2018}, typically attributed to charge trapping and defect formation at the semiconductor–insulator interface and in the adjacent dielectric region~\cite{illarionov2020insulators,AngTDMR2008,knob_iedm_2023,waldhoer2022comphy,WaldhoerTED2021,GHOSH2025112333}.

Overcoming these challenges requires careful interface engineering as conventional van der Waals (vdW) interfaces introduce low-permittivity gaps that act as parasitic capacitances, while covalently bonded interfaces often degrade electronic quality~\cite{pourfathScience}. Zipper material systems have recently emerged as a promising alternative: quasi-covalent bonding eliminates the vdW gap without introducing dangling bonds, combining the electrostatic advantages of 2D semiconductors with high-quality oxide interfaces~\cite{pourfathScience}. A particularly attractive realization is the high-mobility (up to $\SI{812}{\mathrm{cm^2/Vs}}$) semiconductor bismuth oxyselenide (\bissc{})~\cite{peng_vdw,peng_sc,wu2017high,khakbazACS} together with its native high-$\kappa$ ($\SI{20}{}\!\sim\!\SI{35}{}$) oxide bismuth oxoselenate (\bisox{})~\cite{peng_beta,khakbazACS,highKpeng}. This system has already demonstrated excellent interface quality, scalability, and device performance~\cite{knob_iedm_2023,11041216}, making it a compelling platform for assessing the true limits of 2D transistor technology.

Despite these promising results, no study has yet addressed the combined performance, stability, and reliability of \bissc/\bisox{} transistors under realistic scaling conditions. Here, we present a comprehensive study on the performance and the reliability of four generations of \bissc{}$/$\bisox{} transistor types, three of those featuring sub-nanometer equivalent oxide thicknesses (EOTs), including planar and finFETs, as well as gate-all-around geometries.
We first performed electrical characterization, including measurements of gate currents as well as transfer and output characteristics at various conditions, and quantified the hysteresis over a wide range of sweep frequencies and temperatures. 
Subsequently, we employed density functional theory (DFT) calculations to extract intrinsic material and defect properties, such as electronic bands and trap levels. 
These parameters were then used as input for technology computer-aided design (TCAD) simulations to calibrate the electrostatics under operating conditions. Such a multiscale modeling approach~\cite{waldhoer2022comphy,mars_iedm2024} enables quantitative comparison between measured and simulated device characteristics, facilitating the extraction of key performance metrics such as channel electron mobility and contact resistance. In particular, it also allows us to estimate the true performance potential for ultrascaled devices aimed at the IRDS 2035 node and beyond.

In addition, we investigated the most important non-idealities in this novel material system by providing atomistic insight into the nature of point defects that affect their stability.
This is done by thoroughly modeling charge transfer processes using the nonradiative multi-phonon (NMP) framework~\cite{goes2018,GrasserMicRel2012}
to extract the distribution of defect parameters. For this we employ the effective single defect decomposition (ESiD) method~\cite{WaldhoerTED2021,waldhoer2022comphy} integrated with transient TCAD simulations for varying sweeping rates and temperatures. 
This methodology allows us to extract a physically consistent trap distribution that reproduces the experimentally observed hysteresis in \bissc{}$/$\bisox{} FETs, not only at specific current readout (\ith{}) levels but across the whole operating range.
In parallel, defect parameters for various potential defect candidates were calculated using DFT and compared with experimentally extracted trap distributions to identify the defects that are most likely responsible for hysteresis and potentially for parasitic gate
currents in devices employing a \bisox{} dielectric. 
The complete workflow used to evaluate the performance and to identify the point defects that govern the reliability of \bissc{}$/$\bisox{} transistors is illustrated in the schematic diagram shown in \fig{flowchart}.

\section{Results and Discussion}
\subsection{The \bissc{}$/$\bisox{} Material System}
\setlength{\tabcolsep}{2pt}  
\renewcommand{\arraystretch}{0.1}
\begin{figure}[!bt]
{\captionsetup[subfigure]{skip=-1pt}\begin{subfigure}[b]{\linewidth}
\includegraphics[width=1\linewidth]{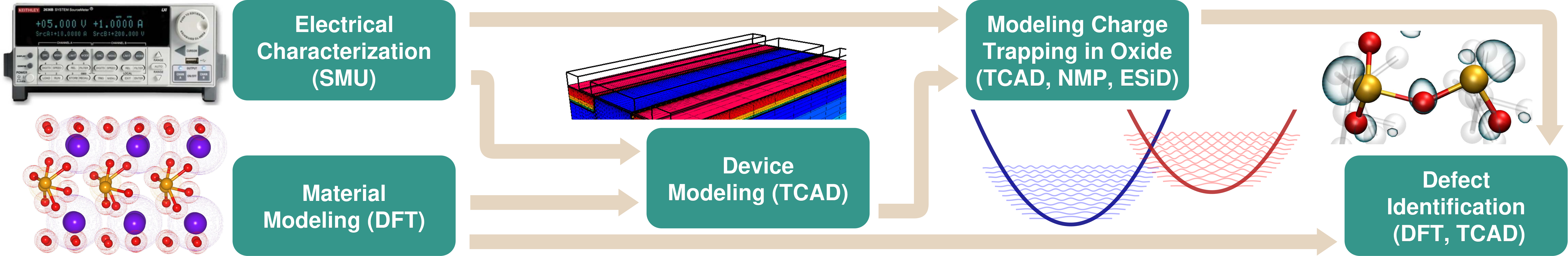}
\caption{}
\label{flowchart}
\end{subfigure}}
{\captionsetup[subfigure]{skip=-12pt}\begin{subfigure}[b]{.34\linewidth}
\includegraphics[width=1.0\linewidth,trim={0cm 0cm 0cm 1.1cm},clip]{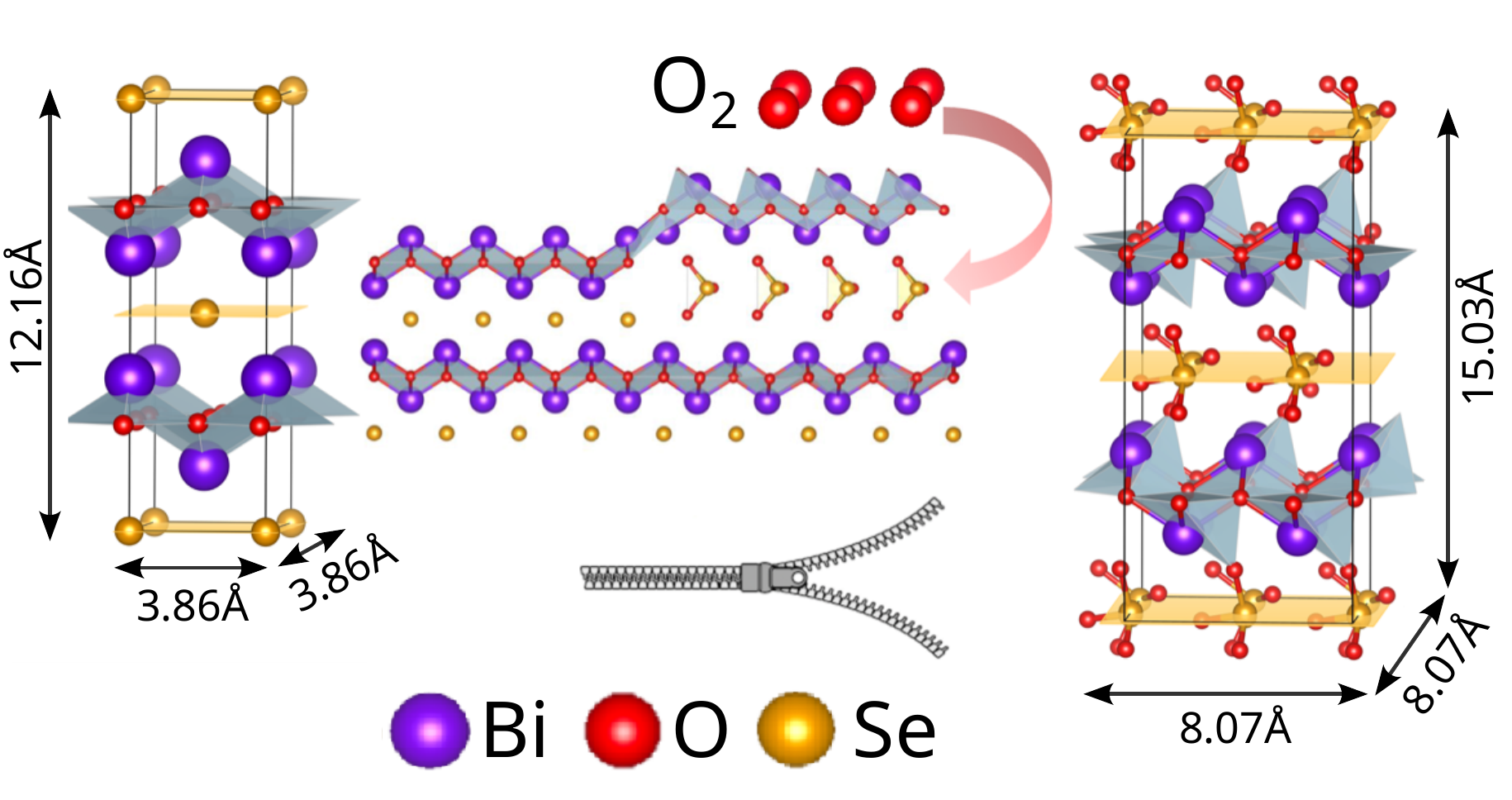} 
\caption{}
\label{zip}
\end{subfigure}}
{\captionsetup[subfigure]{skip=-12pt}\begin{subfigure}[b]{.45\linewidth}
\includegraphics[width=1.0\linewidth,trim={0cm 0cm 0cm 0.25cm},clip]{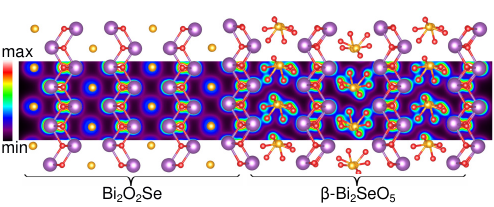} 
\caption{}
\label{hetro}
\end{subfigure}}
{\captionsetup[subfigure]{skip=-12pt}\begin{subfigure}[b]{.19\linewidth}
\includegraphics[width=1.0\linewidth]{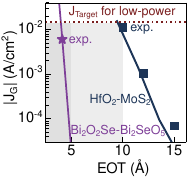} 
\caption{}
\label{tsu}
\end{subfigure}}
\centering\begin{tabular}{@{}pfsg@{}}
\figheader{}\begin{subfigure}[b]{.245\linewidth}
\includegraphics[width=\linewidth,trim={11cm 0cm 24cm 14.5cm},clip]{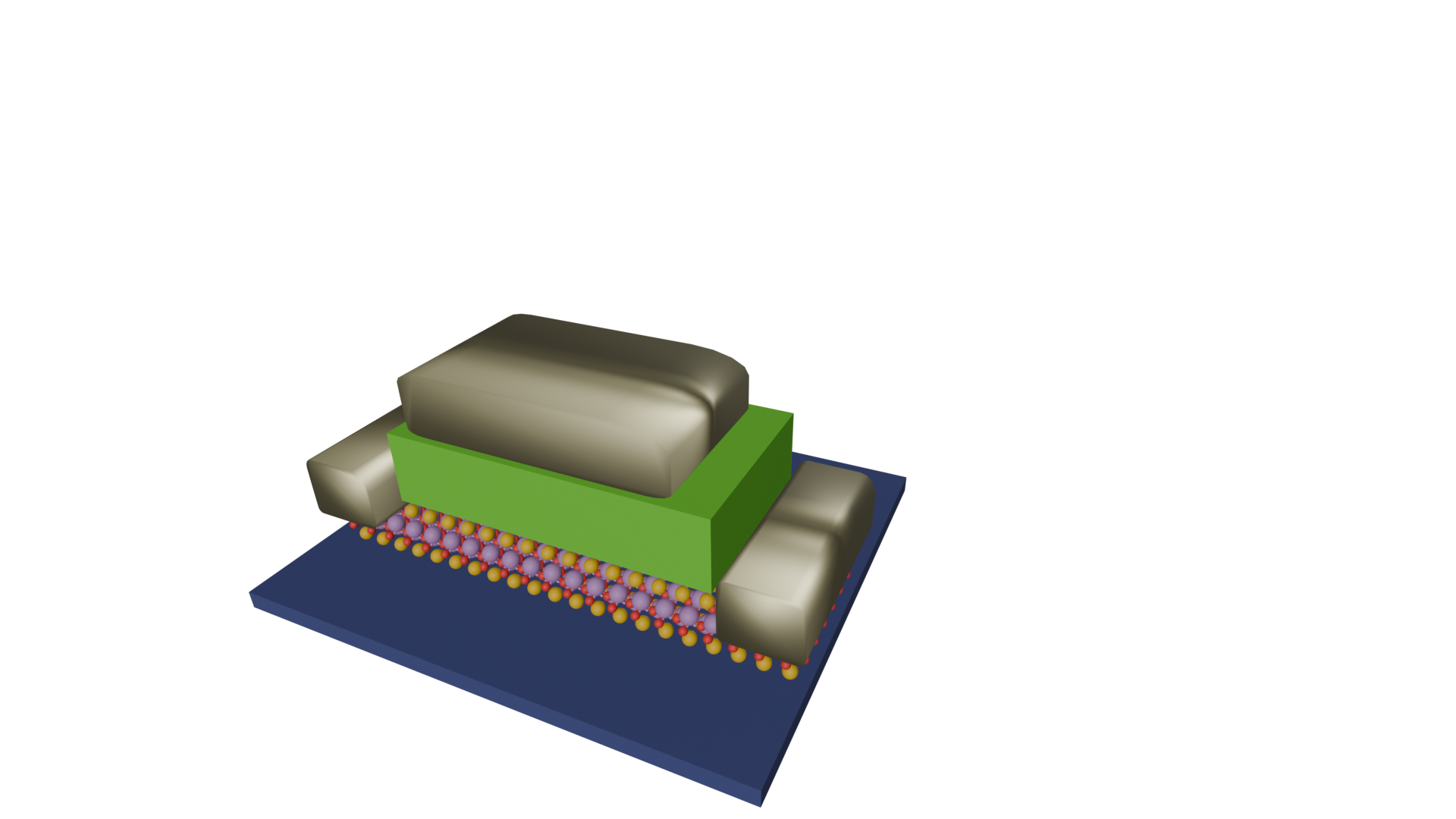}
\caption{}
\label{planar_schem}
\end{subfigure}&
\begin{subfigure}[b]{.245\linewidth}
\includegraphics[width=\linewidth,trim={17cm 9.2cm 17cm 2.5cm},clip]{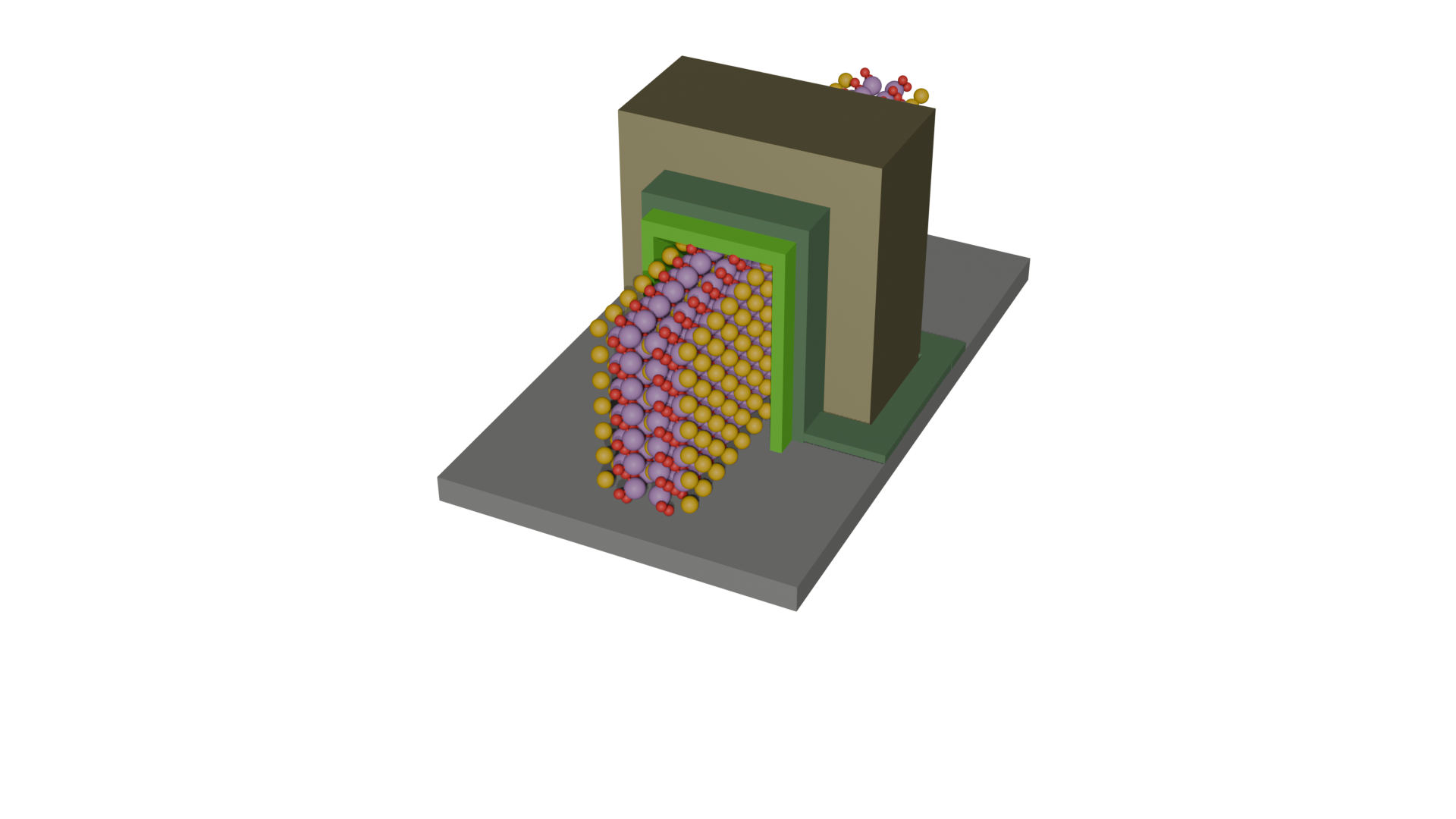}
\caption{}
\label{finfet_schem}
\end{subfigure}&
\begin{subfigure}[b]{.245\linewidth}
\includegraphics[width=\linewidth,trim={6cm 2.5cm 5cm 1cm},clip]{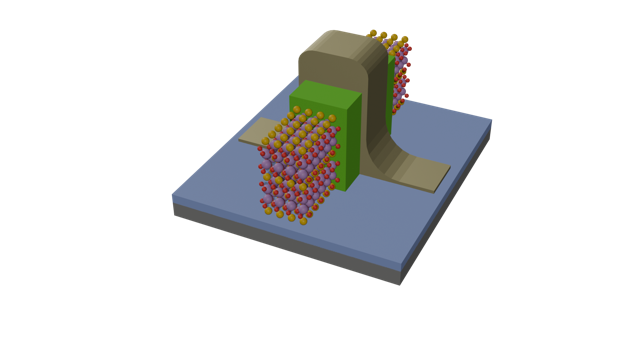}
\caption{}
\label{sgaa_schem}
\end{subfigure}&
\begin{subfigure}[b]{.245\linewidth}
\includegraphics[width=\linewidth,trim={5.8cm 1.7cm 5cm 1cm},clip]{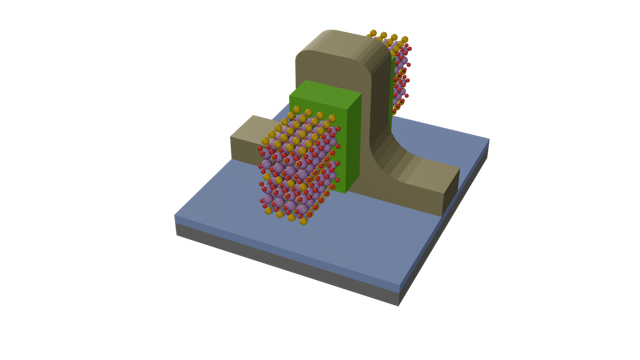}
\caption{}
\label{gaa_schem}
\end{subfigure}\\
{\captionsetup[subfigure]{skip=2pt}\begin{subfigure}[b]{.245\linewidth}
\includegraphics[width=\linewidth,trim={0cm 0cm 0cm 0cm},clip]{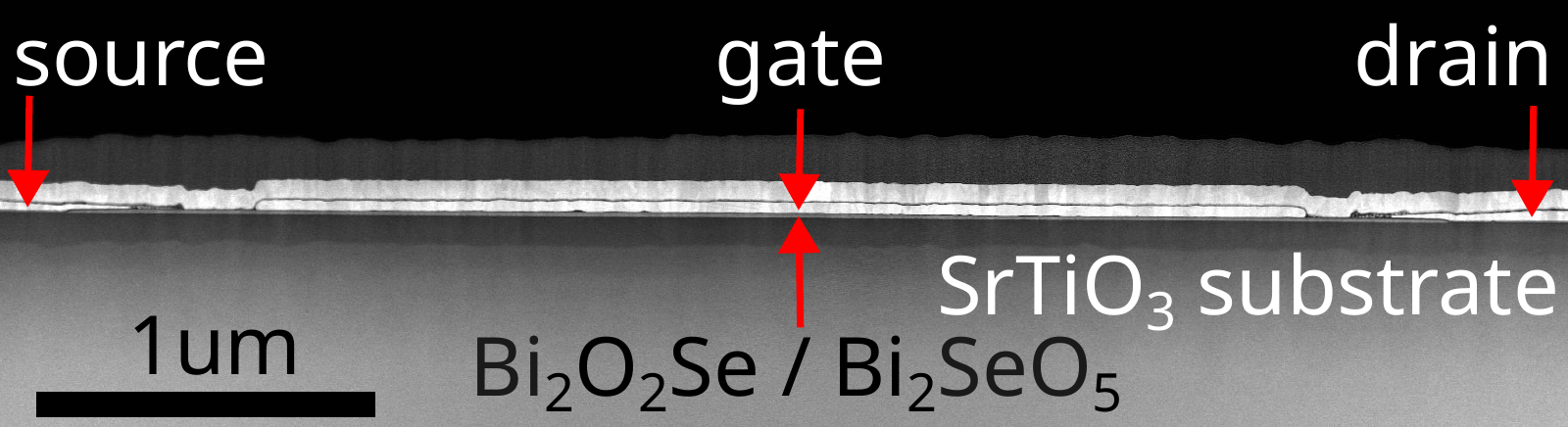}
\includegraphics[width=\linewidth,trim={0cm 0cm 0cm 0cm},clip]{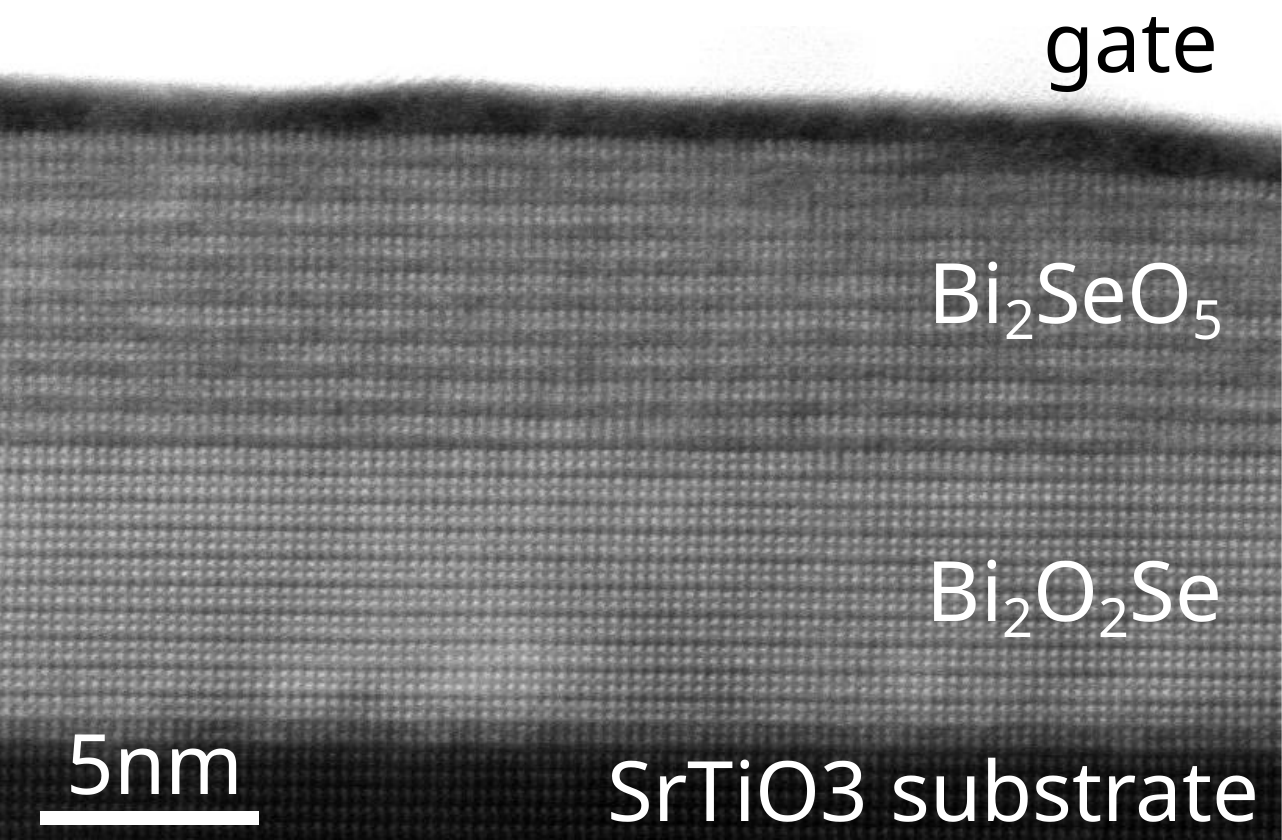}
\caption{}
\label{planar_TEM}
\end{subfigure}}&
{\captionsetup[subfigure]{skip=2pt}\begin{subfigure}[b]{.245\linewidth}
\includegraphics[width=\linewidth,trim={0cm 1.1cm 0cm 0cm},clip]{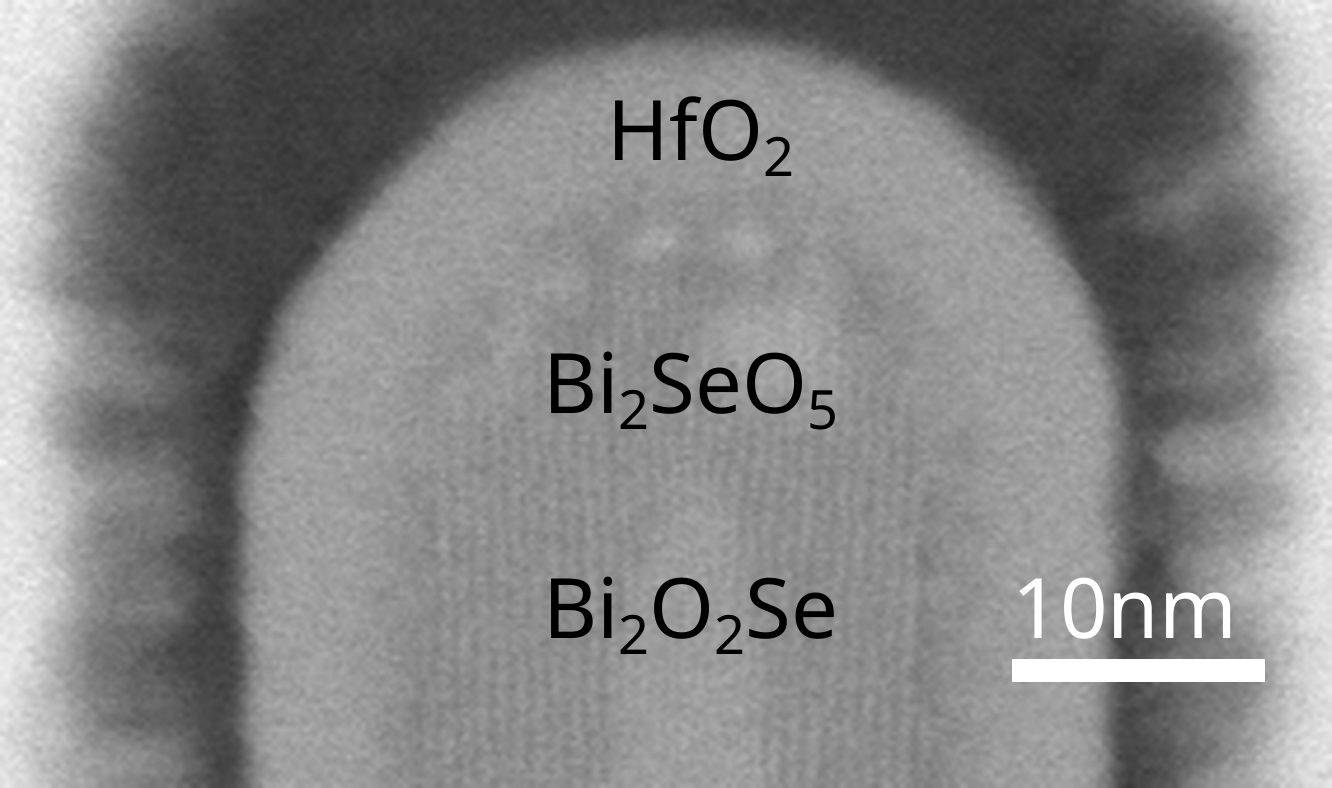}
\includegraphics[width=\linewidth,trim={0cm 0cm 0cm 22cm},clip]{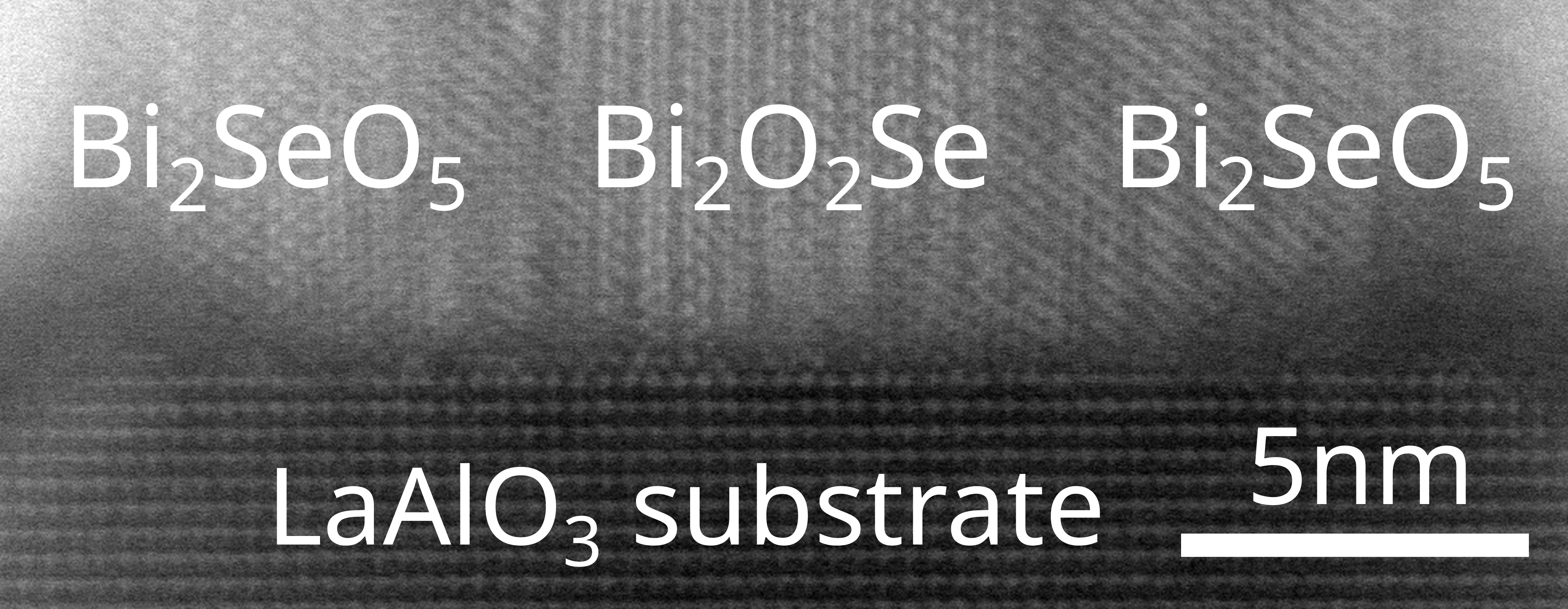}
\caption{}
\label{finfet_TEM}
\end{subfigure}}&
{\captionsetup[subfigure]{skip=2pt}\begin{subfigure}[b]{.245\linewidth}
\includegraphics[width=\linewidth,trim={0cm 0cm 0cm 3.3cm},clip]{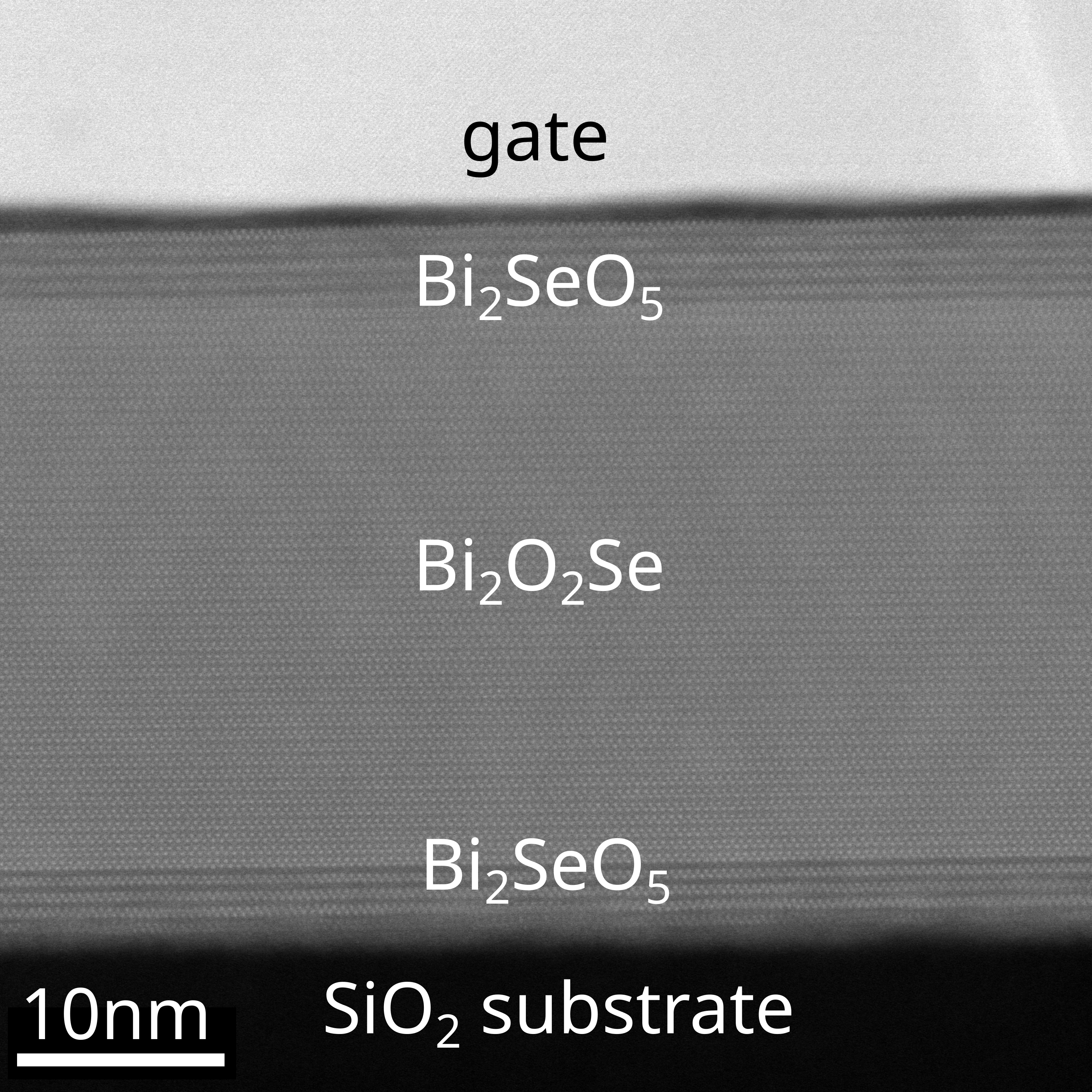}
\caption{}
\label{sgaa_TEM}
\end{subfigure}}&
{\captionsetup[subfigure]{skip=2pt}\begin{subfigure}[b]{.245\linewidth}
\includegraphics[width=\linewidth,trim={0cm 4.3cm 0cm 3cm},clip]{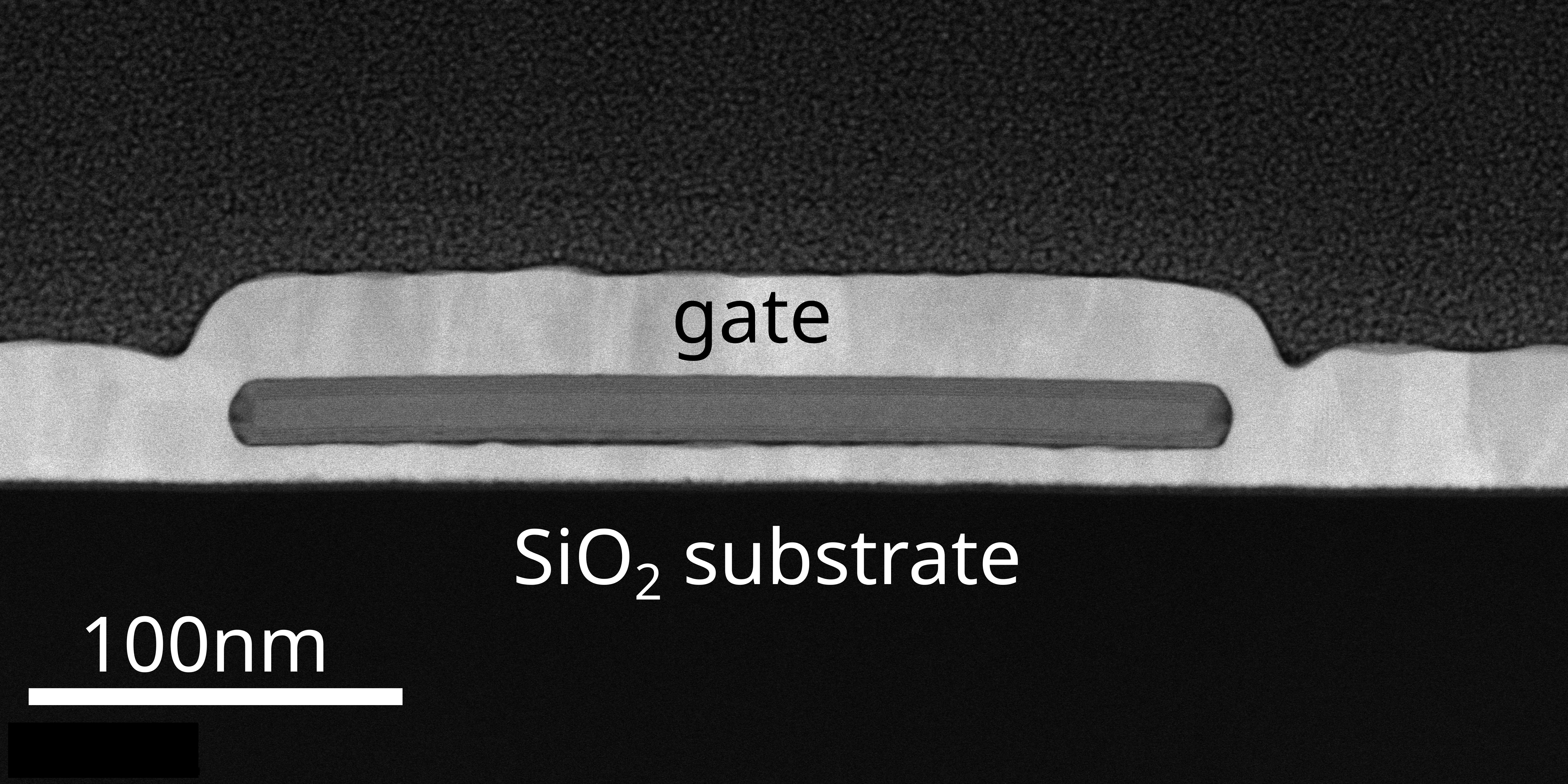}
\includegraphics[width=\linewidth,trim={0cm 0cm 0cm 0cm},clip]{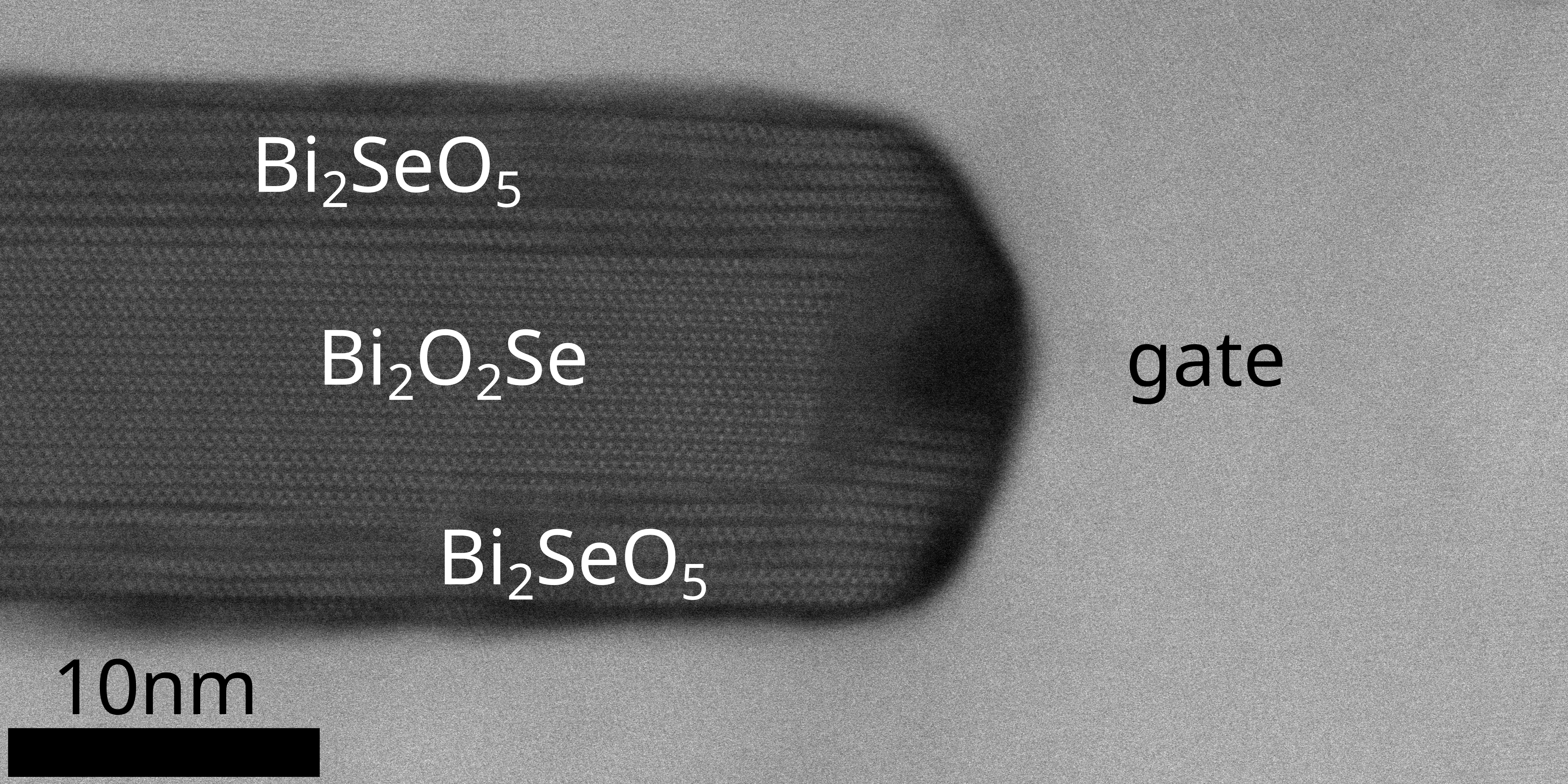}
\caption{}
\label{gaa_TEM}
\end{subfigure}}\\
\end{tabular}
\caption{Multiscale investigation of \bissc{}$/$\bisox{} transistors.
\textbf{(a)} A flowchart illustrating the analysis steps and the flow of data.
\textbf{(b)-(d)} The promising \bissc{}$/$\bisox{} material system.
\textbf{(b)} Schematic illustration of the intercalative oxidation process. UV assisted injection of oxygen atoms between the \ce{[Bi2O2]$^{2n+}$} layers of \bissc{} leads to formation of the layered zipper oxide.
\textbf{(c)} The resulting smooth \bissc{}$/$\bisox{} interface (ideal condition) embedded on the charge density distribution. Covalent, directional bonding is preserved within each constituent, while across the interface it is neither purely covalent (no dangling-bond) nor purely van der Waals (sub-vdW separation).
\textbf{(d)} Gate tunneling current density as a function of EOT for \bissc{}/\bisox{} compared to \ce{HfO2}/\ce{MoS2}. Lines and symbols show Tsu–Esaki calculation results and experimental data ~\cite{peng_beta,Li2019} respectively. Unlike \ce{HfO2}/\ce{MoS2}, \bissc{}/\bisox{} meets the IRDS low-power target (dashed line).
\textbf{(e)–(l)} Investigated \bissc{}$/$\bisox{} transistors, ordered by generation, along with representative minimalistic icons on top, consistently used throughout all device-scale figures in combination with the presented color-coded scheme.
\textbf{(e)–(h)} Schematic illustration of the device geometry (the ball‑and‑sticks, light and dark green layers representing \bissc{}, \bisox{}, and \ce{HfO2} layers respectively) for:
\textbf{(e)} The long channel \plan{}.
\textbf{(f)} The \fin{} with thin fins and an extra \ce{HfO2} layer between the gate and \bisox{} layer.
\textbf{(g)} The \sgaa{} with a gate spreading over the sides of the transistor without extending beneath it.
\textbf{(h)} The \gaa{}, which is the latest and most advanced investigated device sample, exhibiting the lowest EOT.
\textbf{(i)–(l)}: Cross-sectional HAADF-STEM images of:
\textbf{(i)} The \plan{} at two different magnifications showing the grown MOS (\ce{Au}$/$\bisox{}/\bissc{}) stack on the underlying \ce{SrTiO3} substrate.
\textbf{(j)} The \fin{} at two vertical levels of a fin: at topmost level showing the fin cap (top), and at bottom showing a \bisox{}$/$\bissc{}$/$\bisox{} heterostructure on the underlying \ce{LaAlO3} film (bottom) \cite{peng_fin}.
\textbf{(k)} The \sgaa{}, showing the gate on top (the light region).
\textbf{(l)} The \gaa{} at two different magnifications showing the \bisox{}$/$\bissc{}$/$\bisox{} stack surrounded by the metallic gate contact (the light regions).}
\end{figure}
\setlength{\tabcolsep}{6pt}  
\renewcommand{\arraystretch}{1}
Previous studies have shown that the insulator can exist in at least two distinct structural forms, depending on the fabrication technique: \bisalpha{}, synthesized via thermal oxidation and powder sintering of \bissc{}, crystallizing in the $Abm2$ space group; and the layered, single-crystalline \bisbeta{}, synthesized through UV-assisted intercalative oxidation of the semiconductor which shares the same $I4/mmm$ space group as \bissc{}, thereby minimizing lattice mismatch~\cite{khakbazACS,highKpeng,peng_beta}. 

The $\beta$-phase is the structure of interest in this work as it should be able to serve as an ultra-clean gate dielectric and has been predicted to have an even higher permittivity than the $\alpha$ phase~\cite{khakbazACS}. Therefore, \bisbeta{} is used in all investigated device samples in this study and we will use \bisox{} and \bisbeta{} interchangeably in the following. 
\fig{zip} illustrates how \bisbeta{} is formed through the layer-by-layer oxidation of the underlying \bissc{}. This process ideally yields an atomically flat and defect-free interface (see \fig{hetro}) between the semiconductor and its native oxide, a condition under which charge trapping at defects is assumed to be minimal~\cite{peng_beta,khakbazACS}.
In addition to the large dielectric constant of the oxide and the high carrier mobility of the semiconductor, the existence of this native oxide is considered to be one of the most promising features of this material system. This is reminiscent of the successful \ce{Si}/\ce{SiO2} structure, where the clean interface can be considered the hallmark feature of its success.
Moreover, \bissc{} is a chemically stable compound, unlike most other 2D semiconductors with native oxides, such as \ce{HfSe2}, which tend to oxidize rapidly and uncontrollably when exposed to ambient conditions~\cite{Mleczko2017}. 

As discussed earlier, an additional benefit results from the zippered interface between \bissc{} and \bisox{}, which is about half-way in strength between a conventional covalently bonded system such as \ce{Si}/\ce{SiO2} and a weakly bonded vdW system. Both extremes pose challenges as covalent bonding tends to introduce strain, dangling bonds and a high density of interface traps, whereas vdW bonding produces a vacuum-like gap with very low permittivity, effectively adding a penalty to EOT.
Our DFT calculations imply that in this zippered \bissc{}/\bisox{} interface~\cite{khakbazACS,pourfathScience} bonding within the oxide and the semiconductor remains covalent, while at the interface the interaction is neither purely covalent nor purely vdW. The interlayer bonding can be quantified as ionic (weak electrostatic) in nature, which makes it stronger than vdW but weaker than covalent, as revealed by the calculated charge density distribution in \fig{hetro}.
Most importantly, no dangling bonds appear at the interface and the separation between the two sides is significantly smaller than a typical vdW gap. Consequently, the interface avoids the dielectric penalty of a vdW gap while simultaneously suppressing trap formation, thus benefiting from the advantages of both covalent and vdW interfaces while being less prone to their drawbacks.
These interfacial properties enable aggressive scaling of \bissc{}/\bisox{} to meet the IRDS targets beyond 2035~\cite{roadmap_BC,roadmap_MM}, which require a minimum EOT below \SI{0.5}{\nano\meter}. \fig{tsu} compares the calculated gate tunneling leakage current density as a function of EOT using the Tsu–Esaki formalism with experimental data from Ref.~\cite{peng_beta}. Both calculated and experimental results consistently show that \bissc{}/\bisox{} maintains low leakage down to EOT values below \SI{0.5}{\nano\meter}.

A recent study has demonstrated that doping \bissc{} with $\ce{Zn}^{2+}$ ions induces \textit{p}-type conductivity, allowing the integration of \textit{p}-type \bissc{} with its native \textit{n}-type form. This enables the fabrication of p–n homojunctions and complementary MOSFET technology~\cite{polarity_modulation}. \bissc{} crystals can also be grown both horizontally and vertically, depending on the growth substrate and technique, enabling the fabrication of FinFETs with ultra-thin fins
as well as other 3D geometries like gate-all-around (GAA) FETs~\cite{peng_fin,peng_gaa}.
Previous device demonstrations based on the \bissc{}/\bisox{} material system provide encouraging evidence for device-level reproducibility. In one study,  a batch of $120$ \gaa{} devices exhibited narrow distributions of sub-threshold slope, on/off drain current ratio, and mobility~\cite{peng_gaa}, while \fin{} arrays showed reproducible transfer characteristics and mobility statistics across $60$ devices~\cite{peng_fin}.

We analyzed four transistor prototype generations consisting of this material system. The first is a top-gated planar MOSFET (\plan{}), grown on a \ce{SrTiO3} substrate using molecular-beam epitaxy (MBE). This is followed by a fin MOSFET (\fin{}) with thin fins grown on a \ce{LaAlO3} film using chemical vapor deposition (CVD). Contrary to the other devices studied here, the \fin{} includes an additional insulating layer (\ce{HfO2}), which encapsulates the \bissc{}$/$\bisox{} stack. 
This coating serves to prevent potential shorting between the contacts as this device design enables compressed arrangement of contacts, minimizing the size of the access region. 
The third device is a semi-gate-all-around MOSFET (\sgaa{}), in which the gate wraps around the sides of the channel but does not extend beneath it. The final structure is a full gate-all-around MOSFET (\gaa{}). Both \sgaa{} and \gaa{} were grown vertically on Mica substrates using CVD and then physically transferred on \ce{Si}/\ce{SiO2} substrates.
These device geometries are illustrated schematically in \figs{planar_schem}{gaa_schem}, and their cross-sectional high-angle annular dark-field (HAADF) STEM images are shown in \figs{planar_TEM}{gaa_TEM}. Further details on the fabrication processes can be found in~\cite{peng_planar,peng_beta,peng_fin,peng_gaa}.

\subsection{Performance and Reliability Characterization}\label{sec:characterization}
\setlength{\tabcolsep}{2pt}  
\renewcommand{\arraystretch}{0.1}
\begin{figure}[!bt]
\centering\begin{tabular}{@{}pfsg@{}}
\figheader{}
&&&\\&&&\\
\begin{subfigure}[b]{.245\linewidth}
\includegraphics[width=\linewidth]{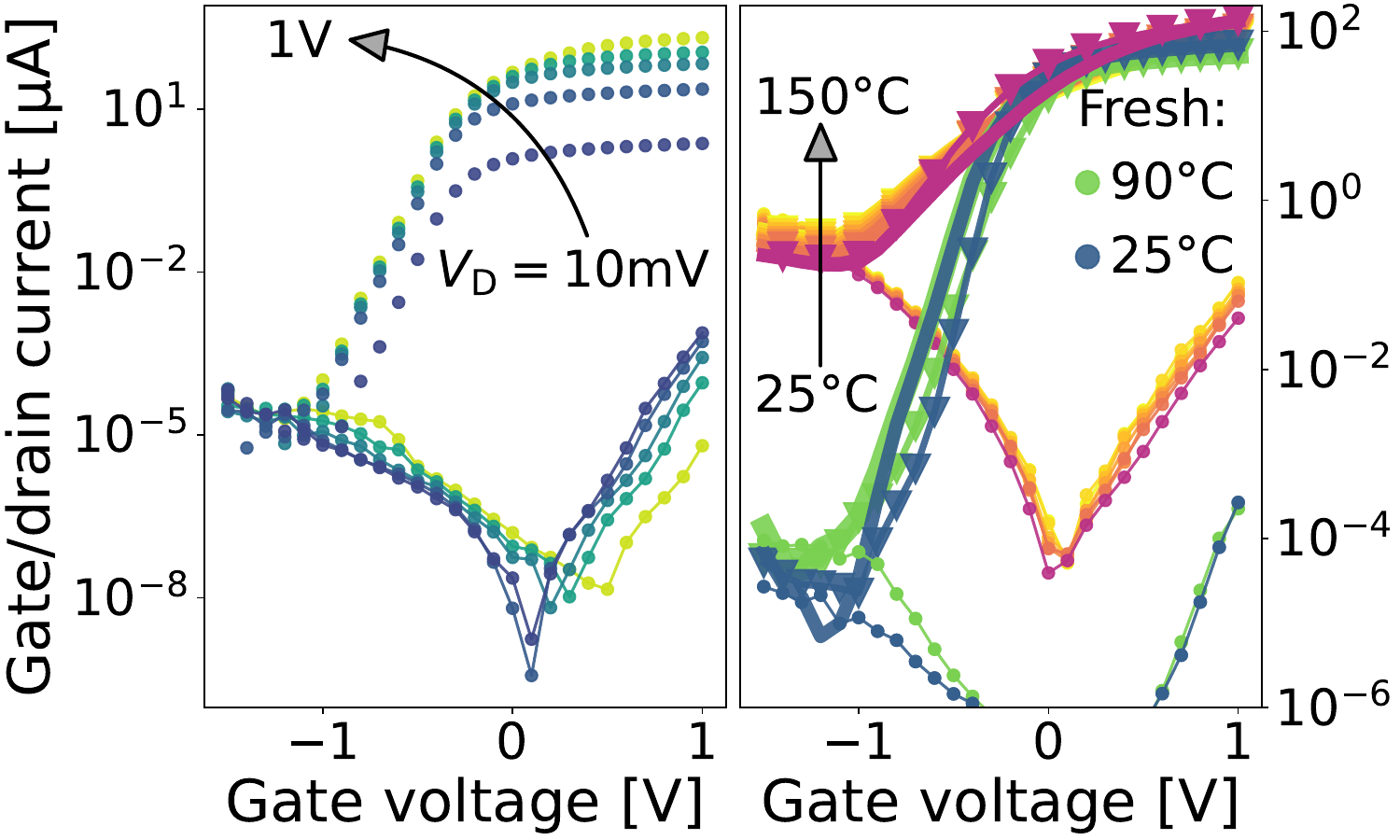}
\caption{}
\label{p_exp_ig}
\end{subfigure}&
\begin{subfigure}[b]{.245\linewidth}
\includegraphics[width=\linewidth]{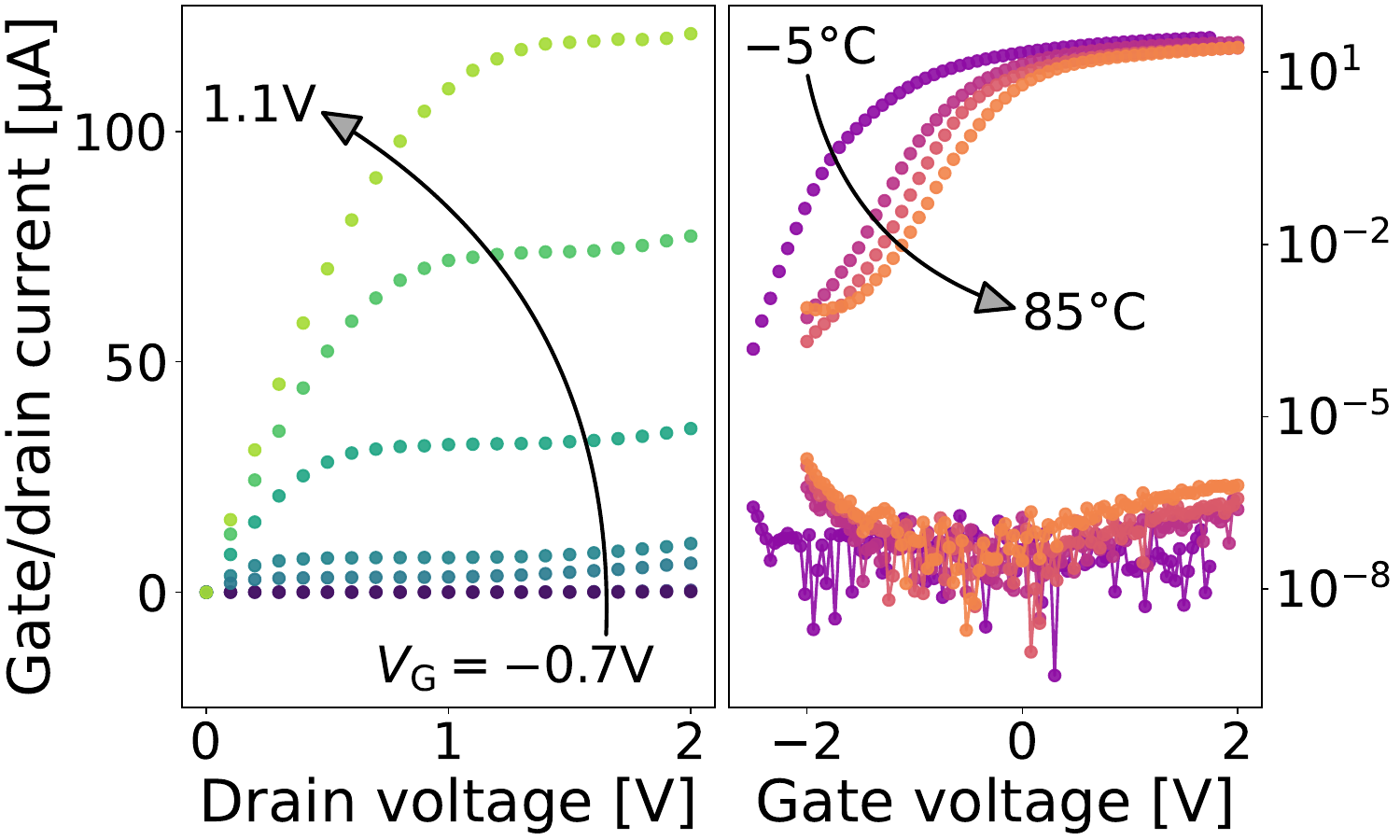}
\caption{}
\label{f_exp_ig}
\end{subfigure}&
\begin{subfigure}[b]{.245\linewidth}
\includegraphics[width=\linewidth]{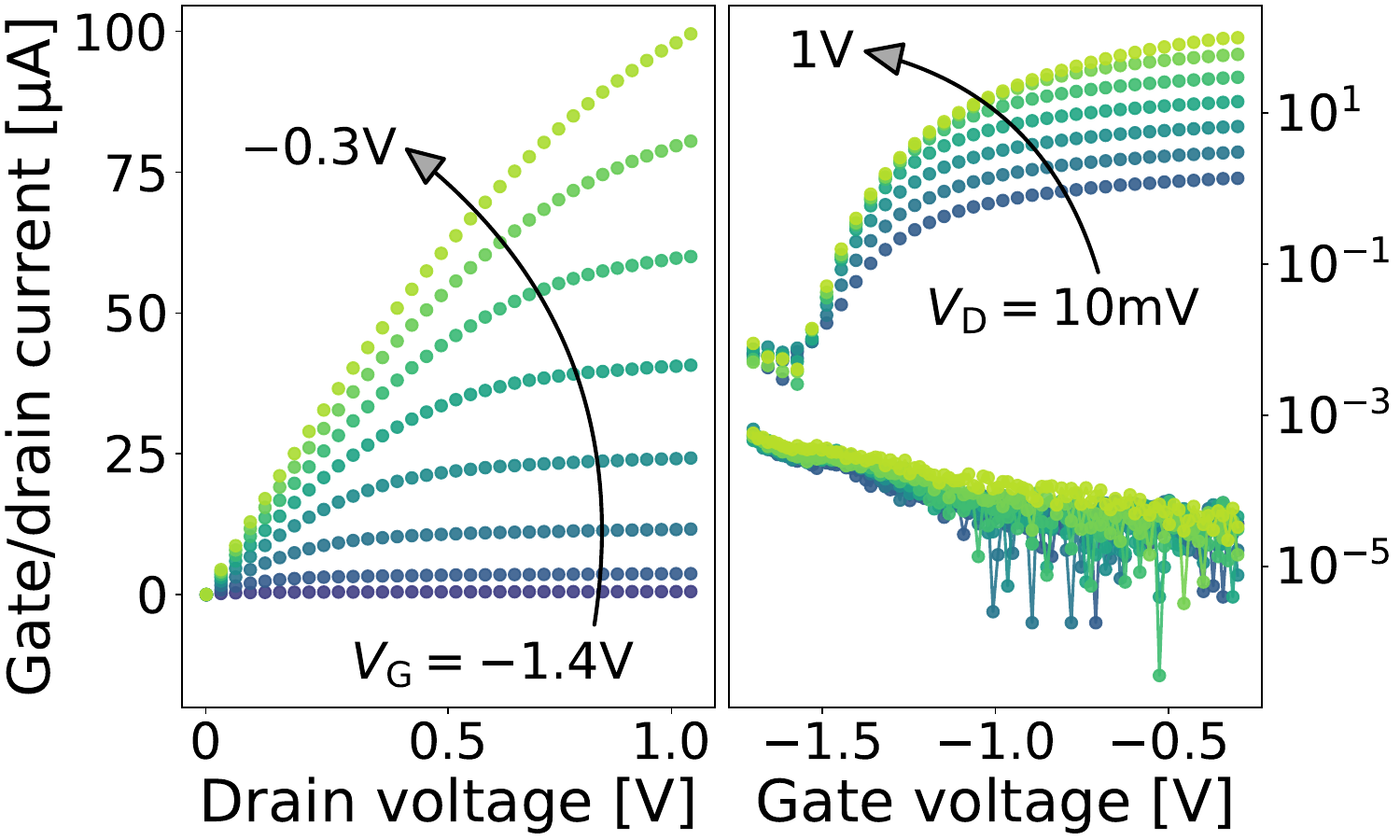}
\caption{}
\label{s_exp_ig}
\end{subfigure}&
\begin{subfigure}[b]{.245\linewidth}
\includegraphics[width=\linewidth]{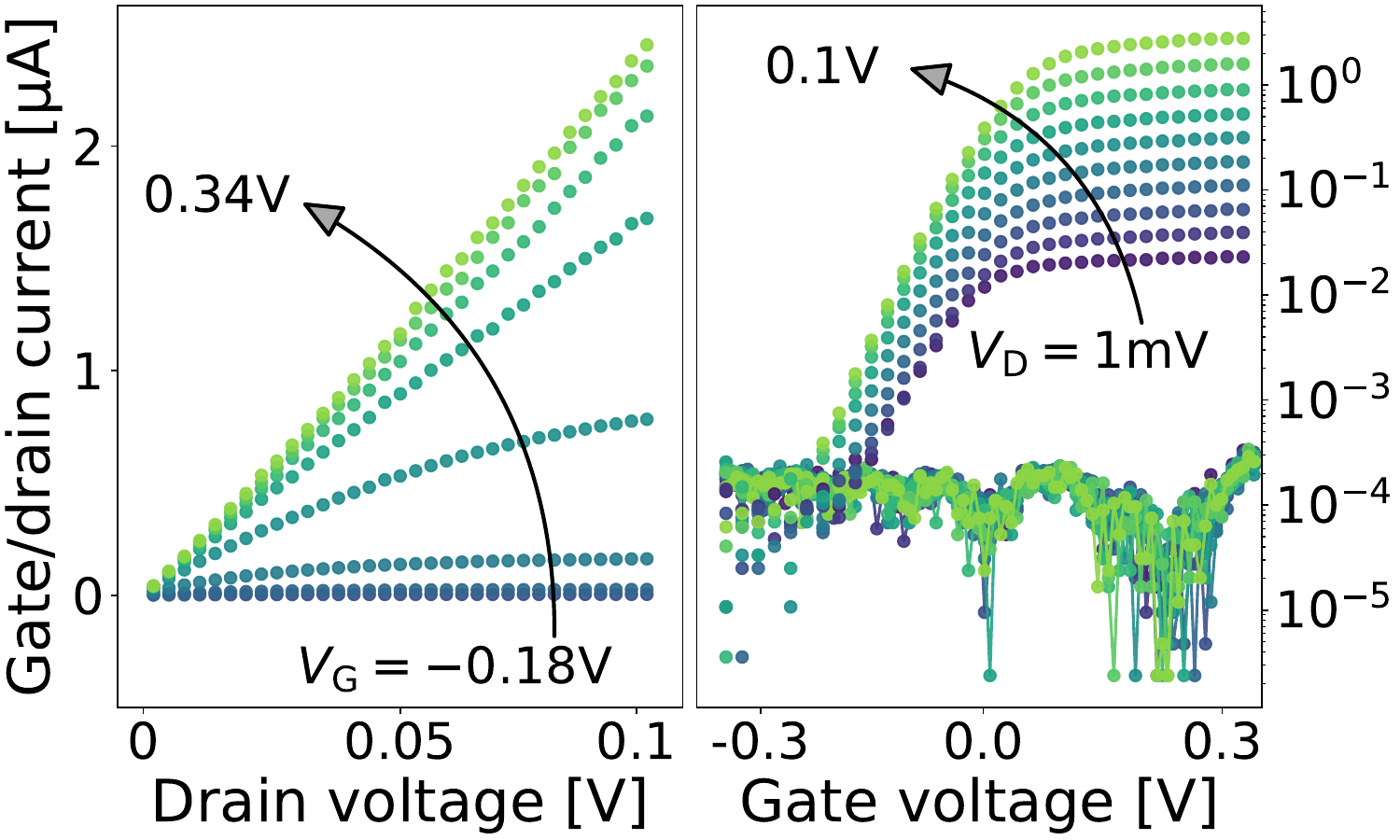}
\caption{}
\label{g_exp_ig}
\end{subfigure}\\
\begin{subfigure}[b]{.245\linewidth}
  \includegraphics[width=\linewidth]{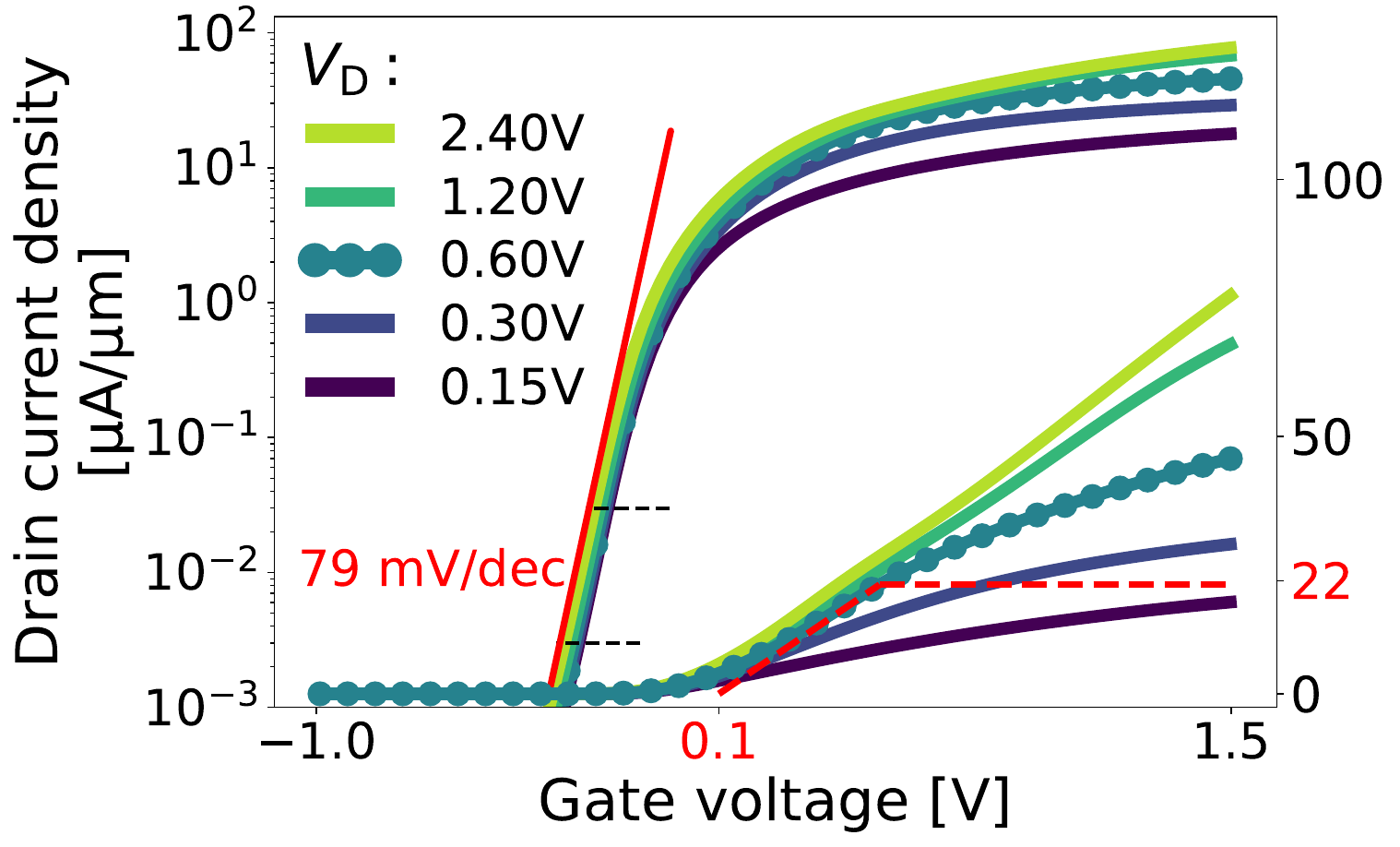}
\caption{}
\label{p_de_embed}
\end{subfigure}&
\begin{subfigure}[b]{.245\linewidth}
  \includegraphics[width=\linewidth]{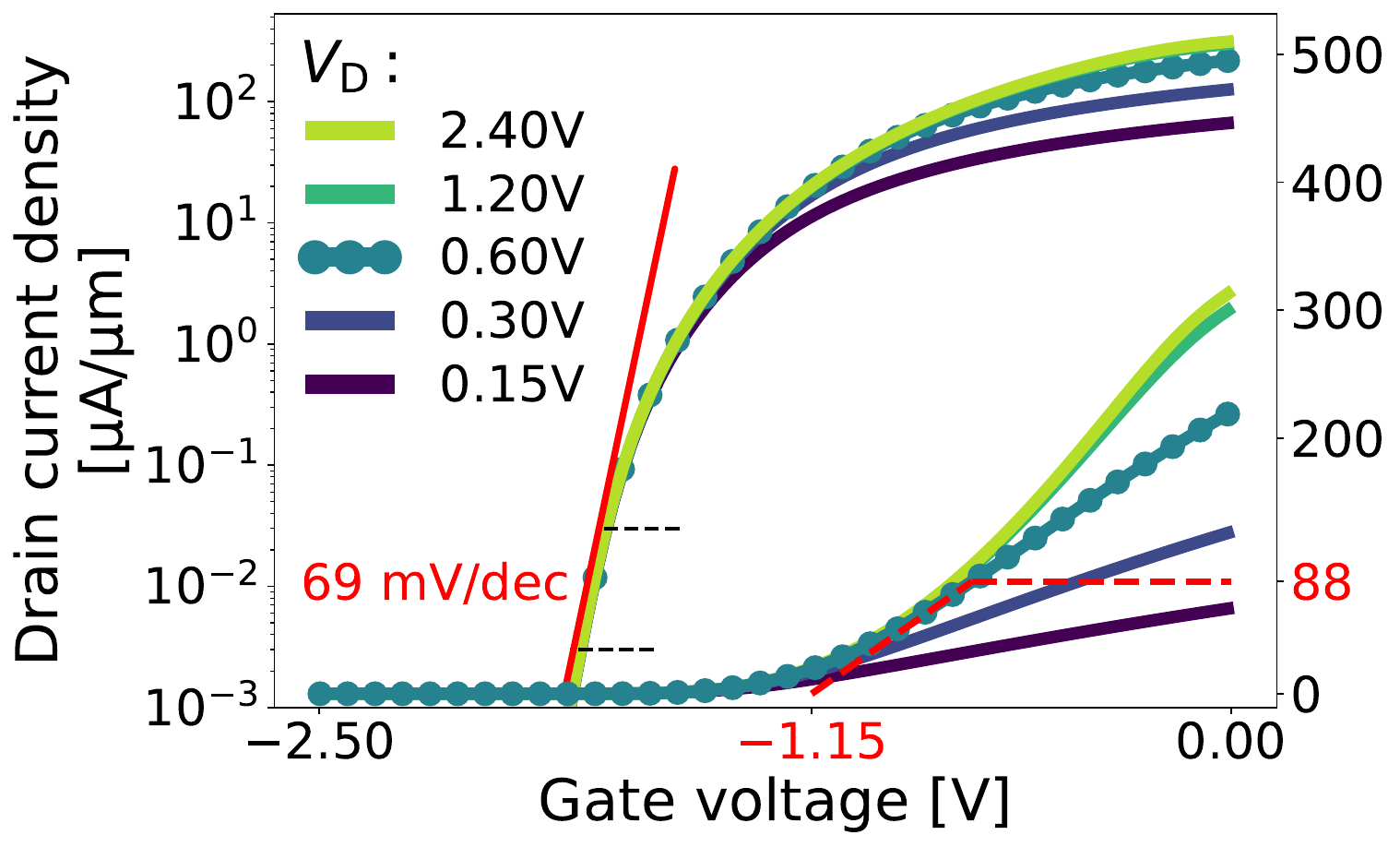}
\caption{}
\label{f_de_embed}
\end{subfigure}&
\begin{subfigure}[b]{.245\linewidth}
  \includegraphics[width=\linewidth]{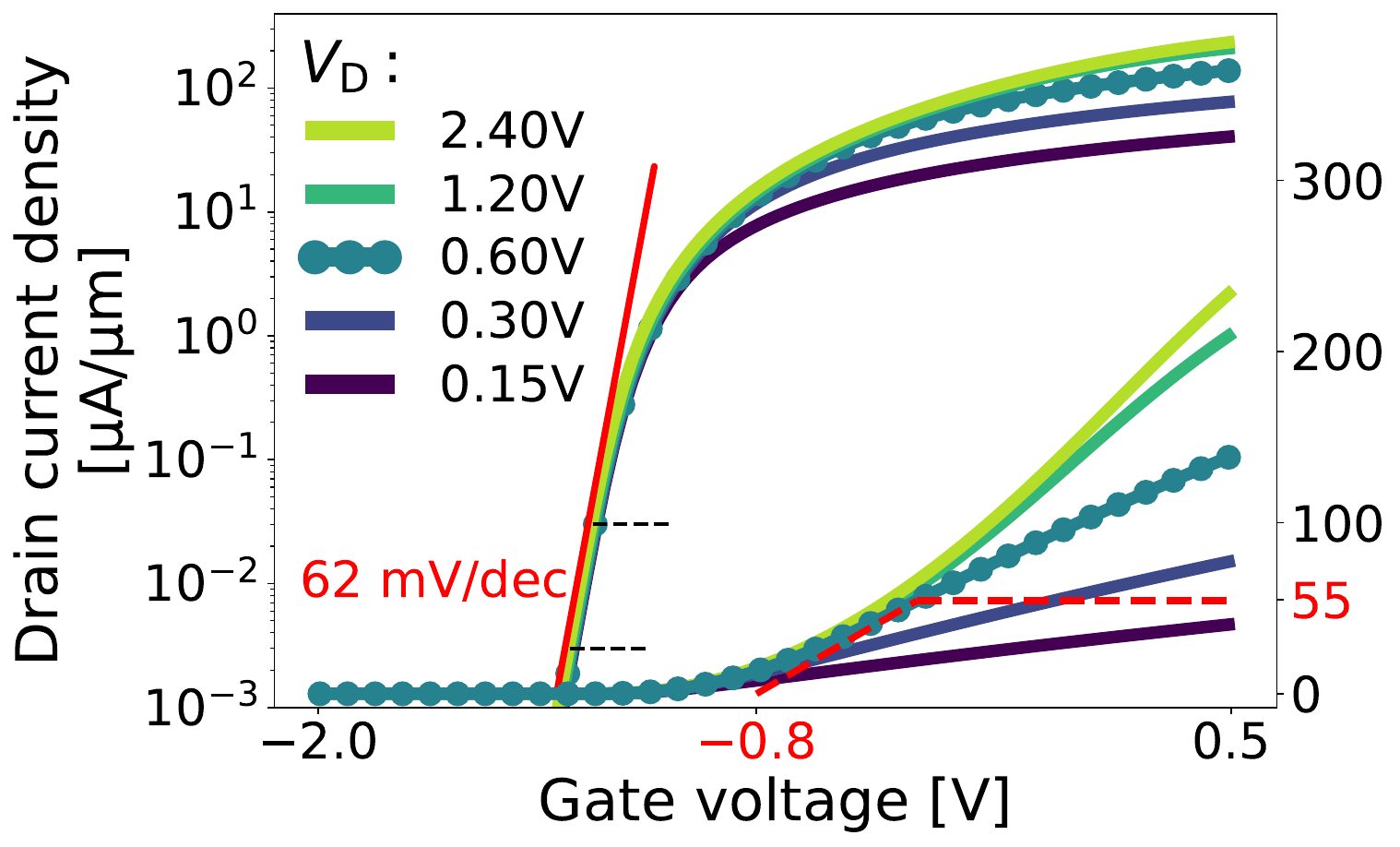}
\caption{}
\label{s_de_embed}
\end{subfigure}&
\begin{subfigure}[b]{.245\linewidth}
  \includegraphics[width=\linewidth]{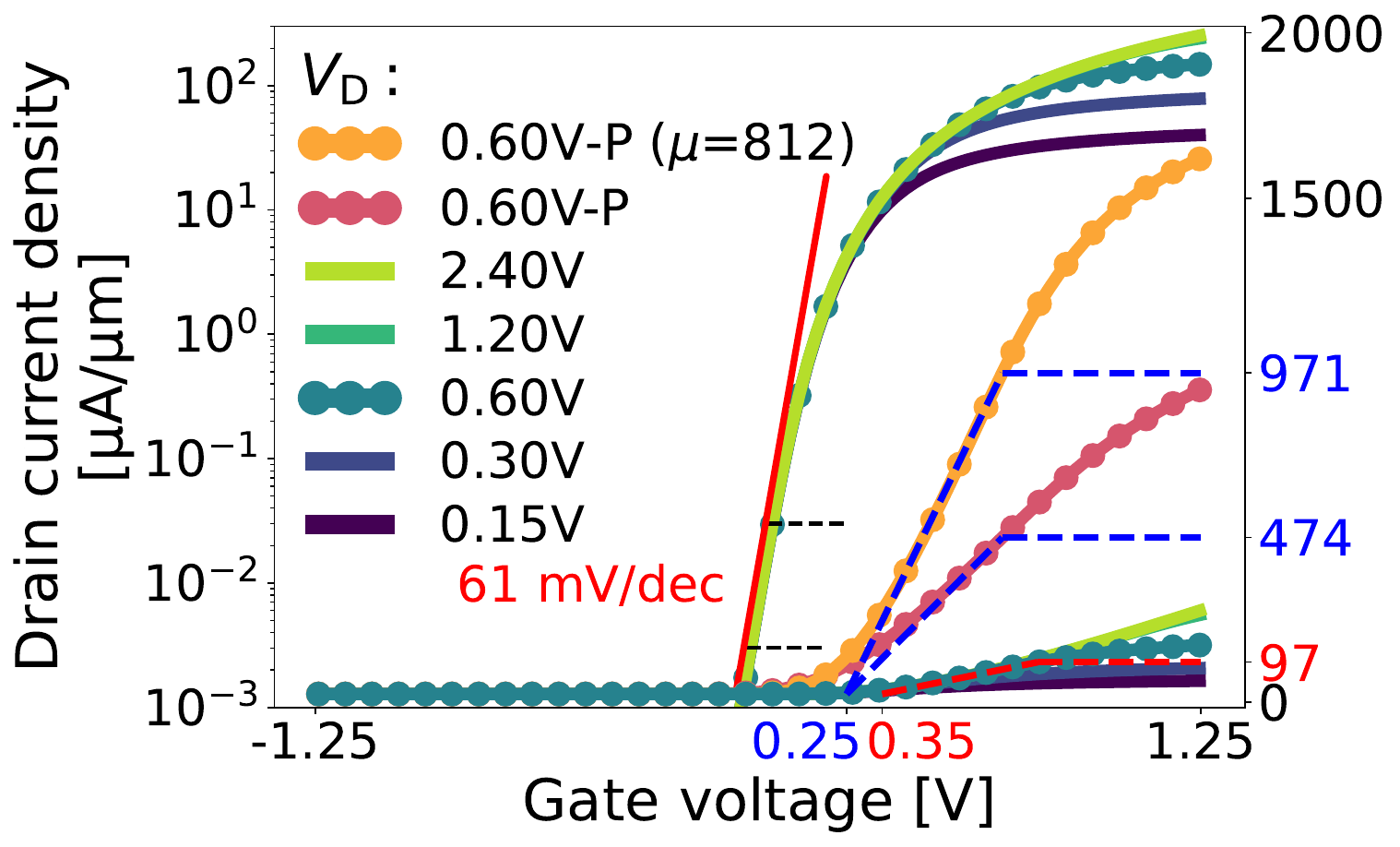}
\caption{}
\label{g_de_embed}
\end{subfigure}\\
\multicolumn{4}{@{}g@{}}{
\begin{subfigure}[b]{\linewidth}
\includegraphics[width=\linewidth]{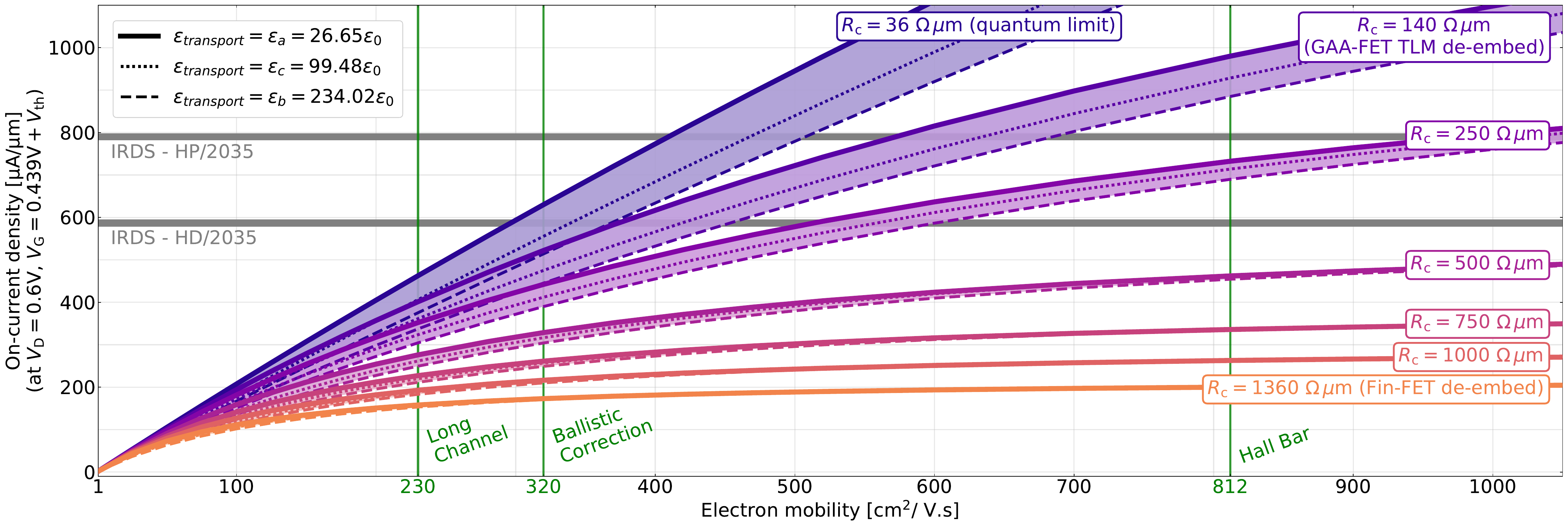}
\caption{}
\label{irds_projection}
\end{subfigure}}\\
\end{tabular}
\caption{Characterization of device performance.
\textbf{(a)-(d)} The measured gate currents (lines and circles) and drain currents (circles, except in the right figure of \textbf{(a)}) of:
\textbf{(a)} The \plan{} as a function of gate voltage at various \vd{}'s at room temperature (left) and at multiple temperatures after degradation (the solid lines represent up-sweeps and the lines and diamonds represent down-sweeps) compared to measurements before degradation (right), note the counter-clockwise hysteresis after degradation (i.e. down-sweep left to the up-sweep).
\textbf{(b)} The \fin{}  output characteristics at various gate voltages (left) and transfer characteristics at multiple temperatures (right).
\textbf{(c)} The \sgaa{} output (left) and transfer (right) characteristics at various gate and drain voltages.
\textbf{(d)} The \gaa{} output (left) and transfer (right) characteristics at various gate and drain voltages.
\textbf{(e)–(h)} 
Transfer characteristics of the de-embedded devices plotted on both logarithmic and linear scales at various drain voltages. The curves highlighted with circular markers correspond to \vd{}$=\SI{0.6}{V}$ and are used to extract the on-state drain current, subthreshold slope, and threshold voltage. The subthreshold slope is extracted from the inverse slope of $\log_{10}$(\id{}) over the fixed one-decade current-density window $\SIrange{3e-3}{3e-2}{\micro A\per\micro m}$, indicated by the black dashed lines. The threshold voltage is obtained from the intercept of a linear fit to the on-state transfer curve on the linear-current scale, and the on-current is evaluated at $V_\mathrm{G}=V_\mathrm{th}+\SI{0.439}{V}$ following the IRDS roadmap~\cite{roadmap_BC,roadmap_MM}. While \textbf{(e)}, \textbf{(f)}, and \textbf{(g)} respectively show results for the de-embedded \plan{}, \fin{}, and \sgaa{}, \textbf{(h)} presents both the de-embedded and IRDS HP/2035-projected curves for the \gaa{}.
\textbf{(i)} Sensitivity analysis of the projected $\SI{12}{\nano\meter}$ \gaa{} on-current (evaluated at $V_\mathrm{D}=\SI{0.6}{V}$ and $V_\mathrm{G}=V_\mathrm{th}+\SI{0.439}{V}$).
Colored curves correspond to different effective contact/access resistances, ranging from the estimated ideal 2D quantum-contact limit, $R_\mathrm{c}=\SI{36}{\ohm\micro\meter}$, to the conservative de-embedded value, $R_\mathrm{c}=\SI{1360}{\ohm\micro\meter}$, including the experimentally reported \gaa{} TLM value, $R_\mathrm{c}=\SI{140}{\ohm\micro\meter}$~\cite{peng_gaa}.
For each $R_\mathrm{c}$, the different line styles correspond to the three transport-direction relative permittivities considered, taken from the anisotropic semiconductor permittivity components. 
Vertical guide lines mark the calibrated long-channel mobility, the ballistic-mobility estimate for the $\SI{12}{\nano\meter}$ gate-length device, and the reported Hall mobility. 
Horizontal guide lines indicate the IRDS 2035 targets for high-performance and high-density nodes.
The ideal 2D quantum-contact limit is obtained from $R_\mathrm{Q}W=(h/2e^2)\sqrt{\pi/(2g_v n_\mathrm{s})}\approx 51/\sqrt{g_v \tilde{n}_\mathrm{2D}}~\si{\ohm\micro\meter}$, with $\tilde{n}_\mathrm{2D}=n_\mathrm{s}/\SI{1e13}{cm^{-2}}$~\cite{quantum_limit_2d_contacts}, using the carrier density reported for the \gaa{} contact, $n_\mathrm{s}=\SI{2e13}{cm^{-2}}$~\cite{peng_gaa}, and the single-valley character of \bissc{}, $g_v=1$~\cite{peng_rashba_effect,peng_nanoribbon}.
}
\end{figure}
\setlength{\tabcolsep}{6pt}
\renewcommand{\arraystretch}{1}

As a first step in evaluating the performance of these devices, we characterize them at various temperatures and record the transfer characteristics at different \vd{} and output characteristics at various \vg{}, along with the gate leakage currents. 
As the first technology investigated, the \plan{} was subjected to temperatures up to \SI{150}{\degree C}. As shown in \fig{p_exp_ig}, while the gate current remains low at the \SI{25}{\degree C}$-$\SI{90}{\degree C} measurements, after the \SI{105}{\degree C}$-$\SI{150}{\degree C} measurements, a pronounced degradation was observed in the measurements taken at all temperatures during cool-down as well as one day later, with an increase in the gate current by almost four orders of magnitude at all subsequent temperatures. In comparison, the devices which were not subjected to such high temperatures showed stable and moderate gate‑leakage current levels, as can be seen in \figs{f_exp_ig}{g_exp_ig}. The large leakage currents observed in these first generation devices may be related to the formation of traps at elevated temperatures that can then initiate trap-assisted tunneling (TAT) leading to an increase in gate current~\cite{khakbazACS,illarionov2020insulators}. Our measurements show that temperatures above roughly \SI{90}{}$-$\SI{100}{\degree C} caused irreversible or long-lasting performance degradation. 
An overview of the device dimensions and experimentally evaluated performance (including mobilities and effective contact/access resistances extracted through TCAD calibration as described in Section~\ref{sec:DFTTCAD}) is given in \tab{tab:metrics}, providing a benchmark for comparisons with the IRDS High Power (HP) node of the year 2035~\cite{roadmap_BC,roadmap_MM}.
The values reported in this study are not statistical population averages, since the analysis focuses on representative devices from each generation within a consistent multiscale workflow. Published device statistics for the same \bissc{}/\bisox{} platform are available in~\cite{peng_fin,peng_gaa}.

\subsubsection{De-embedding and Projection of Device Performance}
\tab{tab:metrics} also includes the projected maximum achievable performance of these four designs. 
Notably, the access region length in all fabricated devices, except the \fin{}, is large, especially in the \gaa{}, where it exceeds the gate length.
By assuming the lowest access-parasitic condition within the present device set, represented by a \fin{}-derived effective specific series resistivity rescaled to each device geometry and the \fin{} access length ($2\times\SI{7.5}{\nano\meter}$), we de-embedded the devices from the detrimental impact of long access regions, which otherwise act as series resistors. With this reduced access length, the remaining effective series resistance is treated as contact-dominated within the de-embedding model.
This allowed us to extract the intrinsic device performance in the absence of unnecessary parasitic resistances (see \figs{p_de_embed}{g_de_embed}). As can be seen, the largest performance enhancement occurs in the \gaa{}, which is the most advanced geometry and also the one with the highest access-to-gate length ratio.
It is worth noting that the current densities obtained for the de-embedded \gaa{} devices, reaching \SI{97}{\micro A \micro m^{-1}}, are in agreement with experimentally reported values for short-channel \gaa{} transistors with a gate length of \SI{30}{nm} when compared at a supply voltage of \SI{0.6}{V}~\cite{peng_gaa}. 
This consistency confirms that our thoroughly calibrated TCAD framework, established using long-channel test devices with comprehensive \idvg{} and \idvd{} measurements across a wide range of bias and temperature conditions, can reliably reproduce the behavior of experimentally reported short-channel devices and provides an essential intermediate validation step before extending the study to the ultra-scaled regime.

To evaluate whether this material system could potentially meet the IRDS - HP/2035 requirements, we also projected the \gaa{} to the roadmap dimensions ($\SI{12}{\nano\meter}$ gate length, $L_{\mathrm{G}}/4=\SI{3}{\nano\meter}$ channel thickness, and $2\times\SI{4}{\nano\meter}$ access-region length).
We then performed a sensitivity analysis spanning conservative to upper-bound parameter choices for related transport parameters (see \fig{irds_projection}).
At the projected $L_\mathrm{G}=\SI{12}{nm}$ node, finite-length and quasi-ballistic transport effects are expected to influence the drain current, whereas the present TCAD framework is based on local drift-diffusion transport. Previous studies have shown that such effects become relevant in deeply scaled transistors and can be treated through ballistic-mobility, kinetic-velocity, or related effective transport corrections~\cite{DD_balistic,LundstromSispad}. To estimate the expected magnitude, we evaluate the ballistic-mobility scale $\mu_\mathrm{B}=v_\mathrm{T}L_\mathrm{G}/(2k_\mathrm{B}T/q)$, with $v_\mathrm{T}=\sqrt{2k_\mathrm{B}T/(\pi m^*)}$. For \bissc{}, using $m^*=0.15m_0$~\cite{khakbazACS} and $L_\mathrm{G}=\SI{12}{nm}$ gives $\mu_\mathrm{B}\approx\SI{320}{\centi\meter\squared\per\volt\per\second}$. We therefore use an effective mobility of $\mu=\SI{320}{\centi\meter\squared\per\volt\per\second}$ in the drift-diffusion TCAD projection. This value should be interpreted as a compact transport parameter representing finite-length and quasi-ballistic contributions within the present model, not as an enhanced intrinsic material mobility or as a complete quasi-ballistic transport calculation.
Our calculations suggest that with these geometry modifications and using the experimentally reported \gaa{} contact resistance of $\SI{140}{\ohm\micro m}$~\cite{peng_gaa}, the on-current can exceed $\SI{474}{\micro A/\micro\meter}$,
which is already more than half of the roadmap target of $\SI{790}{\micro A/\micro\meter}$ (note that rather than reporting the highest achievable current, the drain current is evaluated at $V_{\mathrm{D}} = 0.6\mathrm{V}$ and $V_{\mathrm{G}} = 0.439\mathrm{V} + V_{\mathrm{th}}$, in strict accordance with the IRDS evaluation standard).
By accounting for the non-idealities in our devices and further increasing the mobility to the upper-bound value of $\SI{812}{\mathrm{cm^2/V\cdot s}}$ as extracted from Hall bar measurements~\cite{peng_vdw}, the projected on-current reaches $\SI{971}{\micro A/\micro\meter}$, thereby surpassing the IRDS HP/2035 node requirement. \fig{g_de_embed} illustrates the transfer characteristics of this projected device as well.

\renewcommand{\arraystretch}{1.5}
\begin{table}[!t]
\caption{Dimensions and performance of the investigated devices at room temperature}
\footnotesize\begin{tabular}{@{}l@{}c@{}|p|p|p|f|f|s|s|g|g|g|c@{}}
\bottomrule
 Technology&& \multicolumn{3}{p|}{\plan{}} & \multicolumn{2}{f|}{\fin{}} & \multicolumn{2}{s|}{\sgaa{}} & \multicolumn{3}{g|}{\gaa{}} & IRDS \\
 \hline
 Representation\footnotemark[1]&& degraded & meas. & de-em. & meas. & de-em. & meas. & de-em.  & meas. & de-em. & proj. & HP/2035\\
\hline
$W$&$[\SI{}{\micro \meter}]$ 
  & \multicolumn{3}{p|}{$\SI{7.9}{}$} 
  & \multicolumn{2}{f|}{$\SI{0.4}{}$} 
  & \multicolumn{2}{s|}{$\SI{1.4}{}$} 
  & \multicolumn{3}{g|}{$\SI{12.4}{}$}  
  & $\leq\SI{0.19}{}$ \\ \hline
$L_{\mathrm{S-D}}$&$[\SI{}{\micro \meter}]$
  & \multicolumn{2}{p|}{$\SI{4.9}{}$} & $\SI{3.2}{}$
  & \multicolumn{2}{f|}{$\SI{1.5}{}$} 
  & $\SI{3}{}$ & $\SI{2.5}{}$
  & $\SI{3}{}$ & $\SI{1.4}{}$ & $\SI{0.02}{}$ 
  & $\leq\SI{0.02}{}$ \\ \hline
$L_{\mathrm{G}}$&$[\SI{}{\micro \meter}]$
  & \multicolumn{3}{p|}{$\SI{3.2}{}$} 
  & \multicolumn{2}{f|}{$\SI{1.5}{}$} 
  & \multicolumn{2}{s|}{$\SI{2.5}{}$} 
  & \multicolumn{2}{g|}{$\SI{1.4}{}$} & $\SI{0.012}{}$ 
  & $\leq\SI{0.012}{}$ \\ \hline
$L_{\mathrm{Acc}}$&$[\SI{}{\nano\meter}]$
  & \multicolumn{2}{p|}{$\SI{850}{\!\times2}$} & $\SI{7.5}{\!\times2}$
  & \multicolumn{2}{f|}{$\SI{7.5}{\!\times2}$} 
  & $\SI{250}{\!\times2}$  & $\SI{7.5}{\!\times2}$
  & $\SI{800}{\!\times2}$ & $\SI{7.5}{\!\times2}$ & $\SI{4}{\!\times2}$
  & $\leq\SI{4}{\!\times2}$\\ \hline
$t_{\mathrm{CH}}$&$[\SI{}{\nano\meter}]$
  & \multicolumn{3}{p|}{$\SI{5.5}{}$} 
  & \multicolumn{2}{f|}{$\SI{15}{}$} 
  & \multicolumn{2}{s|}{$\SI{5.5}{}$} 
  & \multicolumn{2}{g|}{$\SI{15}{}$} & $\SI{3}{}$ 
  & $\leq\SI{3}{}$ \\ \hline
$t_{\mathrm{vdW}}$&$[\SI{}{\nano\meter}]$
  & \multicolumn{3}{p|}{$\SI{0}{}$} 
  & \multicolumn{2}{f|}{$\SI{0}{}$} 
  & \multicolumn{2}{s|}{$\SI{0}{}$} 
  & \multicolumn{3}{g|}{$\SI{0}{}$}
  & $\leq\SI{0.3}{}$ \\ \hline
$t_{\mathrm{OX}} (\bisox{})$&$[\SI{}{\nano\meter}]$
  & \multicolumn{3}{p|}{$\SI{5.5}{}$} 
  & \multicolumn{2}{f|}{$\SI{7.5}{}$} 
  & \multicolumn{2}{s|}{$\SI{5.5}{}$} 
  & \multicolumn{3}{g|}{$\SI{4}{}$}
  & $-$\\ \hline
$t_{\mathrm{OX}} (\ce{HfO_2})$&$[\SI{}{\nano\meter}]$
  & \multicolumn{3}{p|}{$-$} 
  & \multicolumn{2}{f|}{$\SI{7.5}{}$} 
  & \multicolumn{2}{s|}{$-$} 
  & \multicolumn{3}{g|}{$-$} 
  & $-$\\ \hline
$\mathrm{EOT}$\footnotemark[2]&$[\SI{}{\nano\meter}]$
  & \multicolumn{3}{p|}{$[\SI{0.6}{}-\SI{1.1}{}]$} 
  & \multicolumn{2}{f|}{$[\SI{2.0}{}-\SI{2.6}{}]$} 
  & \multicolumn{2}{s|}{$[\SI{0.6}{}-\SI{1.1}{}]$} 
  & \multicolumn{3}{g|}{$[\SI{0.45}{}-\SI{0.8}{}]$}
  & $\leq\SI{0.5}{}$\\ \hline
$I_\mathrm{D}^{\mathrm{on}}$\footnotemark[3]&$[\SI{}{\micro A/\micro \meter}]$ 
  & $-$ & $\SI{8.6}{}$ & $\SI{22}{}$
  & \multicolumn{2}{f|}{$\SI{88}{}$}
  & $\SI{26.4}{}$ & $\SI{55}{}$  
  & $\SI{1.5}{}$ & $\SI{97}{}$ & $\SI{971}{}$
  & $\geq\SI{790}{}$ \\ \hline
$\mathrm{SS}$\footnotemark[4]&$[\SI{}{\mV/dec}]$ 
  & $\SI{310}{}$ & $\SI{79}{}$ & $\SI{79}{}$ 
  & \multicolumn{2}{f|}{$\SI{69}{}$}
  & $\SI{62}{}$ & $\SI{62}{}$
  & $\SI{66}{}$ & $\SI{61}{}$ & $\SI{76}{}$
  & $\leq\SI{65}{}$ \\ \hline
$\mu$\footnotemark[5]&$[\mathrm{cm^2\!/V\!\cdot{\!s}}]$
  & $-$ & \multicolumn{2}{p|}{$\SI{140}{}$} 
  & \multicolumn{2}{f|}{$\SI{275}{}$}
  & \multicolumn{2}{s|}{$\SI{155}{}$} 
  & \multicolumn{2}{g|}{$\SI{230}{}$} & $\SI{812}{}$~\cite{peng_vdw}
  & $\geq\SI{60}{}$ \\ \hline
$R_{\mathrm{c}}$\footnotemark[6]&$[\SI{}{k\Omega\!\cdot{\!\micro \meter}}]$
  & $-$ & $\SI{11.1}{}$ & $\SI{0.09}{}$
  & \multicolumn{2}{f|}{$\SI{0.13}{}$}
  & $\SI{2.66}{}$ & $\SI{0.13}{}$
  & $\SI{186}{}$ & $\SI{1.36}{}$ & $\SI{0.14}{}$~\cite{peng_gaa}
  & $\leq\SI{0.21}{}$ \\ \hline
$J_\mathrm{G}^{\mathrm{on}}$&$[\SI{}{A/c\meter^2}]$
  & $\SI{1.5}{}$ & \multicolumn{2}{p|}{$3\!\times\!10^{-3}$} 
  & \multicolumn{2}{f|}{$4\!\times\!10^{-4}$} 
  & \multicolumn{2}{s|}{$2\!\times\!10^{-3}$} 
  & \multicolumn{2}{g|}{$8.5\!\times\!10^{-4}$} & $-$
  & $\leq8\!\times\!10^{1}$ \\ \hline
$J_\mathrm{G}^{\mathrm{off}}$&$[\SI{}{A/c\meter^2}]$
  & {$1.5\!\times\!10^{-1}$}  & \multicolumn{2}{p|}{$2\!\times\!10^{-4}$} 
  & \multicolumn{2}{f|}{$6.5\!\times\!10^{-4}$} 
  & \multicolumn{2}{s|}{$2.5\!\times\!10^{-2}$} 
  & \multicolumn{2}{g|}{$5.5\!\times\!10^{-4}$} & $-$
  & $\leq8\!\times\!10^{1}$ \\
\toprule
\end{tabular}
\label{tab:metrics}
\footnotetext{Here $W$ denotes device width, $L_{\mathrm{S-D}}$ source–drain spacing, $L_{\mathrm{G}}$ gate length, $L_{\mathrm{Acc}}$ access region length, $t_{\mathrm{CH}}$ channel thickness,
$t_{\mathrm{vdW}}$ van der Waals gap,
$t_{\mathrm{OX}} (\bisox{})$ gate insulator thickness, $t_{\mathrm{OX}} (\ce{HfO_2})$ second gate insulator thickness, $\mathrm{EOT}$ equivalent oxide thickness, $I_\mathrm{D}^{\mathrm{on}}$ on-current, $\mathrm{SS}$ sub-threshold slope, $\mu$ electron mobility in channel, $R_{\mathrm{c}}$ normalized contact resistance, $J_\mathrm{G}^{\mathrm{on}}$ gate leakage current at on-state, and $J_\mathrm{G}^{\mathrm{off}}$ gate leakage current at off-state.
}

\footnotetext[1]{Representation labels for the devices: the labels `meas.' and `de-em.' refer to the measured and de-embedded devices, respectively, while `degraded' denotes the degraded \plan{} device, and `proj.' corresponds to the \gaa{} device projected to IRDS HP/2035 specifications within the bounded scaling analysis.}
\footnotetext[2]{By considering the dielectric constant of \bisox{} to be $20\le\kappa\le35$~\cite{peng_beta,khakbazACS}, and for the complete gate stack (\fin{}).}
\footnotetext[3]{Drain current at $V_{\mathrm{D}}=0.6\mathrm{V}$ and $V_{\mathrm{G}}=0.439\mathrm{V}+V_{\mathrm{th}}$ as required by the IRDS roadmap \cite{roadmap_BC,roadmap_MM}.}
\footnotetext[4]{Subthreshold slope extracted at $V_\mathrm{D}=\SI{0.6}{V}$ over the fixed one-decade current-density window ranging from $\SIrange{3e-3}{3e-2}{\micro A\per\micro m}$.}
\footnotetext[5]{Effective mobility extracted from TCAD calibration.}
\footnotetext[6]{Effective contact/access series resistances from TCAD calibration, including access-region resistance, and other source/drain parasitics.}
\end{table}
\renewcommand{\arraystretch}{1}

One important consideration at the projected $L_\mathrm{G}=\SI{12}{nm}$ node is that physical effects which are less important in the fabricated micrometer-scale devices can become significant. Short-channel electrostatics and drain-induced barrier lowering (DIBL) are included in the TCAD projection through the self-consistent electrostatic solution. In particular, the high permittivity of \bissc{} along all three lattice directions ($\varepsilon_{a}=\SI{26.65}{\varepsilon_0}$, $\varepsilon_{b}=\SI{234.02}{\varepsilon_0}$, $\varepsilon_{c}=\SI{99.48}{\varepsilon_0}$)~\cite{khakbazACS} can enhance drain coupling in ultra-scaled devices, which is reflected in the increased subthreshold slope of the projected device listed in \tab{tab:metrics}. To mitigate this effect, the projected device is oriented such that source--drain transport occurs along the $a$ axis, where the in-plane permittivity is lowest. Velocity saturation and quasi-ballistic injection are not described from first principles but instead by using a conventional drift-diffusion model and are therefore treated through the effective-mobility correction and sensitivity analysis discussed above. Quantum confinement may also modify the band edges and charge distribution when the channel is thinned; however, the projection reduces the channel thickness only from the experimentally demonstrated \SI{4}{nm} \gaa{} channel to \SI{3}{nm}, which still contains several atomic layers and is not in the extreme monolayer limit. We therefore expect confinement-related corrections to be secondary for this projection, although they are not explicitly resolved by the present TCAD model. Direct source--to--drain tunneling is also not captured by drift-diffusion and may affect the off-state and subthreshold characteristics at such short gate lengths, especially for further scaling below \SI{12}{nm}. Thus, the present $L_\mathrm{G}=\SI{12}{nm}$ result should be viewed as a calibrated drift-diffusion projection of electrostatics, access scaling, contact effects, and on-current trends, rather than as a complete quantum-transport calculation. A predictive ultimate-scaling assessment will require quantum-corrected or NEGF-based transport simulations.

Contact-length scalability is another important requirement for the applicability of the \bissc{}/\bisox{} platform to ultra-scaled nodes. This requirement is distinct from achieving a low contact resistance in devices with sufficiently long contacts.
For a finite top contact, the width-normalized contact resistance is described within the distributed-contact model~\cite{dual_tlm,BERGER1972145} by
$R_{\mathrm{C}}(L_{\mathrm{C}})=\sqrt{\rho_{\mathrm{C}}R_{\mathrm{SH-C}}}\coth\left(L_{\mathrm{C}}/L_{\mathrm{T}}\right)$, with $L_{\mathrm{T}}=\sqrt{\rho_{\mathrm{C}}/R_{\mathrm{SH-C}}}$,
where $\rho_{\mathrm{C}}$ is the local specific contact resistivity, $R_{\mathrm{SH-C}}$ is the sheet resistance beneath the contact, and
$L_{\mathrm{C}}$ is the physical contact length. Reported \gaa{} TLM data yield
an apparent transfer length of $L_{\mathrm{T}}^{\mathrm{app}}\approx\SI{45}{nm}$~\cite{peng_gaa}.
Because this estimate relies on conventional TLM assumptions regarding the contact geometry and a comparable sheet resistance in the exposed channel and beneath the contact, it should be interpreted as a contact-scaling indicator rather than as a direct local current-injection length. Nevertheless, it indicates that finite-contact current crowding can become relevant when $L_{\mathrm{C}}$ is reduced to a few tens of nanometers, as also observed in recent contact-scaling studies of aggressively scaled 2D transistors~\cite{sakibNatElectron2025}. Therefore, aggressive contact-length scaling will require co-optimization of $\rho_{\mathrm{C}}$ and $R_{\mathrm{SH-C}}$ through interface engineering, doping, contact geometry, and direct validation using devices with systematically varied physical contact lengths.

As mentioned earlier, eliminating the vdW gap removes the additional series capacitance associated with the low-permittivity interfacial region, whose contribution depends on the local interface separation and can therefore vary with roughness, strain, and processing conditions~\cite{pourfathScience}. Since lateral gate-length scaling does not by itself modify the interfacial bonding, it is not expected to destabilize the zipper bonding. However, variability from oxide thickness, stoichiometry, local strain, oxidation-front nonuniformity, point defects, and edge or curvature effects may still remain. Previous first-principles calculations of two \bissc{}/\bisox{} heterostructures, with the lattice mismatch accommodated either in the oxide or the semiconductor, found type-I band alignment in both cases~\cite{khakbazACS}, indicating that strain does not qualitatively alter the band alignment in these configurations. However, these calculations do not establish the strain range over which this alignment is preserved or rule out local reconstruction or defect formation. Thus, our 12-nm TCAD projection addresses electrostatics and performance, while direct structural/electrical validation under sub-10-nm processing remains future work.

\subsubsection{Hysteresis Assessment}
\setlength{\tabcolsep}{2pt}  
\renewcommand{\arraystretch}{0.1}
\begin{figure}[!bt]
\centering\begin{tabular}{@{}pfsg@{}}
\figheader{}
&&&\\&&&\\
\begin{subfigure}[b]{.245\linewidth}
\includegraphics[width=\linewidth]{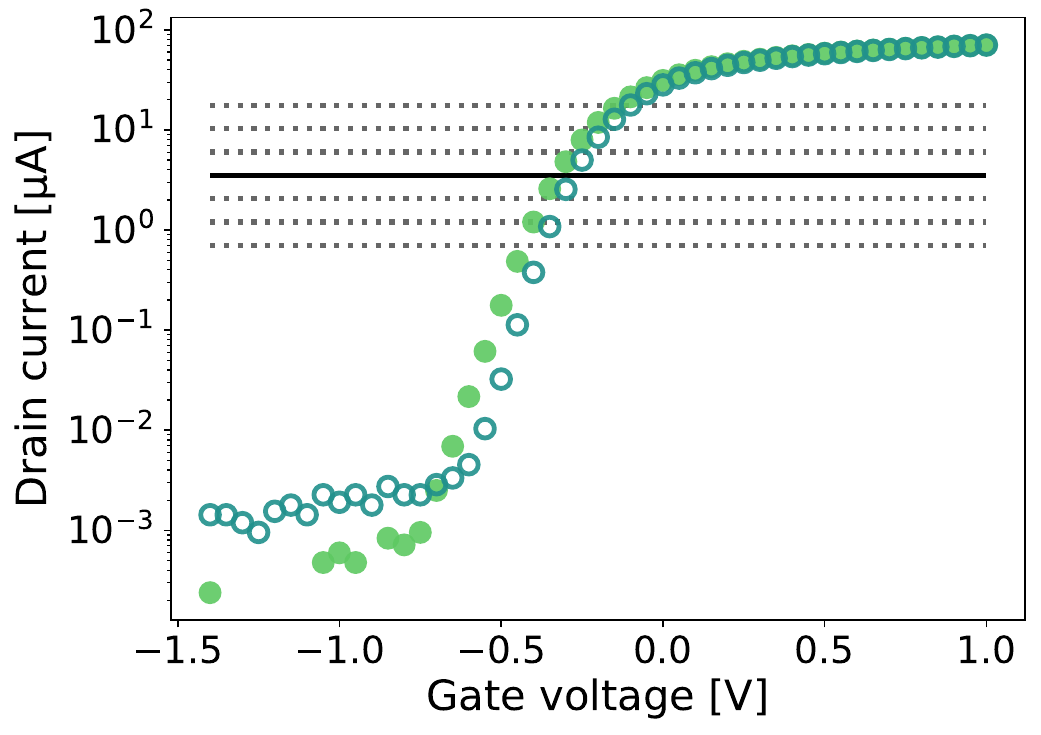}
\caption{}
\label{p_readout}
\end{subfigure}&
\begin{subfigure}[b]{.245\linewidth}
\includegraphics[width=\linewidth]{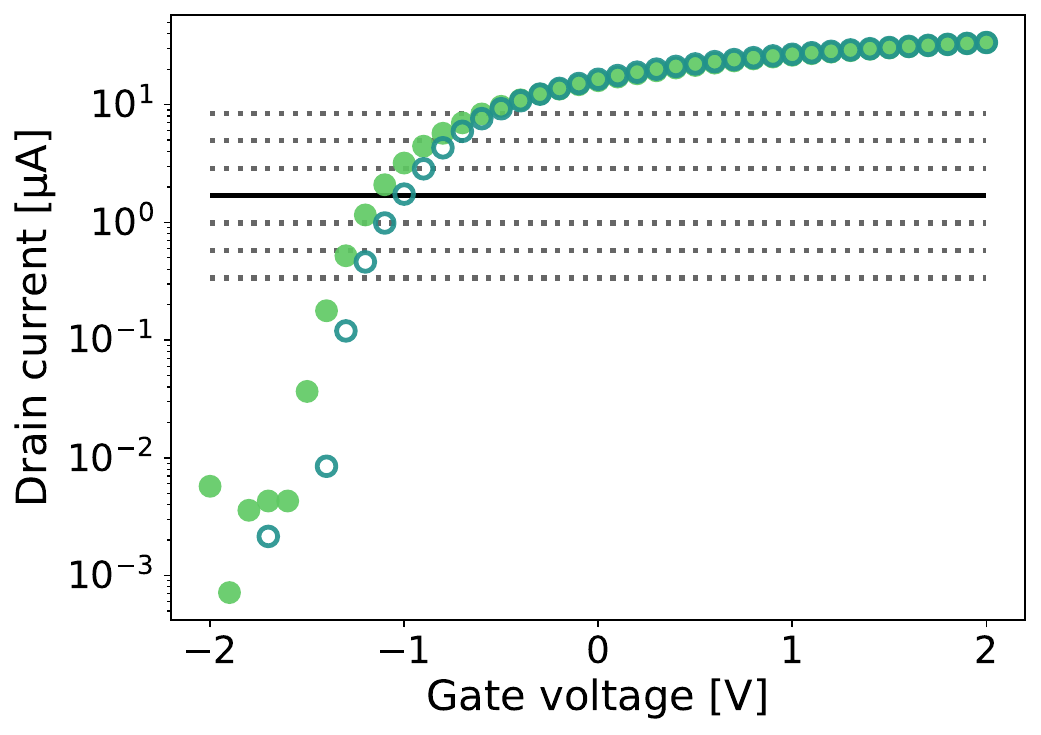}
\caption{}
\label{f_readout}
\end{subfigure}&
\begin{subfigure}[b]{.245\linewidth}
\includegraphics[width=\linewidth]{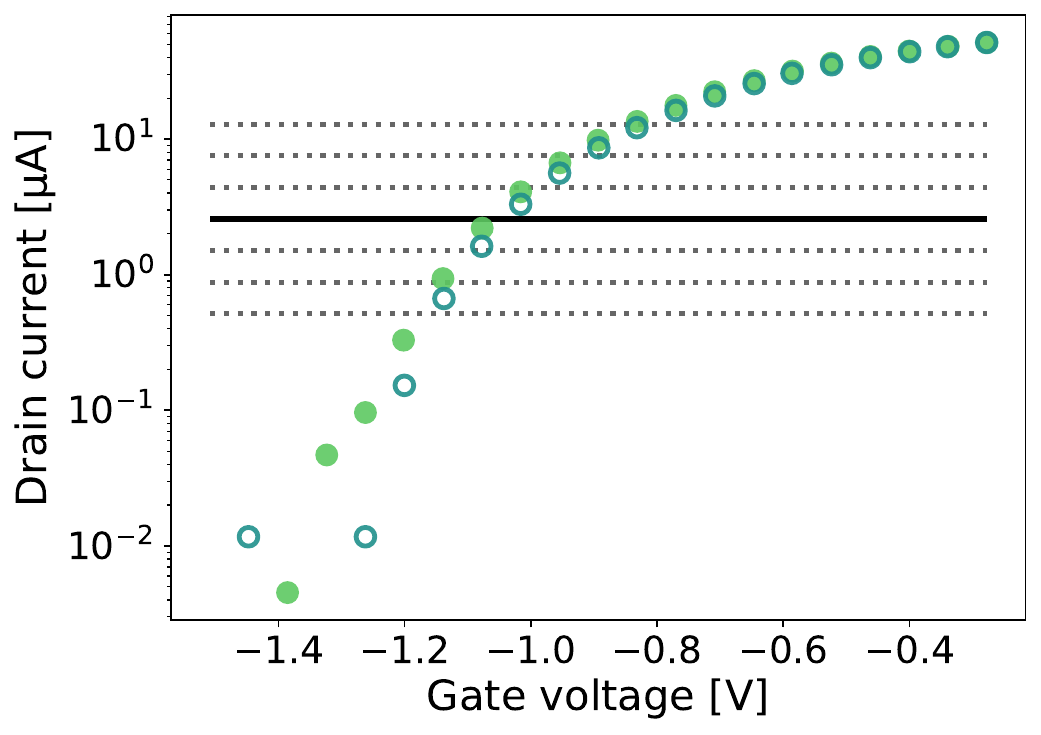}
\caption{}
\label{s_readout}
\end{subfigure}&
\begin{subfigure}[b]{.245\linewidth}
\includegraphics[width=\linewidth]{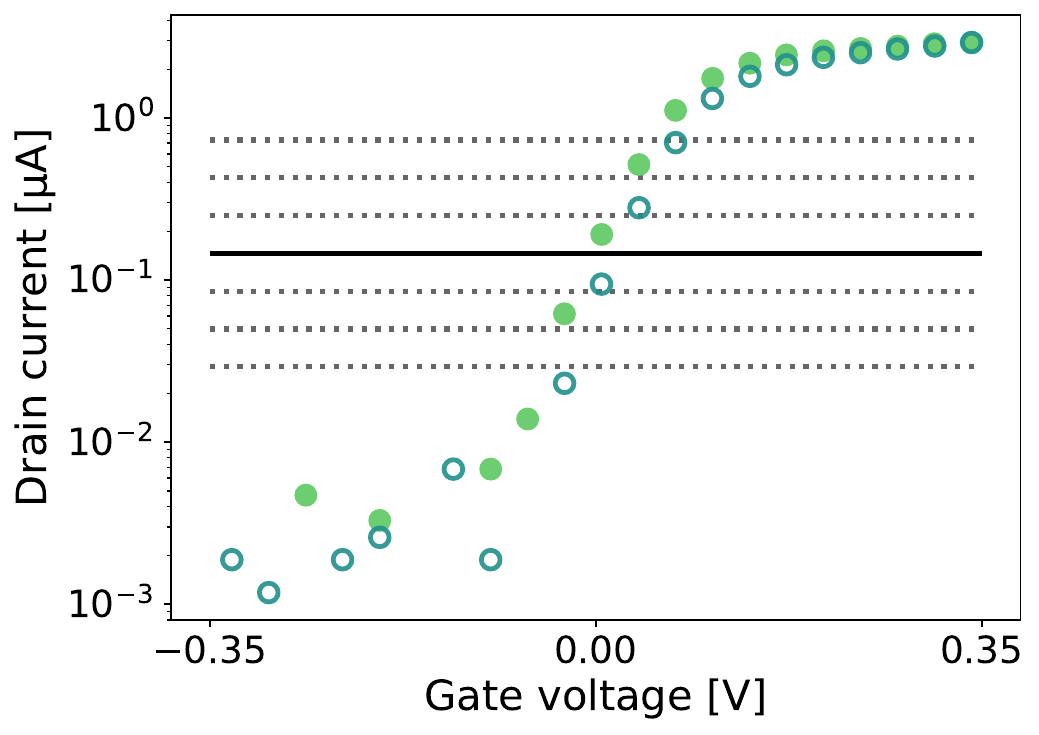}
\caption{}
\label{g_readout}
\end{subfigure}\\
\begin{subfigure}[b]{.245\linewidth}
\includegraphics[width=\linewidth]{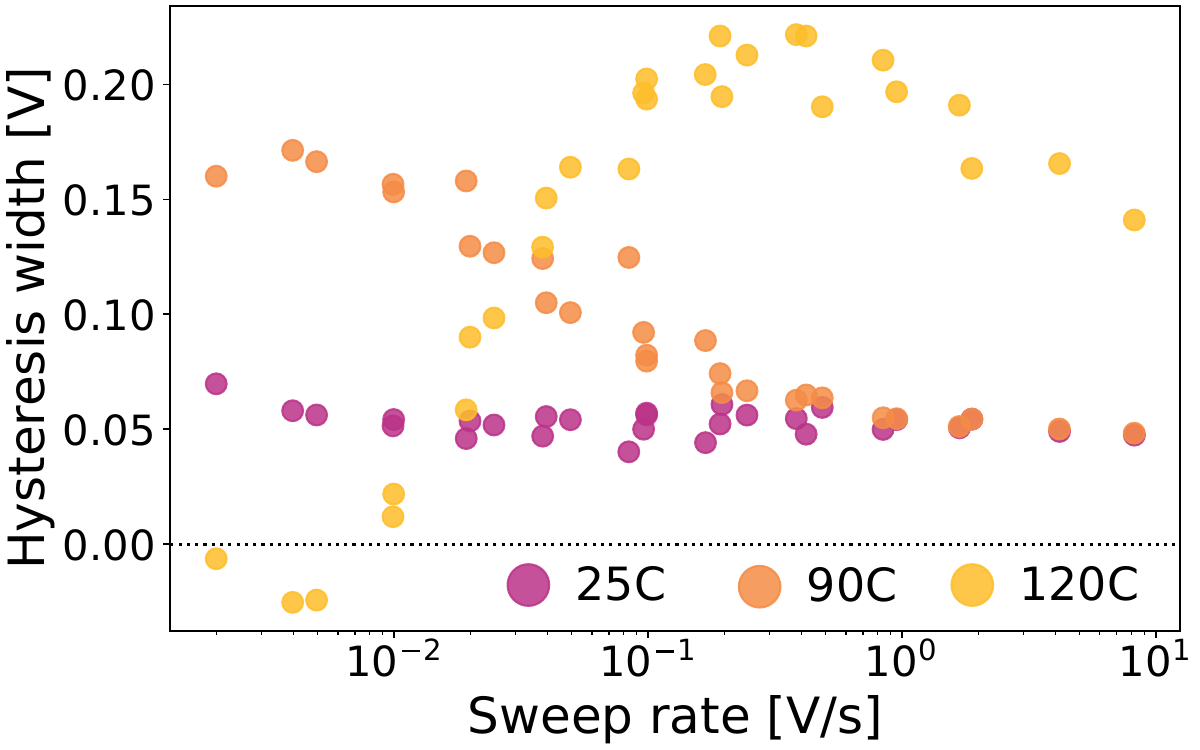}
\caption{}
\label{p_exp}
\end{subfigure}&
\begin{subfigure}[b]{.245\linewidth}
\includegraphics[width=\linewidth]{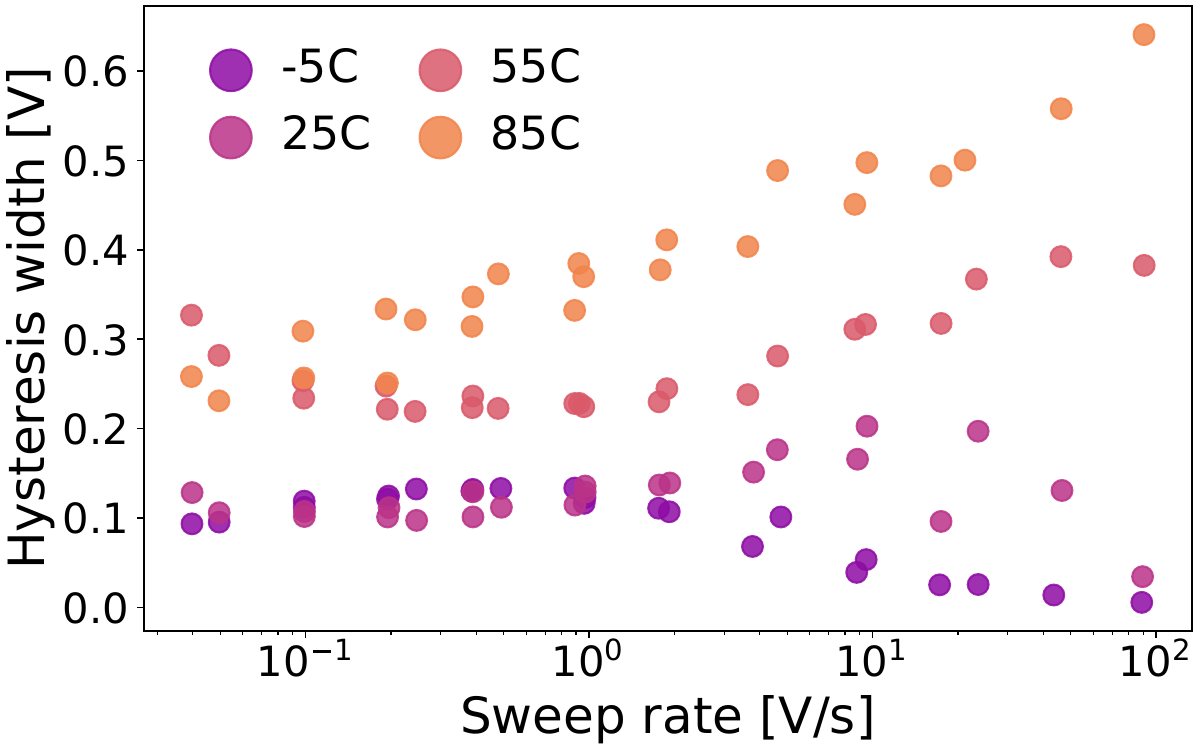}
\caption{}
\label{f_exp}
\end{subfigure}&
\begin{subfigure}[b]{.245\linewidth}
\includegraphics[width=\linewidth]{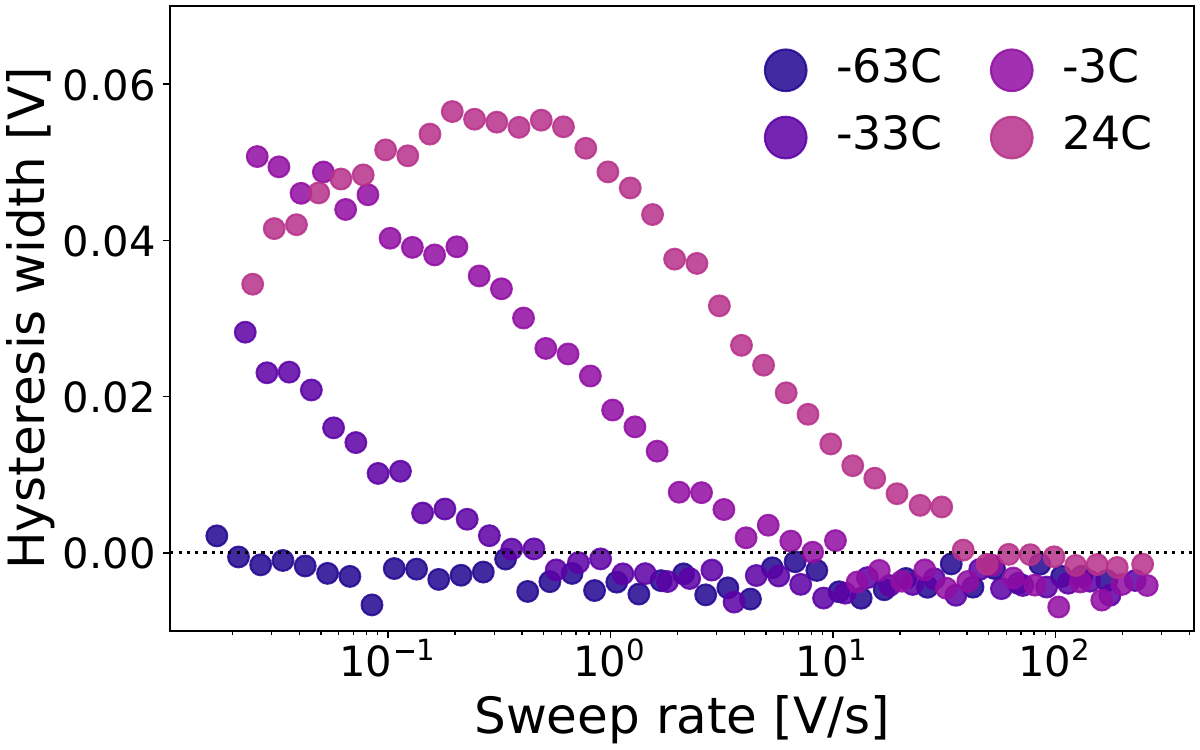}
\caption{}
\label{s_exp}
\end{subfigure}&
\begin{subfigure}[b]{.245\linewidth}
\includegraphics[width=\linewidth]{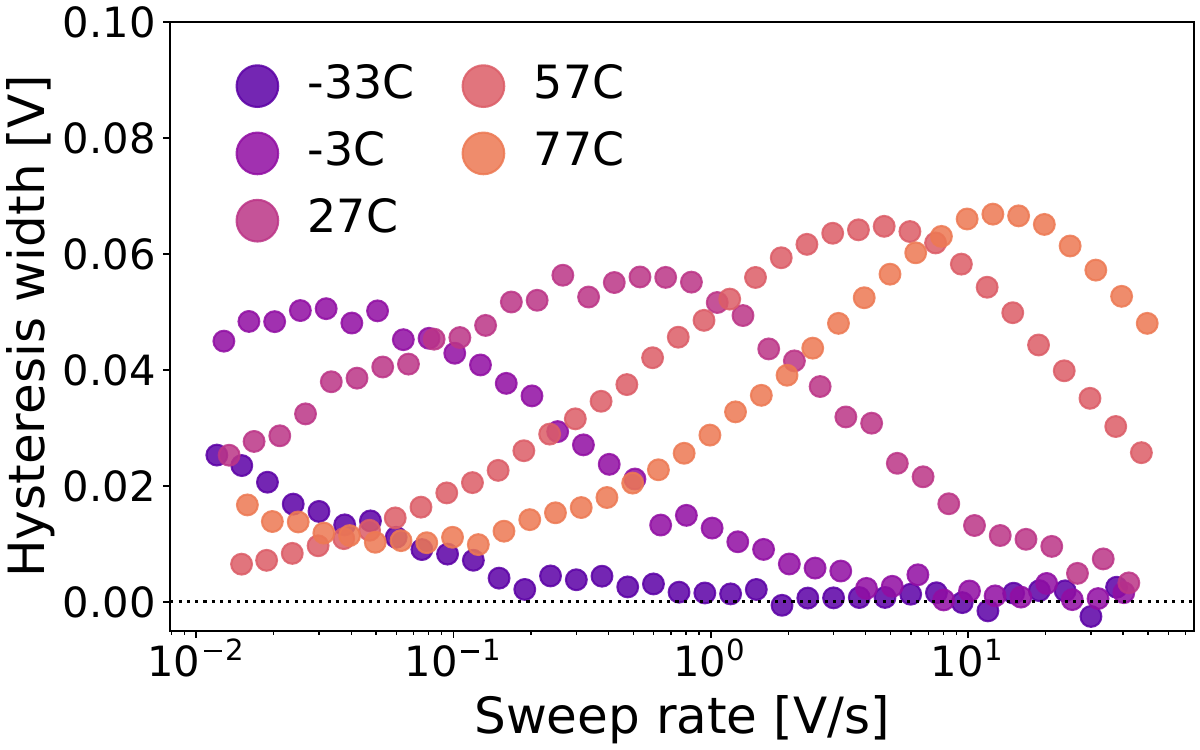}
\caption{}
\label{g_exp}
\end{subfigure}\\
\end{tabular}
\caption{Hysteresis evaluation.
\textbf{(a)-(d)} Positions of the readouts relative to the measured drain currents at room temperature and a sweep-rate of $\SI{5}{V/s}$  with filled and empty circles representing up-sweep and down-sweep data points, respectively (\itr{} is marked in the middle by a solid line) for:
\textbf{(a)} the \plan{},
\textbf{(b)} the \fin{},
\textbf{(c)} the \sgaa{},
and \textbf{(d)} the \gaa{}.
\textbf{(e)-(h)} Evaluated hysteresis at \itr{} as a function of sweep-rate at various temperatures for:
\textbf{(e)} the \plan{},
\textbf{(f)} the \fin{},
\textbf{(g)} the \sgaa{},
and \textbf{(h)} the \gaa{}.
}
\end{figure}
\setlength{\tabcolsep}{6pt}
\renewcommand{\arraystretch}{1}

Next, we investigate the reliability of these devices by evaluating the hysteresis in the transfer characteristics. 
Hysteresis is particularly important because it reflects how consistently and reliably a device can operate under switching conditions. In real-world applications, transistors switch frequently at varying rates and often experience a wide range of temperatures. Therefore, it is critical to assess whether a device can consistently follow the same transfer curve during operation. Failure to do so may result in erratic behavior, which is detrimental in power and logic circuits.
While bias temperature instability (BTI) measurements, which likely have the same microscopic origin as hysteresis~\cite{GrasserMicRel2012}, reveal the overall voltage drifts in transfer curves, those results mostly show long term degradation and can be highly susceptible to subsequent bias-unrelated slow device drifts. 
In contrast, hysteresis is related to the difference in voltage shift between a forward (ramp-up) and reverse (ramp-down) voltage sweep. This approach allows us to isolate and analyze the underlying atomistic phenomena that occur during switching, while minimizing the influence of external drifts, e.g. due to adsorbates. 
This effect is especially pronounced in prototype devices fabricated in university settings, which are typically developed for the purpose of investigating cutting-edge technologies and novel materials without being fully encapsulated. 
However, in hysteresis analysis, the large shifts caused by such slow device drifts tend to cancel out, enabling a more fundamental investigation of the intrinsic reliability under varying operation conditions.

By applying triangular voltage signals to the gate contact across a wide frequency range (spanning up to five orders of magnitude) while measuring the drain current during both the ramp-up and ramp-down phases, the corresponding threshold voltage drift along with the hysteresis is captured:
\dvth{}=$\Delta V(I_\mathrm{th})$=$V(I_\mathrm{th})_\mathrm{down}$-$V(I_\mathrm{th})_\mathrm{up}$.
We further extend the analysis by examining the voltage shift at multiple drain current criteria (\ith{}) levels, rather than focusing solely on the threshold current, in order to probe voltage drifts caused by different types of defects. Measuring the hysteresis widths at several current readouts
($\Delta V($\ith{}$)$=$V($\ith{}$)_\mathrm{down}$-$V($\ith{}$)_\mathrm{up}$)
and across various switching rates enables access to a broader range of defect capture and emission time constants. 
Since charge trapping is a thermally activated process, performing measurements at different temperatures is required to properly de-correlate the extracted defect parameters.

The maximum possible range for [min(\ith{}), max(\ith{})] possessing interpolatable data points at all sweep-rates and temperatures in all devices was found to be $[0.01,0.25]\times I_\mathrm{D}^{\mathrm{on, RT}}$, with the logarithmic midpoint at $0.05\times I_\mathrm{D}^{\mathrm{on, RT}}$. As shown in \figs{p_readout}{g_readout}, we define seven logarithmically spaced read‑out levels within this range and use the central level as the reference readout (\itr{}). As illustrated in \figs{p_exp}{g_exp}, the measured hysteresis is predominantly clockwise (the ramp‑down curve lies to the right of the ramp‑up curve, i.e. \dvth{}$>0$).
On the \plan{} after heating the device above \SI{90}{}$-$\SI{100}{\degree C}, the hysteresis becomes partially counter-clockwise (CCW). In contrast, the other three devices consistently show clockwise hysteresis which can be attributed to charge trapping at defects in the gate insulator~\cite{knob_iedm_2023,mos2hys,GHOSH2025112333}. 

\subsection{Microscopic Nature of Defects Causing the Hysteresis}\label{sec:DFTTCAD}
\setlength{\tabcolsep}{2pt}  
\renewcommand{\arraystretch}{0.1}
\begin{figure}[!bt]
{\captionsetup[subfigure]{skip=-2pt}\begin{subfigure}[b]{.27\linewidth}
\includegraphics[width=1.0\linewidth]{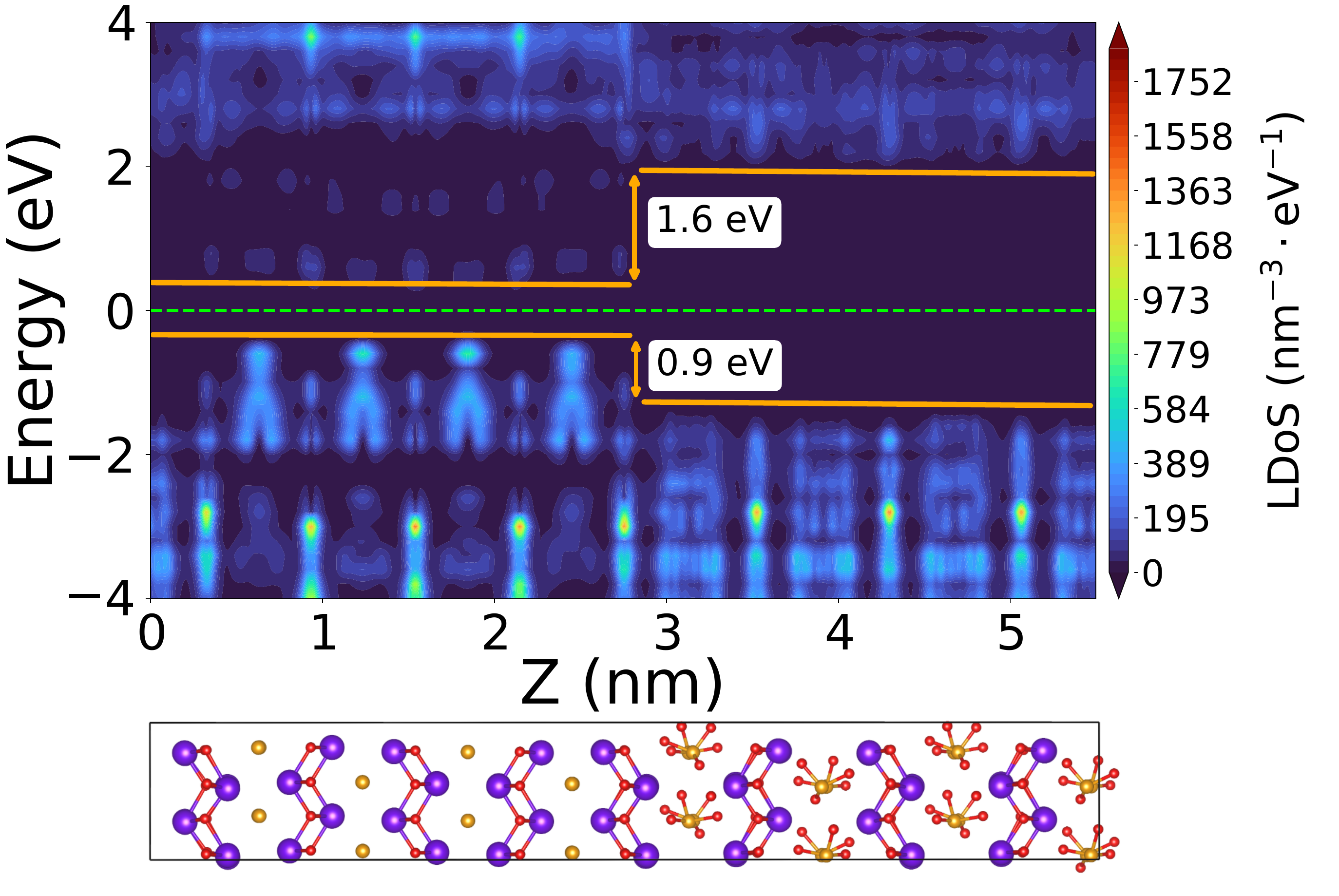}
\caption{}
\label{LDOS}
\end{subfigure}}
{\captionsetup[subfigure]{skip=-2pt}\begin{subfigure}[b]{0.7\linewidth}
\includegraphics[width=1.0\linewidth]{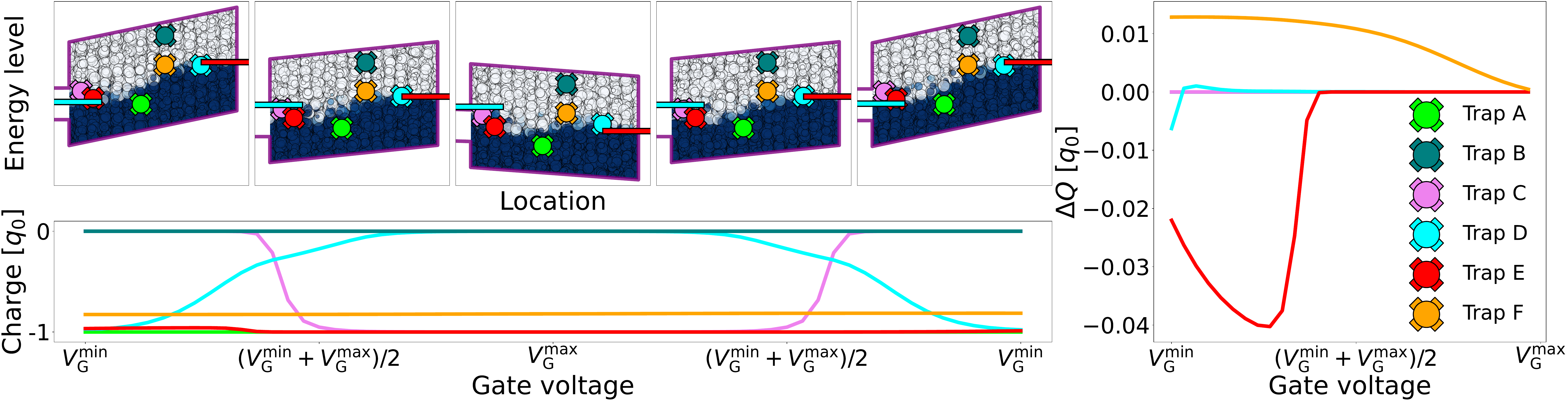}
\caption{}
\label{NMPtraps}
\end{subfigure}}
\centering\begin{tabular}{@{}pfsg@{}}
\figheader{}
&&&\\&&&\\
\begin{subfigure}[b]{.245\linewidth}
\includegraphics[width=1.0\linewidth]{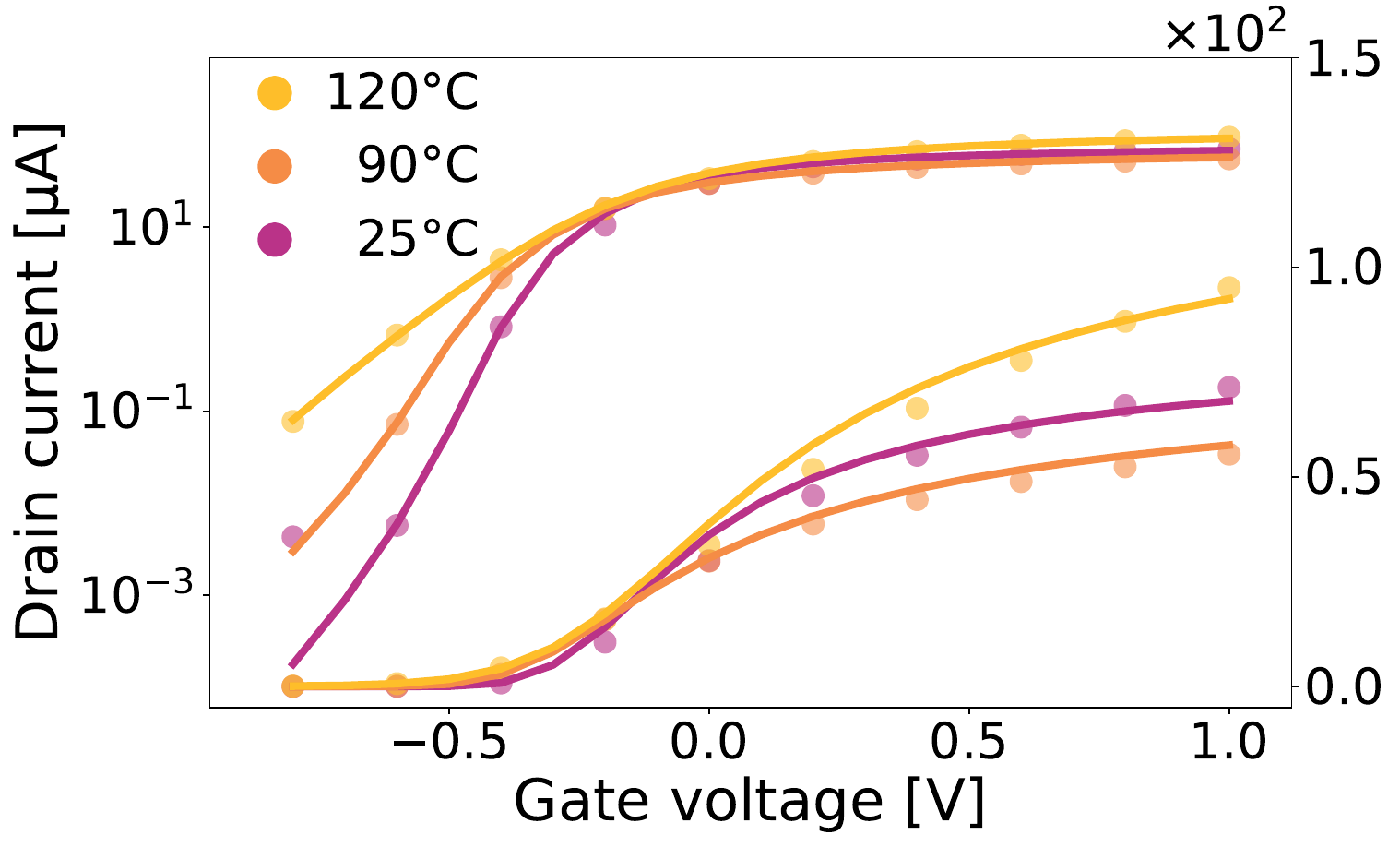} 
\caption{}
\label{p_elecstat}
\end{subfigure}&
\begin{subfigure}[b]{.245\linewidth}
\centering\includegraphics[width=1.0\linewidth]{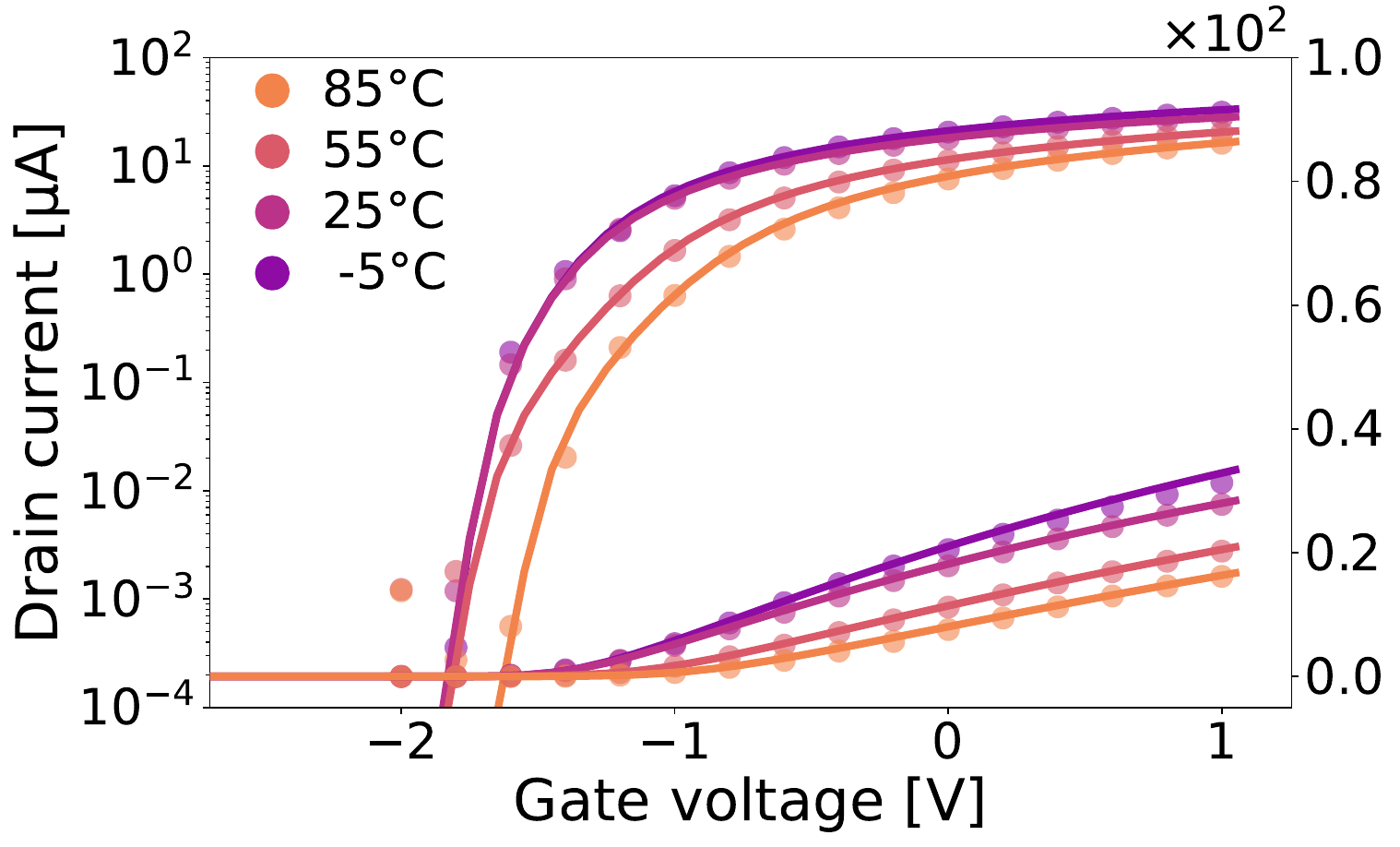} 
\caption{}
\label{f_elecstat}
\end{subfigure}&
\begin{subfigure}[b]{.245\linewidth}
\includegraphics[width=1.0\linewidth]{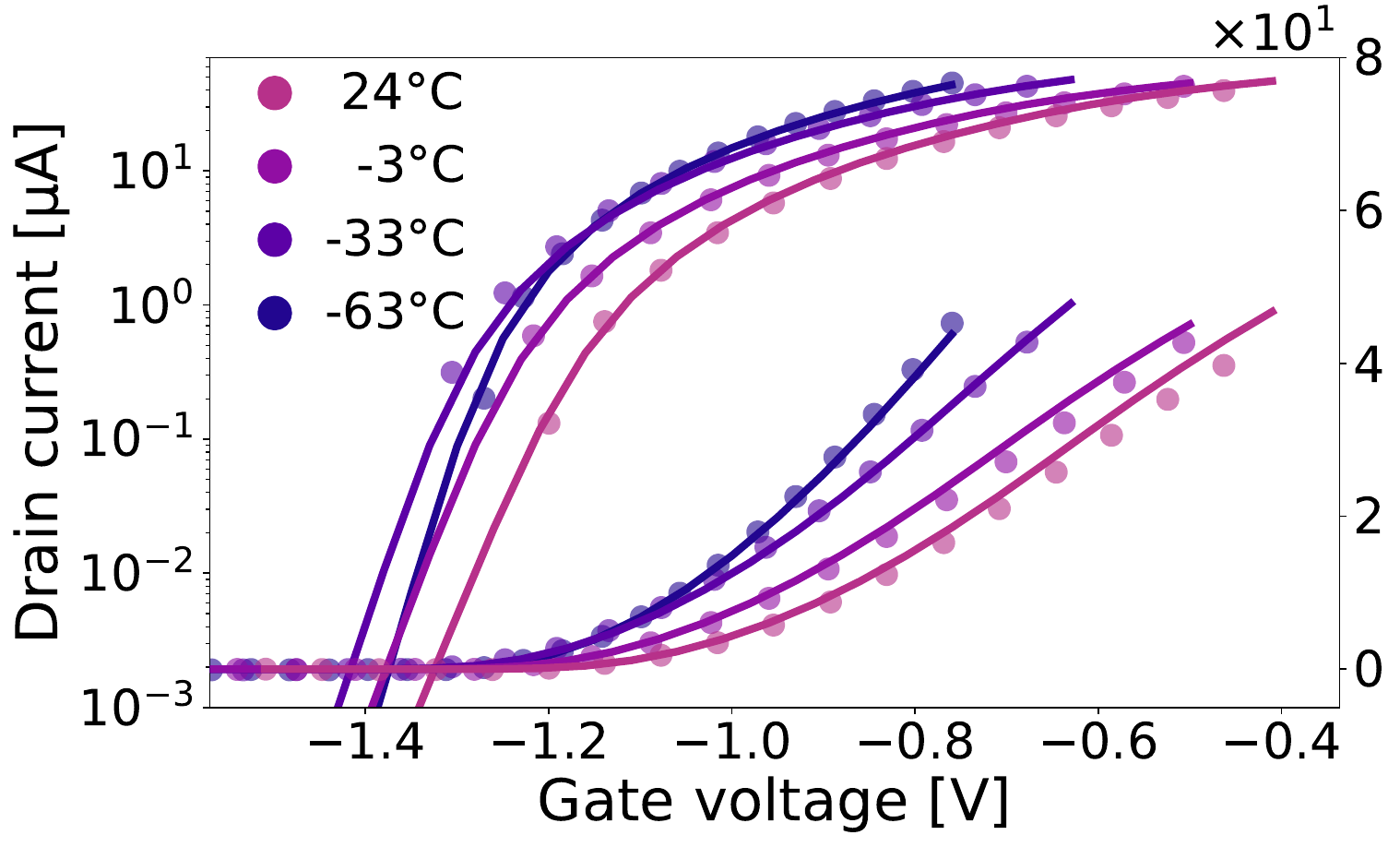} 
\caption{}
\label{s_elecstat}
\end{subfigure}&
\begin{subfigure}[b]{.245\linewidth}
\includegraphics[width=1.0\linewidth]{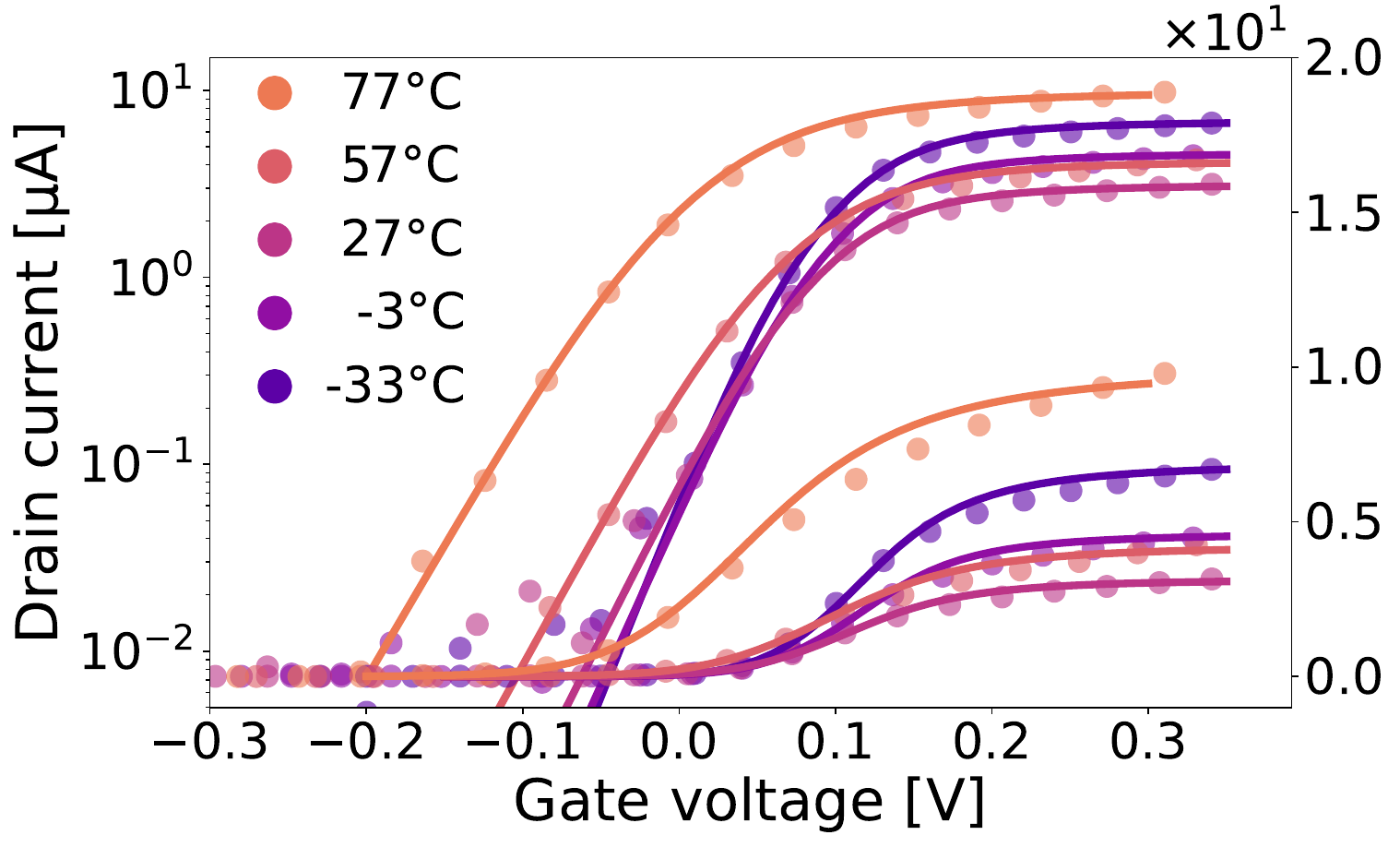} 
\caption{}
\label{g_elecstat}
\end{subfigure}\\
\begin{subfigure}[b]{.245\linewidth}
\includegraphics[width=1.0\linewidth,trim={0.3cm 0.3cm 0.3cm  0.3cm},clip]{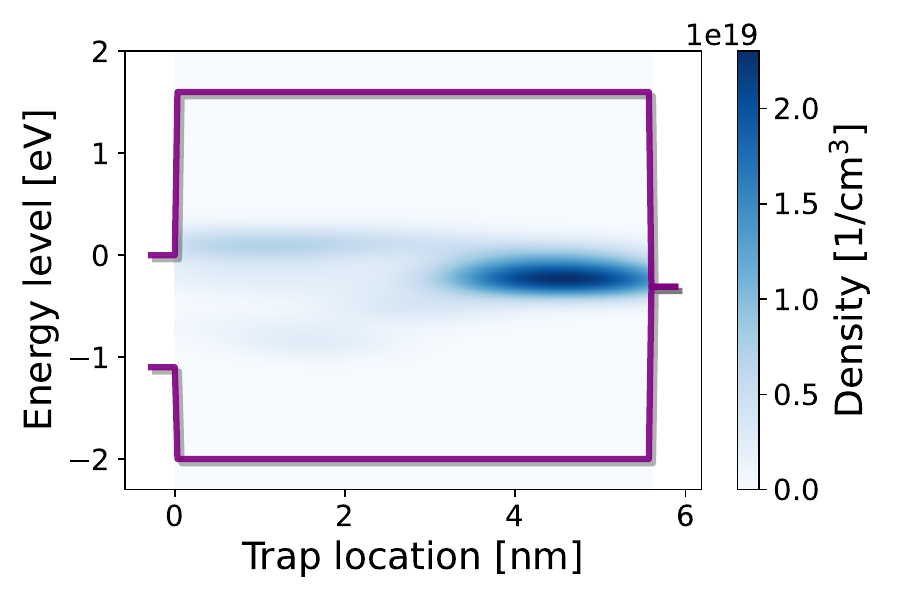}
\caption{}
\label{p_xtet}
\end{subfigure}&
\begin{subfigure}[b]{.245\linewidth}
\includegraphics[width=1.0\linewidth,trim={0.3cm 0.3cm 0.3cm  0.3cm},clip]{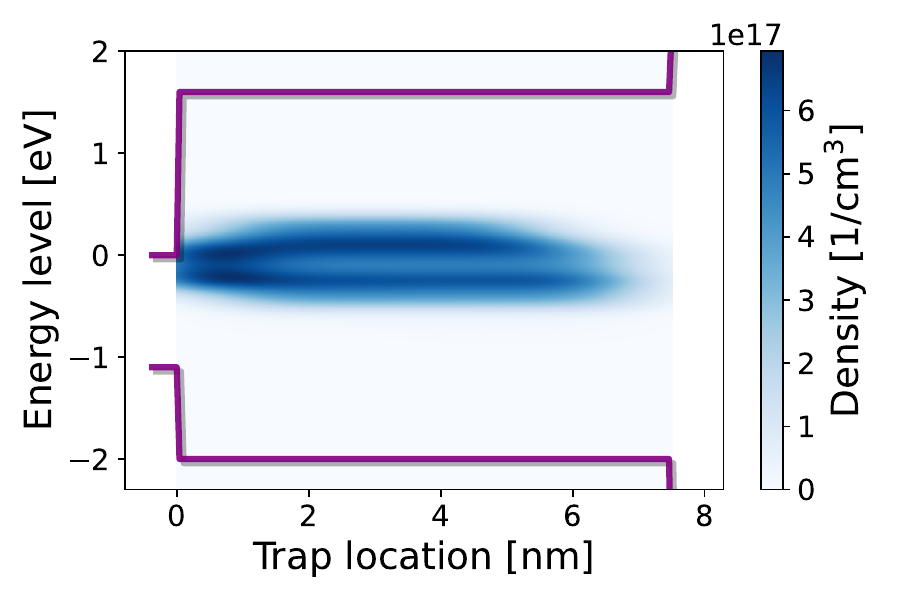}
\caption{}
\label{f_xtet}
\end{subfigure}&
\begin{subfigure}[b]{.245\linewidth}
\includegraphics[width=1\linewidth,trim={0.3cm 0.3cm 0.3cm 0.3cm},clip]{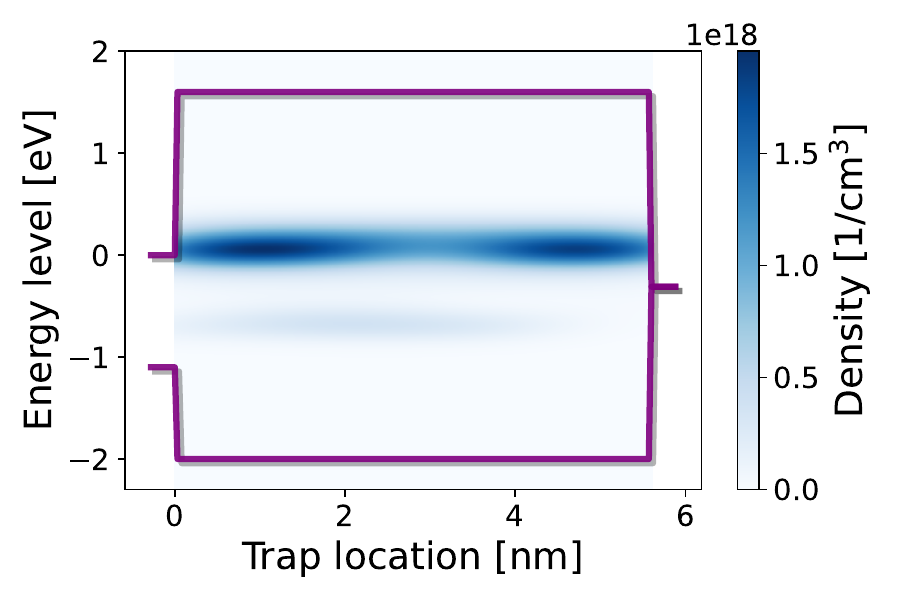}
\caption{}
\label{s_xtet}
\end{subfigure}&
\begin{subfigure}[b]{.245\linewidth}
\includegraphics[width=1.0\linewidth,trim={0.3cm 0.3cm 0.3cm 0.3cm},clip]{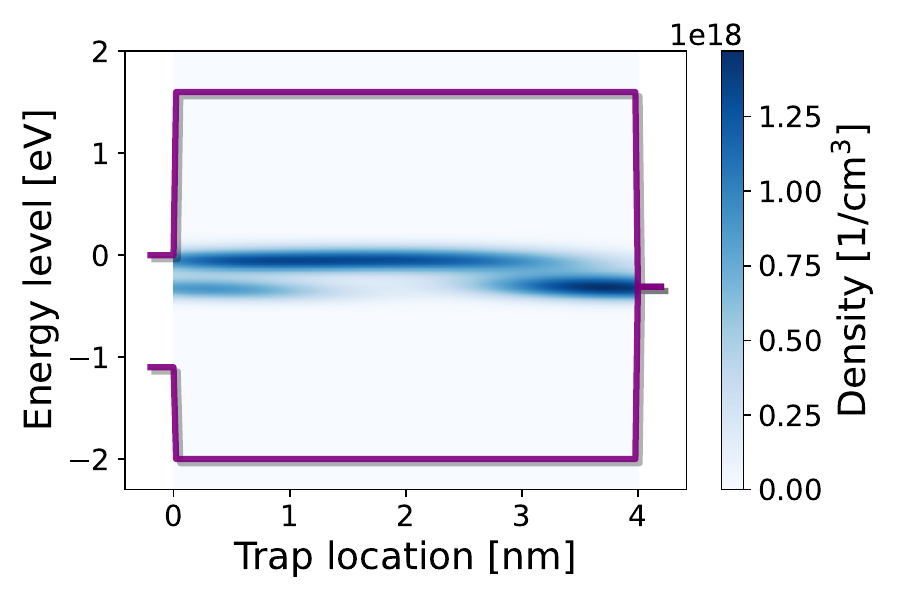}
\caption{}
\label{g_xtet}
\end{subfigure}\\
\begin{subfigure}[b]{.245\linewidth}
\includegraphics[width=1.0\linewidth,trim={0cm 0cm 0cm 16cm},clip]{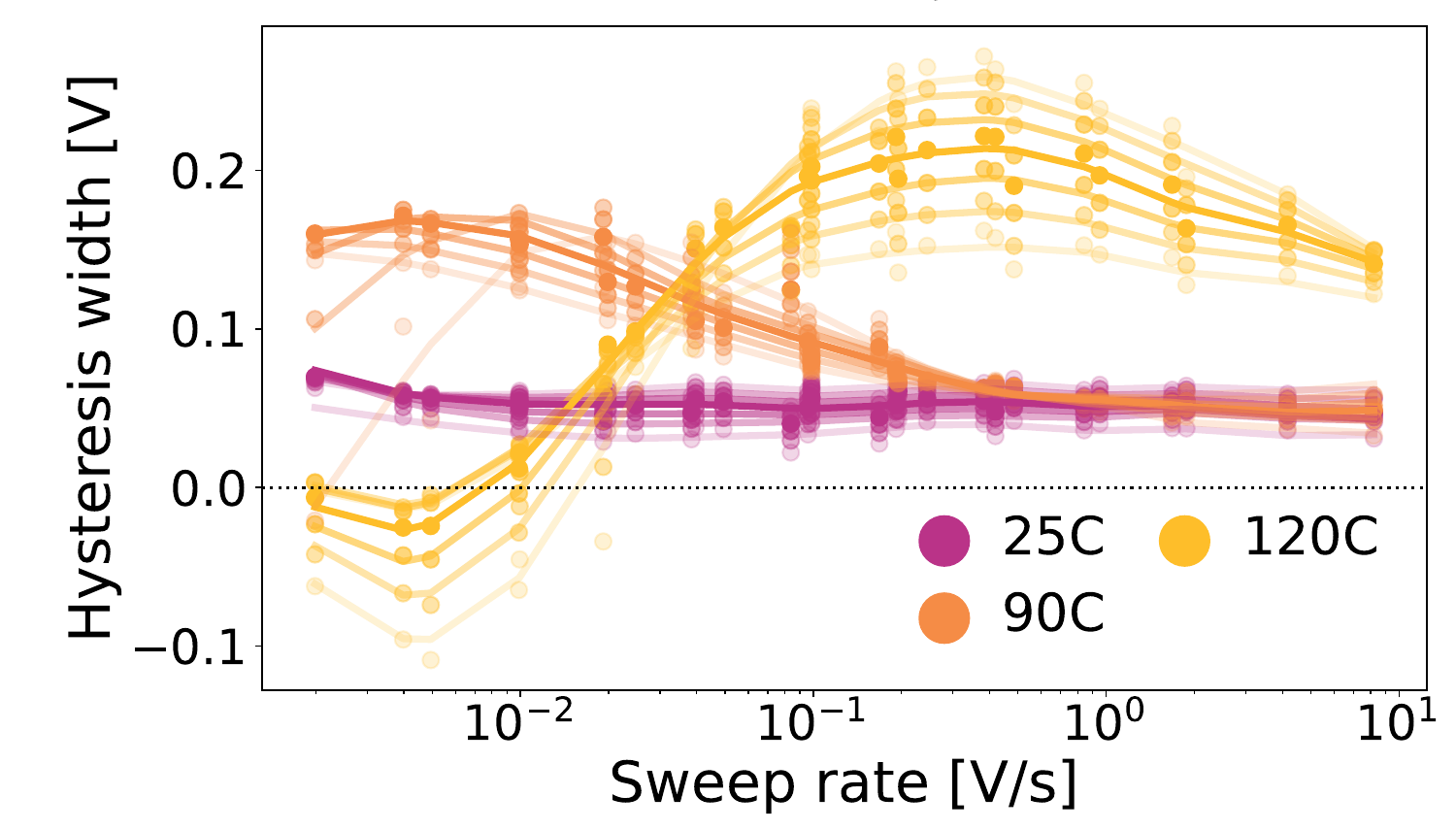}
\caption{}
\label{p_sim_hys}
\end{subfigure}&
\begin{subfigure}[b]{.245\linewidth}
\includegraphics[width=1.0\linewidth,trim={0cm 0cm 0cm 16cm},clip]{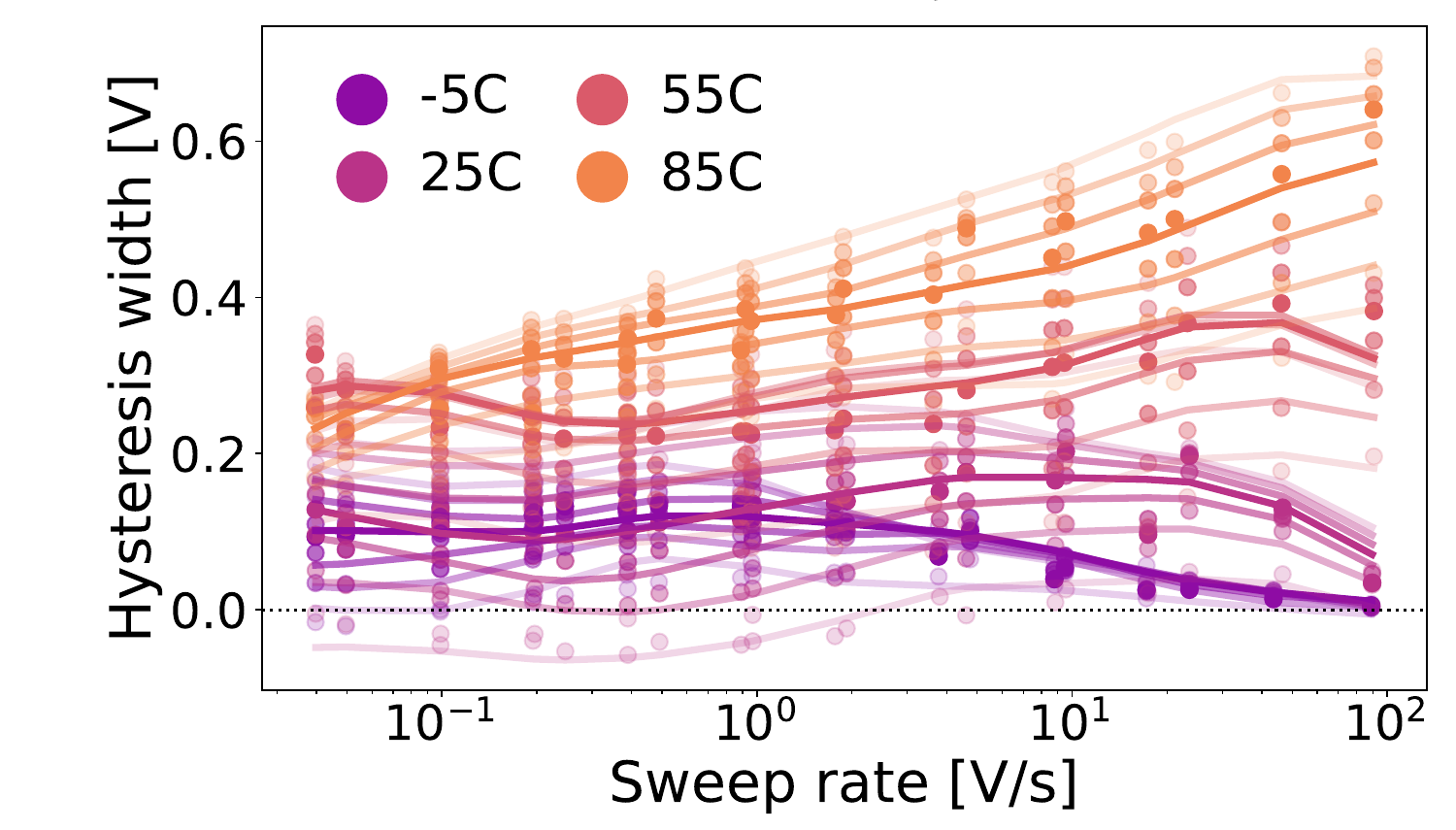} 
\caption{}
\label{f_sim_hys}
\end{subfigure}&
\begin{subfigure}[b]{.245\linewidth}
\includegraphics[width=1.0\linewidth,trim={0cm 0cm 0cm 16cm},clip]{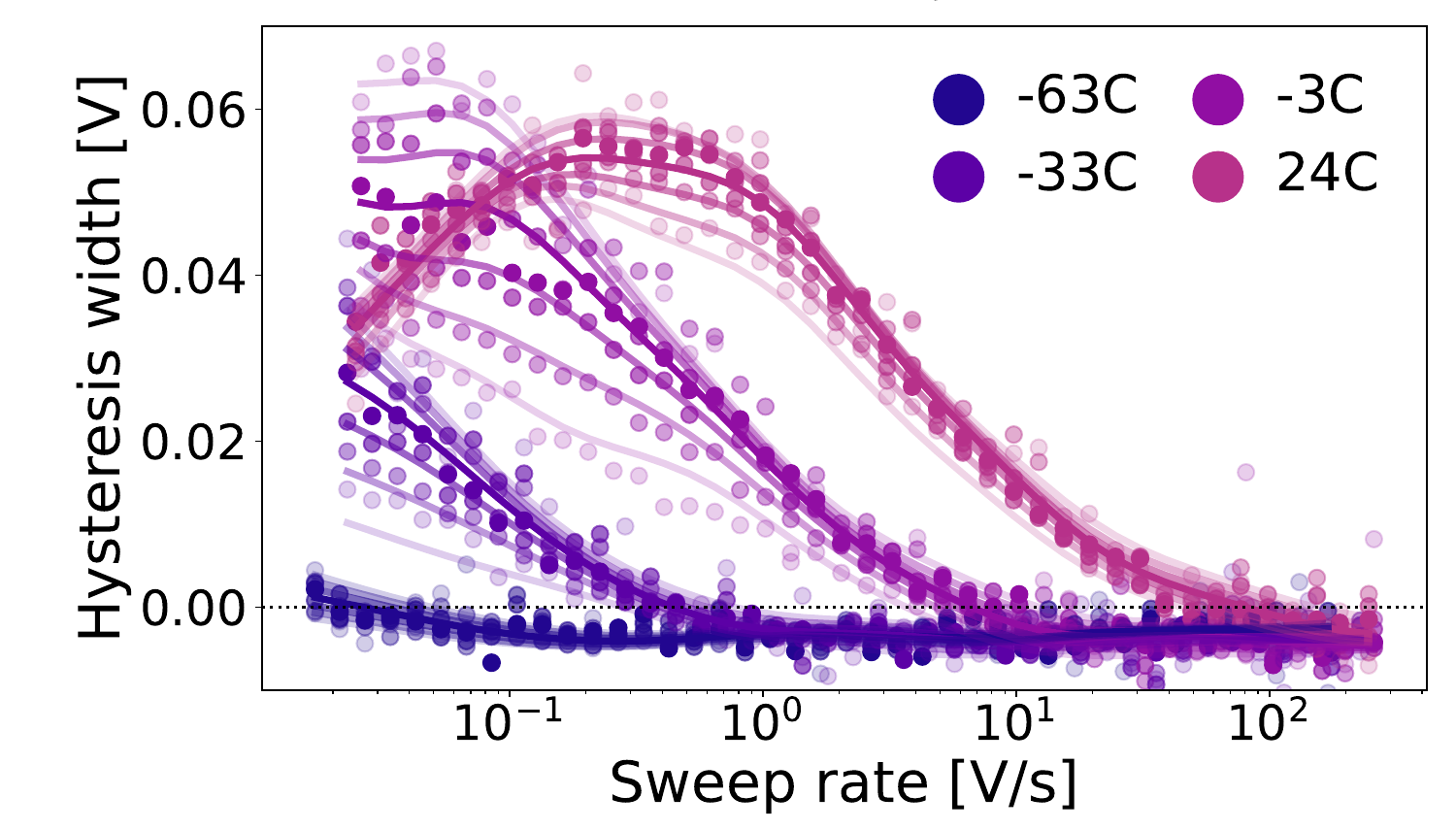} 
\caption{}
\label{s_sim_hys}
\end{subfigure}&
\begin{subfigure}[b]{.245\linewidth}
\includegraphics[width=1.0\linewidth,trim={0cm 0cm 0cm 16cm},clip]{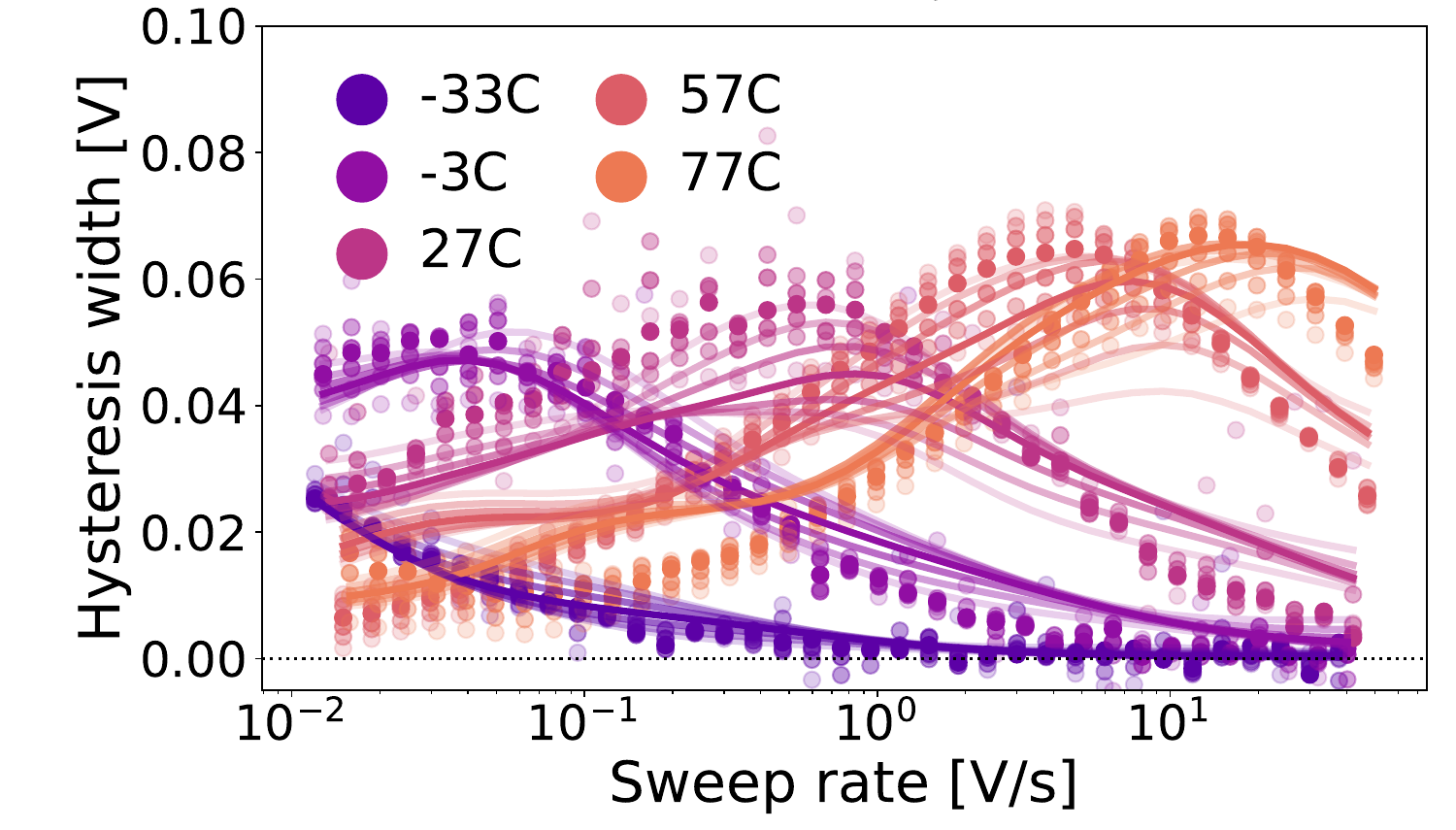} 
\caption{}
\label{g_sim_hys}
\end{subfigure}\\
\end{tabular}
\caption{Modeling the \bissc{}$/$\bisox{} material system and associated transistors.
\textbf{(a)} The local density of states (LDoS) of \bissc{}$/$\bisox{} along the out‑of‑plane axis (top), paired with a ball‑and‑stick representation (bottom), revealing the band offsets. 
\textbf{(b)} The band diagram plots of the position and energy of the initial traps at five cycle phases with \vg{}=[$V_{\mathrm{min}}$, $(V_{\mathrm{max}}+V_{\mathrm{min}})/{2}$, $V_{\mathrm{max}}$, $(V_{\mathrm{max}}+V_{\mathrm{min}})/{2}$, $V_{\mathrm{min}}$]. The blue and red lines as well as the color of the traps represent the Fermi level at the channel and gate interfaces as well as the trapped charges, respectively. Six exemplary traps (A–F) are tracked: their charge evolution over one up-/down-sweep cycle is shown (bottom), as well as the difference in charge between the two phases of the cycle at $\theta$ and $-\theta$ (right). This charge difference leads to a voltage drop that depends on the trap’s distance from the channel interface. 
While traps A and B are located far from the Fermi level and remain fixed, the other four traps exhibit varying charge states. Traps C and D switch rapidly as both have small \er{} (\SI{0.1}{\electronvolt}) with almost symmetric response and negligible hysteresis. Conversely traps E and F change slowly and dominate the hysteresis, particularly trap F, which contributes across nearly all phases (i.e. all extraction \ith{}'s) due to its very slow and gradual charge evolution.
\textbf{(c)-(f)} The excellent agreement between the measured (circles) and TCAD (solid lines) transfer characteristics at various temperatures for:
\textbf{(c)} the \plan{}, \textbf{(d)} the \fin{}, \textbf{(e)} the \sgaa{}, and \textbf{(f)} the \gaa{}, with the left and right axes showing logarithmic and linear scales.
\textbf{(g)-(j)} The extracted trap profiles for:
\textbf{(g)} the \plan{}, \textbf{(h)} the \fin{}, \textbf{(i)} the \sgaa{}, and \textbf{(j)} the \gaa{}.  
\textbf{(k)-(n)} The modeled (solid lines) and measured (circles) hysteresis at various \ith{}'s as a function of sweep-rate at various temperatures for \textbf{(k)} the \plan{}, \textbf{(l)} the \fin{}, \textbf{(m)} the \sgaa{}, and \textbf{(n)} the \gaa{}. The opacity reflects the proximity of the \ith{} to \itr{}.}
\end{figure}
\setlength{\tabcolsep}{6pt}  
\renewcommand{\arraystretch}{1}

To understand the microscopic nature of the point defects responsible for hysteresis, we combine comprehensive experimental data with multi-scale modeling efforts. As the intrinsic properties of \bissc{} and \bisox{} are not well established, we used DFT to investigate the structural, electronic and dielectric properties of the semiconductor and oxide. Based on crystal structures consistent with crystallographic data (e.g. electron microscopy and photoelectron spectroscopy images), we modeled the semiconductor-oxide heterostructures and extracted the electronic properties (see \fig{LDOS}); a detailed discussion of this analysis is provided in~\cite{khakbazACS}. 
The derived parameters, including band gaps, band offsets, and effective masses were subsequently incorporated into Minimos-NT~\cite{MINIMOS-NT,manualGTS}, a commercial TCAD tool to model device-level electrostatics. 
However, inherent and unclear experimental non-idealities (such as distribution of strain, layer thickness, defectivity, and interface quality),  cannot be fully captured by DFT calculations. 
To ensure consistency with experimental observations, we adopted intermediate values for the band gaps and offsets of the semiconductor and oxide, reflecting their sensitivity to strain and interface composition. 
Moreover, since the permittivity of the oxide is strongly influenced by thickness~\cite{peng_vdw}, defect density, and interface quality, we used a permittivity value aligned with experimentally reported capacitance measurements~\cite{peng_beta,peng_vdw}. 
These considerations allowed us to build a consistent device model that bridges atomic-scale insights with experimentally relevant behavior.

\begin{table}[!t]
\caption{Material and device parameters at room temperature (TCAD calibration/DFT)}
\renewcommand{\arraystretch}{1.25}
\begin{tabular}{@{}l|l@{}c|pfsg|c@{}}
\toprule
\multicolumn{3}{l|}{~~~~~Physical quantity} & \plan{} & \fin{} & \sgaa{} & \gaa{} & DFT\cite{khakbazACS,minaACS} \\
\midrule
\multirow{6}{*}{\rotatebox[origin=c]{90}{\bissc{}}}
&Electron mobility $(\mu)$ & $[\mathrm{cm^2\!/V\!\cdot{\!s}}]$ & $\SI{140}{}$ & $\SI{275}{}$ & $\SI{155}{}$ & $\SI{230}{}$ & $-$  \\
&Out-of-plane permittivity $(\varepsilon_{\mathrm{SC}}^{\bot})$ & $[\varepsilon_0]$& \multicolumn{4}{c|}{$\SI{99.5}{}$} & $\SI{99.5}{}$ \\
&Band gap $(E_{\mathrm{g,SC}})$ & $[\mathrm{eV}]$ & \multicolumn{4}{c|}{$\SI{1.1}{}$} & $[\SI{1.04}{}-\SI{1.26}{}]$\\
&Electron affinity $(\xi_{\mathrm{SC}})$ & $[\mathrm{eV}]$ & \multicolumn{4}{c|}{$\SI{4.3}{}$} & $\SI{4.3}{}$ \\
&Electron effective mass $(m_\mathrm{e}^{x}, m_\mathrm{e}^{y})$ & $[m_0]$ & \multicolumn{4}{c|}{$(\SI{0.15}{}, \SI{0.15}{})$} & $(\SI{0.15}{}, \SI{0.15}{})$ \\
&Interface traps energy level\footnotemark[1] $(E_{\mathrm{T}})$ & $[\mathrm{eV}]$ & $\SI{-0.095}{}$ & $\SI{0.0}{}$ & $\SI{0.0}{}$ & $\SI{-0.01}{}$ & $\SI{-0.04}{}$  \\
\midrule
\multirow{3}{*}{\rotatebox[origin=c]{90}{Contacts\footnotemark[2]}}
&Schottky barrier height $(\phi_{\mathrm{b}})$ & $[\mathrm{eV}]$ & $\SI{0.05}{}$ & $\SI{0.014}{}$ & $\SI{0.0}{}$ & $\SI{0.0}{}$ & $-$  \\
&Specific resistivity $(\rho_{\mathrm{c}})$ & $[\mathrm{\Omega\!\cdot{\!cm^2}}]$ & $\SI{1.7}{\times10^{-4}}$  & $\SI{1.36}{\times10^{-6}}$ & $\SI{2.6}{\times10^{-5}}$ & $\SI{1.8}{\times10^{-4}}$ & $-$ \\
&Normalized resistance $(R_{\mathrm{c}})$ & $[\SI{}{k\Omega\!\cdot{\!\micro \meter}}]$ & $\SI{11.1}{}$ & $\SI{0.13}{}$ & $\SI{2.66}{}$ & $\SI{186}{}$ & $-$ \\
\midrule
\multirow{3}{*}{\rotatebox[origin=c]{90}{\bisox{}}}
&Out-of-plane permittivity $(\varepsilon_{\mathrm{OX}}^{\bot})$ & $[\varepsilon_0]$ &  \multicolumn{4}{c|}{$\SI{22}{}$~\cite{peng_beta}} & $\SI{35.3}{}$\\
&Band gap $(E\mathrm{g}_{\mathrm{OX}})$ & $[\mathrm{eV}]$ &  \multicolumn{4}{c|}{$\SI{3.6}{}$} & $[\SI{3.16}{}-\SI{3.70}{}]$\\
&Electron affinity $(\xi_{\mathrm{OX}})$ & $[\mathrm{eV}]$ &  \multicolumn{4}{c|}{$\SI{2.7}{}$} & $\SI{2.7}{}$ \\
\botrule
\end{tabular}
\label{tab:elecstat}
\footnotetext[1]{With respect to the \bissc{} conduction band edge.}
\footnotetext[2]{The listed resistances are effective contact/access series resistances and include access-region resistance and other source/drain parasitics. Since the \fin{} device has a very short access length of only $2\times\SI{7.5}{\nano\meter}$, its fitted value is least affected by long-access region parasitics and therefore provides the closest contact-dominated effective series-resistance estimate within the present device set.}
\end{table}
\renewcommand{\arraystretch}{1}

We rigorously calibrated the device model to match all experimental curves simultaneously (including all output and transfer curves at various gate and drain biases as well as temperatures) on both logarithmic and linear scales. 
This approach resulted in excellent agreement of the TCAD model with the experimental data as illustrated in \figs{p_elecstat}{g_elecstat} (see also appendix~\ref{supp:elecstat}).
During this process, we determined the electron mobility in the channel, the Schottky barrier height, and the contact/access series resistance at the source and drain contacts.
A sensitivity analysis of the extracted parameters and a Y-function analysis for mobility extraction are presented in Supplementary Appendix~\ref{supp:sensitivity} and \ref{supp:y_function}, respectively.
Additionally, we identified trap levels of the interface traps, which coincide with the energy levels of semiconductor defects calculated using DFT, see appendix~\ref{supp:sc_defects}.
These interface states have small relaxation energies and short tunneling distances to the channel, leading to rapid charge exchange on the gate-sweep time scale. They therefore mainly affect the quasi-static electrostatics and subthreshold behavior, but do not contribute meaningfully to the observed hysteresis.
\tab{tab:elecstat} summarizes the extracted material and device parameters. Mobility and contact/access series resistance have been extracted during the same procedure but are also listed in the overview \tab{tab:metrics}.

\subsubsection{Modeling Charge Trapping in the Gate Insulator}\label{sec:hysteresis}
As the next step, in order to be able to theoretically describe the experimentally observed hysteresis, we introduced oxide charge traps in our TCAD model. 
For modeling the oxide traps we used an effective 2-state NMP model~\cite{goes2018,GrasserMicRel2012}, in which each trap is represented by two parabolic energy curves corresponding to its charged and neutral state. In the classical limit of NMP theory, traps are characterized by only three parameters: trap location (\xt{}, with respect to the channel–oxide interface), trap energy level (\et{}, referenced to the conduction band edge of \bissc{}), and trap relaxation energy (\er{}).
As depicted in \fig{NMPtraps}, we initially introduced NMP traps uniformly distributed across the oxide and over \et{} and \er{}. 
We then performed corresponding simulations to determine the theoretical hysteresis width caused by each individual trap at the experimental readout currents during an up- and down-sweep at various sweep-rates and temperatures. 
Within the charge sheet approximation~\cite{csa} the threshold voltage change $\delta V_{\text{th},i}$ of the $i$-th trap is only determined by its charge (occupation) $q_i$ and its location according to:
\begin{equation}
\delta V_{\text{th},i} = \frac{q_i}{C}\times(1-X_{\mathrm{T},i}/t_\mathrm{OX})   \quad\text{with}\quad  
 C=\epsilon_0\epsilon_\mathrm{ox}\frac{A}{t_\mathrm{OX}}\ ,
\label{eq1}
\end{equation}
\noindent where $A$ denotes the sheet area. Examples of such individual contributions during a full sweep cycle are shown in \fig{NMPtraps} for six specific traps sampled from the full distribution. For reasonably small defect concentrations with negligible interactions between traps (non-selfconsistent case), the overall \dvth{} caused by the whole defect ensemble can then be approximated by the sum of all $\delta V_{\text{th},i}$.
In order to decompose the measured hysteresis into individual defect contributions, we employed the ESiD~\cite{WaldhoerTED2021,waldhoer2022comphy} algorithm, which at its core solves a non-negative least squares (NNLS) problem to determine the best weight factors for each trap~\cite{WaldhoerTED2021}. Note that the non-negativity constraint is essential, as the weight factors resulting from the optimization correspond to the trap density. 
This optimization technique enables us to find a density distribution $N(p)$ in the defect parameter space that reproduces the experimental voltage shifts for each experiment and for each of the seven readout currents of the hysteresis width, as specified in Section~\ref{sec:characterization}. 
The experimental \dvth{} is generated by this density function through 
\begin{equation}
\Delta V_{\mathrm{th}}(t) = \sum_i N(p_i) \delta V_{\mathrm{th},i}(t)\,,
\label{eq2}
\end{equation}
\noindent where $p_i$ denotes the set of parameters characterizing the $i$-th trap. \figs{p_xtet}{g_xtet} show the band alignment of the devices, including the obtained traps, reproducing the observed hysteresis with excellent agreement, as illustrated in \figs{p_sim_hys}{g_sim_hys}.

\subsubsection{Defect Identification}\label{sec:defects}
\newlength{\legendwidth} 
\settowidth{\legendwidth}{\includegraphics{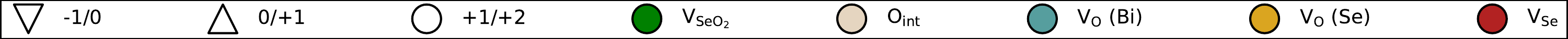}}
\setlength{\tabcolsep}{2pt}  
\renewcommand{\arraystretch}{0.1}
\begin{figure}[!btp]
{\captionsetup[subfigure]{skip=-7pt}\begin{subfigure}[b]{0.97\linewidth}
\includegraphics[width=1.0\linewidth]{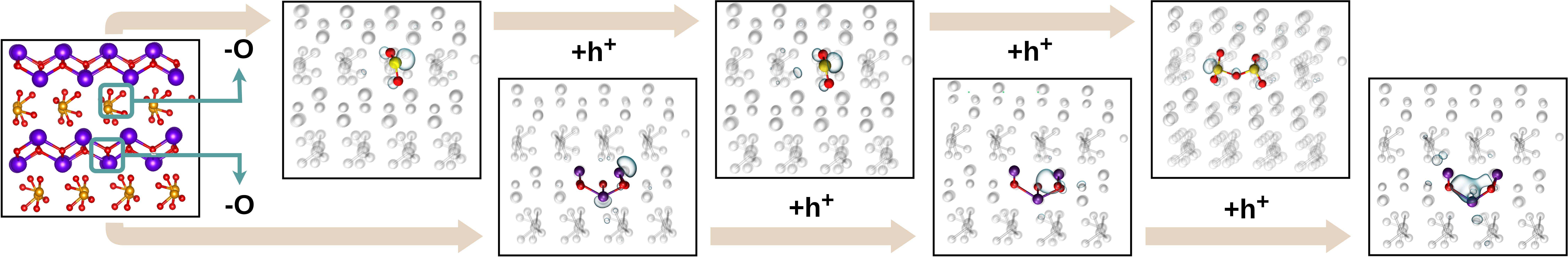} 
\caption{}
\label{dft_vac}
\end{subfigure}}
\centering\begin{tabular}{@{}pfsg@{}}
\figheader{}
&&&\\&&&\\
\multicolumn{4}{c}{ 
\begin{subfigure}[b]{1.0\linewidth} 
\includegraphics[width=1.0\linewidth]{final_figures/vertical_legend.pdf} 
\end{subfigure}}\\
\begin{subfigure}[b]{.245\linewidth}
\includegraphics[width=1.0\linewidth,trim={0cm 4cm 0cm 0cm},clip]{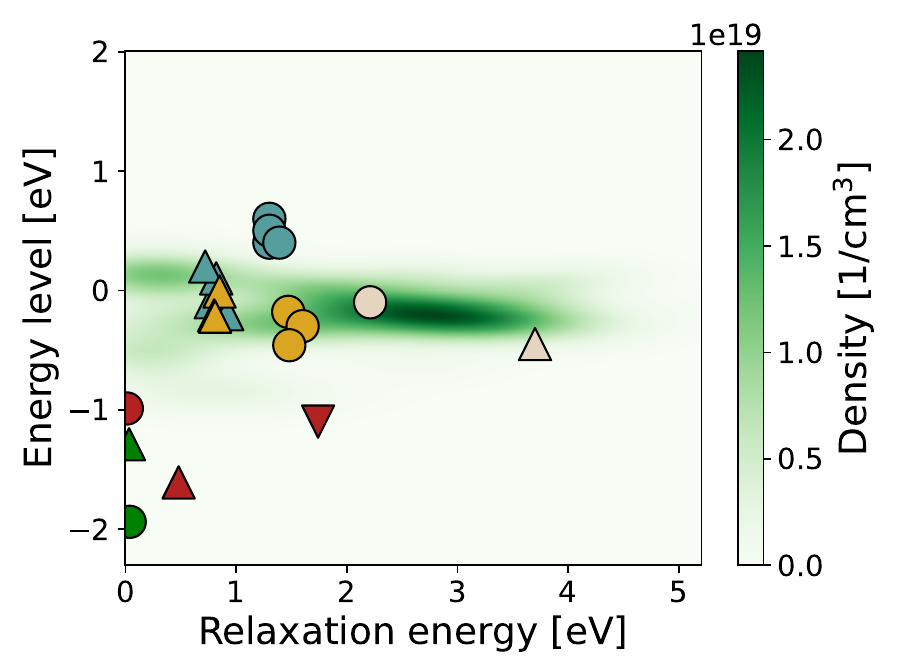}
\caption{}
\label{p_ets}
\end{subfigure}&
\begin{subfigure}[b]{.245\linewidth}
\includegraphics[width=1.0\linewidth,trim={0cm 4cm 0cm 0cm},clip]{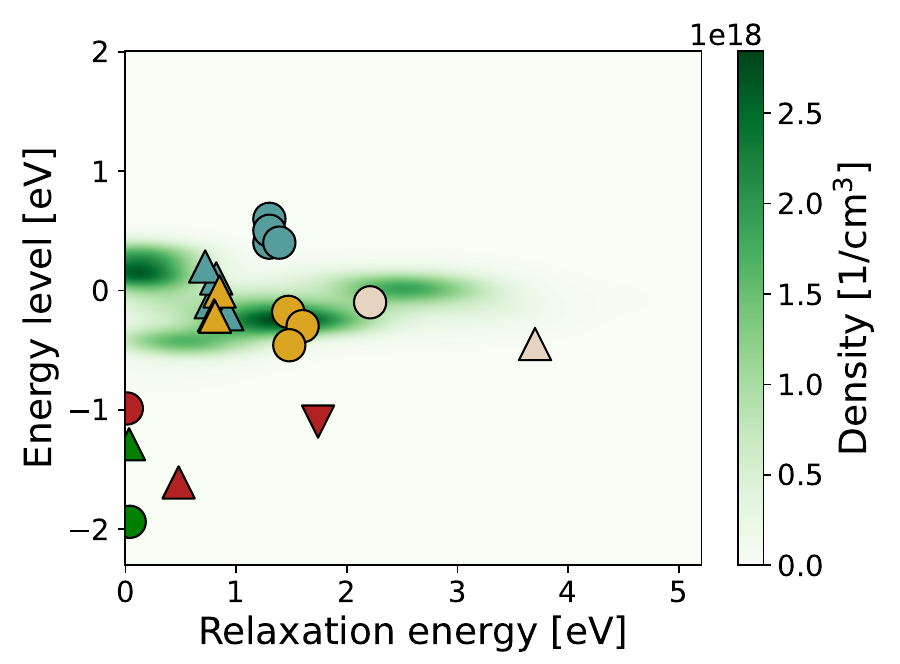}
\caption{}
\label{f_ets}
\end{subfigure}&
\begin{subfigure}[b]{.245\linewidth}
\includegraphics[width=1\linewidth,trim={0cm 4cm 0cm 0cm},clip]{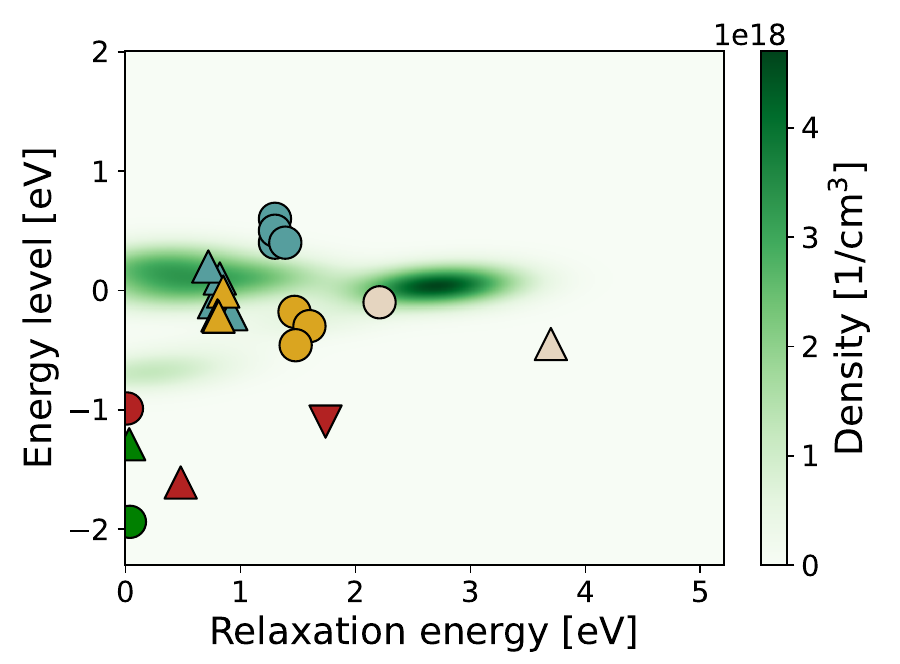}
\caption{}
\label{s_ets}
\end{subfigure}&
{\captionsetup[subfigure]{skip=-9pt}\begin{subfigure}[b]{.245\linewidth}
\includegraphics[width=1\linewidth,trim={0cm 4cm 0cm 0cm},clip]{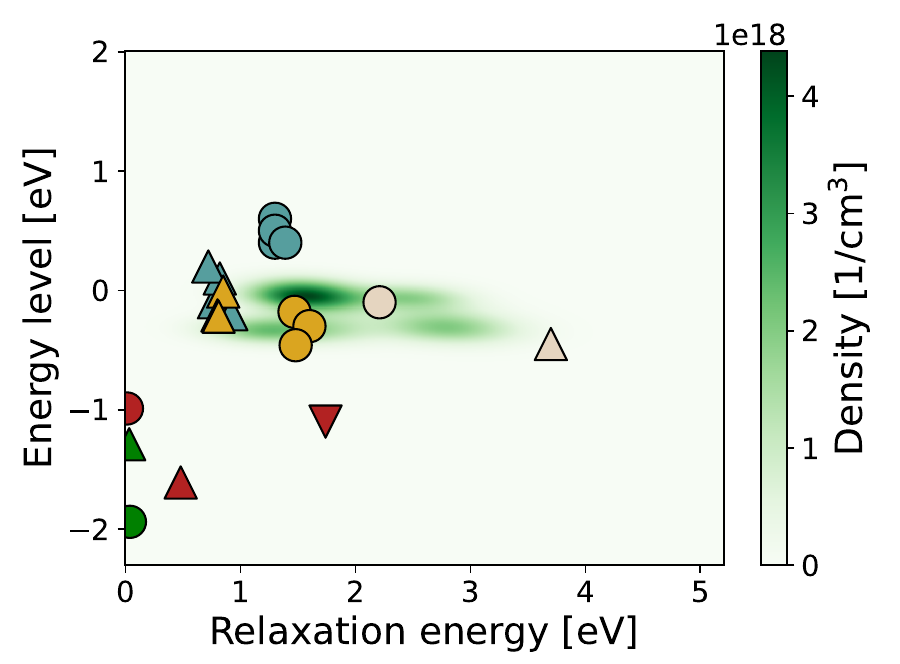}
\caption{}
\label{g_ets}
\end{subfigure}}\\
\begin{subfigure}[b]{.245\linewidth}
\includegraphics[width=1.0\linewidth,trim={1.7cm 0.5cm 0cm 2.2cm},clip]{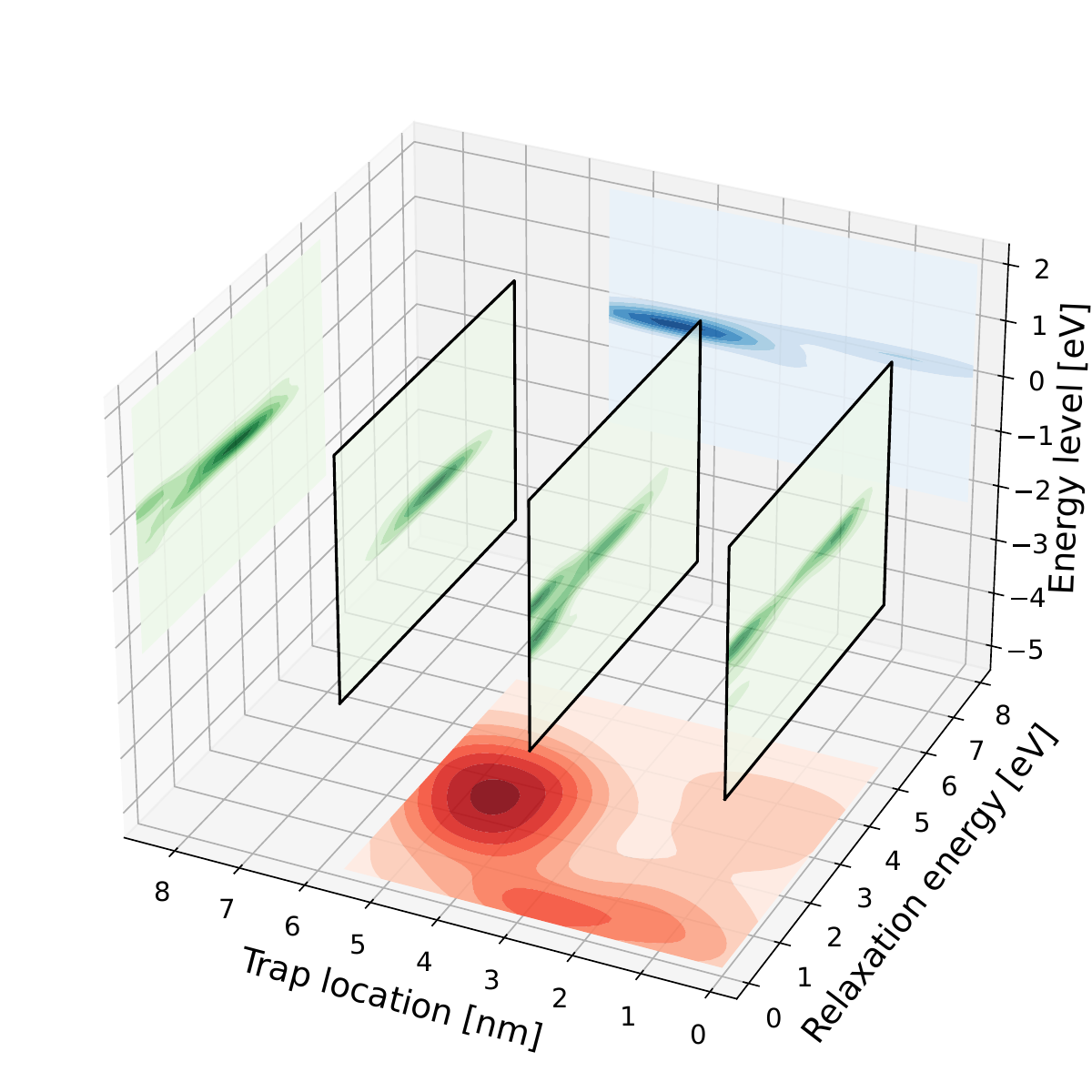} 
\caption{}
\label{p_3d}
\end{subfigure}&
\begin{subfigure}[b]{.245\linewidth}
\includegraphics[width=1.0\linewidth,trim={1.7cm 0.5cm 0cm 2.2cm},clip]{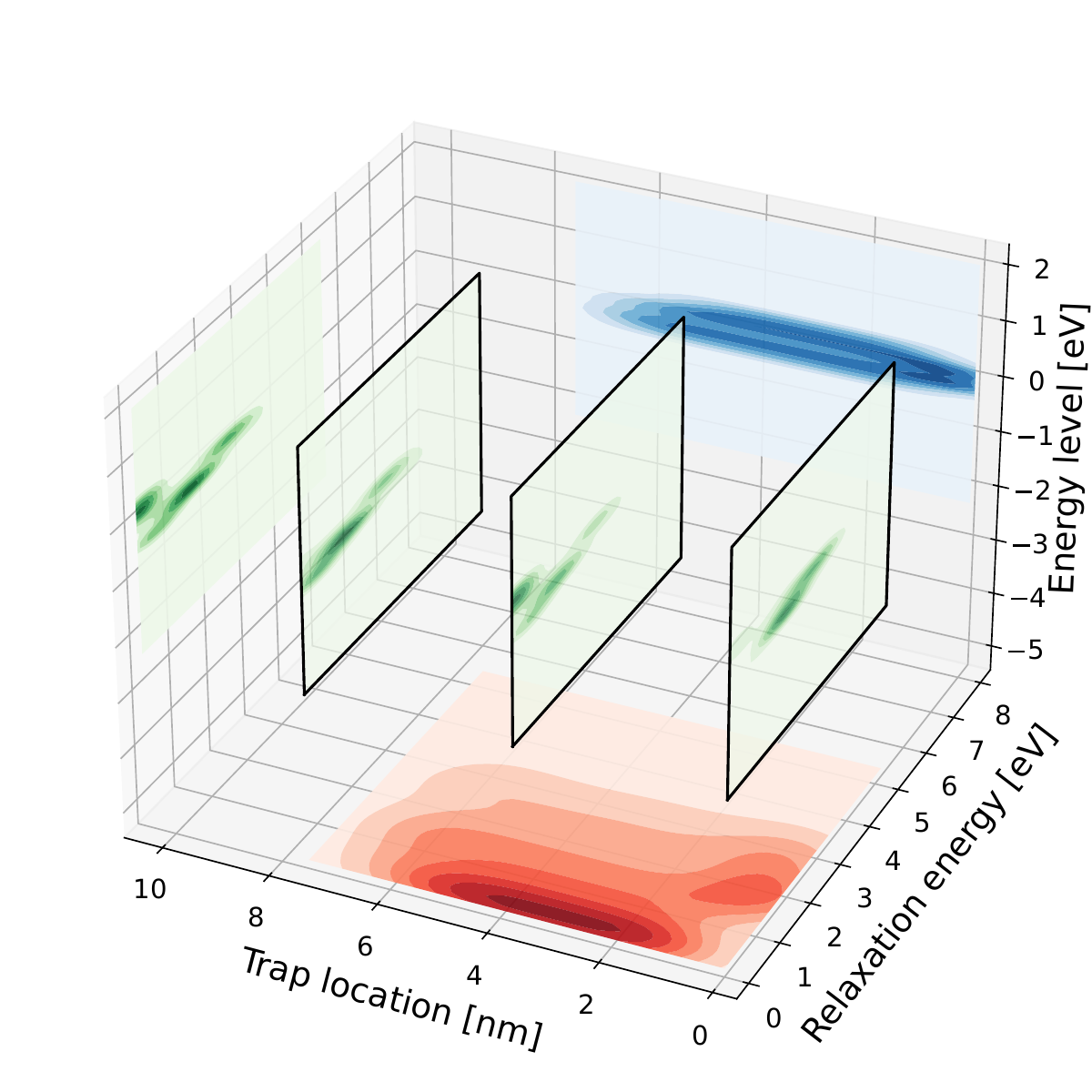} 
\caption{}
\label{f_3d}
\end{subfigure}&
\begin{subfigure}[b]{.245\linewidth}
\includegraphics[width=1.0\linewidth,trim={1.7cm 0.5cm 0cm 2.2cm},clip]{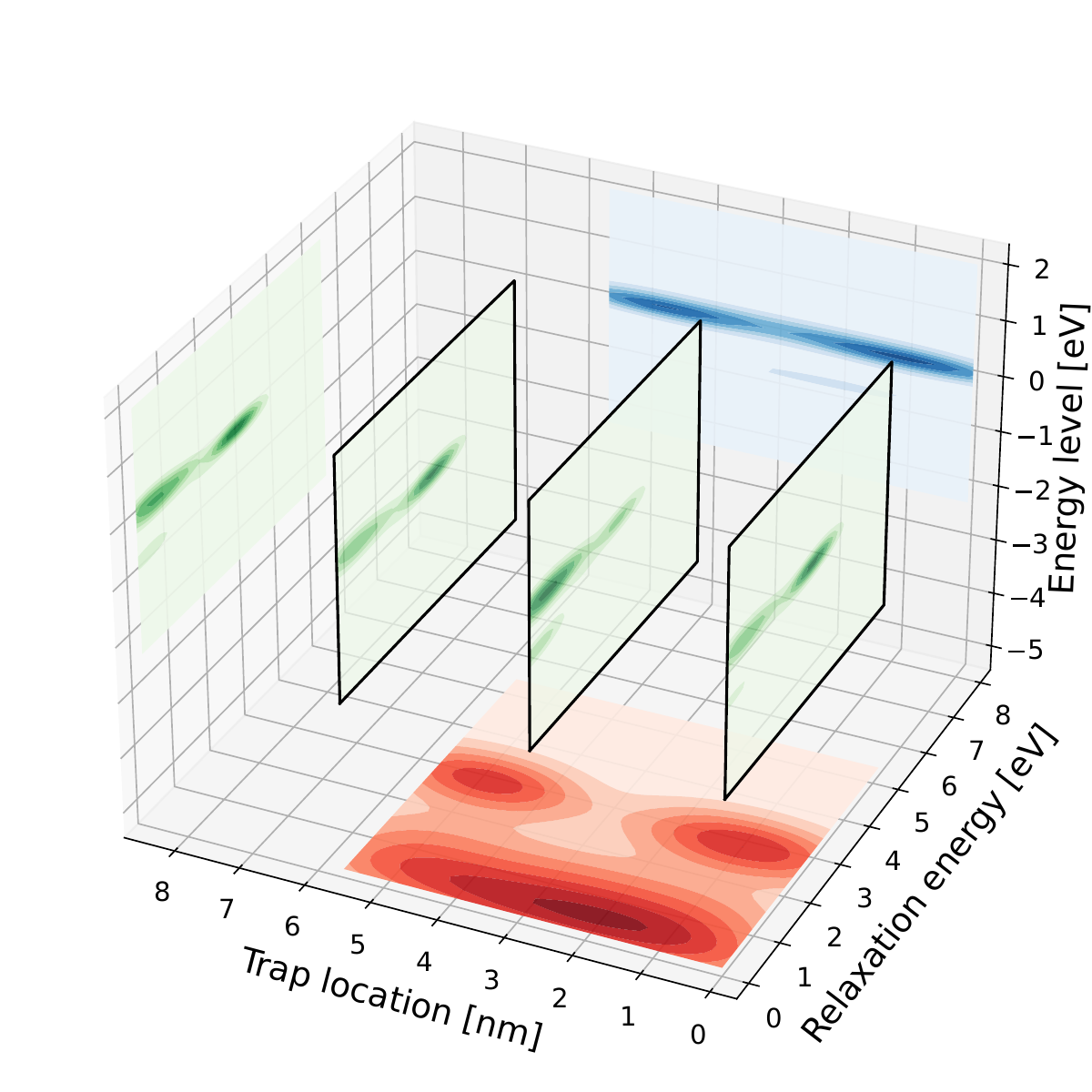} 
\caption{}
\label{s_3d}
\end{subfigure}&
\begin{subfigure}[b]{.245\linewidth}
\includegraphics[width=1.0\linewidth,trim={1.7cm 0.5cm 0cm 2.2cm},clip]{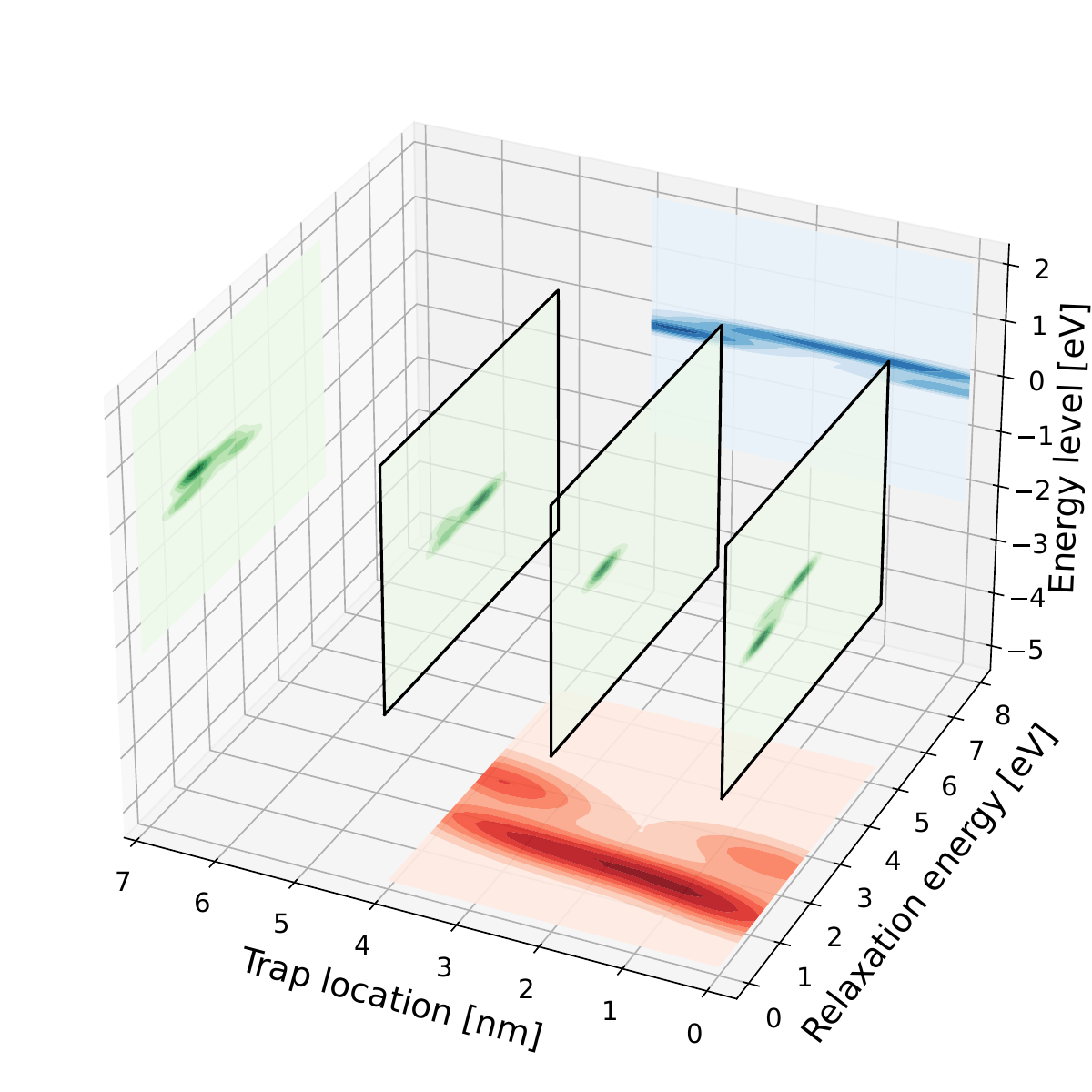} 
\caption{}
\label{g_3d}
\end{subfigure}\\
\end{tabular}
{\captionsetup[subfigure]{skip=-12pt}\begin{subfigure}[b]{1\linewidth}
\includegraphics[width=1.0\linewidth,trim={0cm 0.3cm 0cm 0.8cm},clip]{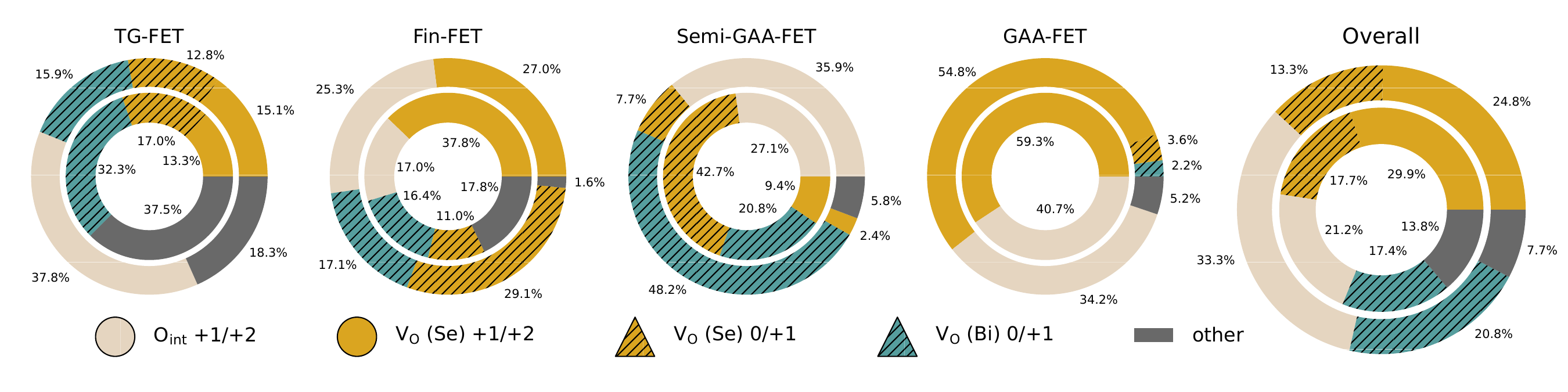}
\caption{}
\label{pie}
\end{subfigure}}
{\captionsetup[subfigure]{skip=2pt}\begin{subfigure}[b]{\linewidth}
\setlength{\tabcolsep}{9pt}  
\renewcommand{\arraystretch}{1.1}
\footnotesize
\begin{tabular}{@{}l|cccccc|}
\hline
Defect & Type & Material & Charge Transition ($Q_1/Q_2$) & Energy level (\et{}) & Relaxation Energy (\er{})\\
\hline
$\mathrm{V_O (Bi)}$ & single atom vacancy & \bisox{} & 0/+1 & \SI{0}{\electronvolt} & \SI{0.81}{\electronvolt}\\
$\mathrm{V_O (Se)}$ & single atom vacancy & \bisox{} & 0/+1 & \SI{-0.15}{\electronvolt} & \SI{0.82}{\electronvolt}\\
$\mathrm{V_O (Se)}$ & single atom vacancy & \bisox{} & +1/+2 & \SI{-0.31}{\electronvolt} & \SI{1.52}{\electronvolt}\\
$\mathrm{O_{int}}$ & interstitial atom & \bisox{} & +1/+2 & \SI{-0.10}{\electronvolt} & \SI{2.21}{\electronvolt}\\
\hline
$\mathrm{V_{O}}$ & single atom vacancy & \bissc{} & +1/+2 & \SI{-0.04}{\electronvolt} &  \SI{0.12}{\electronvolt} \\
\hline
\end{tabular}
\caption{}
\label{tab:defects}
\end{subfigure}}
\caption{Defects in the \bissc{}/\bisox{} material system.
\textbf{(a)} Ball‑and‑stick representations of the oxygen vacancies at the selenium (top) and bismuth (bottom) sites; left to right: the intact structure, after removing one oxygen atom in its neutral, singly, and doubly positively charged states. The blue bubbles represent the highest occupied molecular orbital (HOMO) in the neutral charge state and the lowest occupied molecular orbital (LUMO) in the positive charge state, which corresponds to a localized hole at the defect site.
\textbf{(b)-(e)} Illustration of the energy levels and relaxation energies of different charge transitions of the various theoretically (DFT) predicted structural defects, on top of the experimentally (ESiD) extracted distributions from hysteresis measurements of: \textbf{(b)} the \plan{}, \textbf{(c)} the \fin{}, \textbf{(d)} the \sgaa{}, and \textbf{(e)} the \gaa{}.
\textbf{(f)-(i)} The extracted trap profiles for: \textbf{(f)} the \plan{}, \textbf{(g)} the \fin{}, \textbf{(h)} the \sgaa{}, and \textbf{(i)} the \gaa{}; the blue, red, and green heatmaps on the walls represent the \xt{}/\et{}, \xt{}/\er{}, and \et{}/\er{} correlations, respectively. The characteristic \et{}/\er{} heatmaps at three spatial slices are shown as well.
\textbf{(j)} Contribution of defects to hysteresis; left to right: \plan{}, \fin{}, \sgaa{}, \gaa{}, and the full device set. The outer circles show the weights that best reconstruct the experimentally extracted defect distribution, while the inner circles show the weights obtained by considering only DFT defects.
\textbf{(k)} An overview of the dominant defect charge transitions in the \bissc{}/\bisox{} material system. \et{} and \er{} for the \bisox{} oxygen vacancies ($\mathrm{V_O (Se)}$, $\mathrm{V_O (Bi)}$) correspond to the averaged values of each cluster, with only minor variations as can be seen in \textbf{(b)-(e)}.}
\end{figure}
\setlength{\tabcolsep}{6pt}  
\renewcommand{\arraystretch}{1}
As the final step in the multi-scale modeling chain, we use DFT calculations to determine the characteristic parameters of the most likely point defects in \bisox.
By comparing these DFT parameters to the ESiD extraction, we can identify the most likely defect types responsible for the observed hysteresis. While most material parameters can be derived from the primitive unit cell of the crystal, the study of defect properties requires calculations in larger supercells, followed by geometry relaxation in different charge states.
\fig{dft_vac} shows the atomic structure of representative defects (oxygen vacancies in this case) in the \bisox{} crystal, as well as the structural changes upon electron capture or emission. 
We calculate defect parameters for various defect types in \bisox{} by evaluating charge transitions between different states.
\figs{p_ets}{g_ets} overlay the energetic parameters (\et{} and \er{}) of the theoretically predicted defects on top of the experimentally extracted distribution of the properties of the defects obtained by ESiD. For the spatial distribution of the extracted traps, see \figs{p_3d}{g_3d}, which visualize the trap distribution in three-dimensional parameter space.

By decomposing the experimental distribution into contributions from individual theoretical defects, we determine the relative weight of each defect type, shown in \fig{pie}. These weights represent the dominant contributors rather than the exact defect parameters or generation rates. However, a challenge arises in bridging the discrete nature of DFT-calculated point defects with the continuous energy distributions obtained from the experiment.
To address this, we employ two complementary approaches. In the first approach, we broadened each point defect by assigning a small finite energy broadening parameter to both \et{} and \er{}. This allowed us to create an effective distribution, averaging over different broadening widths ($\delta$E), from which the relative contributions were extracted, represented by the outer circle in \fig{pie}.
In the second approach, we directly used only the DFT-predicted discrete traps in place of the broader empirical parameter space. 
Remarkably, this reduced set of traps reproduces the general hysteresis behavior of the devices (see appendix A, sections \ref{supp:overal_def} and \ref{supp:hys_def}), while the required defect densities are consistent with those derived from the ESiD analysis, as indicated by the inner circles in \fig{pie}.

As the pie charts reveal, the dominant defects are oxygen vacancies ($\mathrm{V_O}$ at both the \ce{Se} and \ce{Bi} sites) and interstitial oxygen atoms ($\mathrm{O_{int}}$). Molecular-dynamics calculations further identify oxygen in the \ce{SeO3} interlayer as the most mobile species in \bisox{}~\cite{khakbazACS}. This interpretation is consistent with the spatially resolved trap distributions of the investigated devices. The unencapsulated \plan{} shows an increased trap density toward the outer, gate-side region of the \bisox{} layer after the highest-temperature measurement sequence. By contrast, the \fin{}, which is the only investigated device coated with an additional (\ce{HfO2}) layer, exhibits a reduced trap population near the outer \bisox{} side compared with the three unencapsulated devices.
Taken together, these results support a central role of oxygen exchange or redistribution in the electrical instability of the present devices and are consistent with a beneficial role of encapsulation. The degradation observed above approximately 90--100~$^\circ$C in the unencapsulated \plan{} should therefore be interpreted as a device-level electrical stability limit of the tested unencapsulated prototype under the applied thermal sequence, rather than an intrinsic thermal limit of the ideal \bissc{}/\bisox{} material system. This electrical degradation must be considered together with the reported transformation of \bisbeta{} toward \bisalpha{} near 380~$^\circ$C~\cite{peng_beta}, which represents a separate structural thermal-budget constraint. The present measurements do not establish whether the electrically observed degradation and the reported phase transition are independent or partially coupled.
Accordingly, the current results do not yet demonstrate compatibility with high-temperature front-end CMOS processing. A more realistic near-term integration route is low temperature mid-end, back-end or monolithic three-dimensional integration, with subsequent processing restricted to a thermally validated window. Low temperature encapsulation or a dense diffusion-barrier layer may improve electrical stability by suppressing oxygen exchange and helping stabilize oxide stoichiometry. A hybrid gate stack that preserves a thin native \bisox{} interfacial layer while adding a more thermally robust high-$\kappa$ cap may also be a beneficial strategy, retaining the chemically matched zipper interface.
An overview of these defects, along with the oxygen vacancy in \bissc{} (which was previously shown in \tab{tab:metrics} to coincide with the extracted interface trap level) is provided in \fig{tab:defects}.

\section{Conclusions}\label{sec:conclusions}
We have thoroughly investigated the performance and reliability potential of transistors built with the layered semiconductor bismuth oxyselenide \bissc{} and its native oxide \bisox{}. 
Our multi-scale study confirms that the \bissc{}$/$\bisox{} material system offers several key advantages such as high carrier mobility, a high‑$\kappa$ native oxide with an atomically clean interface, steep sub‑threshold slopes, and low contact resistance. All four investigated generations exhibit effective ohmic contacts (negligible or zero Schottky barriers) and overall promising performance, though irreversible device degradation may set in above \SI{90}{}$-$\SI{100}{\degree C} if the transistor is not properly encapsulated.

Moreover, by integrating systematic electrical characterization with density-functional theory and TCAD simulations, we linked macroscopic instabilities to atomistic origins.
Specifically, oxygen vacancies and interstitials in the oxide drive hysteresis and recoverable threshold-voltage shifts. These findings identify a critical reliability challenge but also highlight viable mitigation strategies such as \ce{HfO2} encapsulation or oxygen-rich annealing. More broadly, our workflow demonstrates how defect-level modeling can directly inform interface engineering for 2D material systems.
The IRDS-oriented projection indicates the performance potential of optimized \bissc{}/\bisox{} devices, supporting further development of this system for ultra-scaled nanoelectronics.

\section{Methods}
\subsection{TEM Measurements}
Cross-sectional STEM characterization was conducted to investigate the 2D \bissc{}$/$\bisox{} heterostructure along with the metal-gate/dielectric/semiconductor interfaces in all presented device geometries. 
TEM samples were prepared using standard FIB processes with a FEI Scios 2 DualBeam SEM/FIB system. High-resolution STEM imaging was carried out using an aberration-corrected (Cs corrected) FEI Titan Themis G2 60–300 (Cubed) operated at 300 kV.

\subsection{Electrical Characterization}
Electrical characterization of our FETs consisted of measurements of \idvg{} and \idvd{} characteristics in auto-range mode, as well as hysteresis measurements with controlled sampling rates recorded in fixed current-range mode. Hysteresis was analyzed by measuring the double-sweep \idvg{} characteristics using different sweep times ($t_{\mathrm{SW}}$) and sweep ranges. We used a Keithley 2636B, which incorporates two Source-Measure Units (SMUs) and contacted the devices using tungsten tips. During all measurements the devices were placed inside the chamber of a Lakeshore vacuum probe station ($\leq2 \times 10^{-6}$ torr) in complete darkness.

\subsection{DFT Material Simulation}
We perform DFT calculations using a set of computational methods to achieve both precision and efficiency at various length scales. 
For calculations of structural and electronic properties, we use the QuantumATK package~\cite{smidstrup2019quantumatk}, using a localized basis set based on the linear combination of atomic orbitals (LCAO) and PSEUDODOJO pseudopotentials~\cite{van2018pseudodojo}. 
We used a \textit{k}-point sampling density of approximately 8\,\AA\ and applied a density mesh cutoff of 125\,Ha. We employed a non-local hybrid functional for all calculations to minimize errors in the electronic state calculations. Van der Waals (vdW) interactions were included using Grimme's DFT-D3 dispersion corrections~\cite{grimme2011effect}. 
The structural relaxations continued until the interatomic forces were below 0.01\,eV/\AA\ and the pressure of the pristine unit cell pressure was reduced to less than 0.1\,GPa. For calculations involving large supercells, such as those required for defect studies, we used the CP2K software package~\cite{kuhne2020cp2k} due to its superior scalability and performance on large-scale systems. 
 To calculate defect parameters (i.e. \et{} and \er{}) for various defects in the \bisox{} crystal structure through different transitions of charge states, we employed DFT with periodic boundary conditions using the Gaussian Plane Wave (GPW) method as implemented in the CP2K code~\cite{kuhne2020cp2k}.
A double-$\zeta$ Gaussian basis set~\cite{vandevondele2007gaussian} was used in combination with Goedecker-Teter-Hutter (GTH) pseudopotentials~\cite{goedecker1996separable}, and an energy cutoff of 800\,Ry was applied. We employed the non-local PBE0\_TC\_LRC hybrid functional~\cite{guidon_pbe0} for all defect calculations. In addition, the pFIT3~\cite{guidon2010auxiliary} auxiliary basis set was used to accelerate the computation of the Hartree-Fock exchange required for hybrid functionals. Geometry optimizations in various charge states were performed using the Broyden-Fletcher-Goldfarb-Shanno (BFGS) algorithm.

\subsection{TCAD Device Simulation}
To simulate the charge carrier transport, we employed the efficient drift-diffusion model~\cite{dd,LundstromSispad}. This method is well-suited for our devices as it accurately captures the behavior of charge carriers in micrometer-scaled prototype devices~\cite{10319609}, where the transport is diffusive and dominated by scattering at ambient temperatures. All TCAD simulations were performed using Minimos-NT~\cite{MINIMOS-NT,manualGTS} using a constant mobility model and Schottky contact model was used to model the source and drain contacts.

\subsection{Effective Single Defect Decomposition}
To ensure a robust and meaningful estimation of trap densities, a Tikhonov regularization~\cite{tikhonov} term $\gamma^2 \sum_i N(p_i)^2$  is added to the optimization problem~\cite{WaldhoerTED2021}. This regularization enforces a smooth distribution function with low total defect density and prevents overfitting. The regularization parameter $\gamma$ plays a crucial role in determining the strength of the regularization. For our dataset, we identified optimal $\gamma$ values  of [$5\times 10^{-9} - 1\times 10^{-5}$]. By finding a good balance between accuracy and regularization (the correlation between the regularization parameter $\gamma$ and the error norm to the experimental data as well as the corresponding required total trap density), we obtained reliable and informative trap distributions. 

\newpage
\section*{Declaration}

\bmhead{Acknowledgments}
Not applicable.

\bmhead{Funding}
We would like to express our gratitude to Huawei Technologies R\&D Belgium for their generous financial support, which was instrumental in advancing this research. Furthermore, this work was supported by the European Research Council (ERC) under grant agreement no. 101055379 (F2GO) and by the Austrian Science Fund (FWF) under project no. PAT 1601525 (BOLD).

\bmhead{Data availability}
The data that support the findings of this study are available from the corresponding author upon reasonable request.

\bmhead{Author Contributions Statement}
M.R.D. conceived and coordinated the study, performed the experimental data analysis, carried out device modelling and modelling of charge trapping and reliability, and wrote the original draft of the manuscript. M.B. performed the DFT calculations, including defect studies, contributed to data analysis and visualization, and co-wrote the manuscript. A.V. performed electrical characterization and measurements, contributed to the methodology and experimental strategy, contributed to device fabrication, and contributed to manuscript revision. P.K. contributed to the methodology, performed DFT calculations including materials studies, and contributed to manuscript writing and revision. D.W. and T.K. contributed to the methodology, provided guidance, supported model development, and contributed to manuscript revision. M.P. contributed to formal analysis and visualization, supported the analysis and interpretation of the results, and contributed to manuscript revision. A.K. contributed to model development and methodology, and supported the analysis. C.W. provided guidance, supported the analysis and interpretation of the results, and contributed to manuscript revision. Y.I. and S.M.S.-E. performed electrical characterization and measurements. Y.Zhang, J.T., Y.L., C.T. and H.P. contributed to device fabrication. X.G. performed the transmission electron microscopy measurements. A.N., Y.Zheng and C.L. provided input on experimental design and simulation consistency. M.W. supervised the technical experimental work. T.G. devised and supervised the research and co-wrote the manuscript. All authors discussed the results and contributed to the final version of the manuscript.

\bmhead{Competing Interests Statement}
The authors declare no competing interests.

\clearpage
\newpage
\begin{appendices}
\section{Supplementary Information}
\subsection{Measured Hysteresis at Various Readouts}\label{supp:readouts}
\begin{figure}[!h]
\begin{subfigure}[b]{.245\linewidth}
\includegraphics[width=1.0\linewidth]{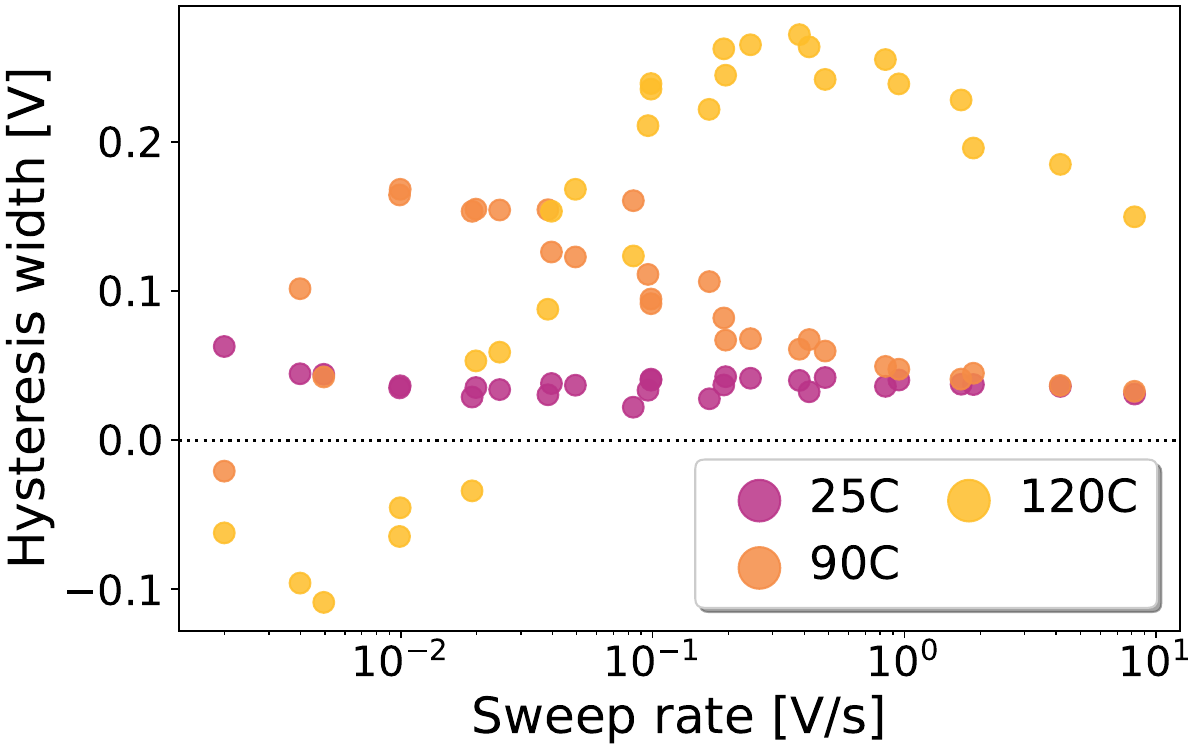}
\end{subfigure}
\begin{subfigure}[b]{.245\linewidth}
\includegraphics[width=1.0\linewidth]{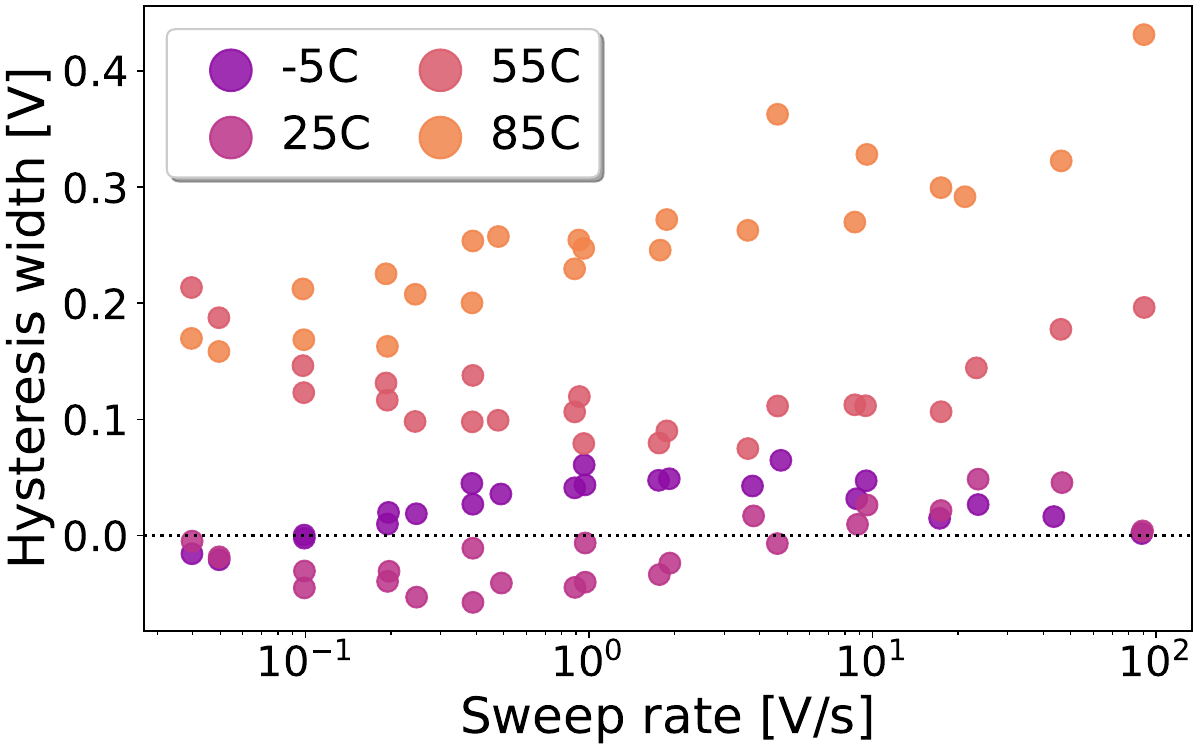}
\end{subfigure}
\begin{subfigure}[b]{.245\linewidth}
\includegraphics[width=1\linewidth]{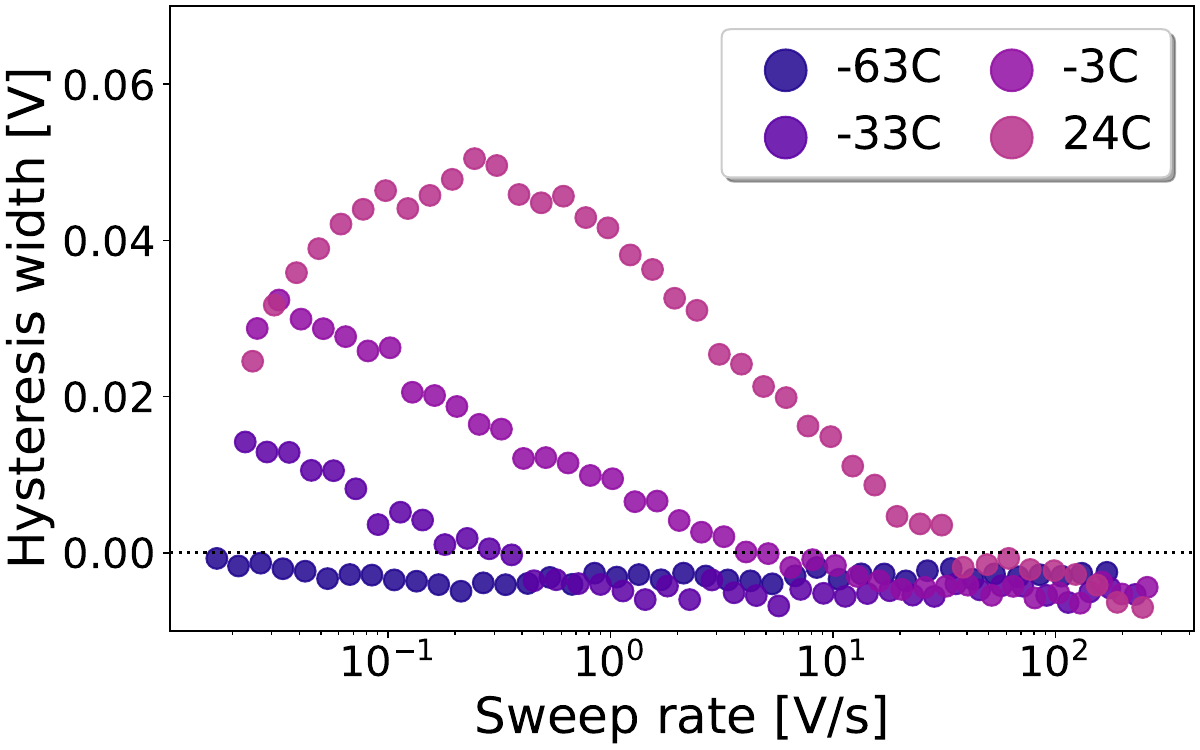}
\end{subfigure}
\begin{subfigure}[b]{.245\linewidth}
\includegraphics[width=1\linewidth]{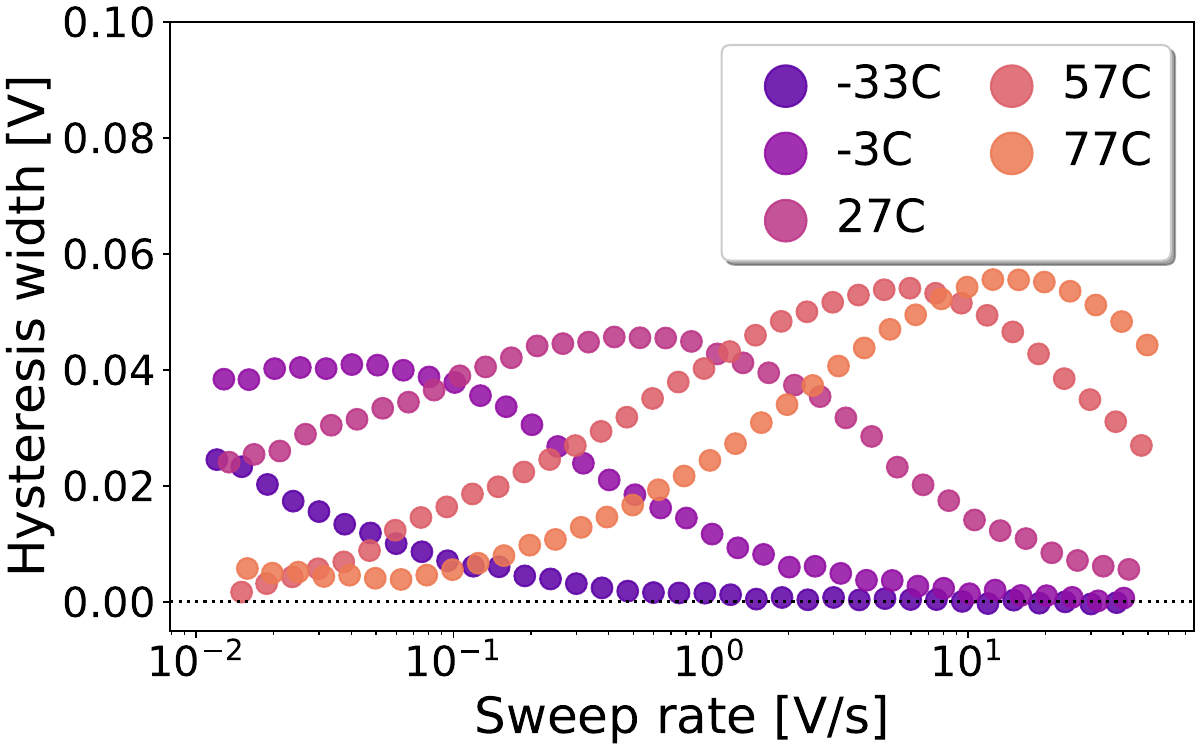}

\end{subfigure}
\begin{subfigure}[b]{.245\linewidth}
\includegraphics[width=1.0\linewidth]{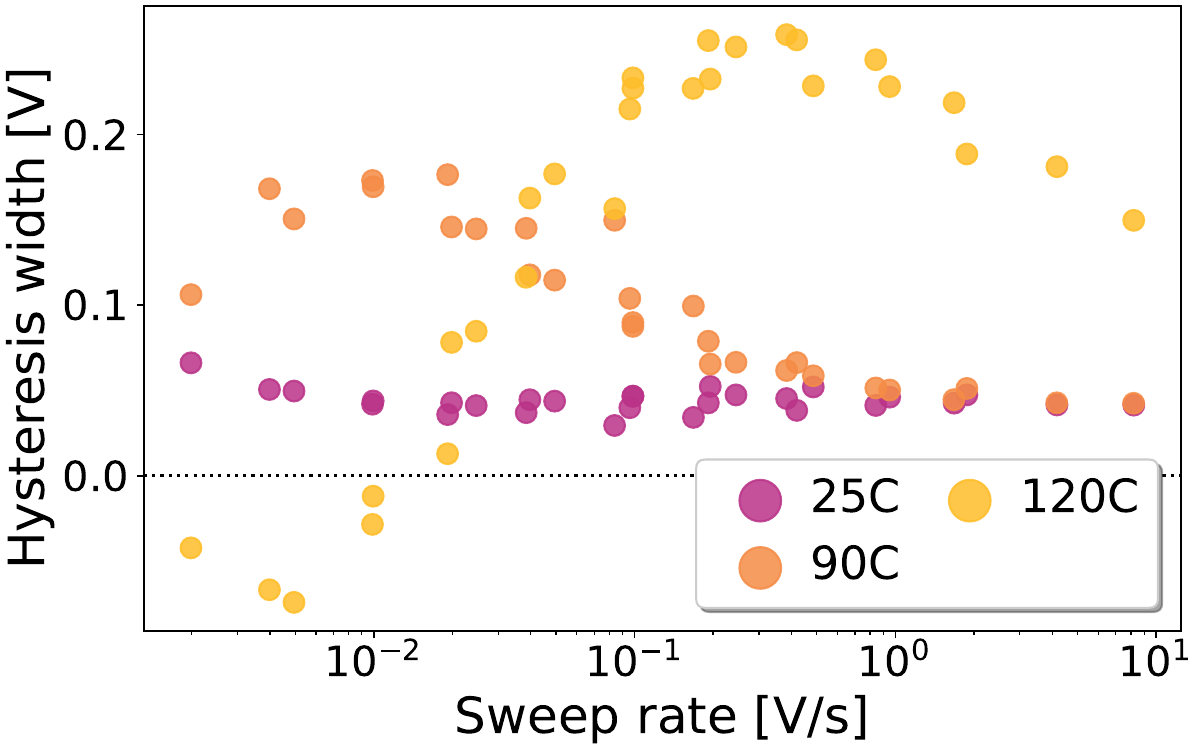}
\end{subfigure}
\begin{subfigure}[b]{.245\linewidth}
\includegraphics[width=1.0\linewidth]{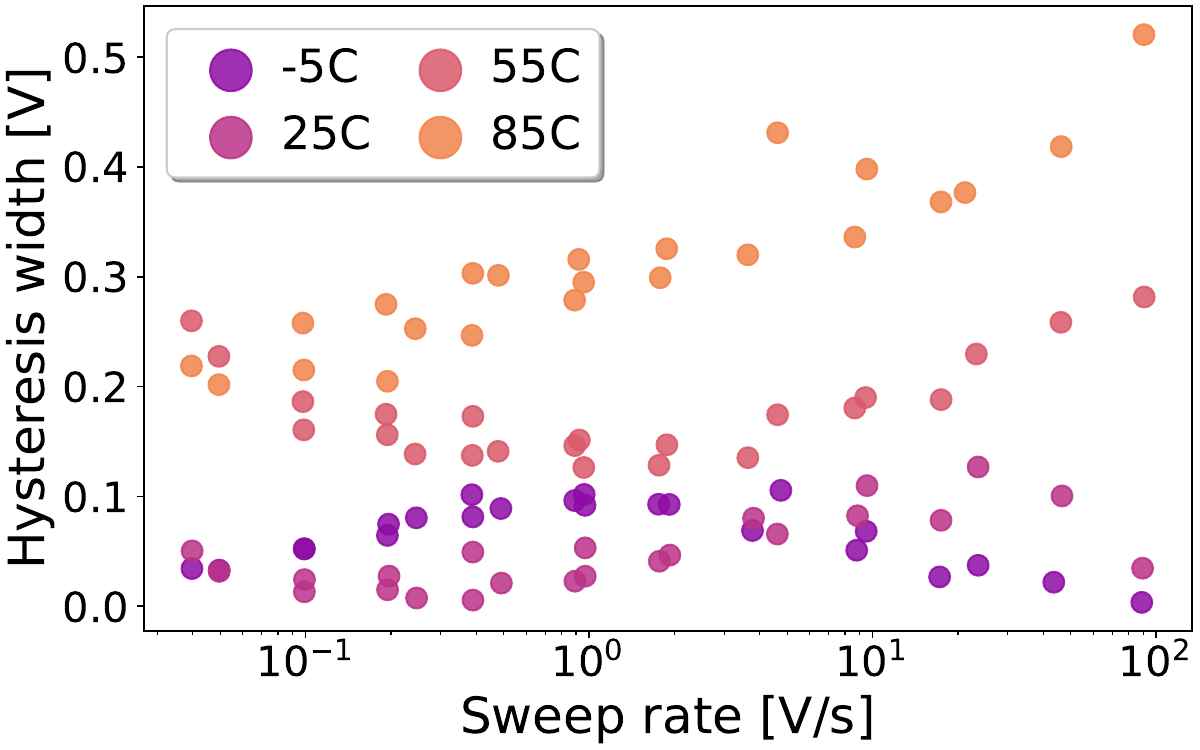}
\end{subfigure}
\begin{subfigure}[b]{.245\linewidth}
\includegraphics[width=1\linewidth]{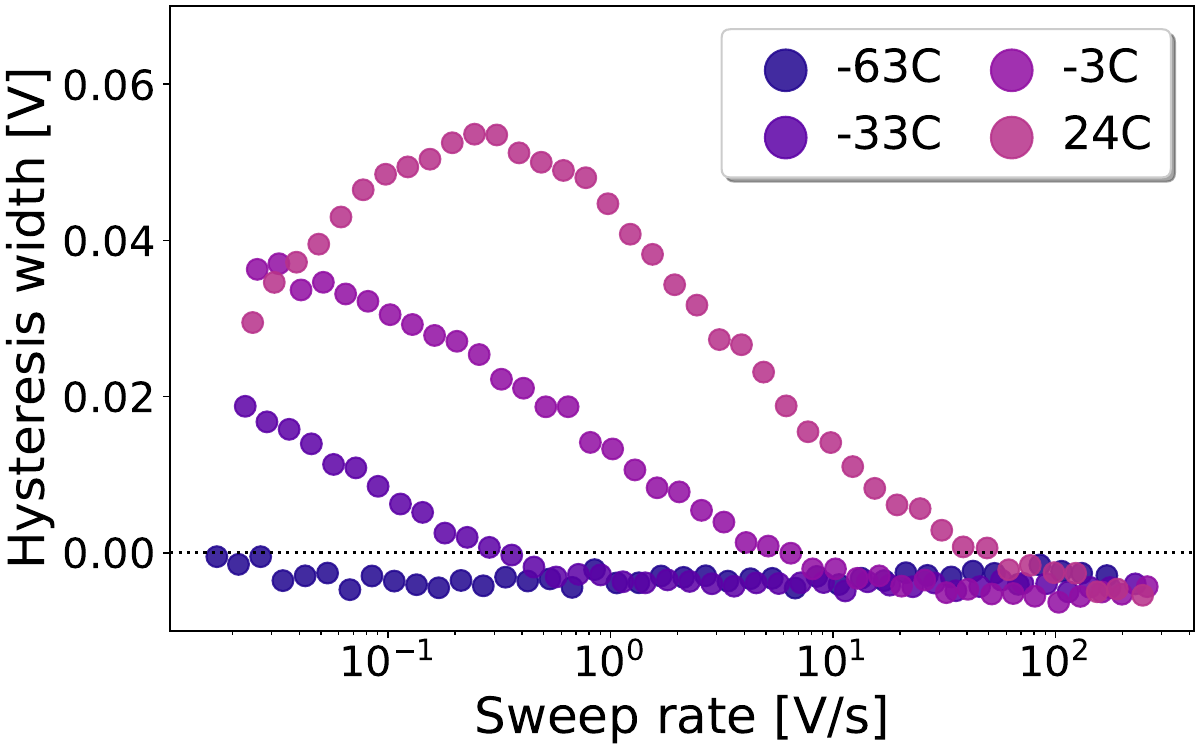}
\end{subfigure}
\begin{subfigure}[b]{.245\linewidth}
\includegraphics[width=1\linewidth]{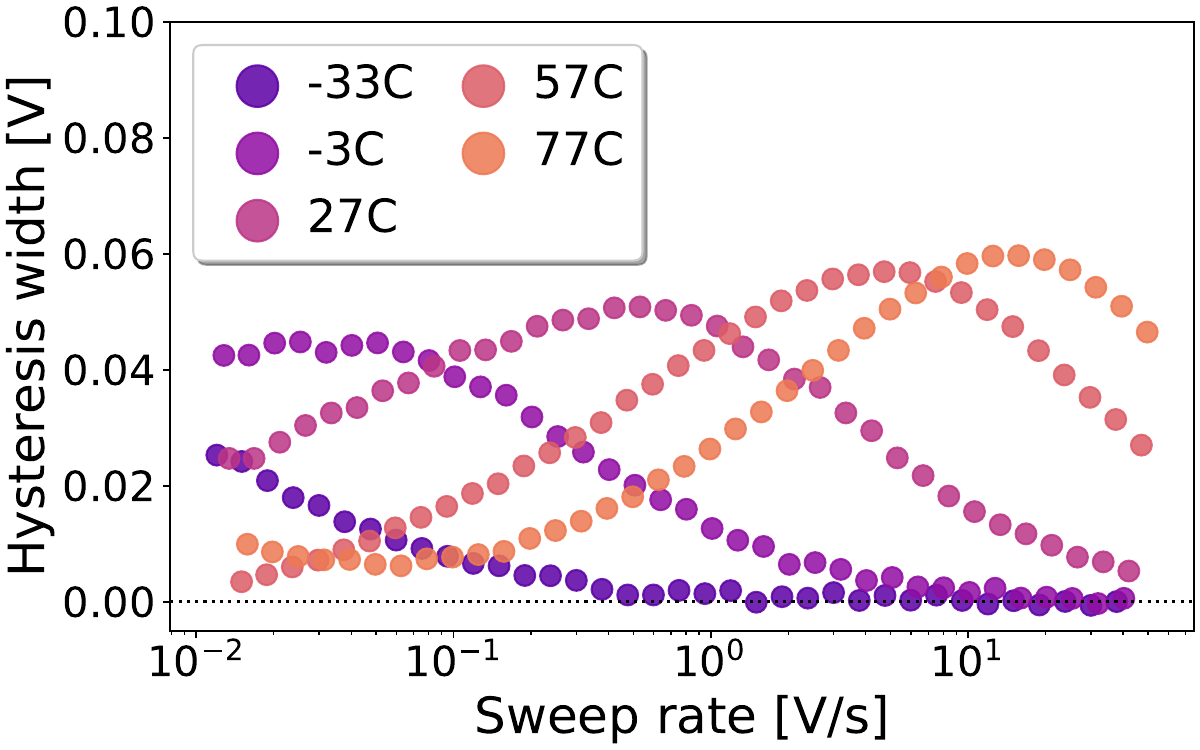}
\end{subfigure}

\begin{subfigure}[b]{.245\linewidth}
\includegraphics[width=1.0\linewidth]{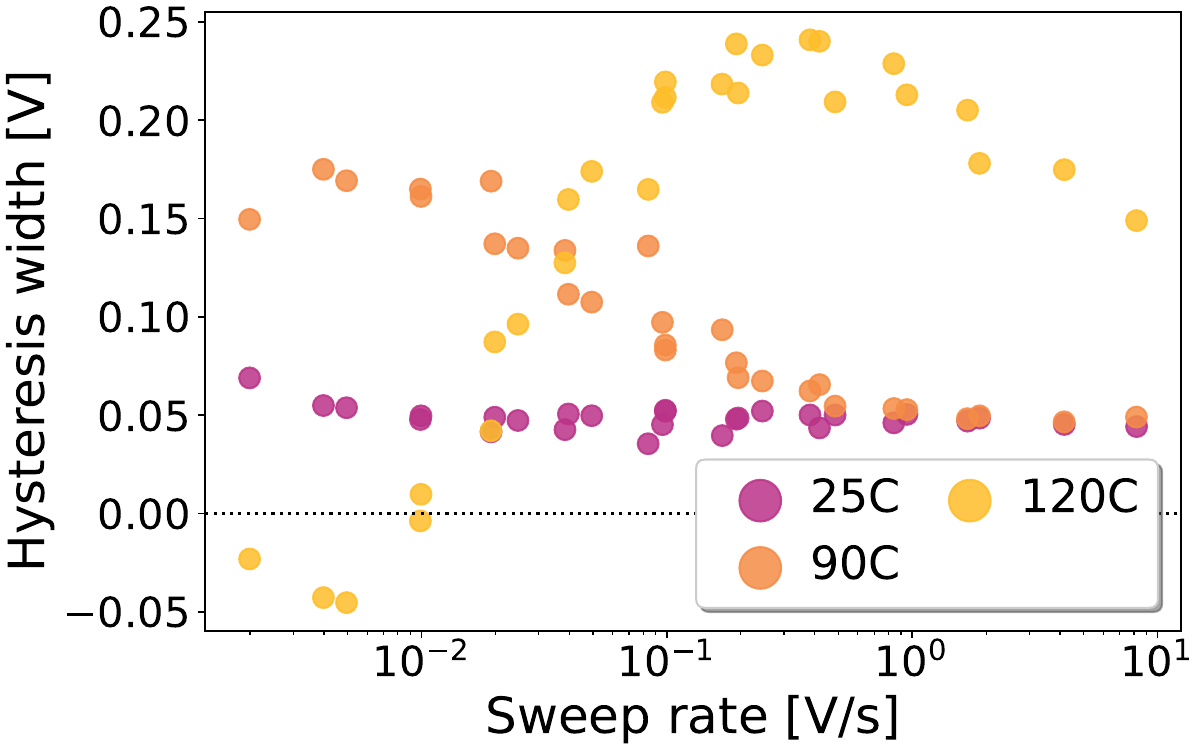}
\end{subfigure}
\begin{subfigure}[b]{.245\linewidth}
\includegraphics[width=1.0\linewidth]{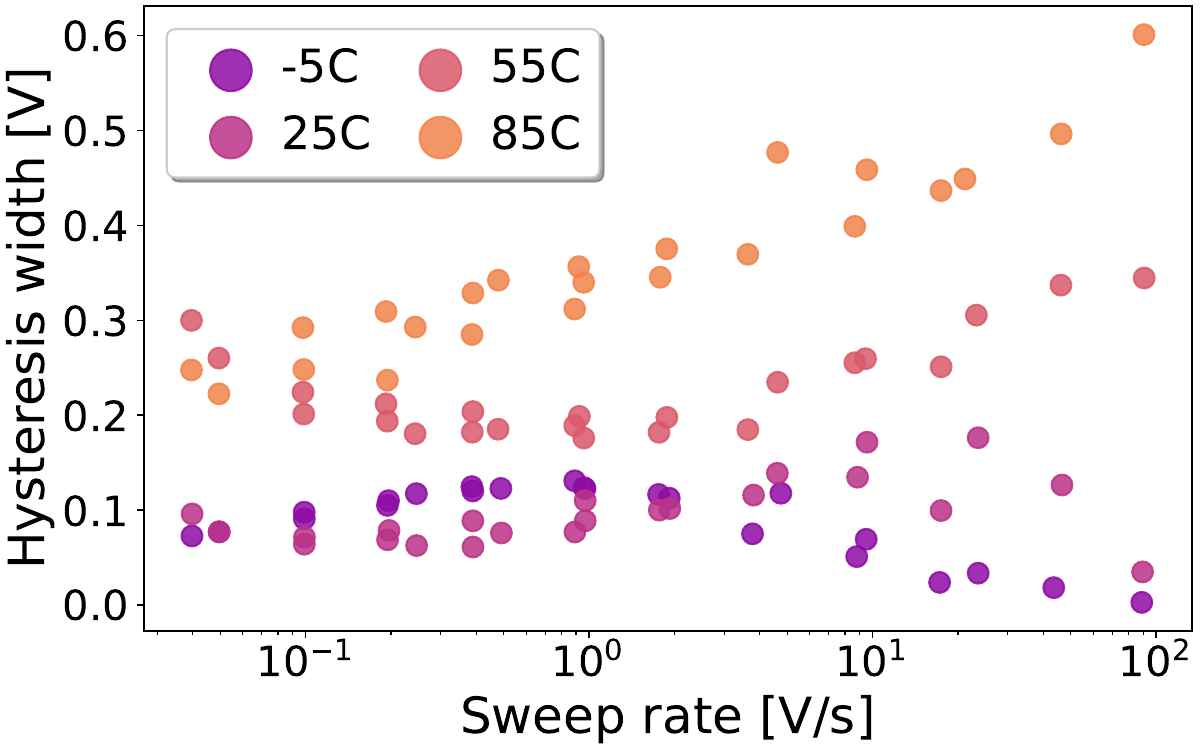}
\end{subfigure}
\begin{subfigure}[b]{.245\linewidth}
\includegraphics[width=1\linewidth]{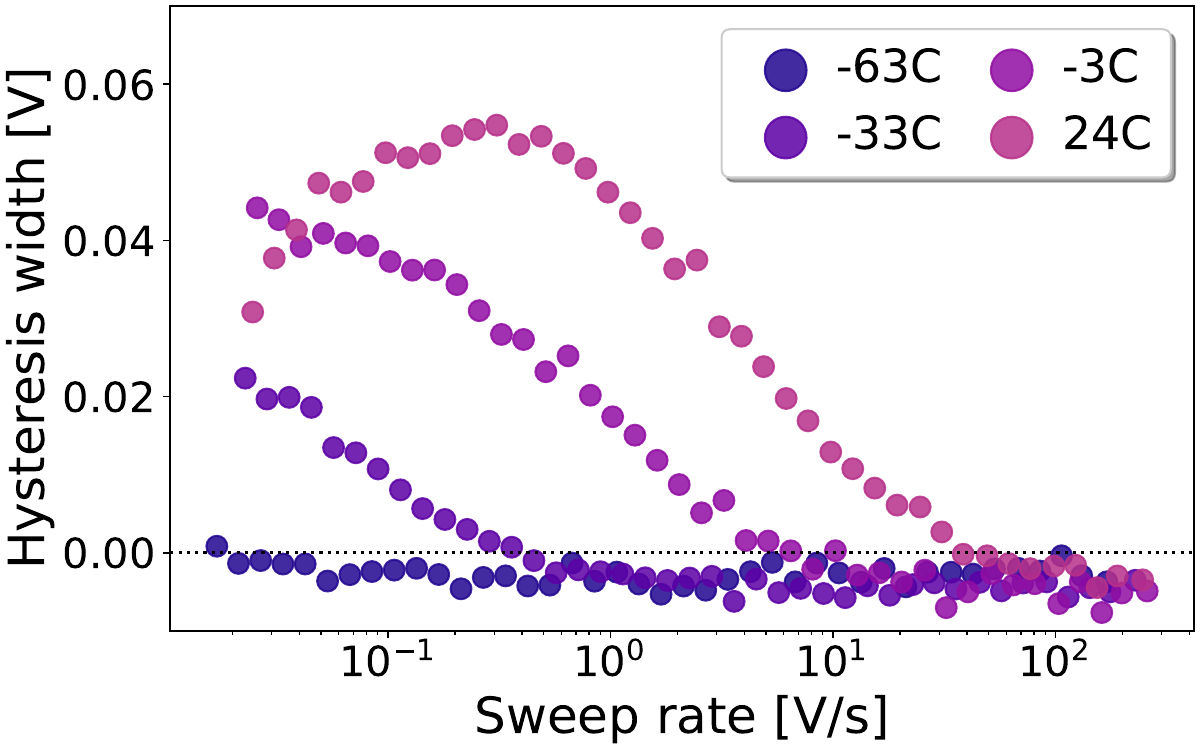}
\end{subfigure}
\begin{subfigure}[b]{.245\linewidth}
\includegraphics[width=1\linewidth]{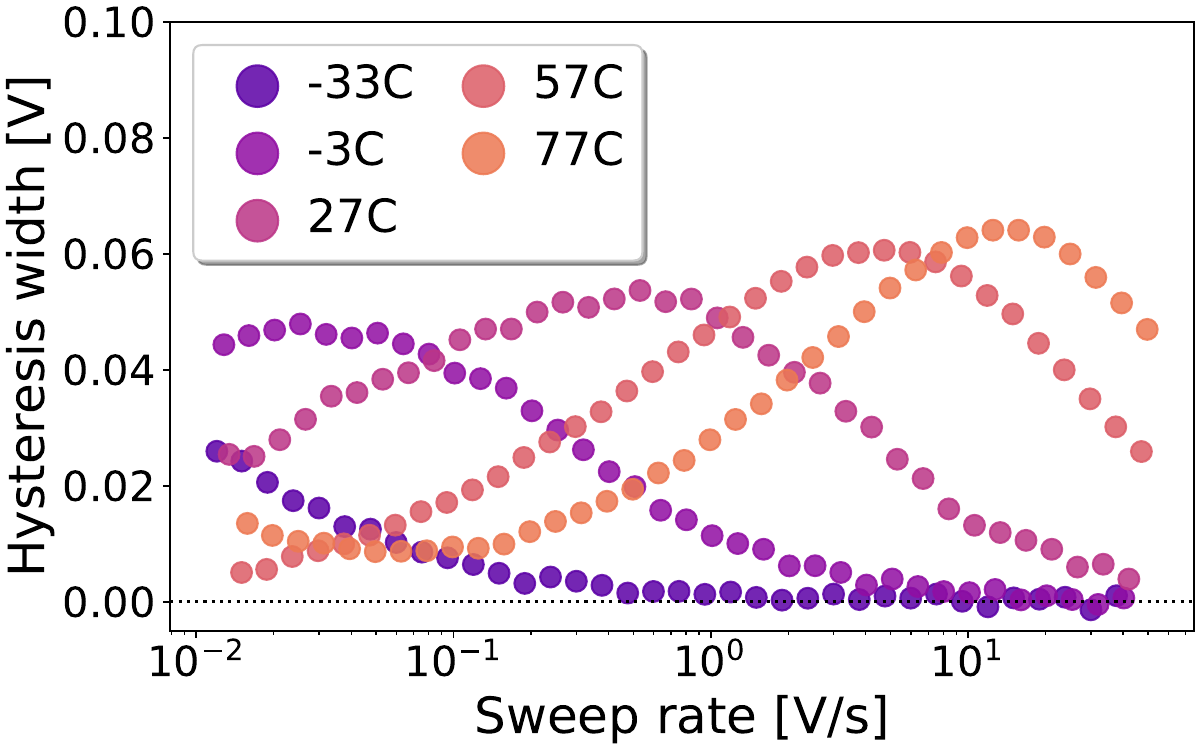}
\end{subfigure}

\begin{subfigure}[b]{.245\linewidth}
\includegraphics[width=1.0\linewidth]{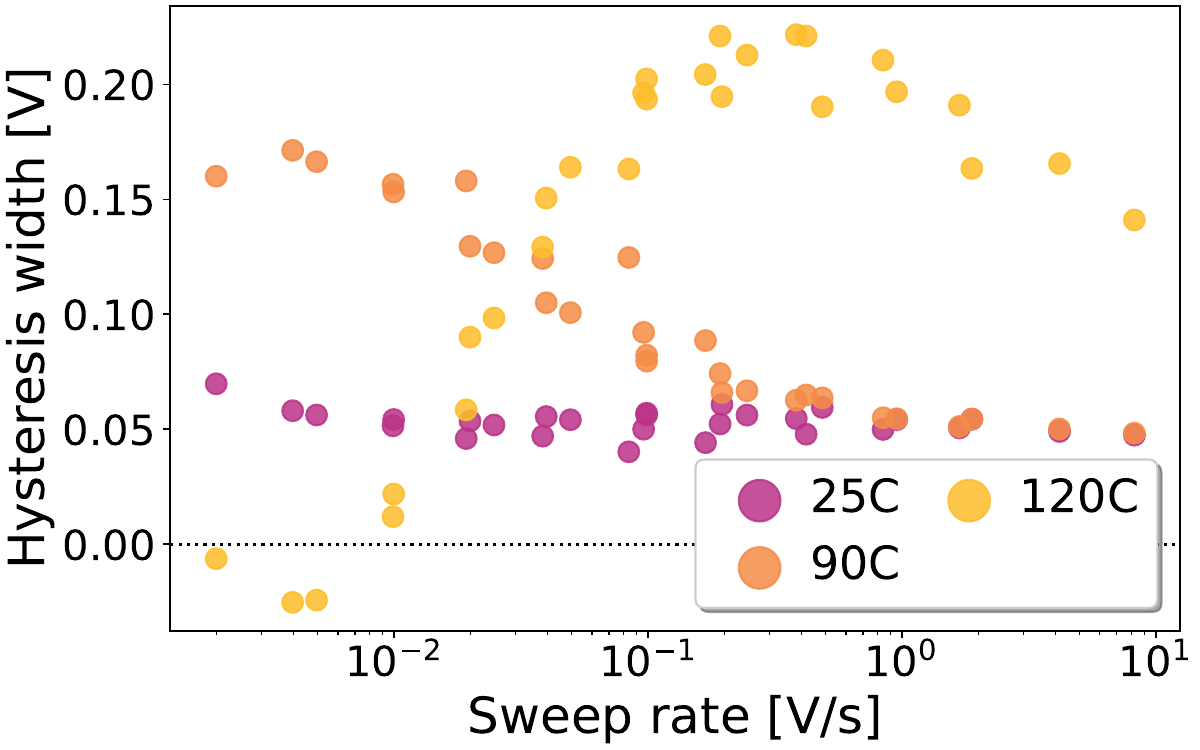}
\end{subfigure}
\begin{subfigure}[b]{.245\linewidth}
\includegraphics[width=1.0\linewidth]{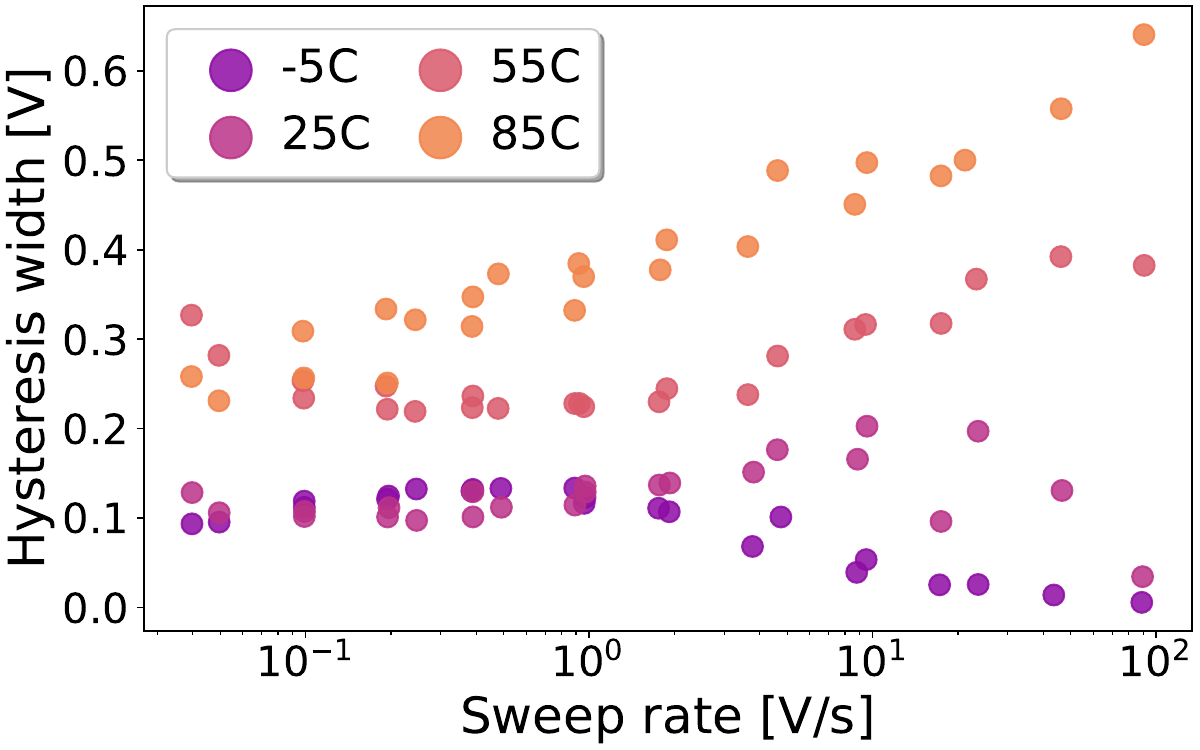}
\end{subfigure}
\begin{subfigure}[b]{.245\linewidth}
\includegraphics[width=1\linewidth]{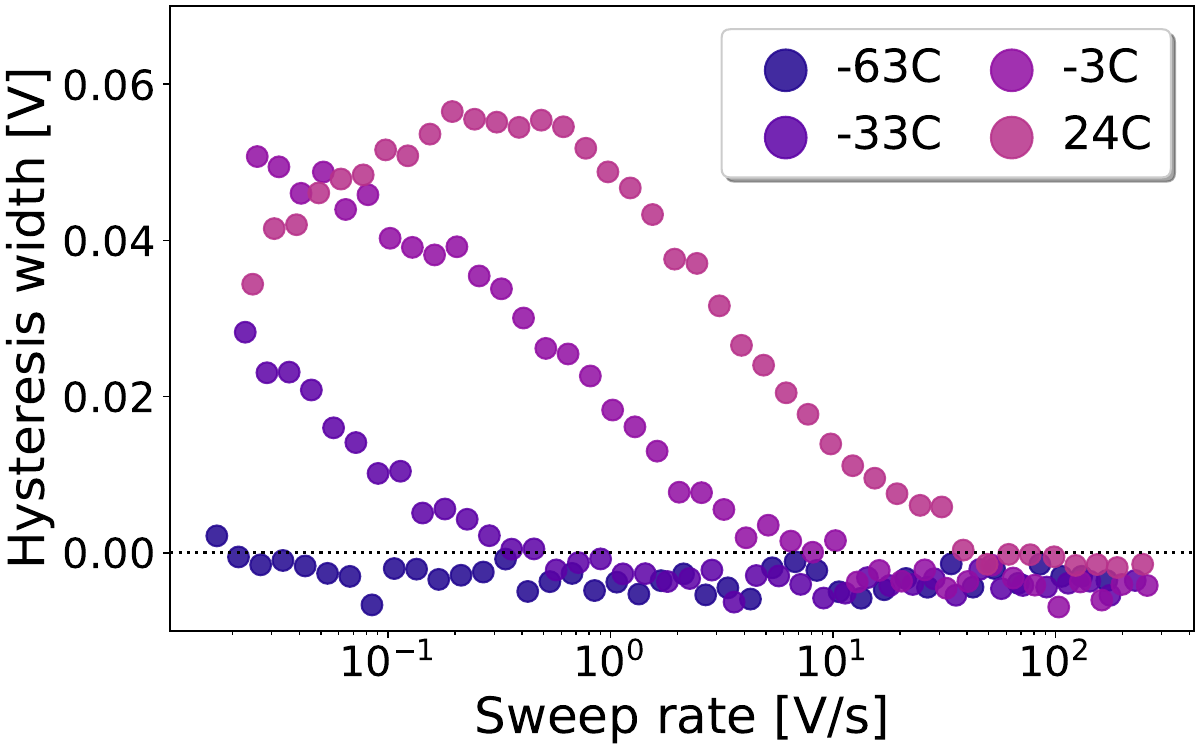}
\end{subfigure}
\begin{subfigure}[b]{.245\linewidth}
\includegraphics[width=1\linewidth]{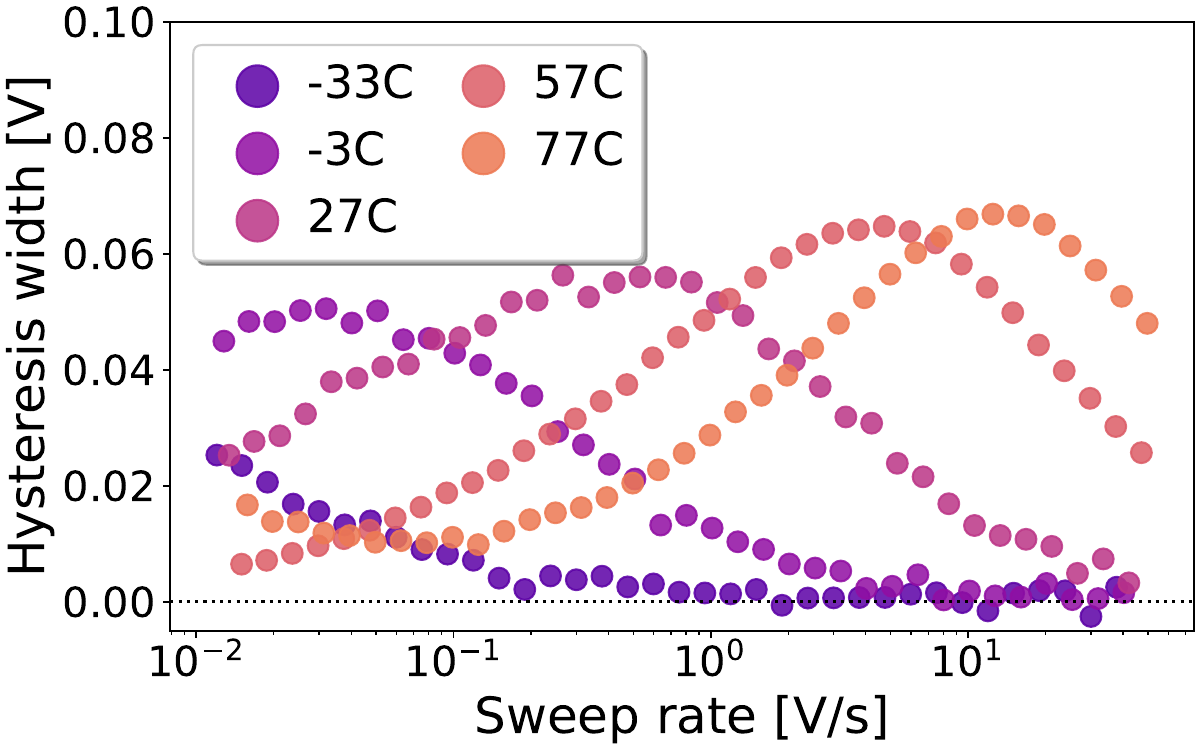}
\end{subfigure}

\begin{subfigure}[b]{.245\linewidth}
\includegraphics[width=1.0\linewidth]{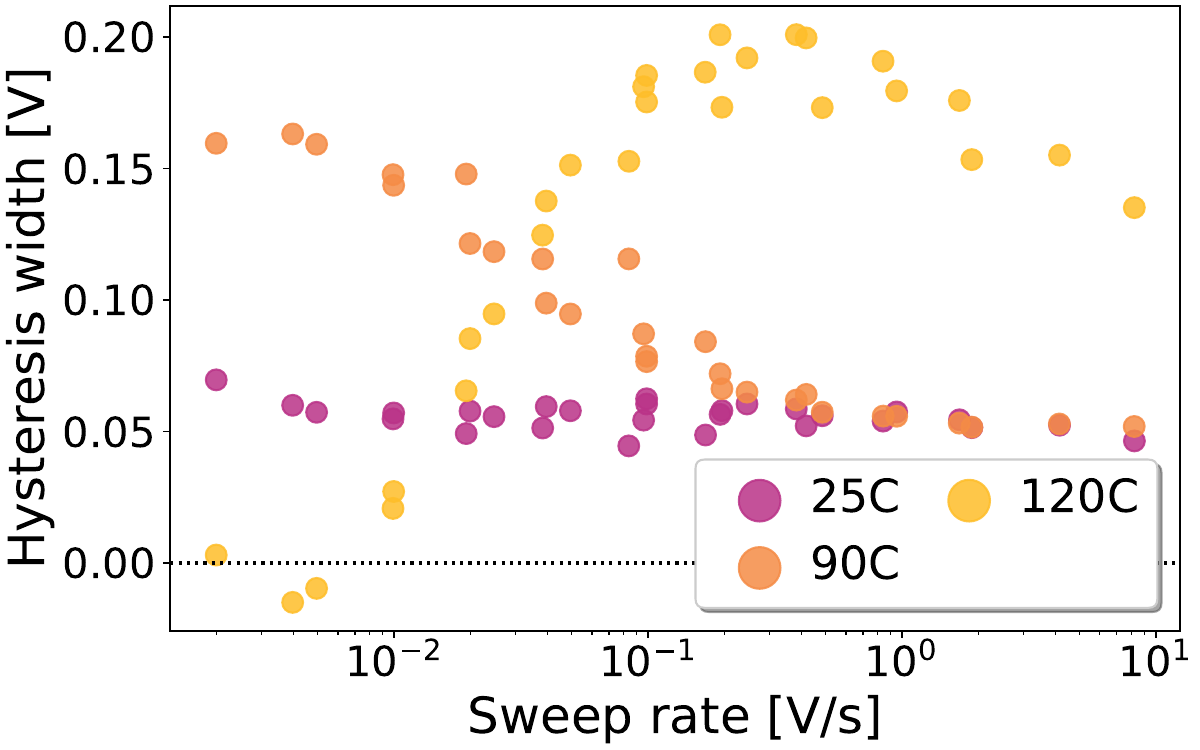}
\end{subfigure}
\begin{subfigure}[b]{.245\linewidth}
\includegraphics[width=1.0\linewidth]{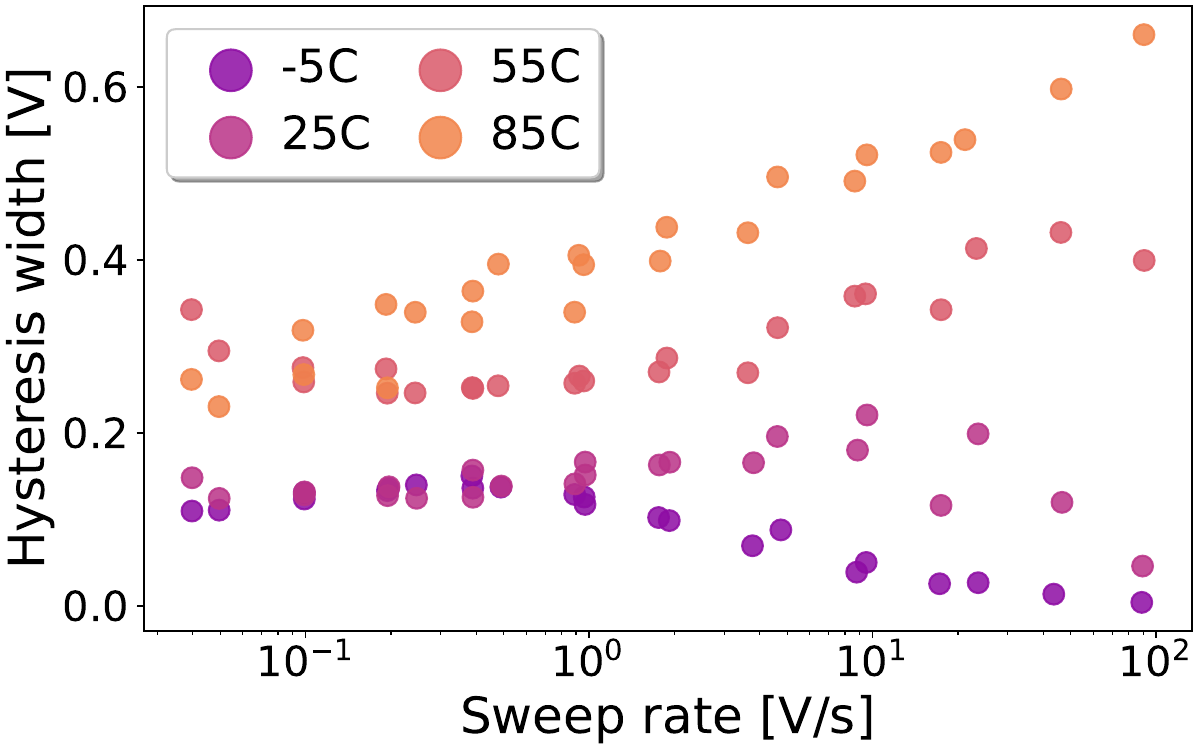}
\end{subfigure}
\begin{subfigure}[b]{.245\linewidth}
\includegraphics[width=1\linewidth]{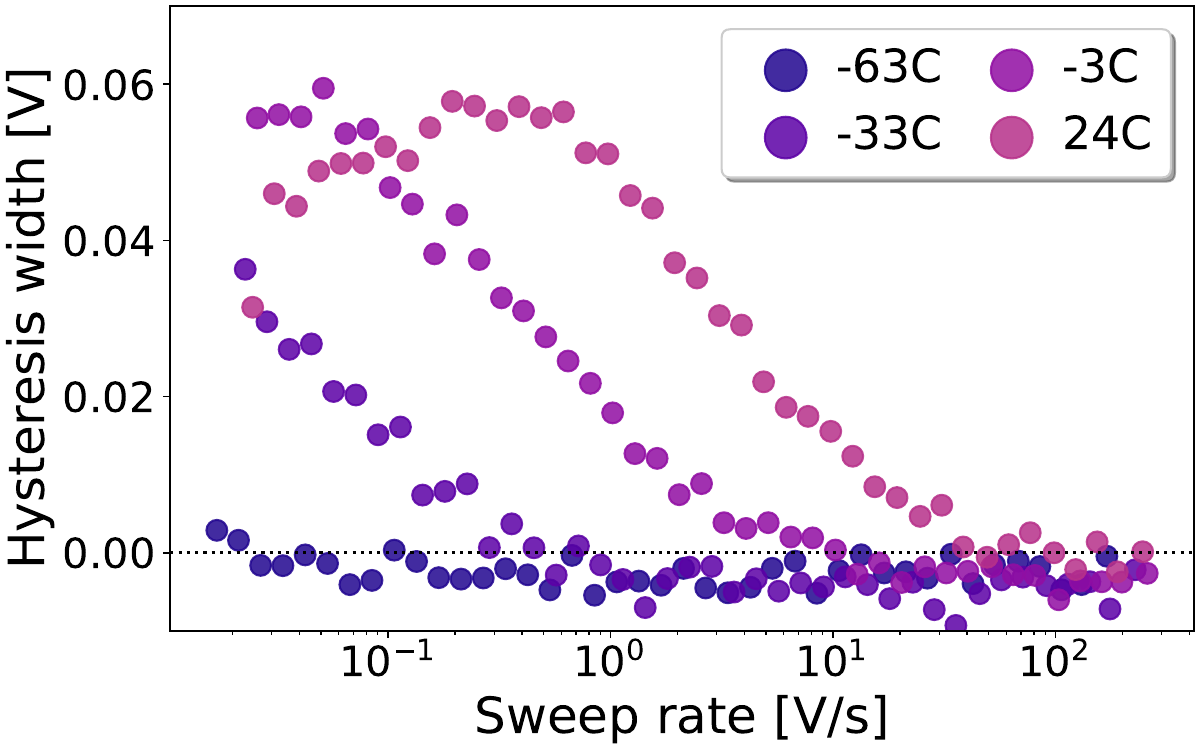}
\end{subfigure}
\begin{subfigure}[b]{.245\linewidth}
\includegraphics[width=1\linewidth]{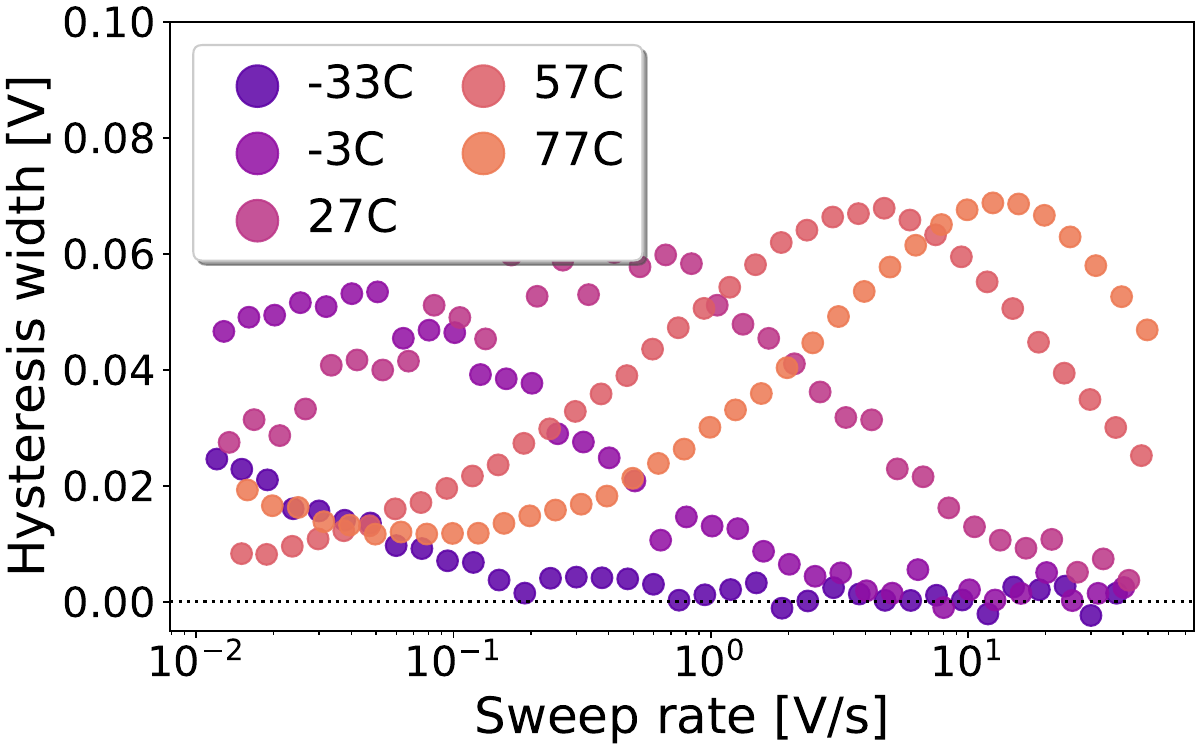}
\end{subfigure}

\begin{subfigure}[b]{.245\linewidth}
\includegraphics[width=1.0\linewidth]{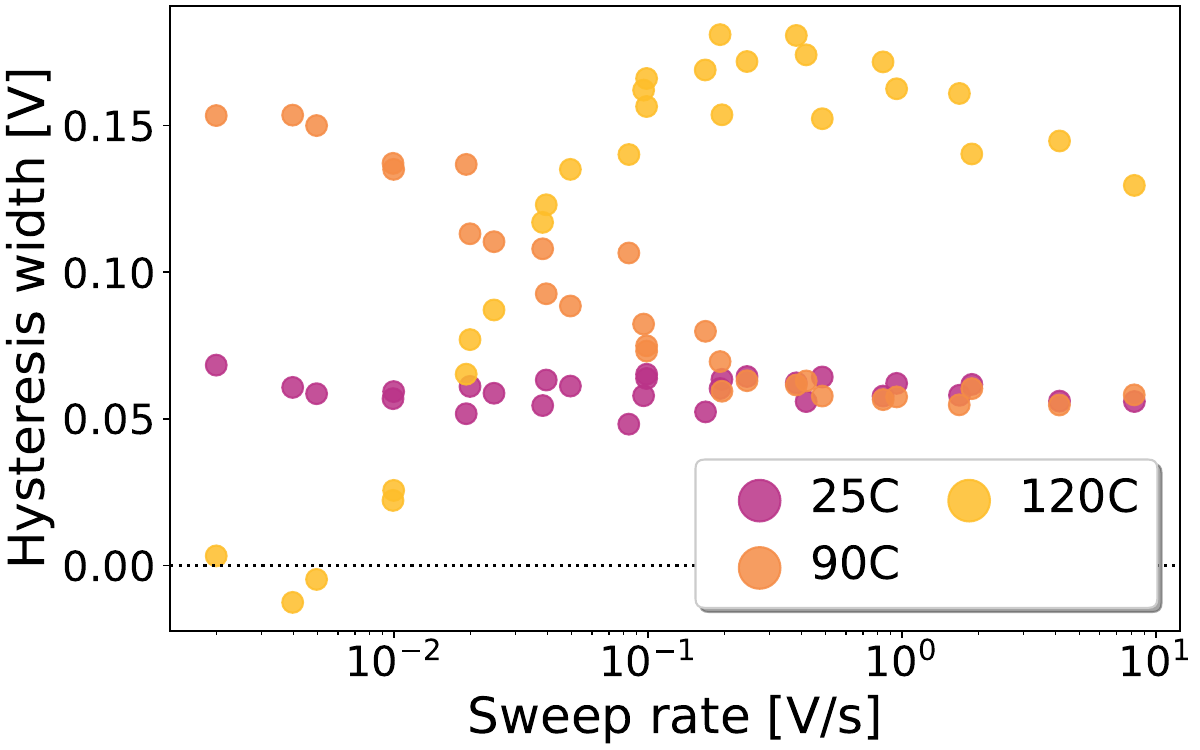}
\end{subfigure}
\begin{subfigure}[b]{.245\linewidth}
\includegraphics[width=1.0\linewidth]{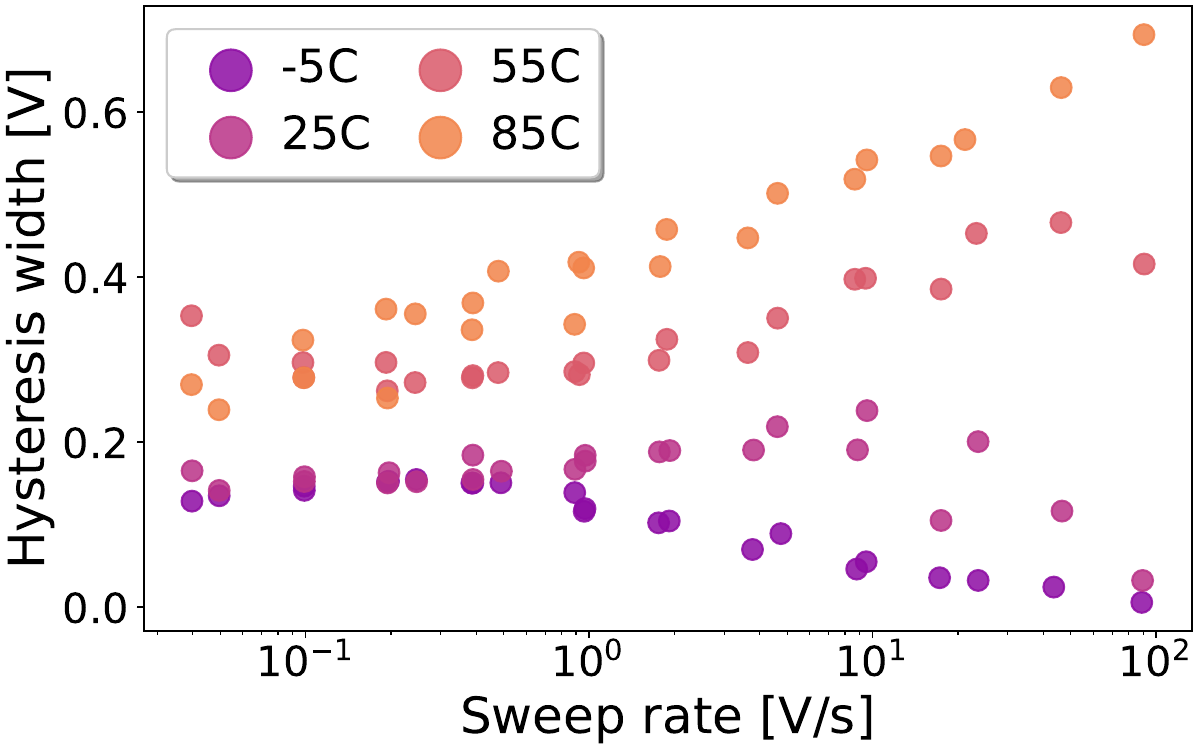}
\end{subfigure}
\begin{subfigure}[b]{.245\linewidth}
\includegraphics[width=1\linewidth]{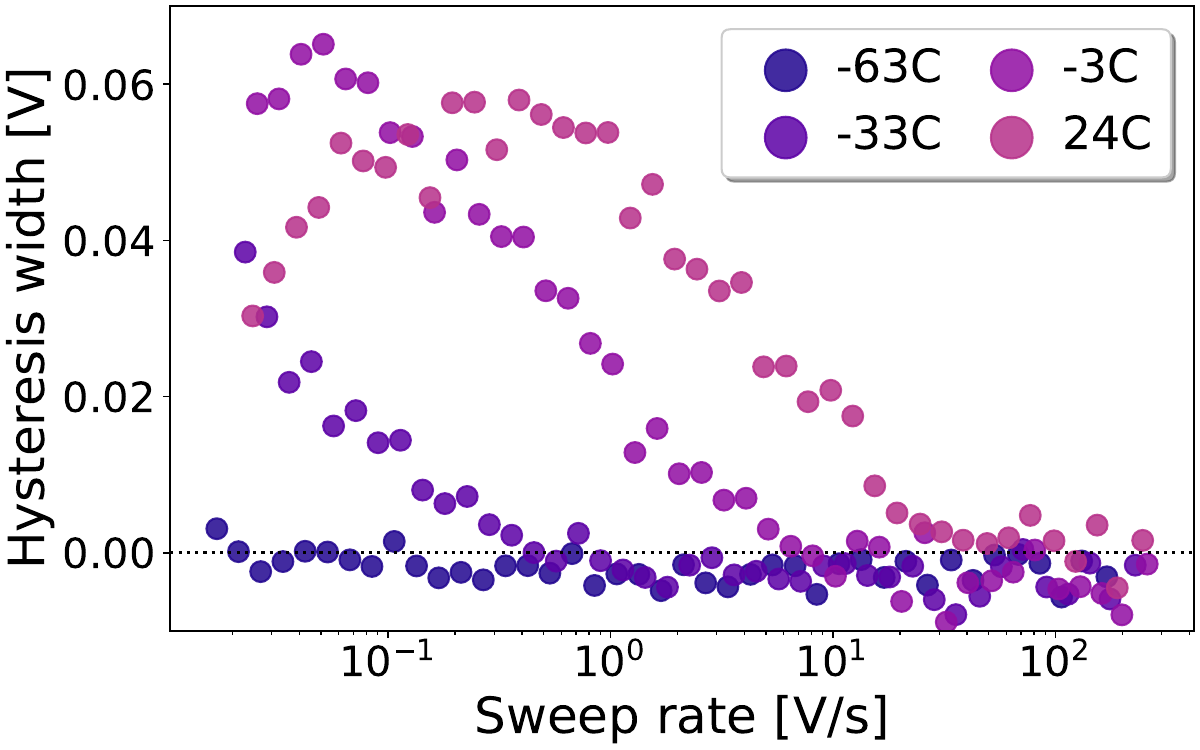}
\end{subfigure}
\begin{subfigure}[b]{.245\linewidth}
\includegraphics[width=1\linewidth]{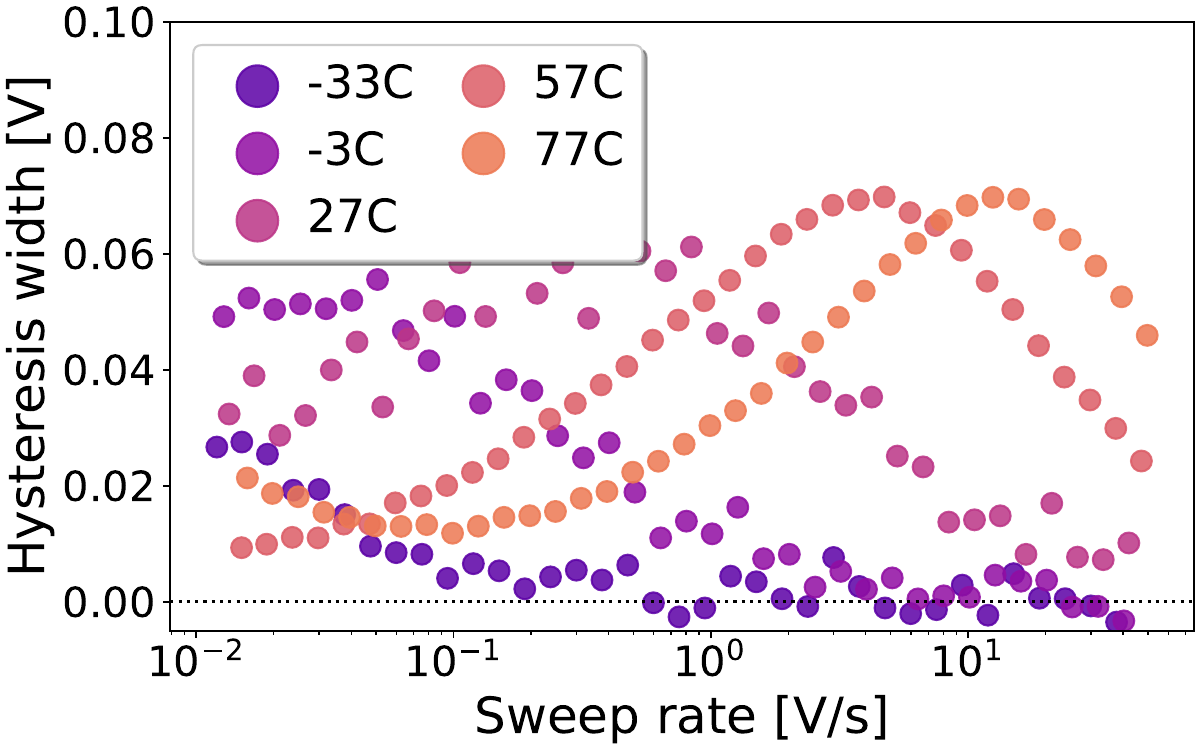}
\end{subfigure}

\begin{subfigure}[b]{.245\linewidth}
\includegraphics[width=1.0\linewidth]{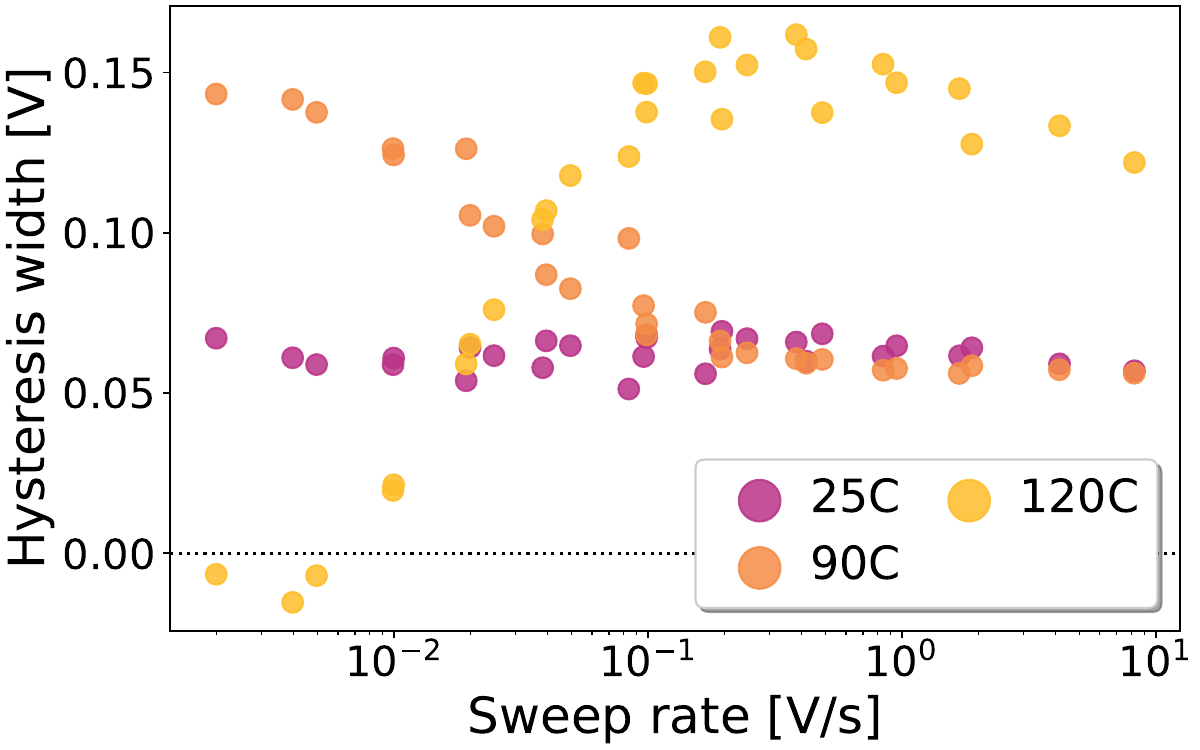}
\caption{}

\end{subfigure}
\begin{subfigure}[b]{.245\linewidth}
\includegraphics[width=1.0\linewidth]{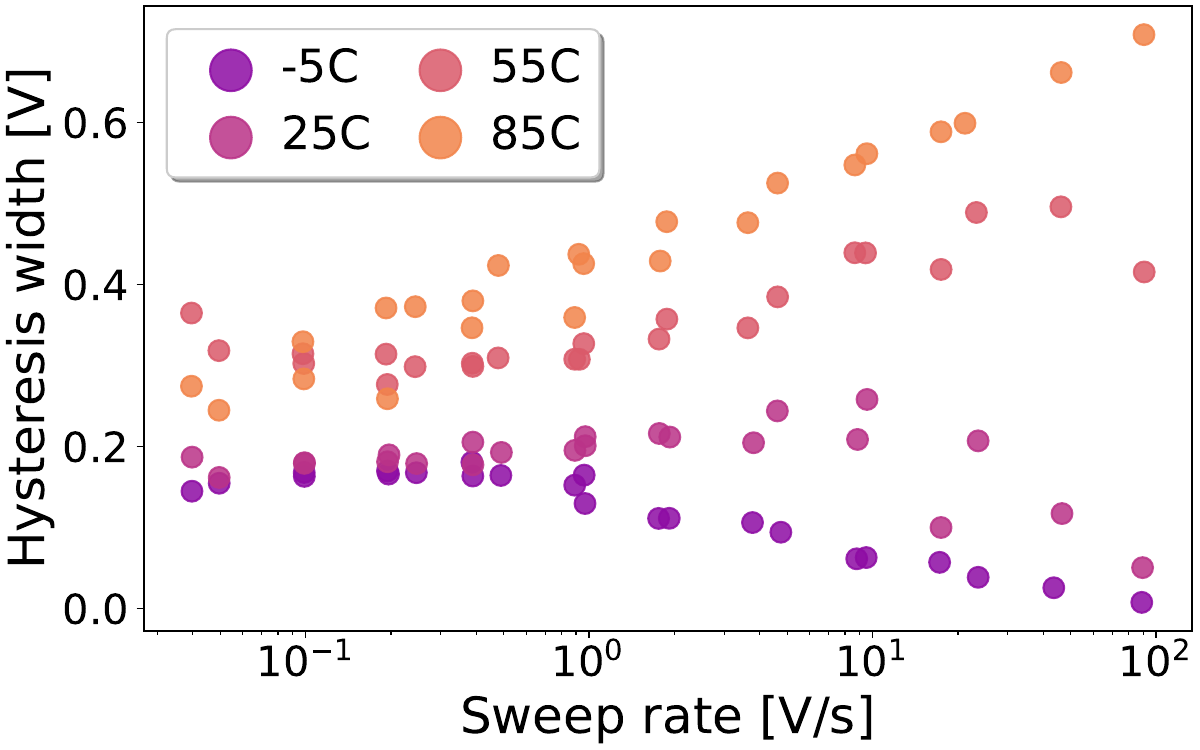}
\caption{}

\end{subfigure}
\begin{subfigure}[b]{.245\linewidth}
\includegraphics[width=1\linewidth]{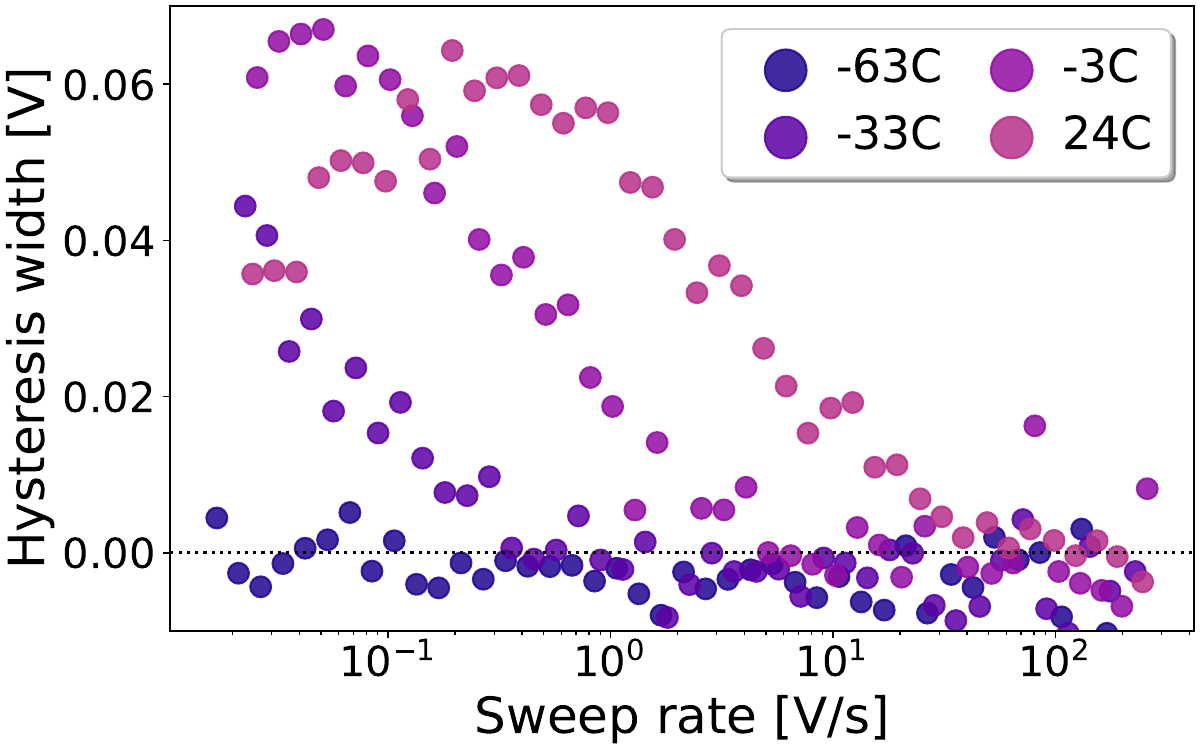}
\caption{}

\end{subfigure}
\begin{subfigure}[b]{.245\linewidth}
\includegraphics[width=1\linewidth]{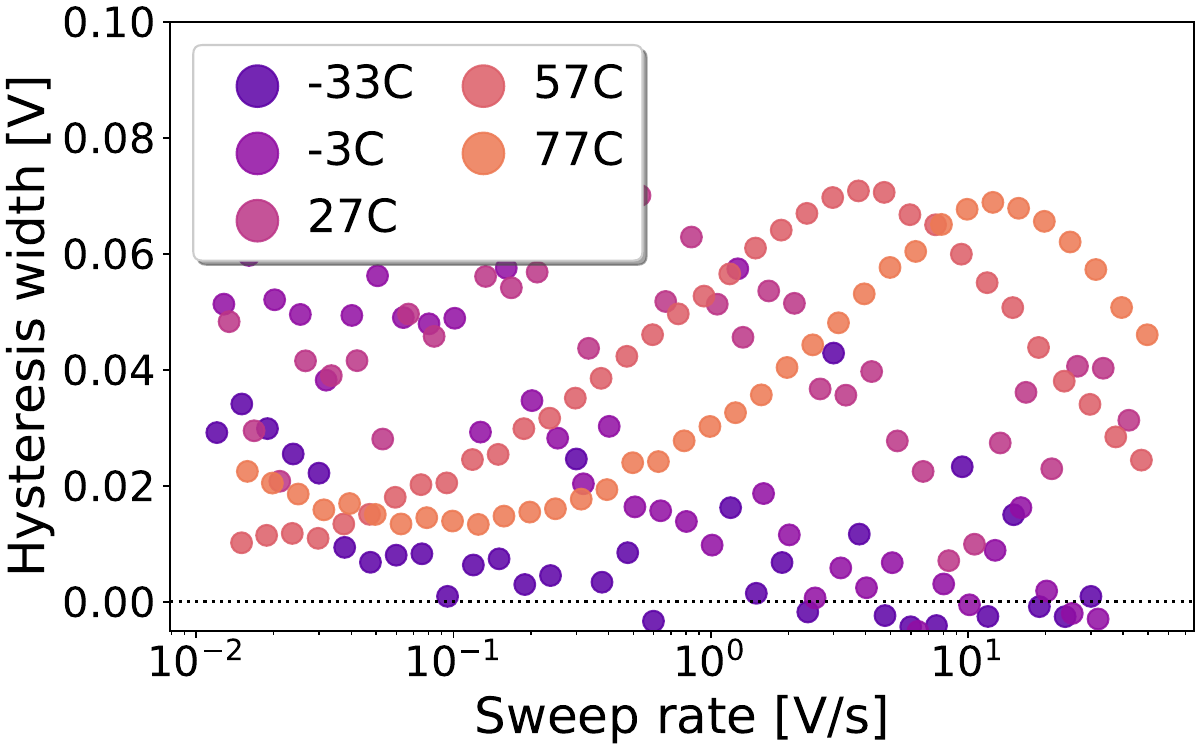}
\caption{}

\end{subfigure}
\caption{Measured hysteresis at different readouts (top to down: the highest to the lowest \ith{}) as a function of sweep-rate at various temperatures for \textbf{(a)} the \plan{}, \textbf{(b)} the \fin{},\textbf{(c)} the \sgaa{}, and \textbf{(d)} the  \gaa{}.}
\end{figure}

\newpage
\subsection{Defect Levels in Semiconducting \bissc{}}\label{supp:sc_defects}
Native point defects in the channel material (\bissc{}) play a critical role in device performance, as they significantly affect its electrical and optical properties. In this part, we analyze the energy levels associated with these intrinsic defects. To accurately reproduce the experimental \idvg{} characteristics in our device simulations, an interface trap density was introduced. To validate the parameters used in these simulations, we performed DFT calculations on \bissc{} to determine the energy levels associated with native point defects. In our study, we investigated both oxygen vacancies (V$_\text{O}$) and selenium vacancies (V$_\text{Se}$), which are likely the most prevalent native point defects in \bissc{}. These defects are of particular interest due to the material's unique layered crystal structure and the differing chemical stabilities of its constituent atoms. 

\bissc{} is characterized as a polar layered compound comprising strongly covalently bonded [\ce{Bi2O2}] layers. These layers are separated and linked by  electrostatic interactions with adjacent planar selenium layers, as opposed to the weaker conventional van der Waals interaction found in 2D materials. The selenium atoms located between the [\ce{Bi2O2}] slabs (interlayer Se) are especially prone to removal or chemical modification, as demonstrated by oxidation reactions that preferentially affect these interlayer Se atoms while largely preserving the integrity of the [\ce{Bi2O2}] framework~\cite{peng_beta}.

\begin{figure}[!htb]
        \centering
        \captionsetup{width=0.9\linewidth}
        \begin{subfigure}{0.5\textwidth}
            \centering
            \includegraphics[width=\textwidth]{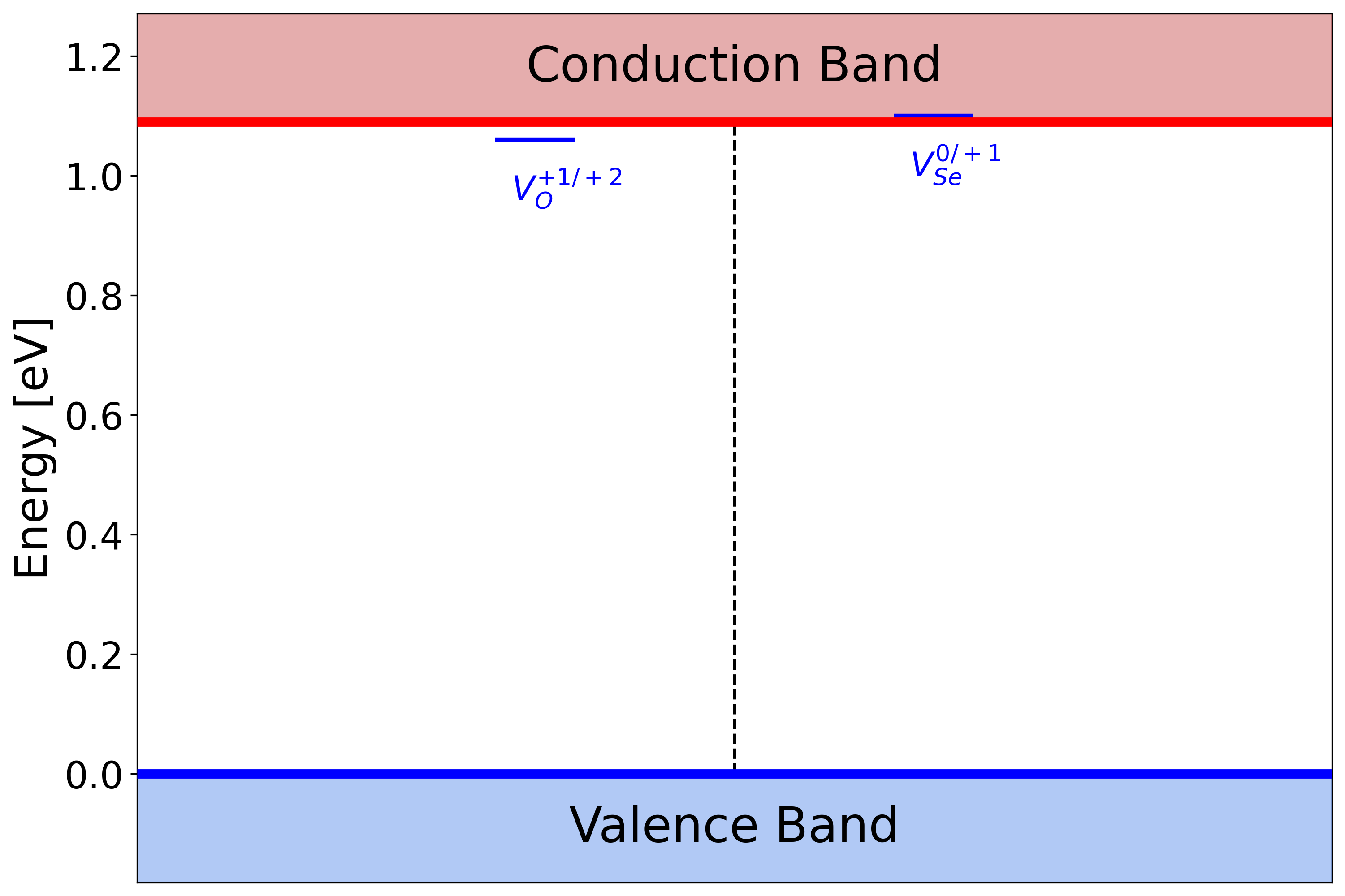}
            \caption {}
        \end{subfigure}
        \begin{subfigure}{0.7\textwidth}
            \centering
            \includegraphics[width=\textwidth]{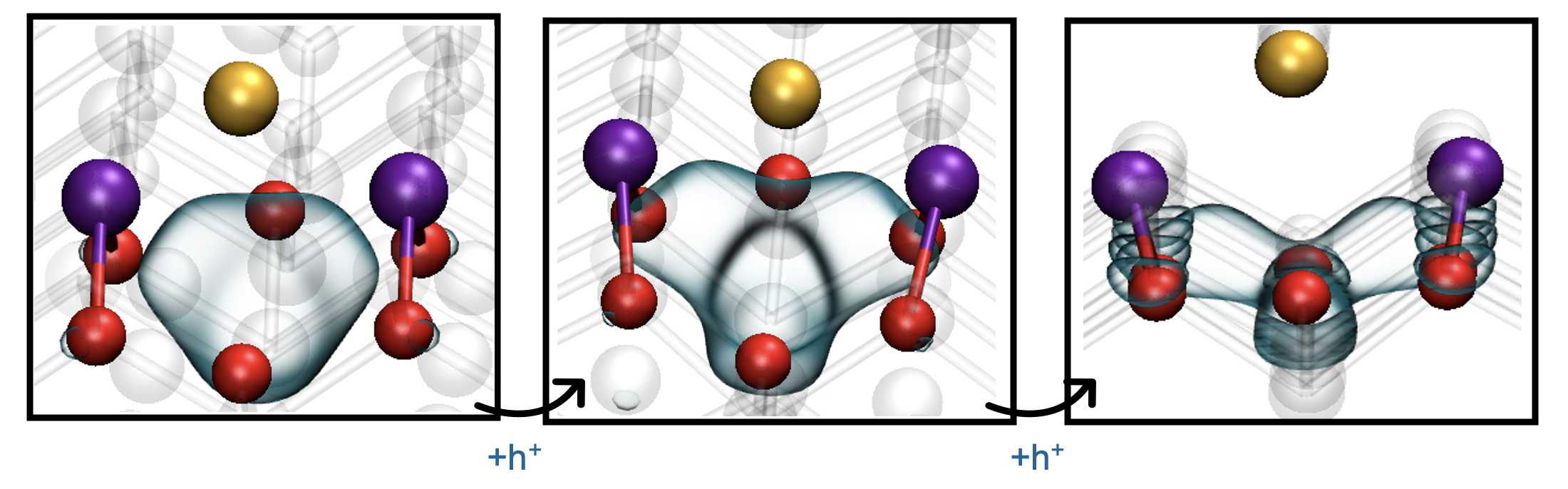}
            \caption {}
        \end{subfigure}
    \caption{Defect levels in the semiconductor.
    \textbf{(a)} The charge transition levels of defect candidates within the band gap of \bissc{}. \textbf{(b)} An example of an oxygen vacancy in different charge states, showing the respective highest occupied molecular orbital (HOMO) for the neutral (left), singly charged, and doubly charged states.}
        
    \label{fig:sc}
\end{figure}

Oxygen vacancies (V$_\text{O}$), alongside selenium vacancies (V$_\text{Se}$), are thus identified as the dominant anion deficiency-related defects in \bissc{}, significantly impacting the material's electronic structure and defect energetics. Hence, these defects explain both the required density of fixed negative charges as well as the density of interface traps used in the TCAD simulations. As shown in the \fig{fig:sc} the trap level for the oxygen vacancy in the \( +1/+2 \) charge transition is consistent with the interface trap levels and fixed charge densities used in our TCAD device simulations.

\newpage
\subsection{Simulated Electrostatics}\label{supp:elecstat}
\begin{figure}[!h]
\begin{subfigure}[b]{.33\linewidth}
\includegraphics[width=1.0\linewidth]{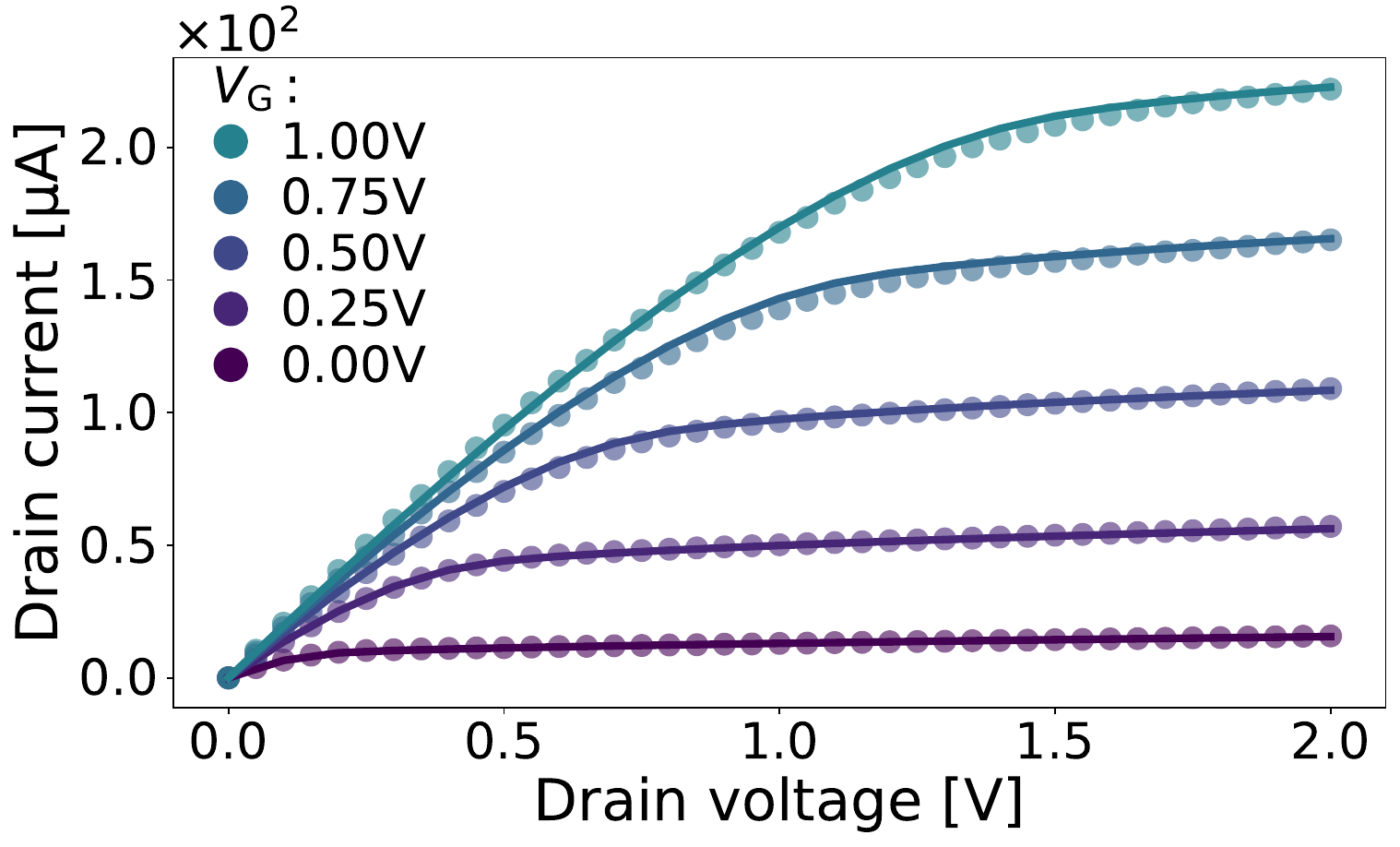}\caption{}
\end{subfigure}
\begin{subfigure}[b]{.33\linewidth}
\includegraphics[width=1.0\linewidth]{final_figures/electrostatics/planar_vgt.pdf}
\end{subfigure}
\begin{subfigure}[b]{.33\linewidth}
\includegraphics[width=1\linewidth]{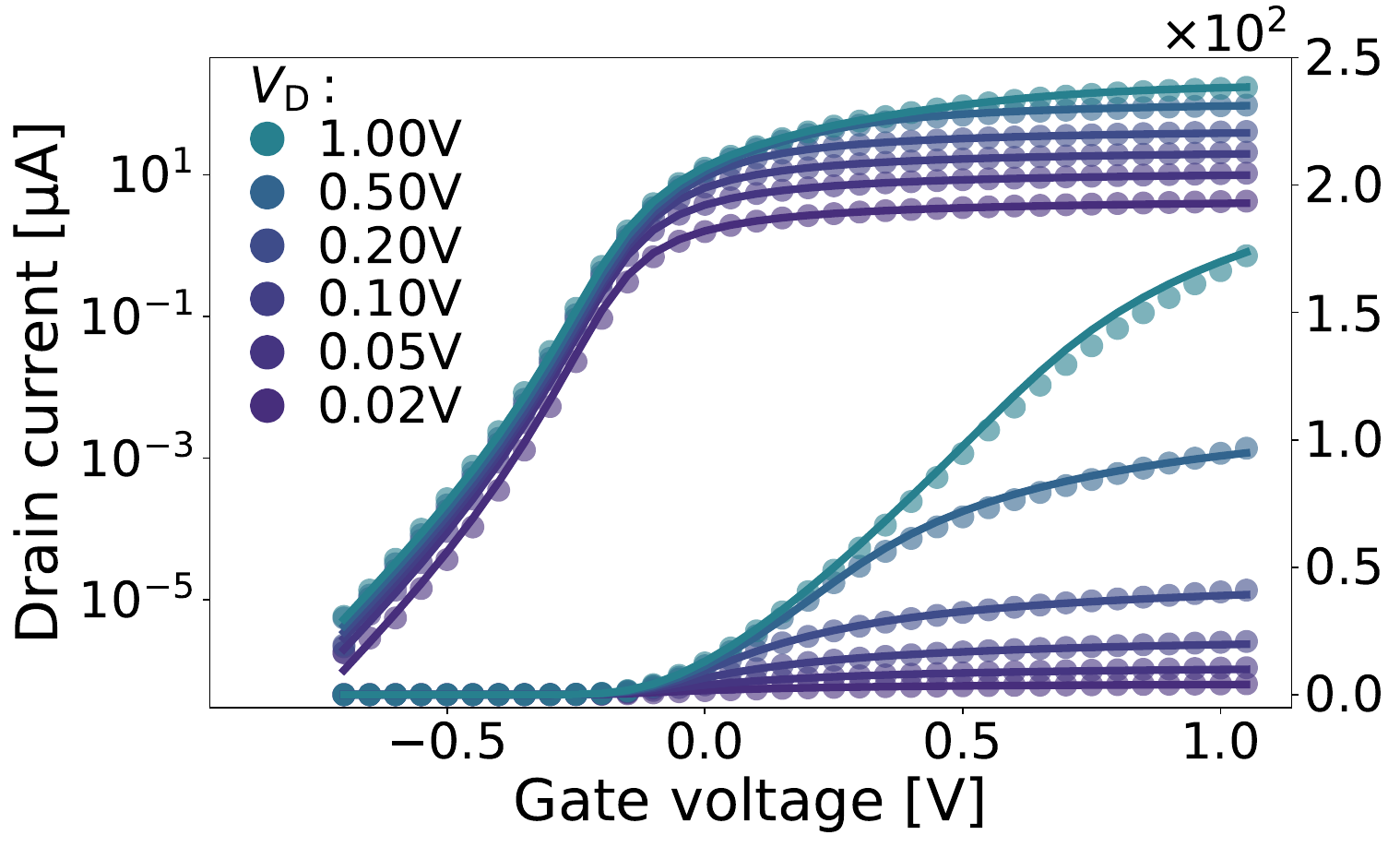}

\end{subfigure}
\begin{subfigure}[b]{.33\linewidth}
\includegraphics[width=1\linewidth]{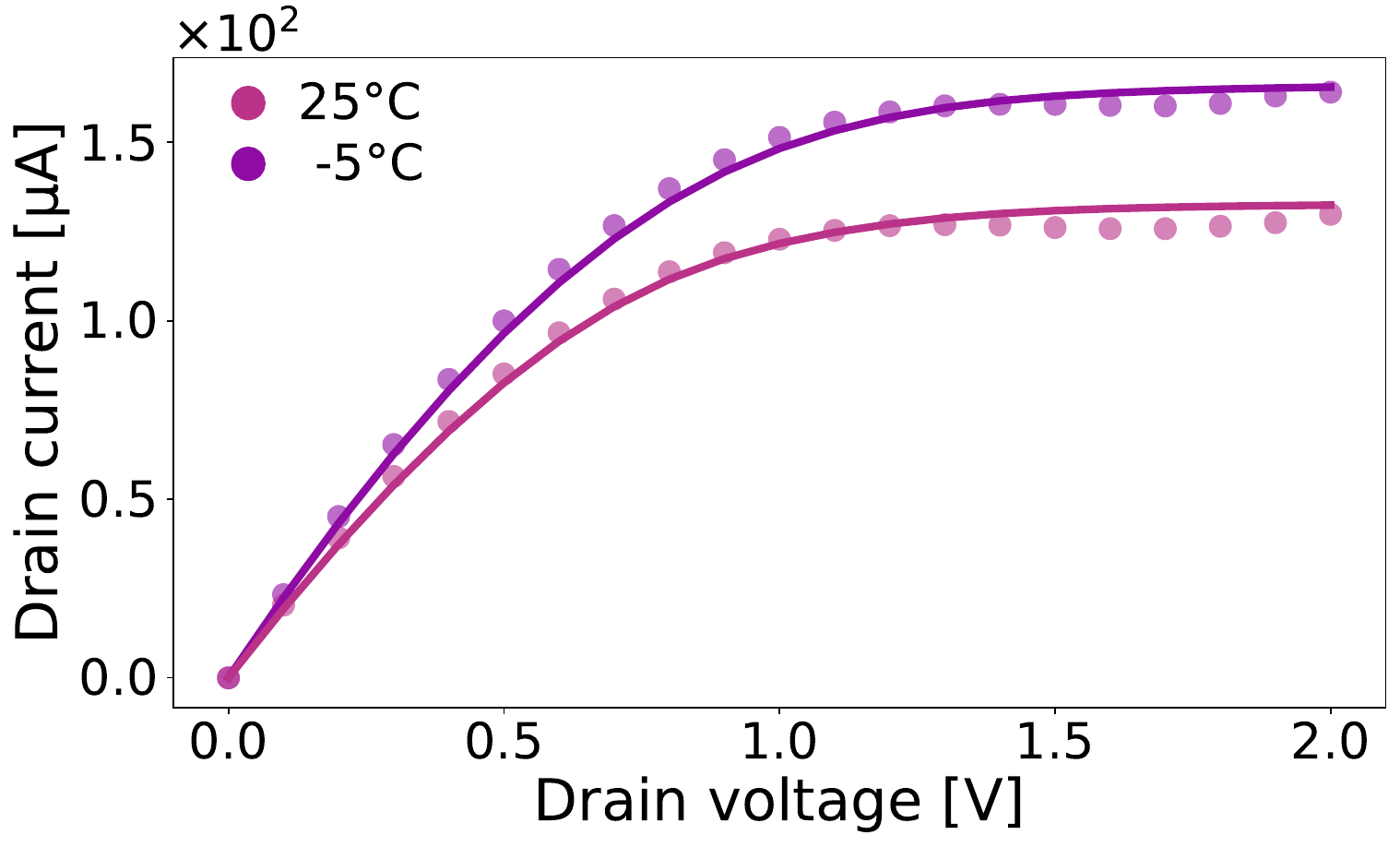}\caption{}
\end{subfigure}
\begin{subfigure}[b]{.33\linewidth}
\includegraphics[width=1.0\linewidth]{final_figures/electrostatics/finfet_vgt.pdf}
\end{subfigure}

\begin{subfigure}[b]{.33\linewidth}
\includegraphics[width=1.0\linewidth]{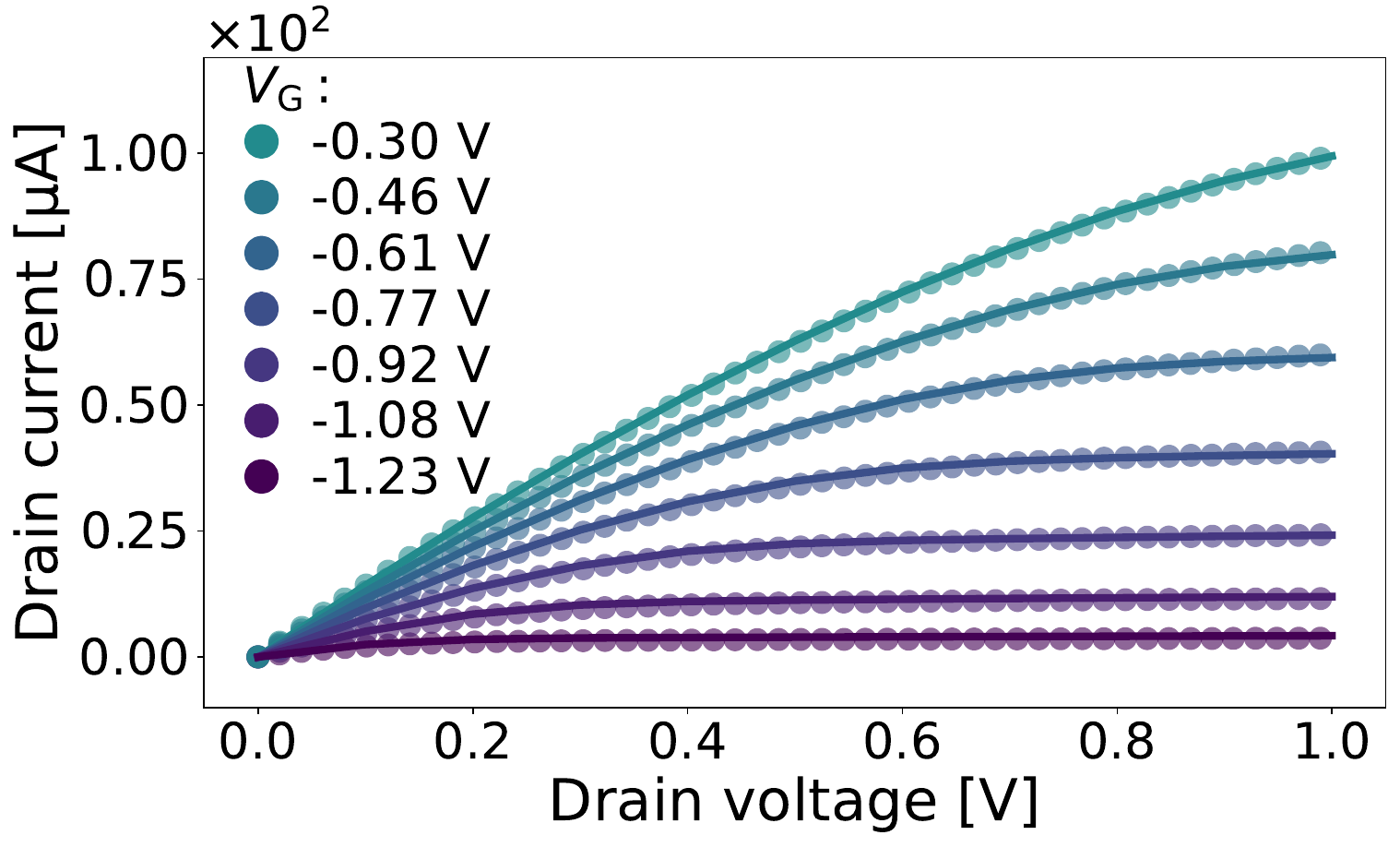}\caption{}
\end{subfigure}
\begin{subfigure}[b]{.33\linewidth}
\includegraphics[width=1\linewidth]{final_figures/electrostatics/sgaa_vgt.pdf}
\end{subfigure}
\begin{subfigure}[b]{.33\linewidth}
\includegraphics[width=1.0\linewidth]{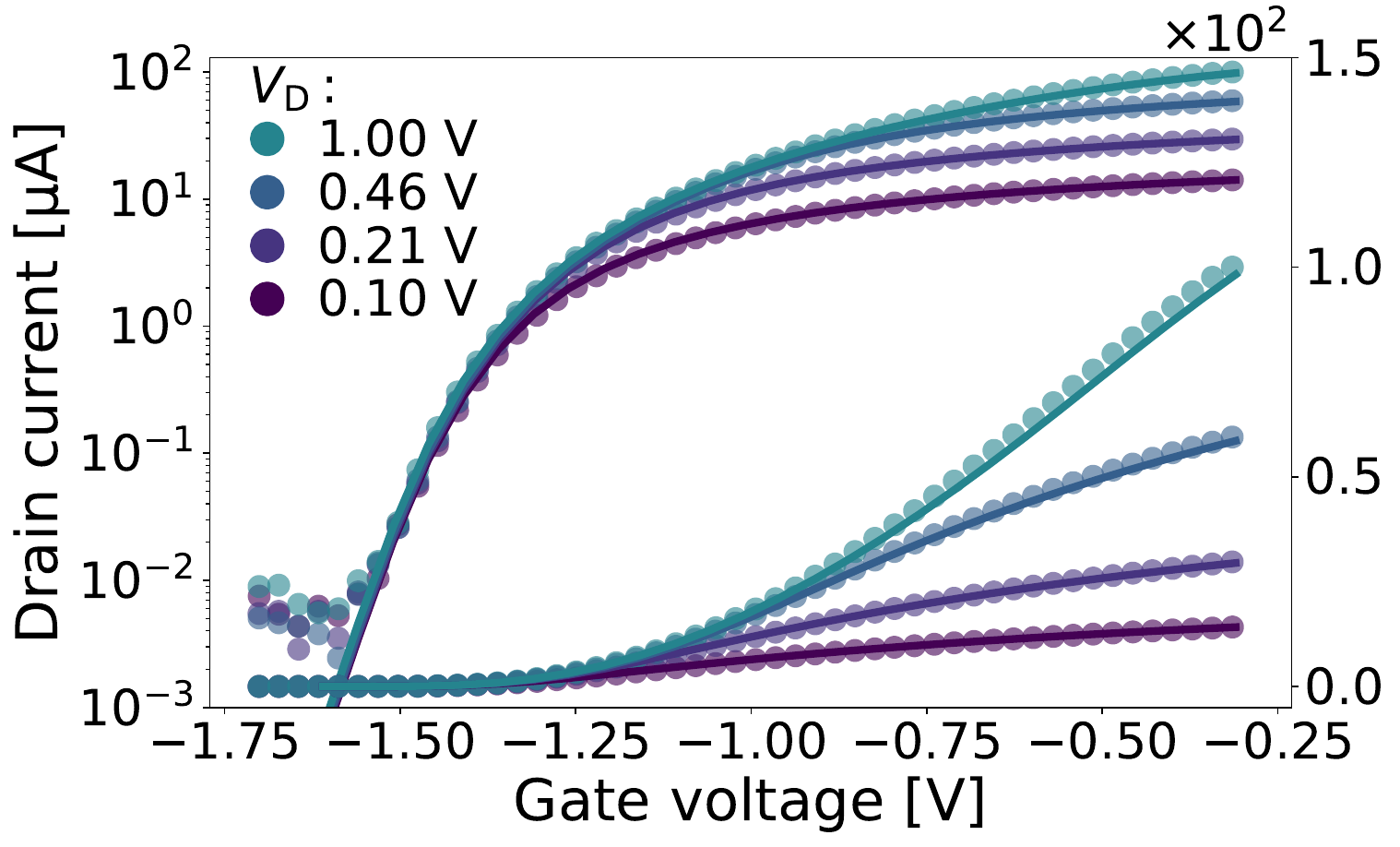}
\end{subfigure}

\begin{subfigure}[b]{.33\linewidth}
\includegraphics[width=1.0\linewidth]{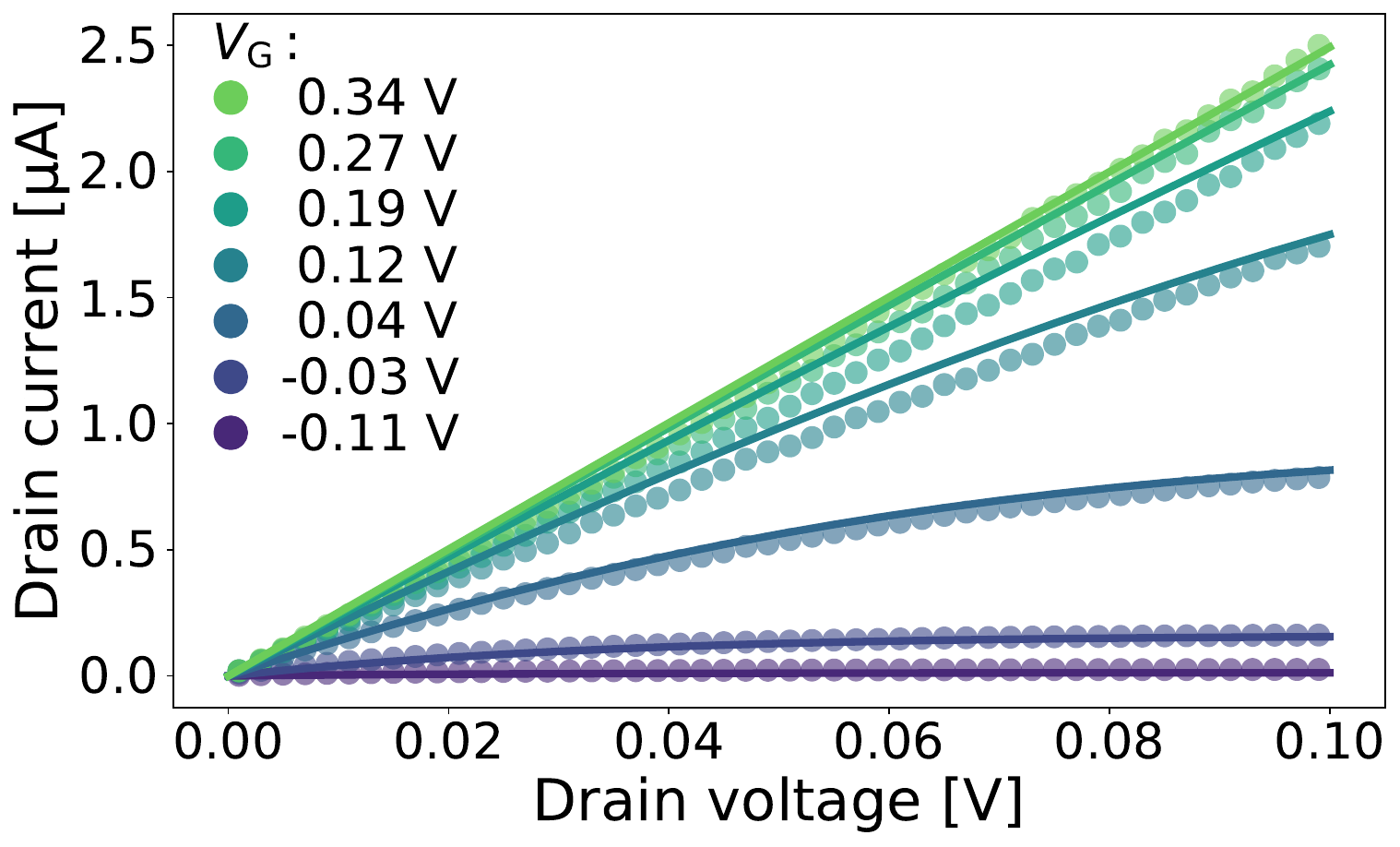}\caption{}
\end{subfigure}
\begin{subfigure}[b]{.33\linewidth}
\includegraphics[width=1\linewidth]{final_figures/electrostatics/gaa_vgt.pdf}
\end{subfigure}
\begin{subfigure}[b]{.33\linewidth}
\includegraphics[width=1\linewidth]{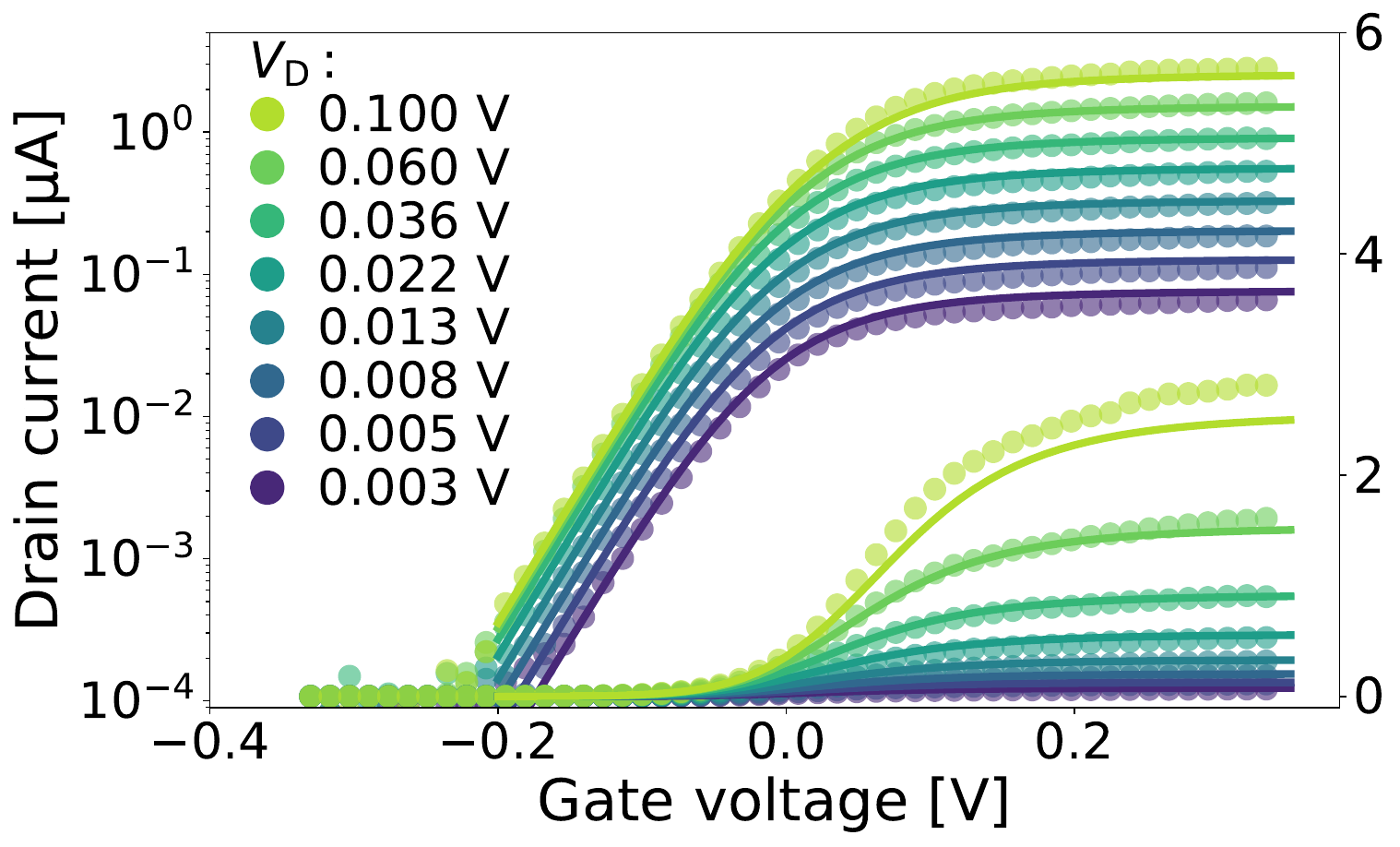}
\end{subfigure}

\caption{Modeling accuracy. The simulated (line) and measured (circles) electrostatics of \textbf{(a)} the \plan{}, \textbf{(b)} the \fin{},\textbf{(c)} the \sgaa{}, and \textbf{(d)} the  \gaa{} on logarithmic (left) and linear (right) scales.}
\end{figure}

\newpage
\subsection{Sensitivity Analysis of TCAD Calibration}\label{supp:sensitivity}
\begin{figure}[!h]
    \centering
    \includegraphics[width=0.9\linewidth]{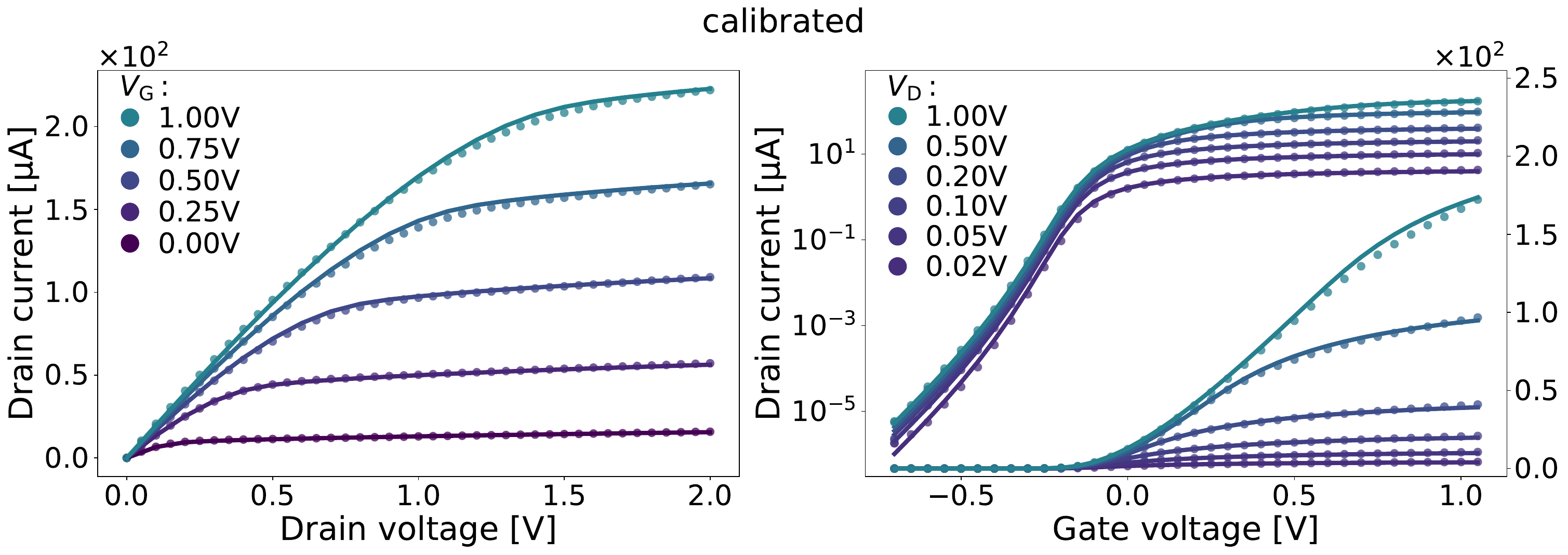}
    \includegraphics[width=0.9\linewidth]{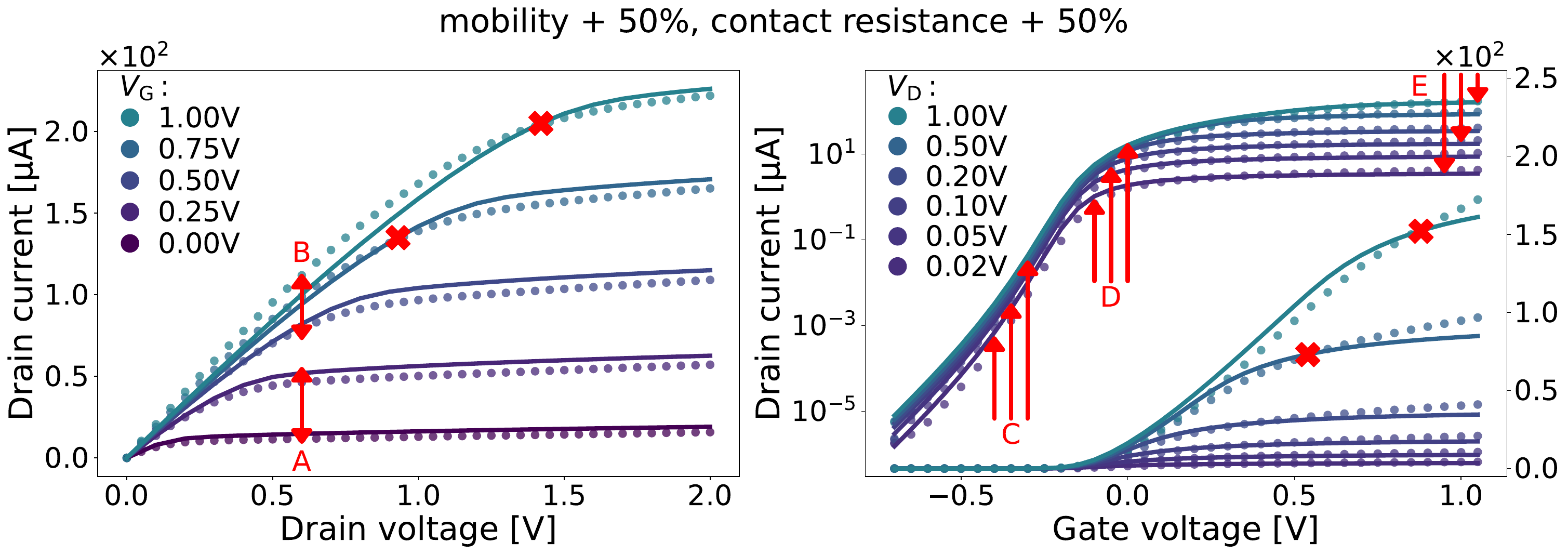}
    \includegraphics[width=0.9\linewidth]{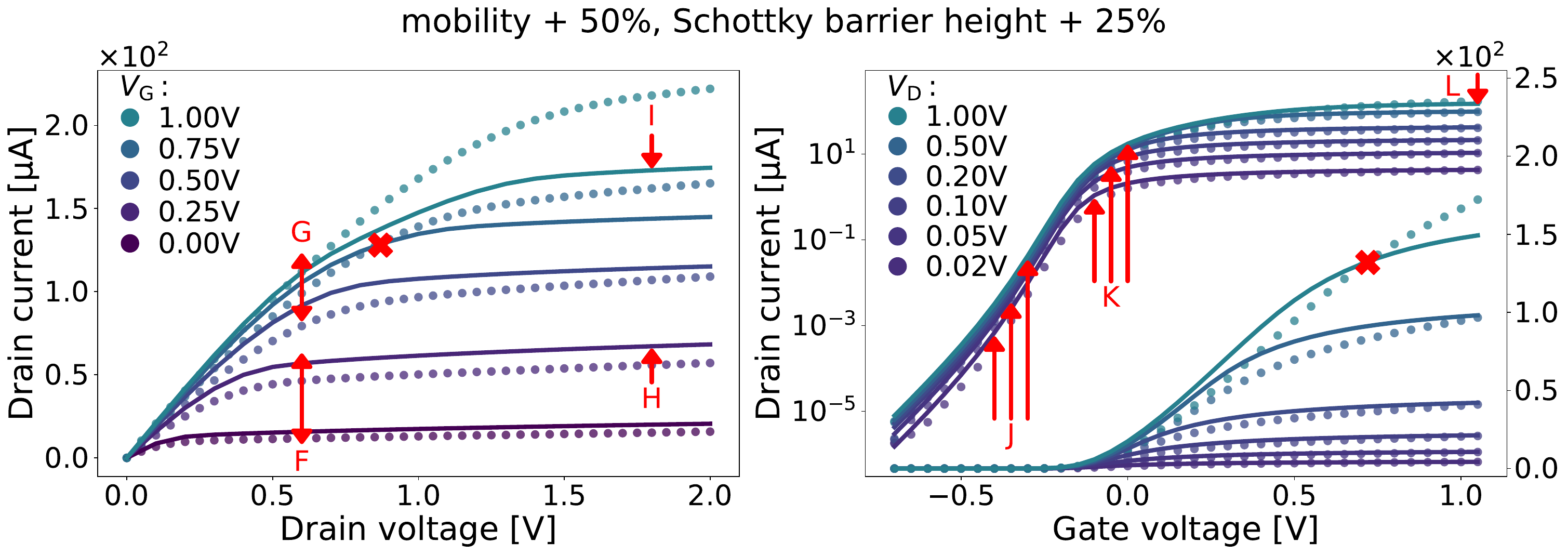}
    \caption{Sensitivity analysis illustrating the robustness of the TCAD mobility calibration. 
    Left panels show the output characteristics, \idvd{}; right panels show the transfer characteristics, \idvg{}, on both logarithmic and linear scales. 
    The top row shows the calibrated TCAD results using one common parameter set, the middle row shows a representative perturbation with mobility and contact resistance both increased by \SI{50}{\percent}, and the bottom row shows mobility increased by \SI{50}{\percent} together with a \SI{25}{\percent} increase in Schottky barrier height. 
    Only a partial compensation between the perturbed parameter sets is possible. Near an optimum region the contributions cancel each other out, whereas away from this point considerable induced errors are observed. 
    In the middle row, the output characteristics show increased spacing between lower $V_\mathrm{G}$ branches in region A but reduced spacing between higher $V_\mathrm{G}$ branches in region B. 
    The corresponding transfer characteristics overestimate the subthreshold and intermediate current regions marked by C and D, while underestimating the current branches in region E. 
    The bottom-row perturbation also shows mismatched trends: at low drain bias, F and G mark opposite spacing changes, while H and I show simultaneous enhancement and suppression of current for different gate biases.
    In the transfer characteristics, J--K versus L show overestimation in the subthreshold/intermediate current regime but underestimation of the high-current branches specially in the highest-$V_\mathrm{D}$ curve. 
    The red crosses mark intersections between perturbed simulations and measured data, showing that the error changes sign across the bias range.
    Symbols denote measured data and solid lines denote TCAD simulations.}
    \label{fig:parameters}
\end{figure}

\newpage
\subsection{Mobility Analysis Using Y-Function}\label{supp:y_function}
\setlength{\tabcolsep}{2pt}  
\renewcommand{\arraystretch}{0.1}
\begin{figure}[!h]
\centering
\begin{tabular}{@{}pfsg@{}}
\figheader{}
&&&\\&&&\\
\begin{subfigure}[b]{.245\linewidth}
\includegraphics[width=\linewidth]{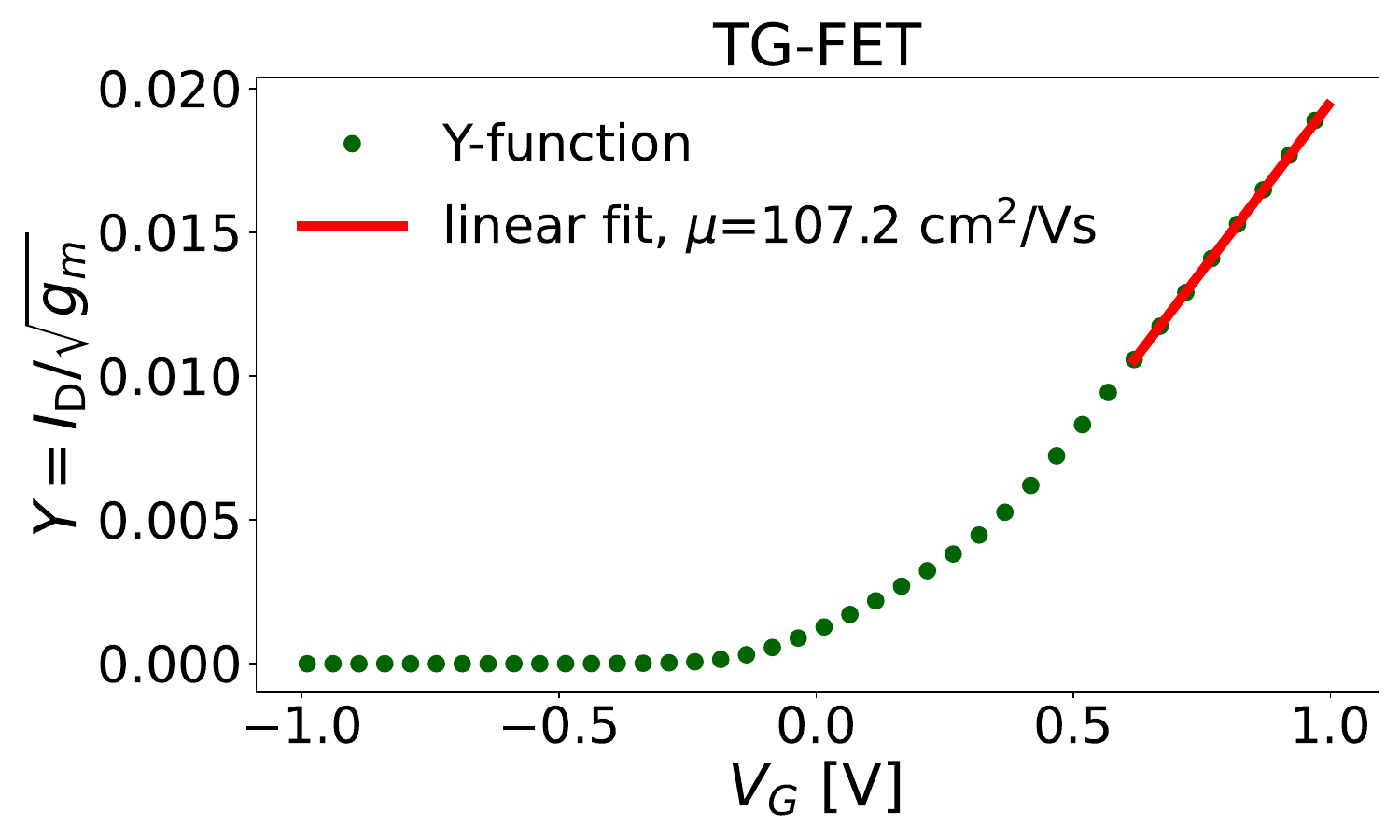}
\label{p_y_function}
\end{subfigure}&
\begin{subfigure}[b]{.245\linewidth}
\includegraphics[width=\linewidth]{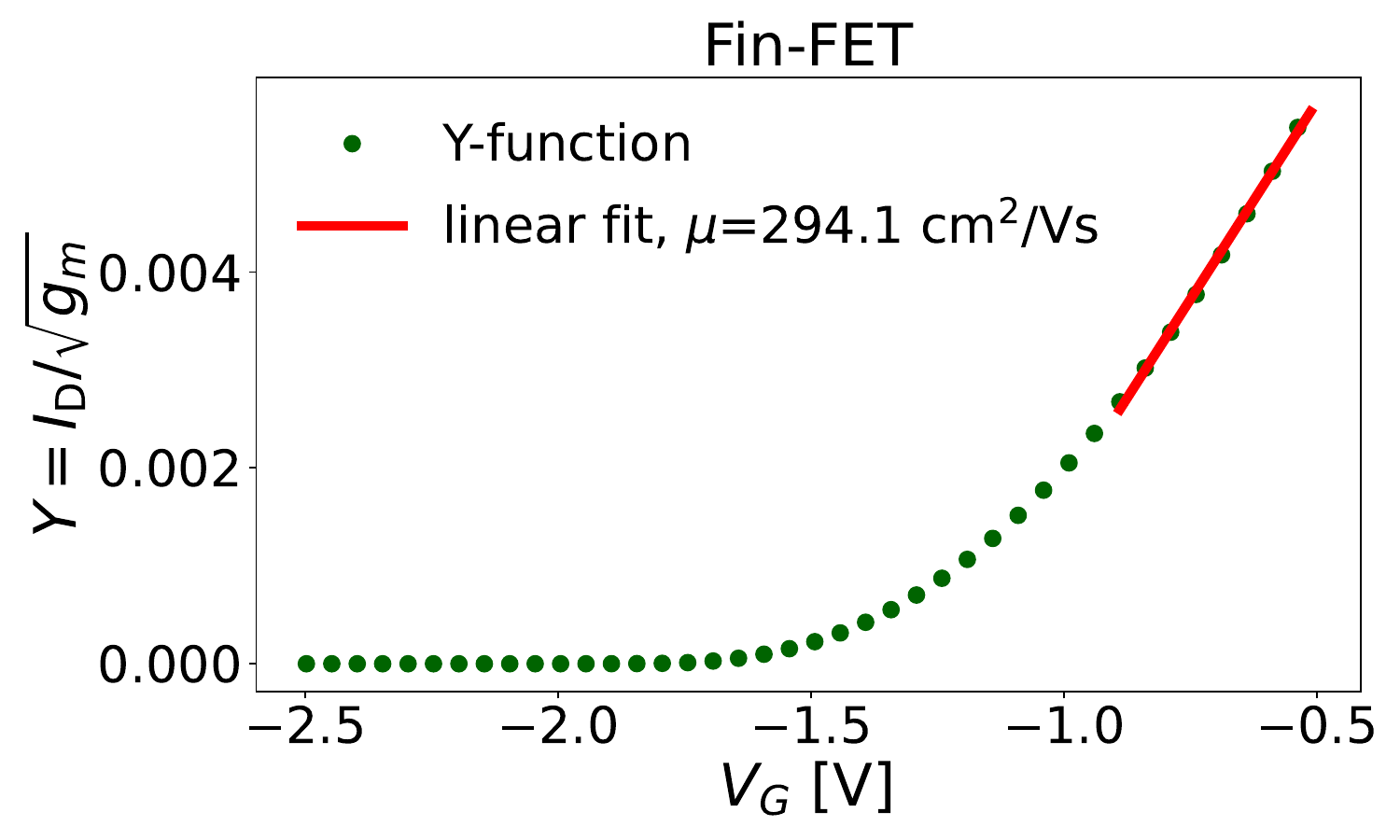}
\label{f_y_function}
\end{subfigure}&
\begin{subfigure}[b]{.245\linewidth}
\includegraphics[width=\linewidth]{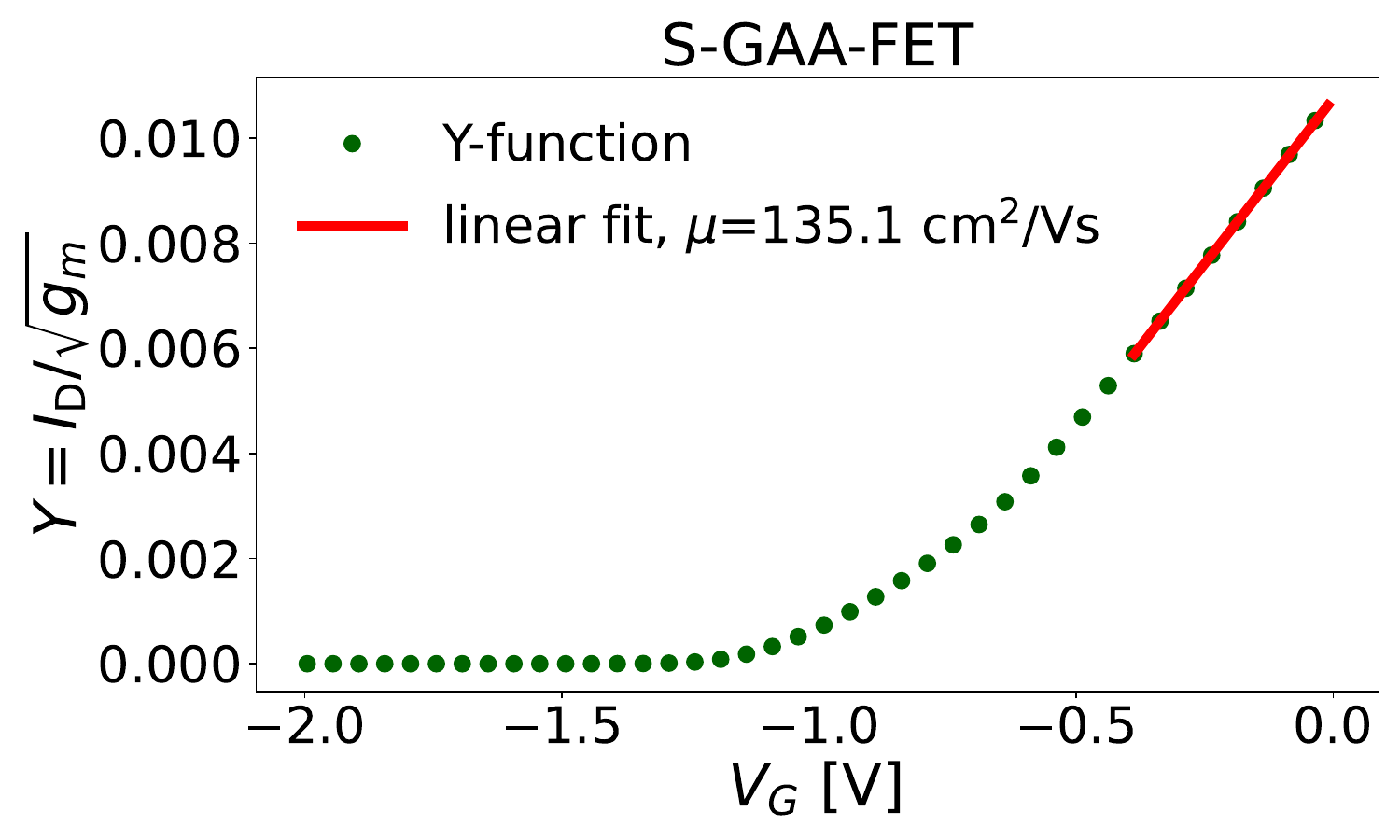}
\label{s_y_function}
\end{subfigure}&
\begin{subfigure}[b]{.245\linewidth}
\includegraphics[width=\linewidth]{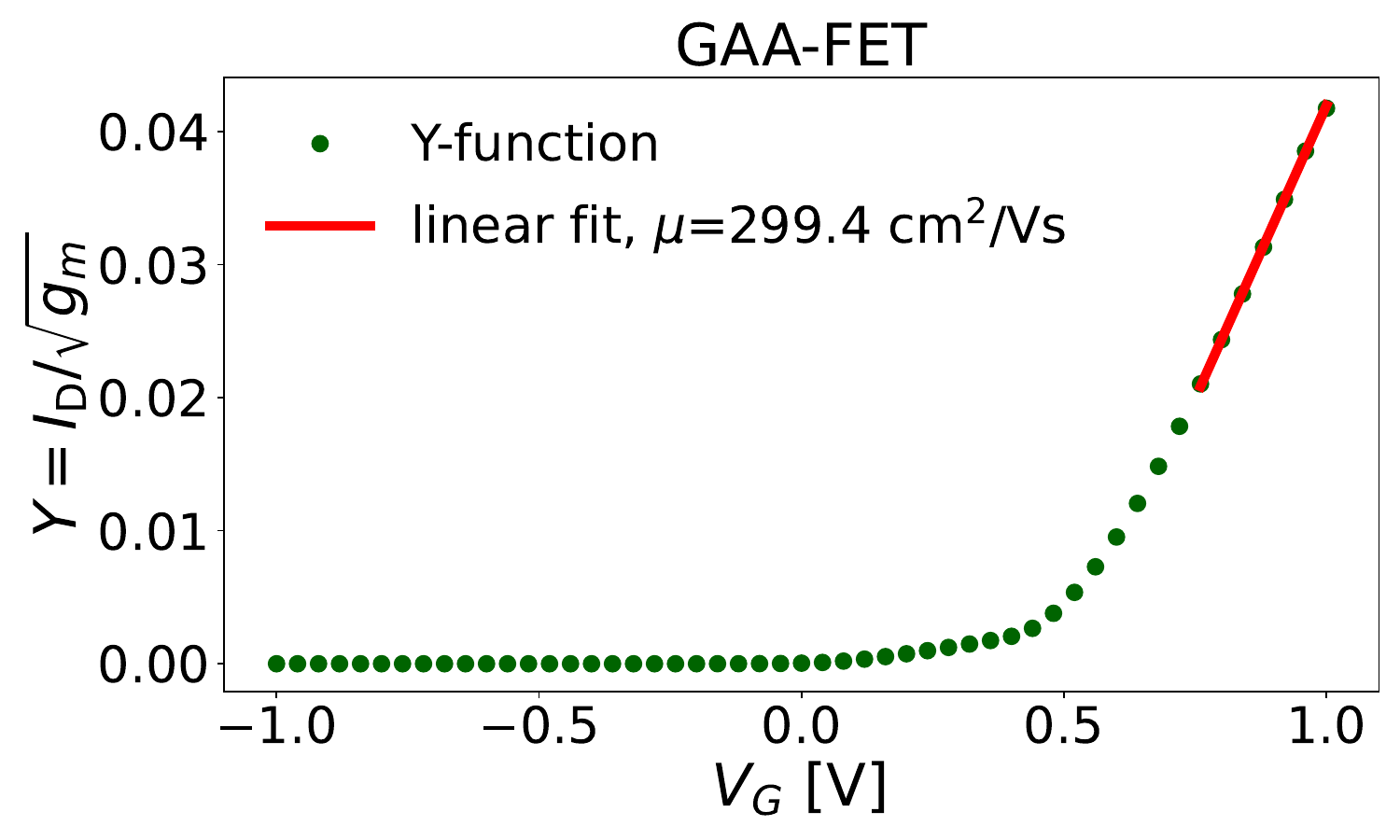}
\label{g_y_function}
\end{subfigure}\\
\end{tabular}
\caption{Y-function mobility calculated for the four device generations. The quantity $Y=I_\mathrm{D}/\sqrt{g_m}$ was constructed from the transfer characteristics and fitted in the approximately linear on-state region. The extracted Y-function mobilities are $\SI{107.2}{cm^2 V^{-1} s^{-1}}$ for \plan{}, $\SI{294.1}{cm^2 V^{-1} s^{-1}}$ for \fin{}, $\SI{135.1}{cm^2 V^{-1} s^{-1}}$ for \sgaa{}, and $\SI{299.4}{cm^2 V^{-1} s^{-1}}$ for \gaa{}.
}
\label{fig:y_function}
\end{figure}
\setlength{\tabcolsep}{6pt}
\renewcommand{\arraystretch}{1}

\newpage
\subsection{Simulated Hysteresis for Individual Readouts}\label{supp:hys_arb}
\begin{figure}[!h]
\begin{subfigure}[b]{.245\linewidth}
\includegraphics[width=1.0\linewidth]{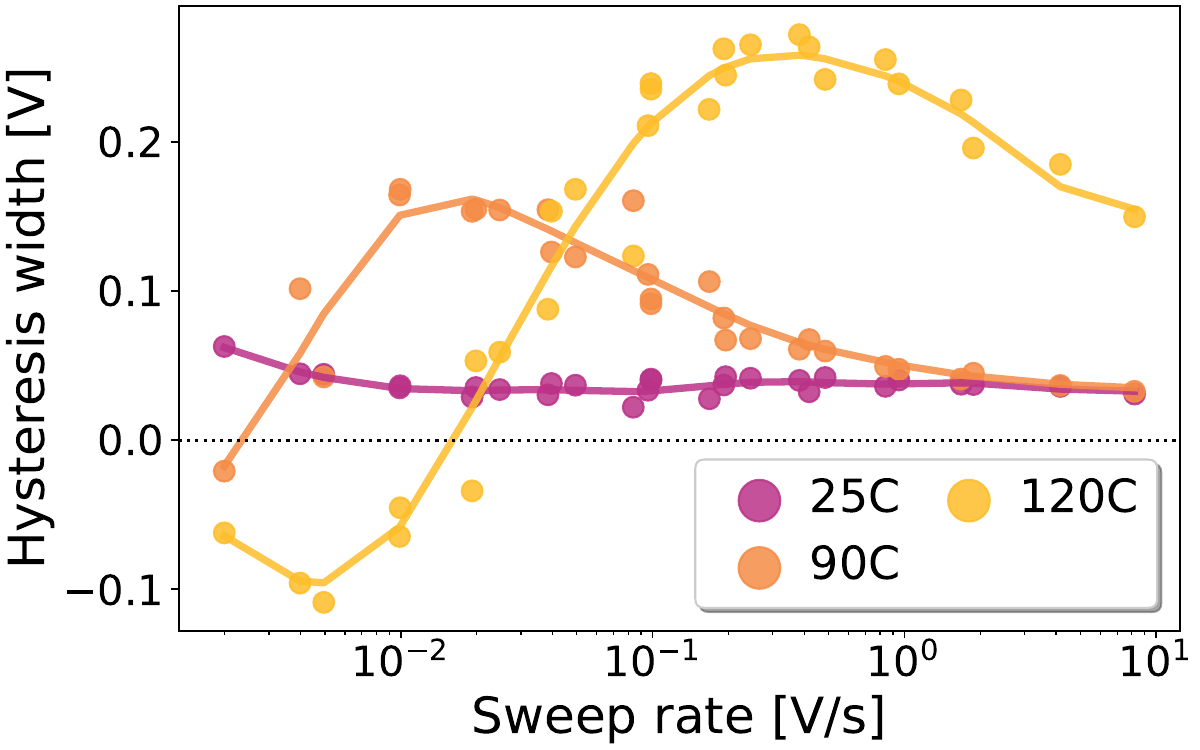}
\end{subfigure}
\begin{subfigure}[b]{.245\linewidth}
\includegraphics[width=1.0\linewidth]{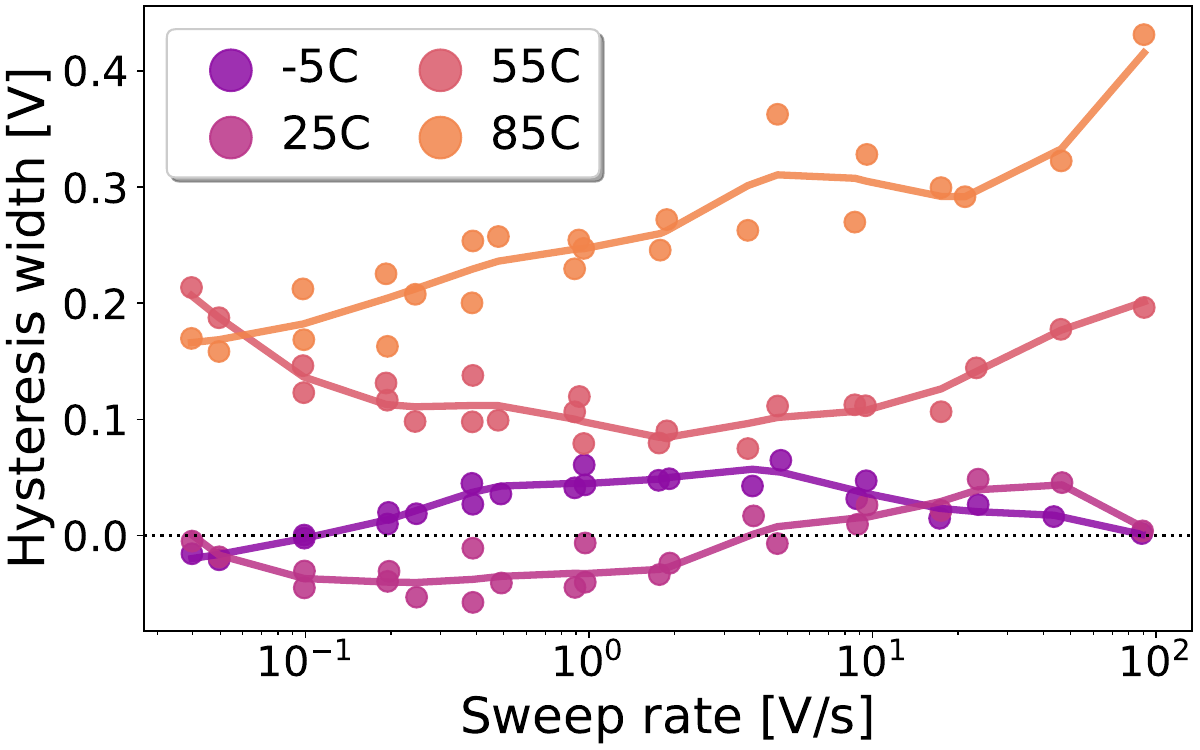}
\end{subfigure}
\begin{subfigure}[b]{.245\linewidth}
\includegraphics[width=1\linewidth]{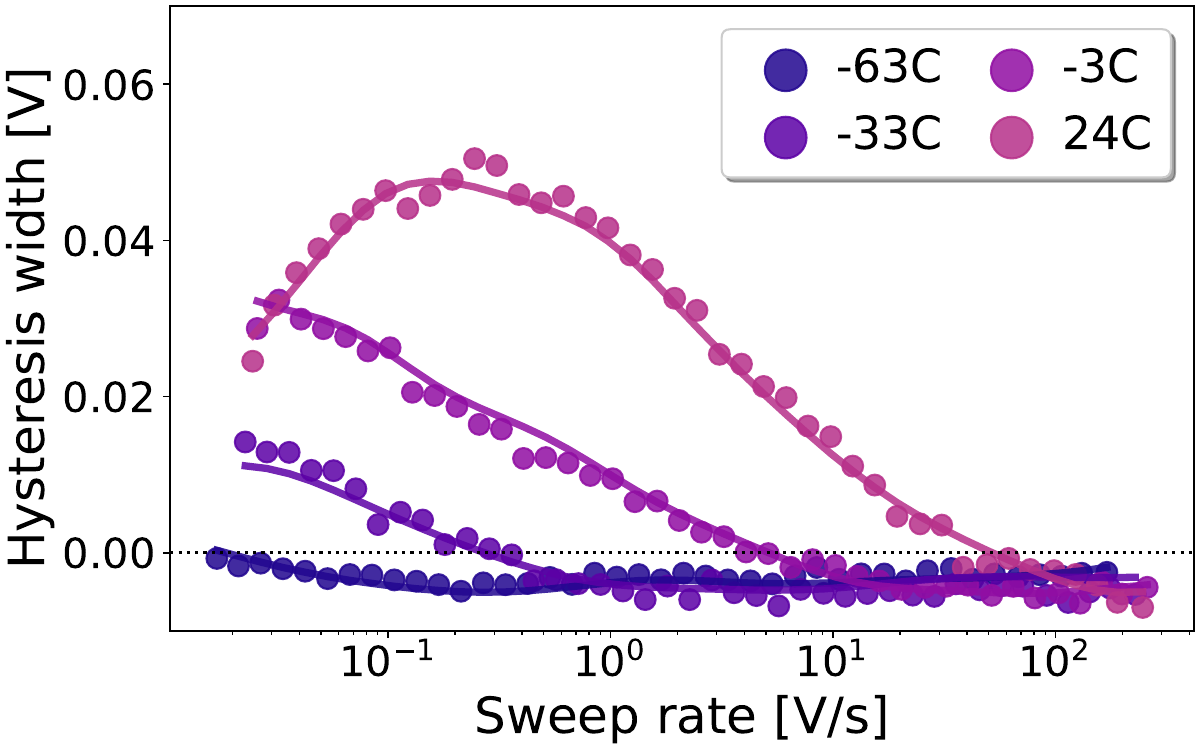}
\end{subfigure}
\begin{subfigure}[b]{.245\linewidth}
\includegraphics[width=1\linewidth]{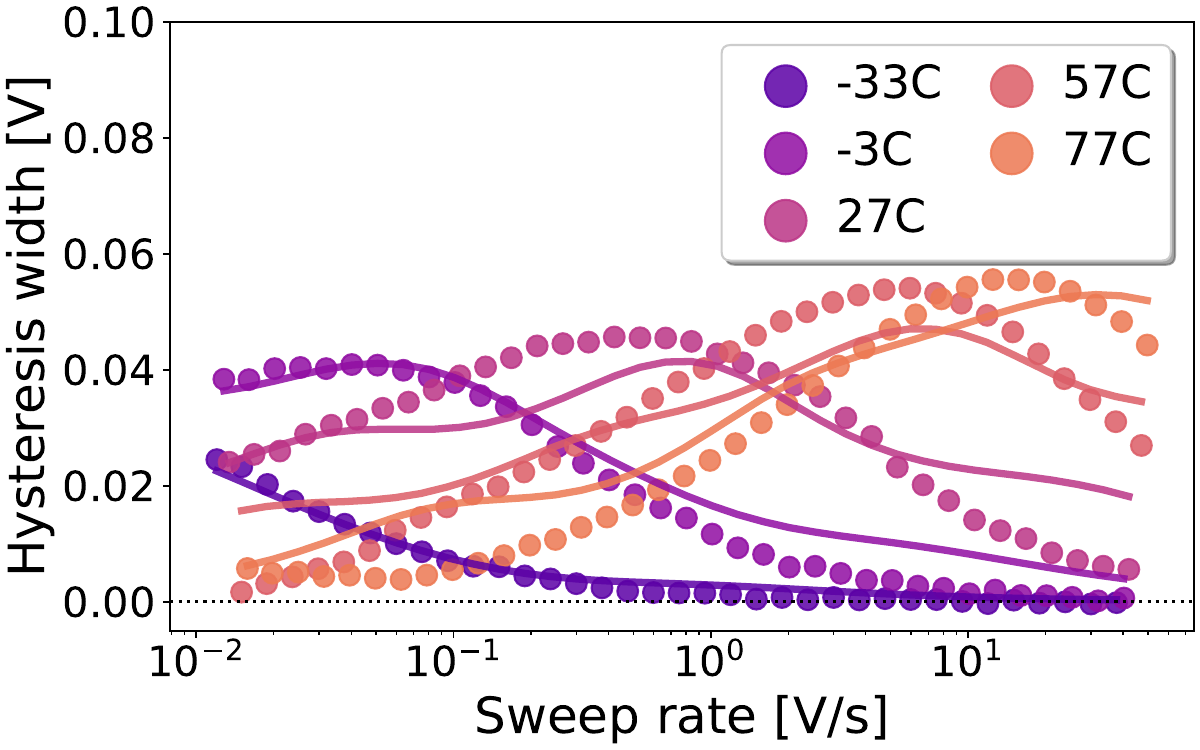}

\end{subfigure}
\begin{subfigure}[b]{.245\linewidth}
\includegraphics[width=1.0\linewidth]{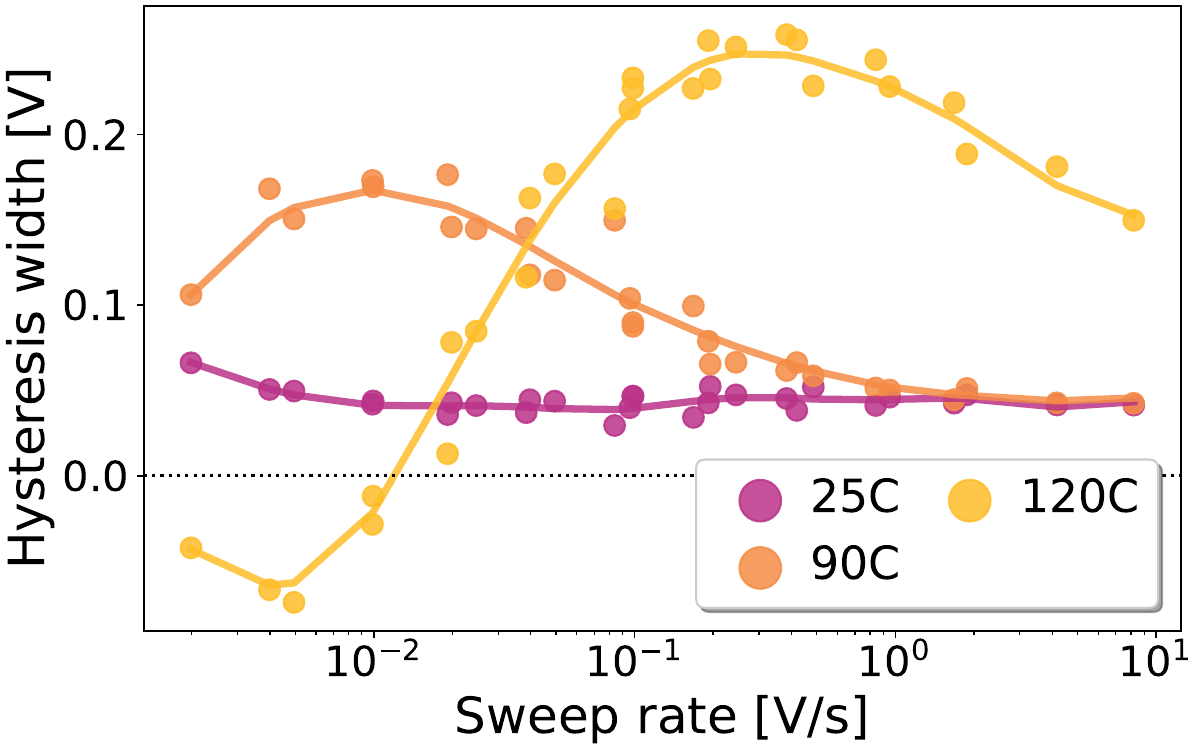}
\end{subfigure}
\begin{subfigure}[b]{.245\linewidth}
\includegraphics[width=1.0\linewidth]{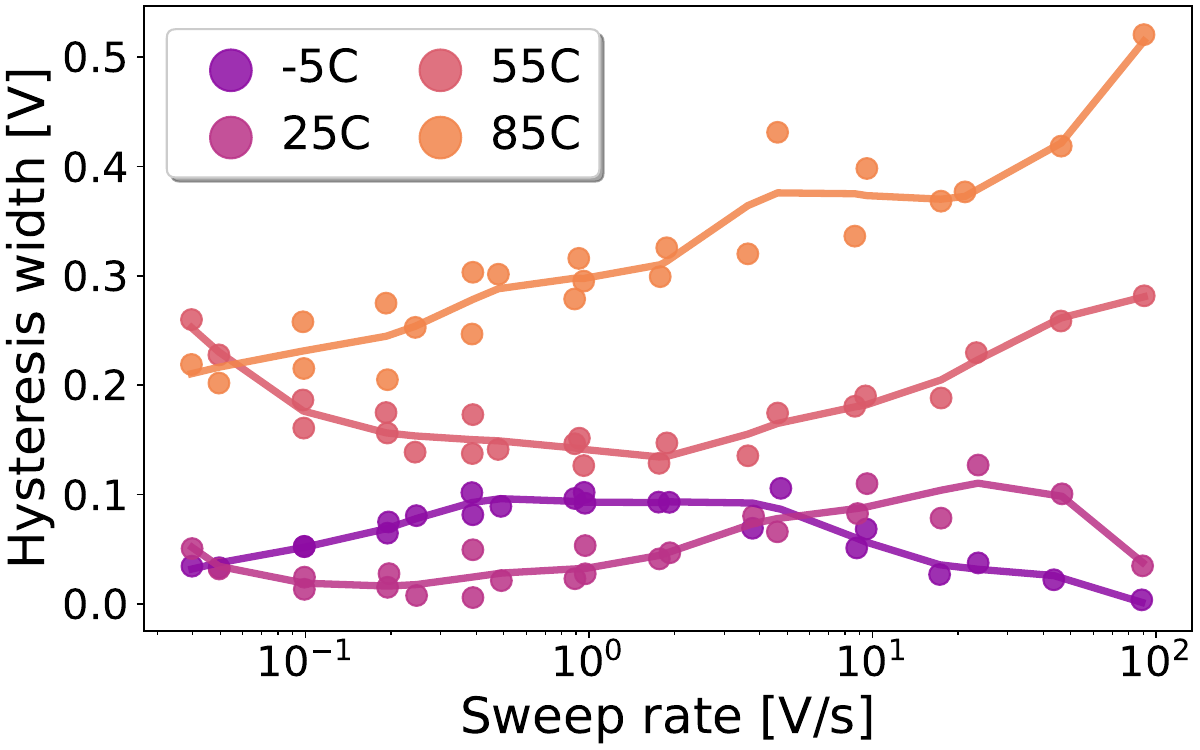}
\end{subfigure}
\begin{subfigure}[b]{.245\linewidth}
\includegraphics[width=1\linewidth]{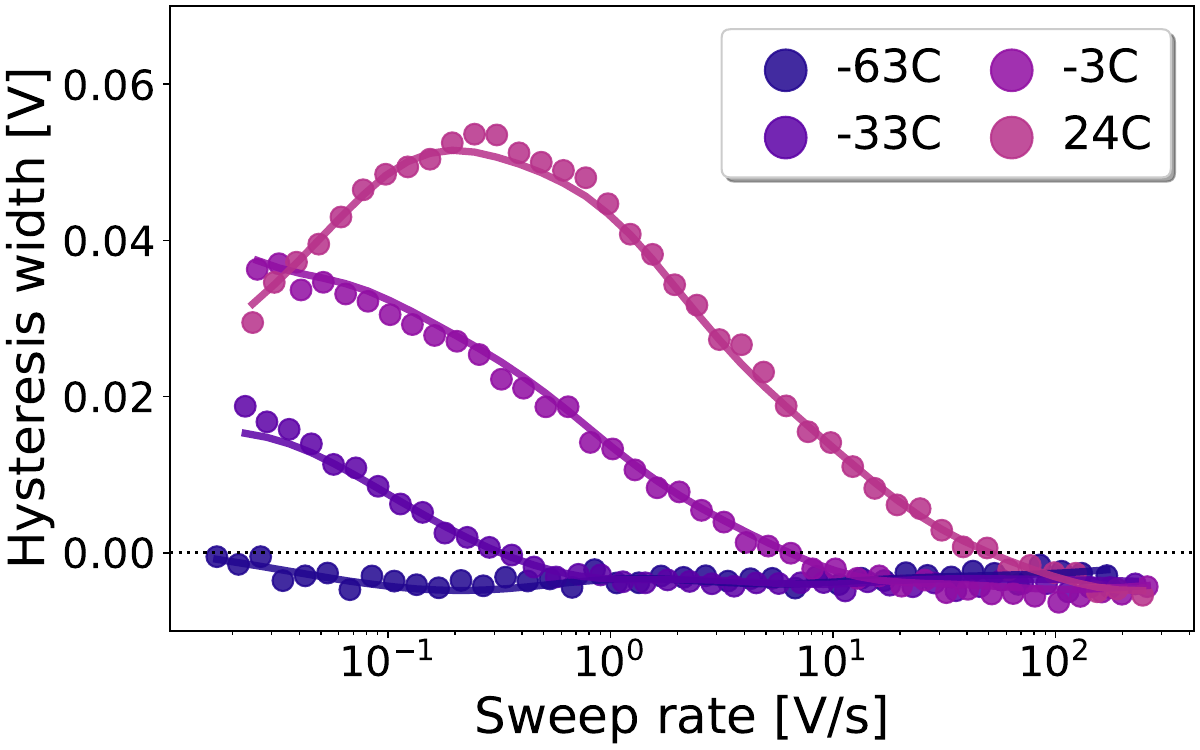}
\end{subfigure}
\begin{subfigure}[b]{.245\linewidth}
\includegraphics[width=1\linewidth]{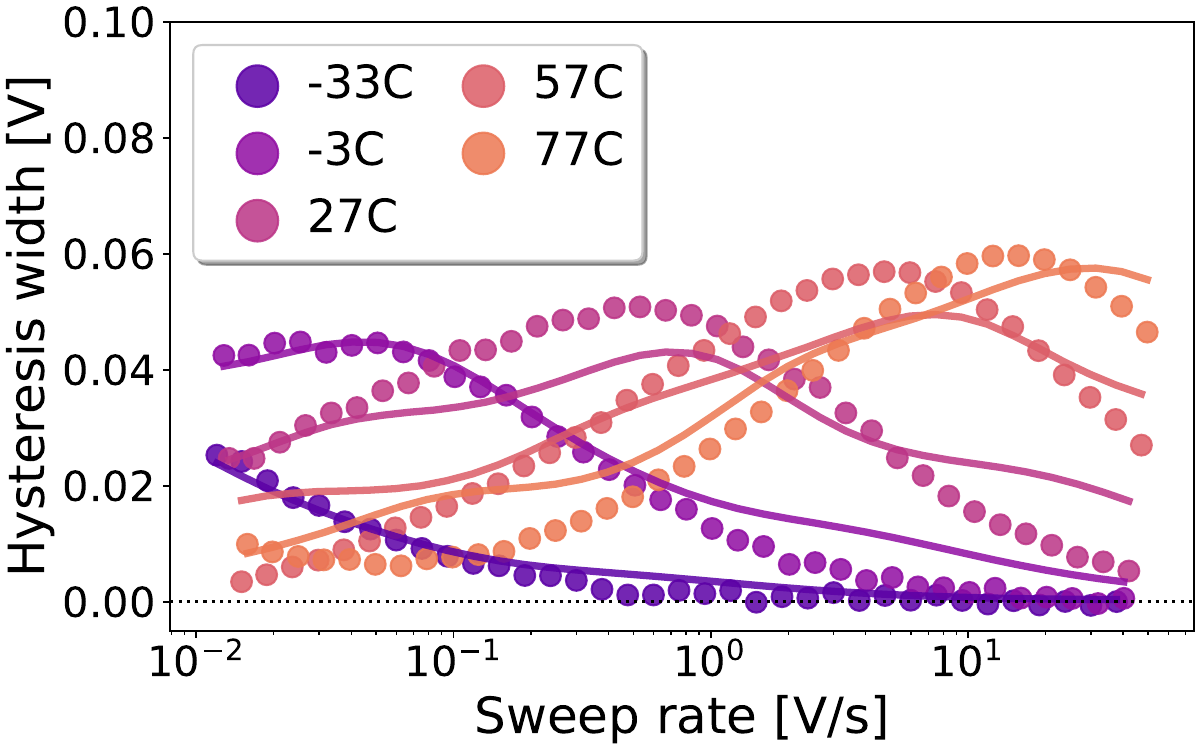}
\end{subfigure}

\begin{subfigure}[b]{.245\linewidth}
\includegraphics[width=1.0\linewidth]{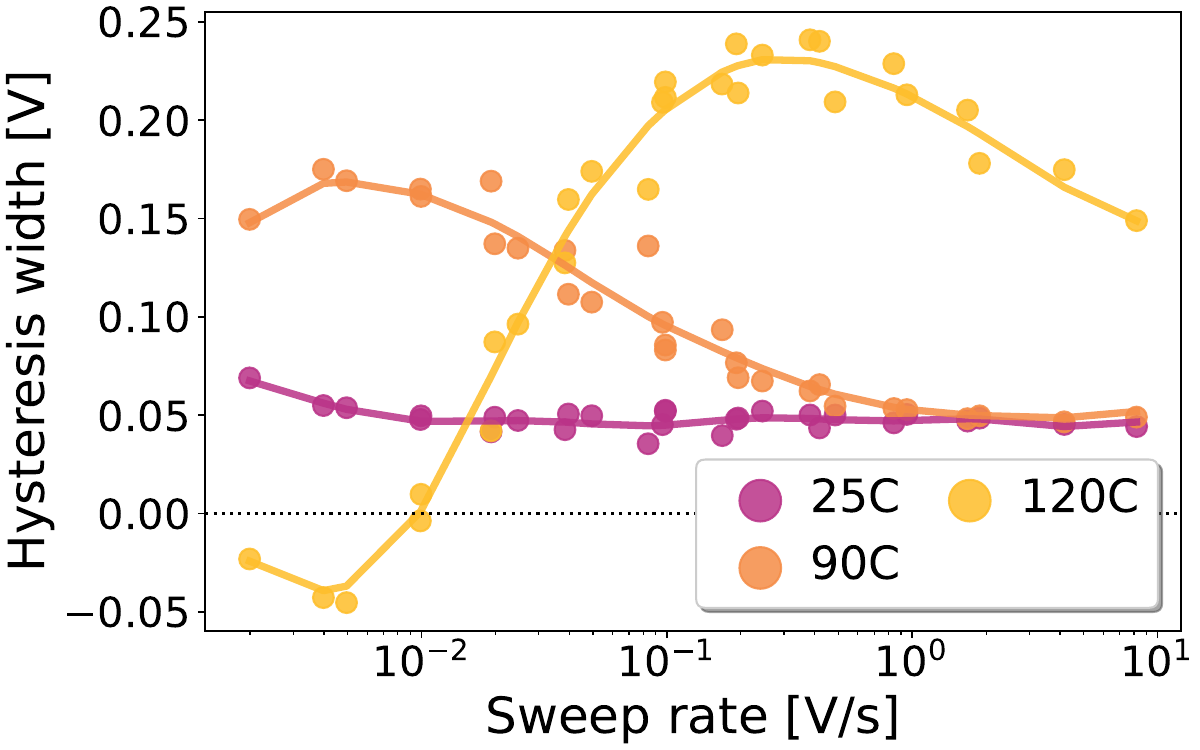}
\end{subfigure}
\begin{subfigure}[b]{.245\linewidth}
\includegraphics[width=1.0\linewidth]{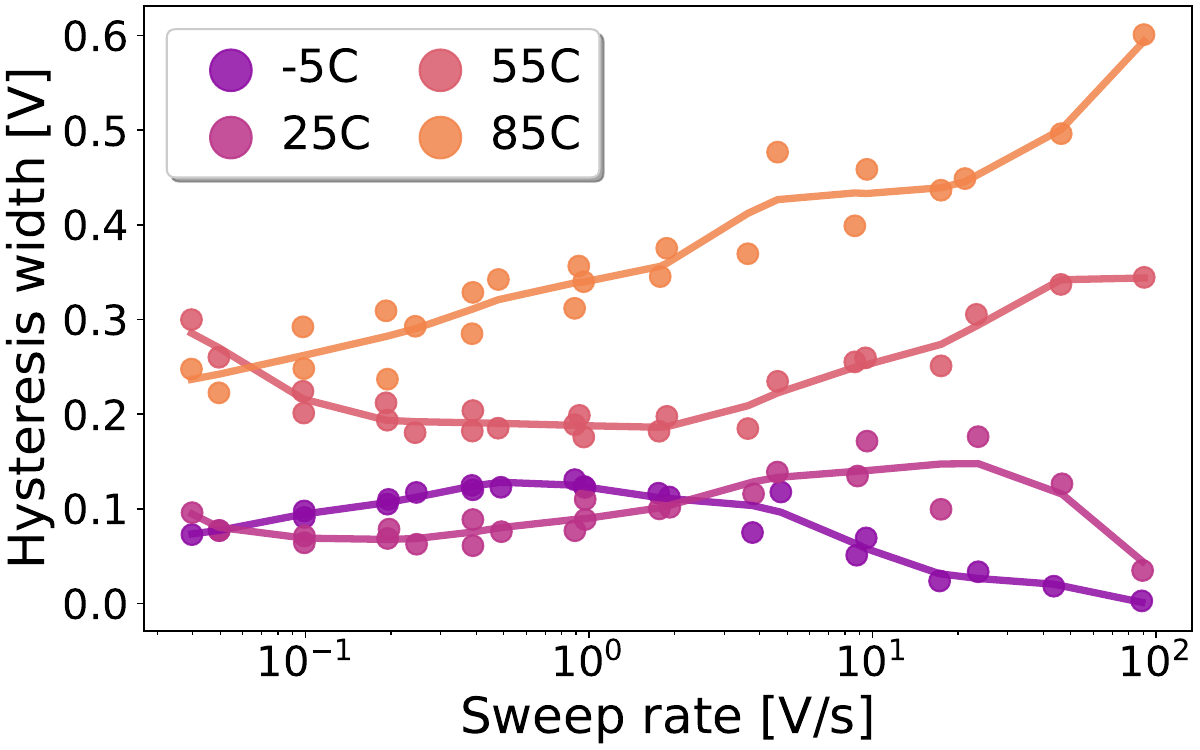}
\end{subfigure}
\begin{subfigure}[b]{.245\linewidth}
\includegraphics[width=1\linewidth]{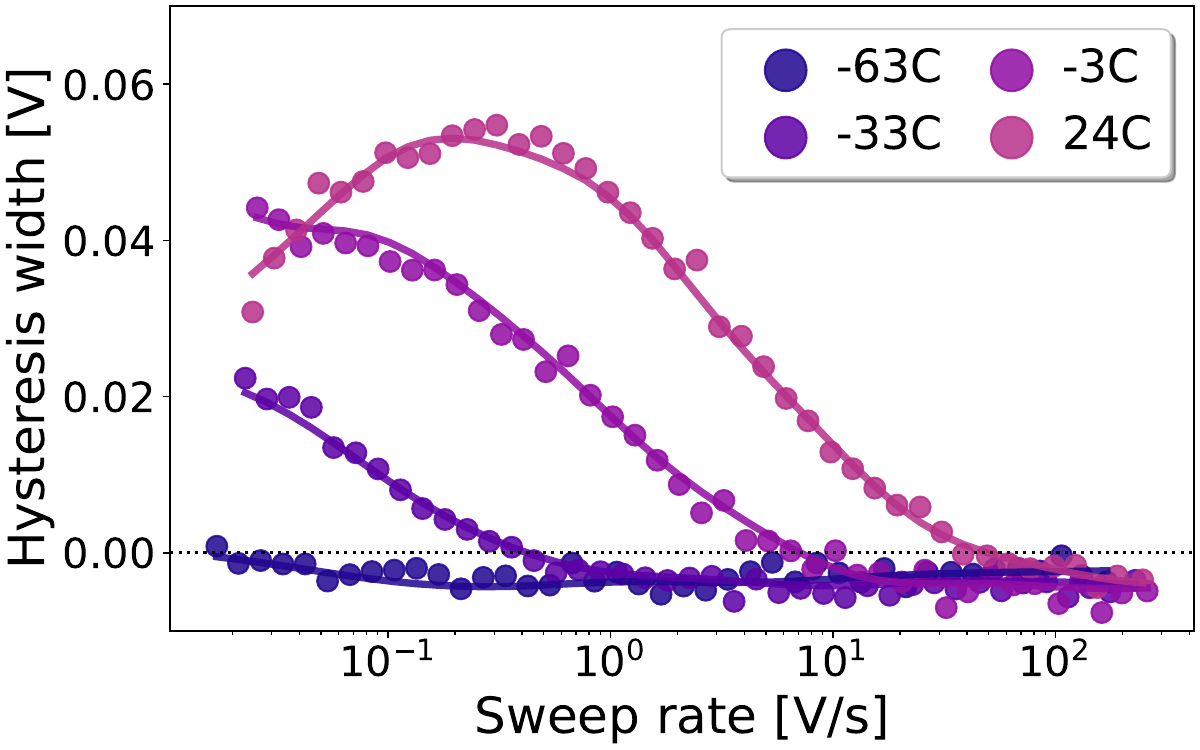}
\end{subfigure}
\begin{subfigure}[b]{.245\linewidth}
\includegraphics[width=1\linewidth]{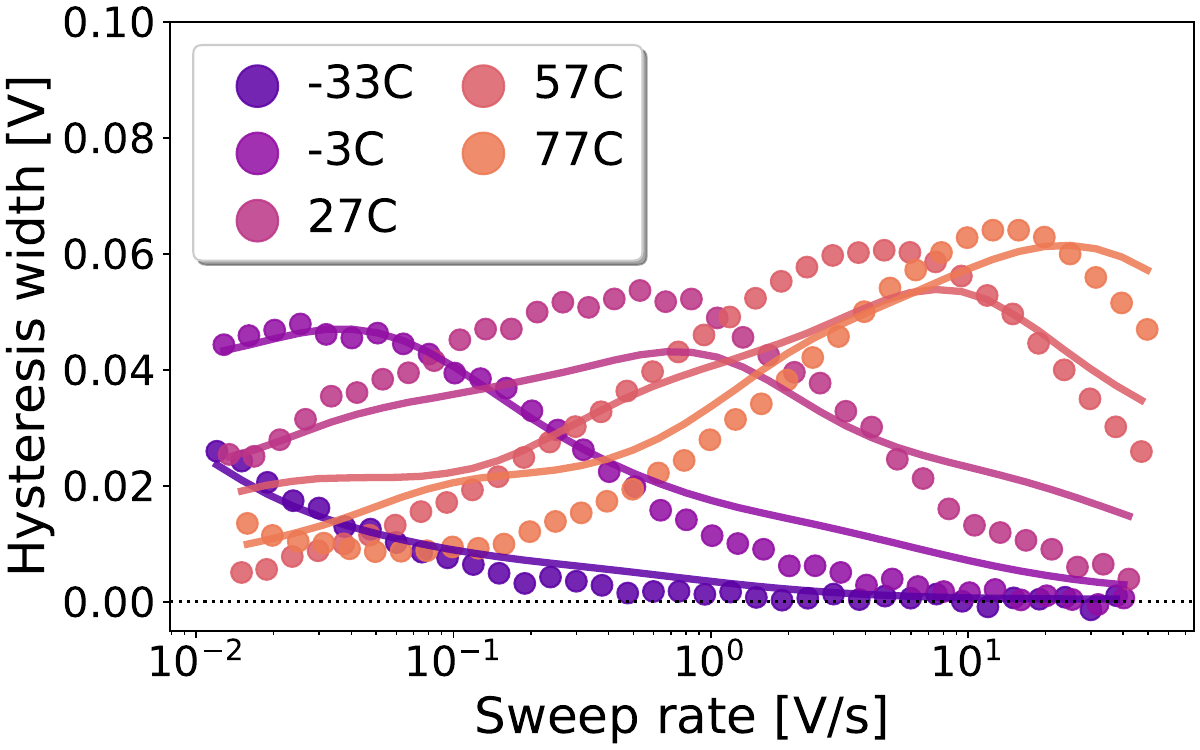}
\end{subfigure}

\begin{subfigure}[b]{.245\linewidth}
\includegraphics[width=1.0\linewidth]{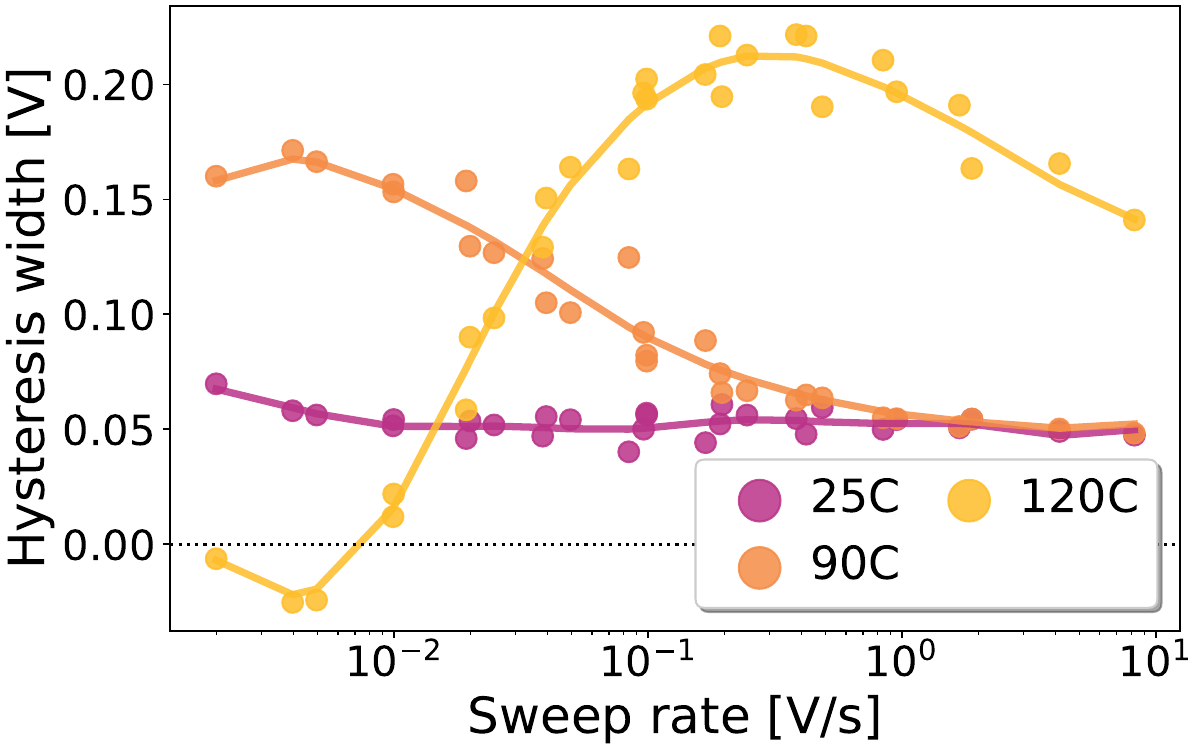}
\end{subfigure}
\begin{subfigure}[b]{.245\linewidth}
\includegraphics[width=1.0\linewidth]{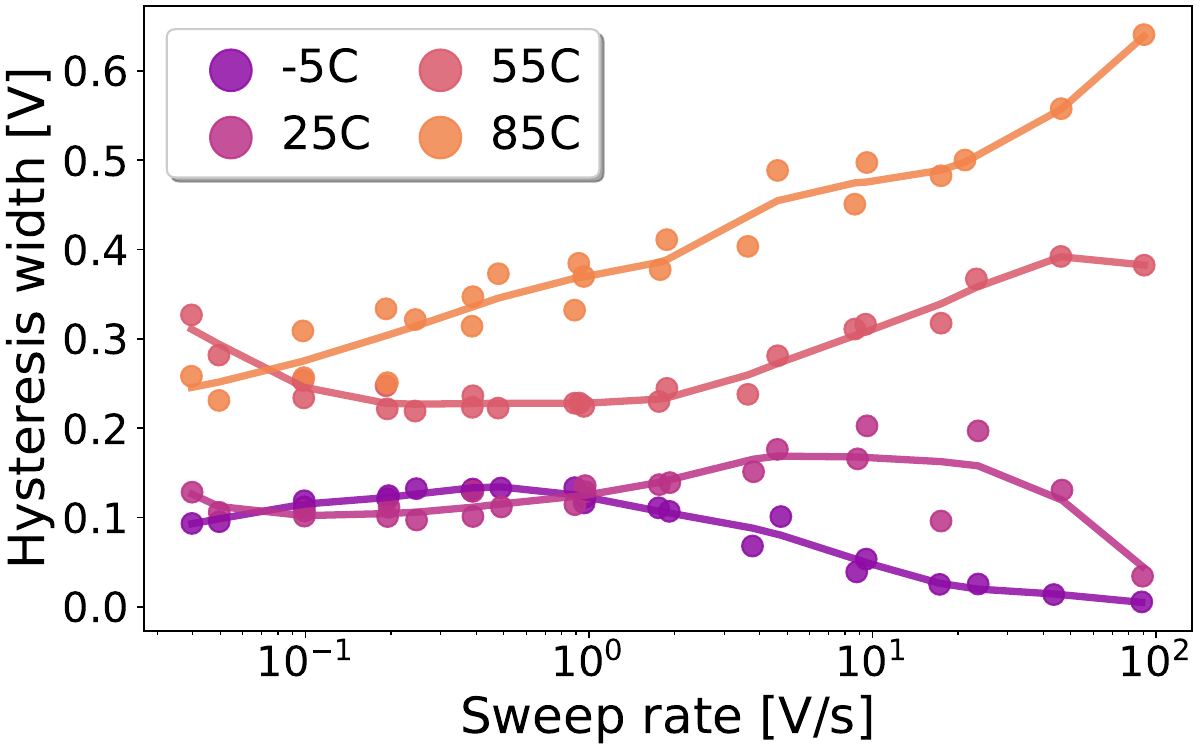}
\end{subfigure}
\begin{subfigure}[b]{.245\linewidth}
\includegraphics[width=1\linewidth]{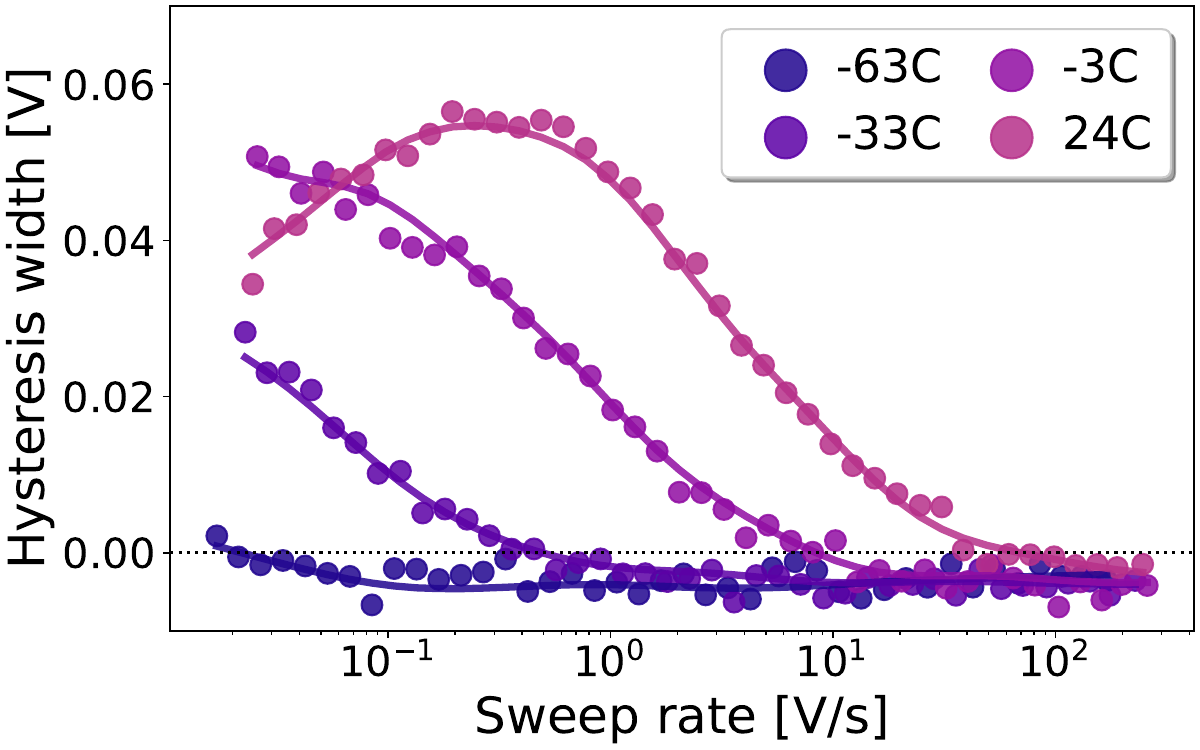}
\end{subfigure}
\begin{subfigure}[b]{.245\linewidth}
\includegraphics[width=1\linewidth]{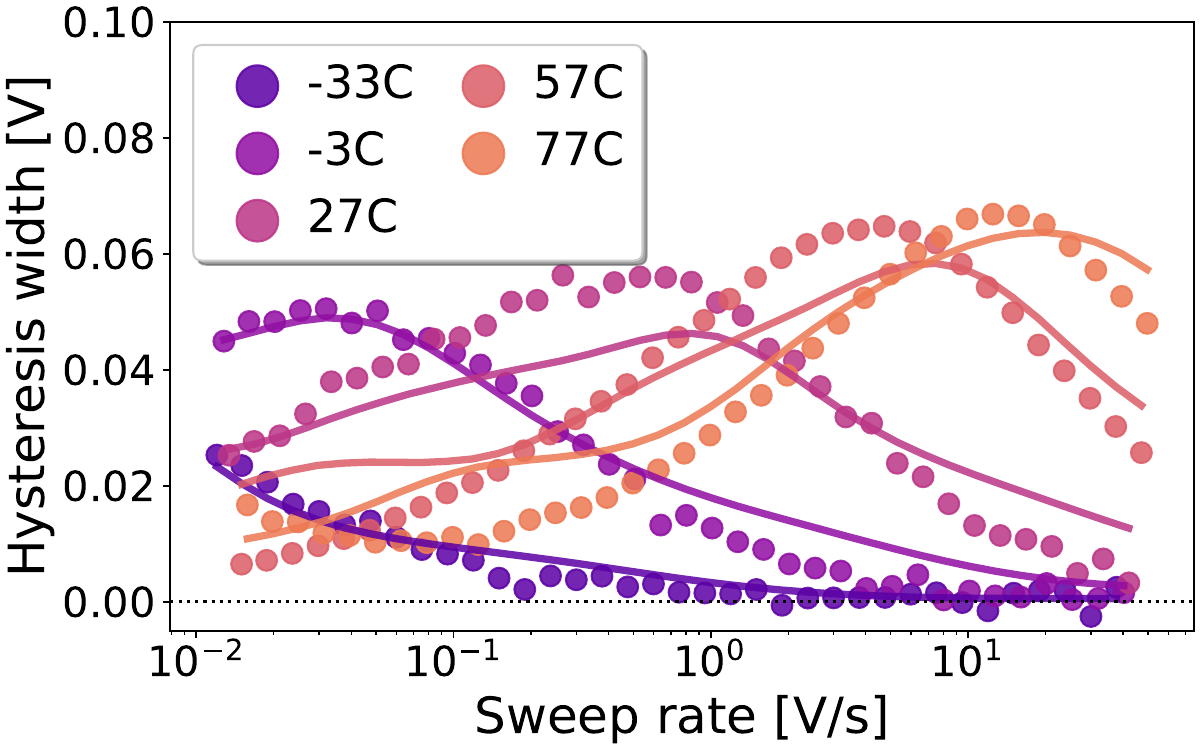}
\end{subfigure}

\begin{subfigure}[b]{.245\linewidth}
\includegraphics[width=1.0\linewidth]{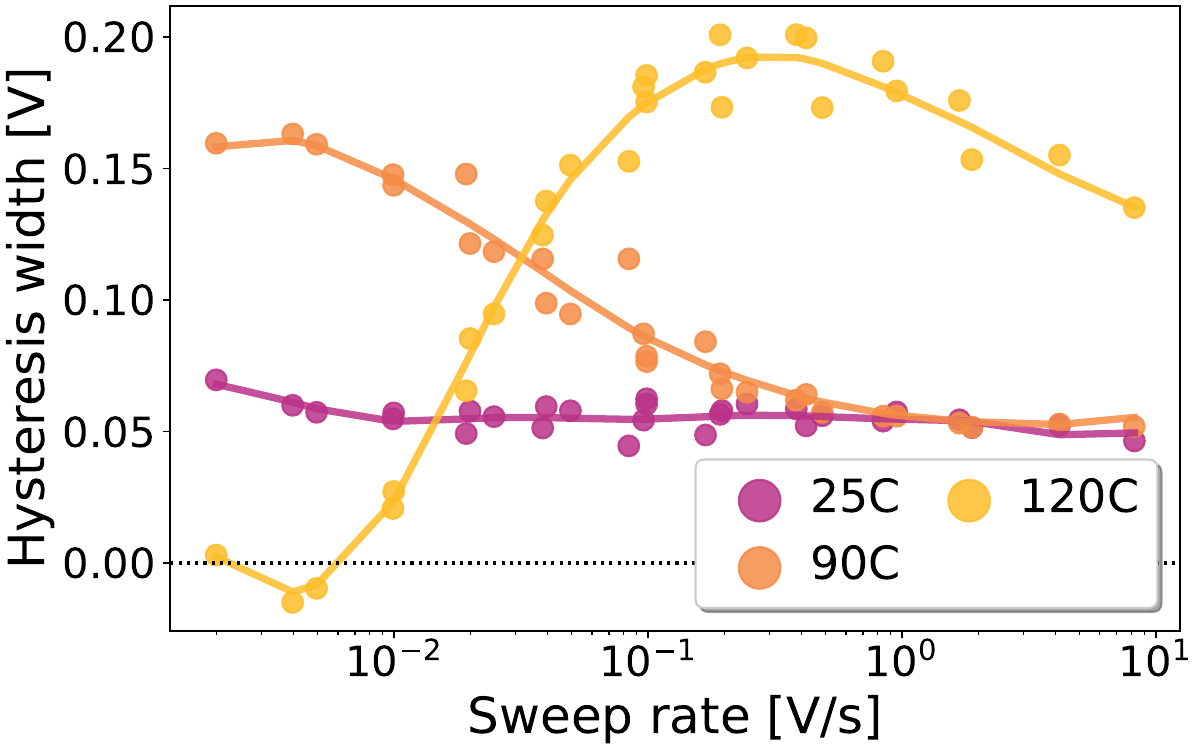}
\end{subfigure}
\begin{subfigure}[b]{.245\linewidth}
\includegraphics[width=1.0\linewidth]{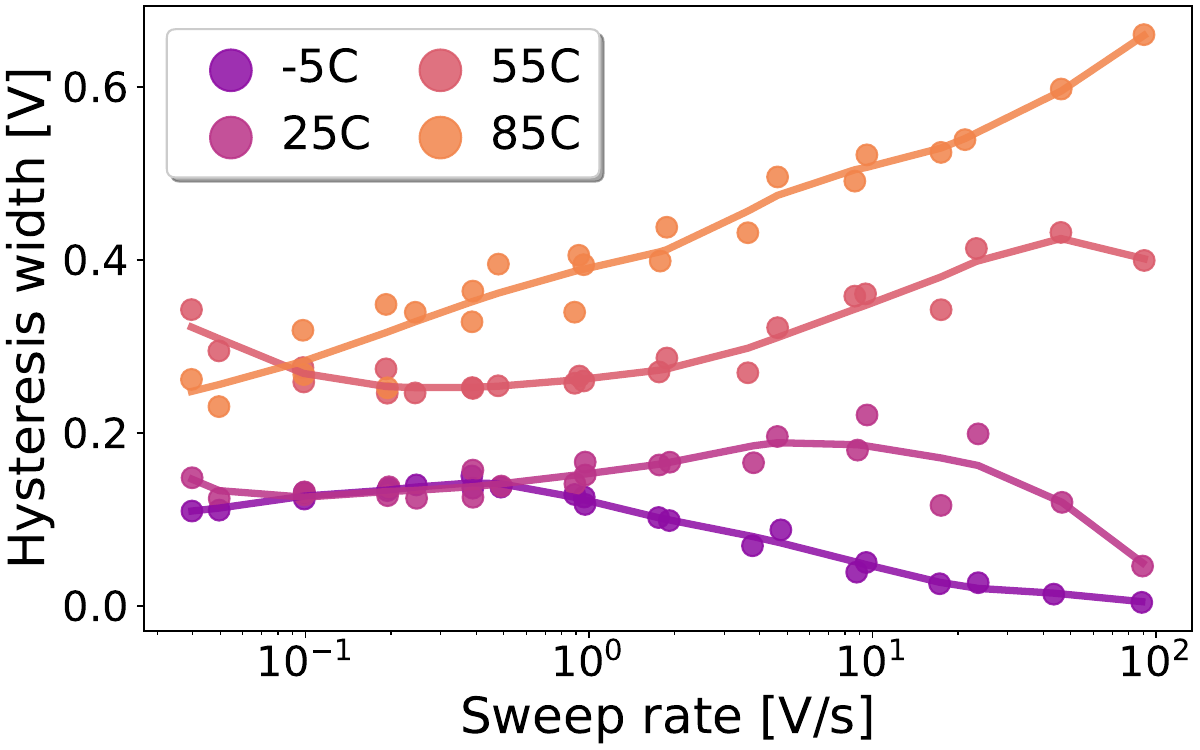}
\end{subfigure}
\begin{subfigure}[b]{.245\linewidth}
\includegraphics[width=1\linewidth]{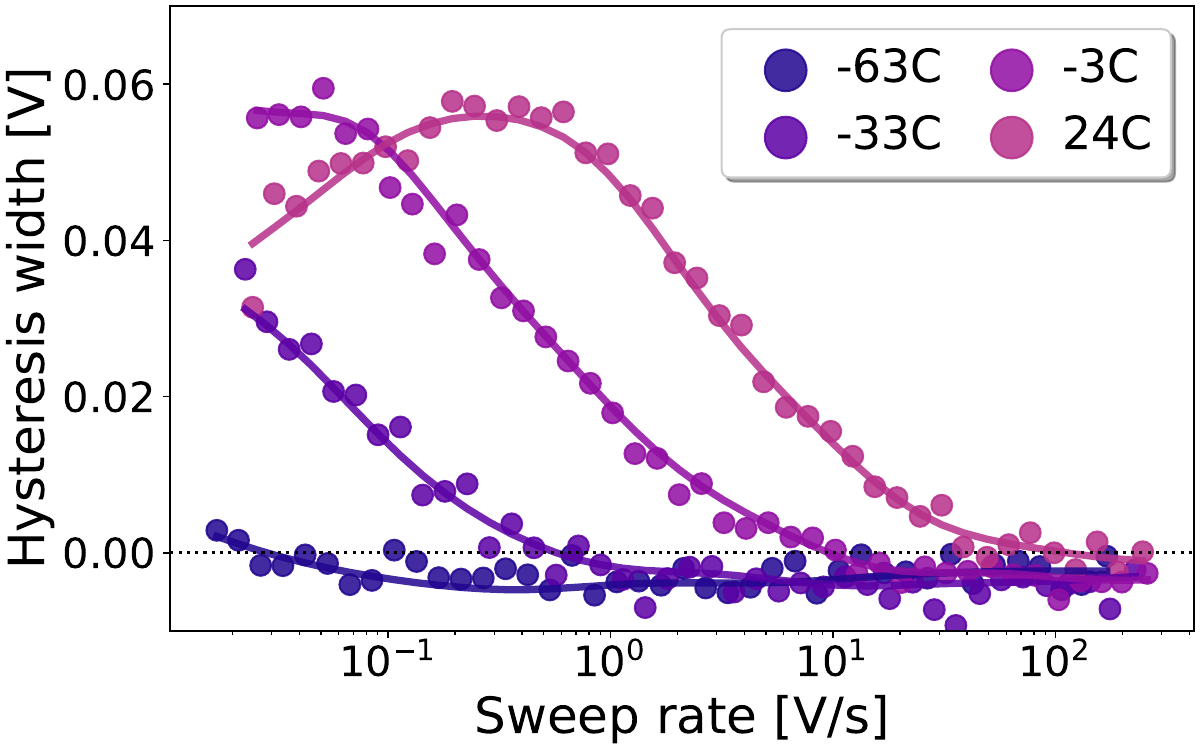}
\end{subfigure}
\begin{subfigure}[b]{.245\linewidth}
\includegraphics[width=1\linewidth]{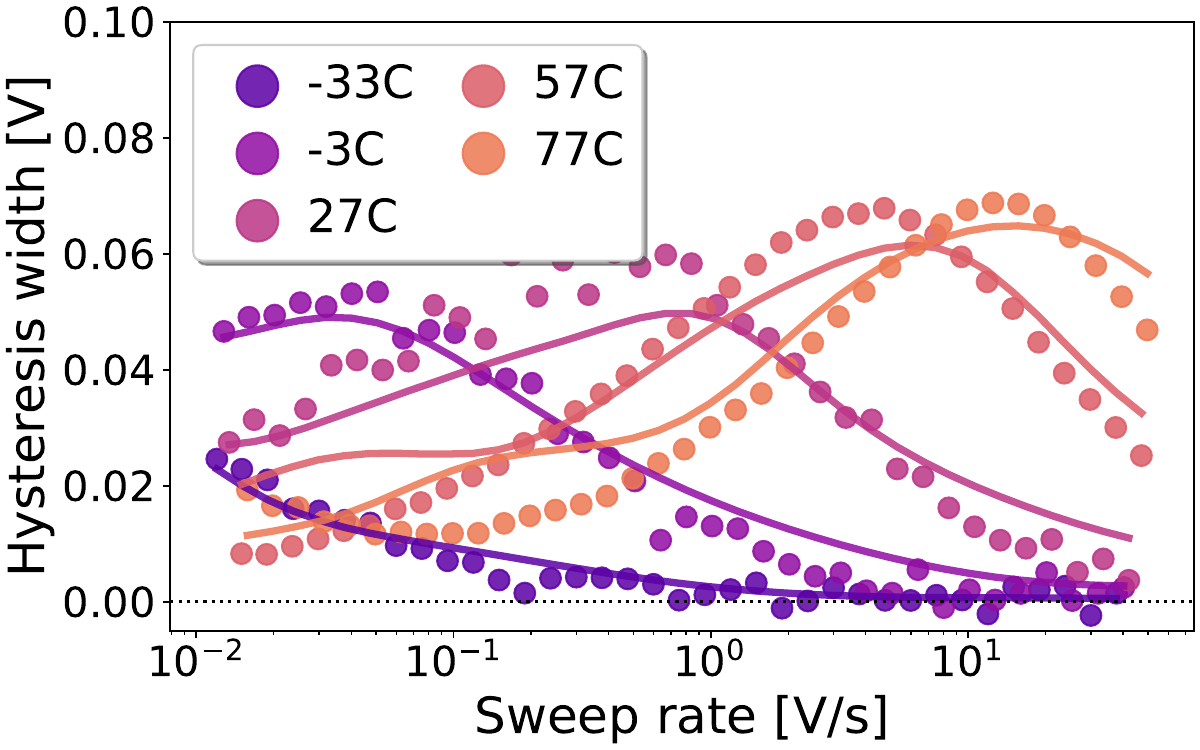}
\end{subfigure}

\begin{subfigure}[b]{.245\linewidth}
\includegraphics[width=1.0\linewidth]{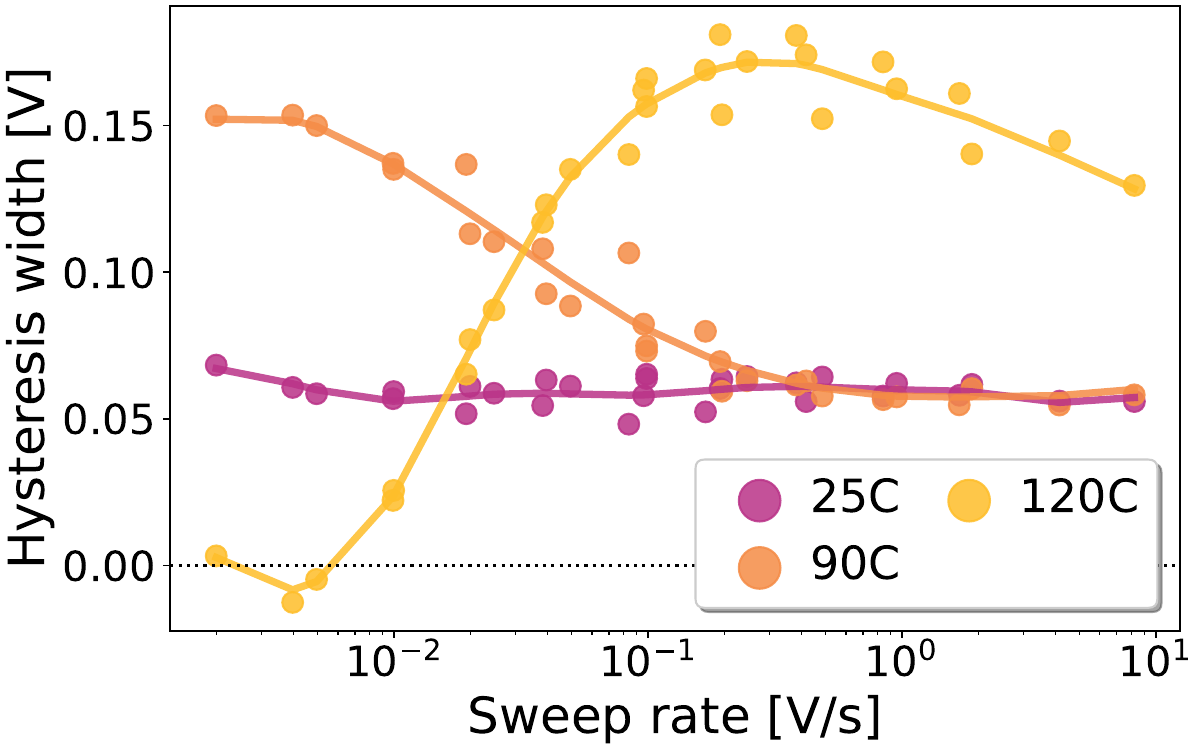}
\end{subfigure}
\begin{subfigure}[b]{.245\linewidth}
\includegraphics[width=1.0\linewidth]{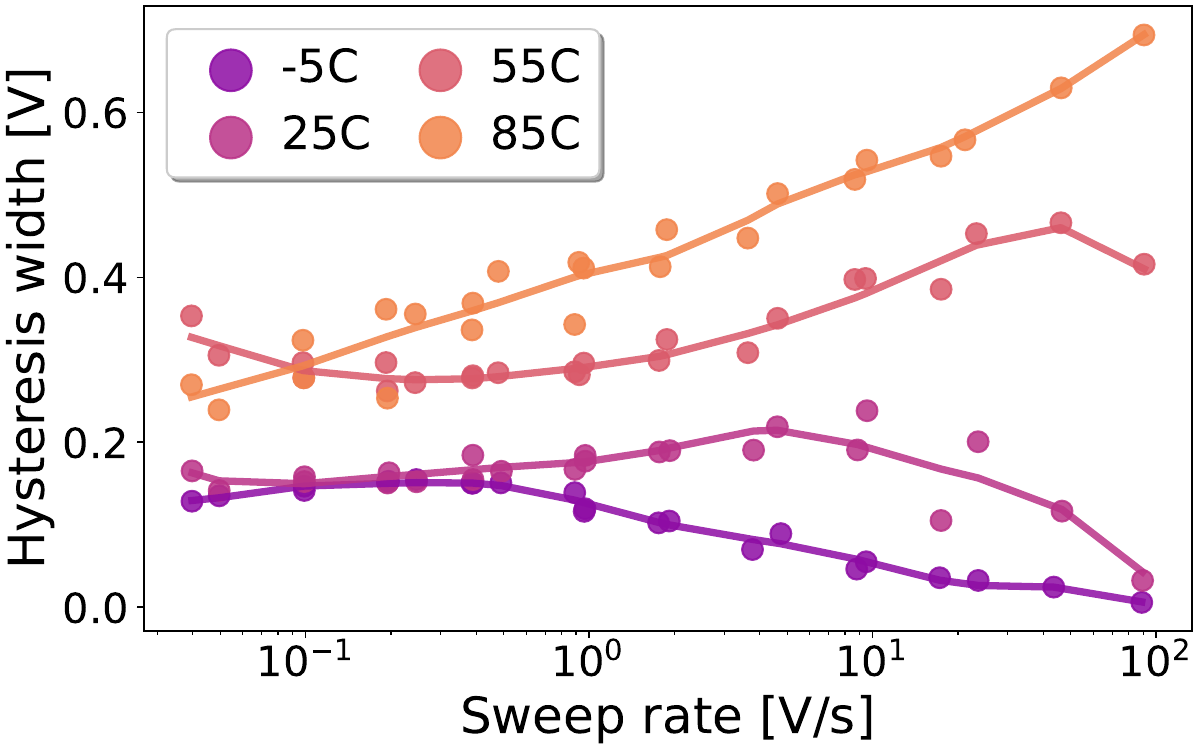}
\end{subfigure}
\begin{subfigure}[b]{.245\linewidth}
\includegraphics[width=1\linewidth]{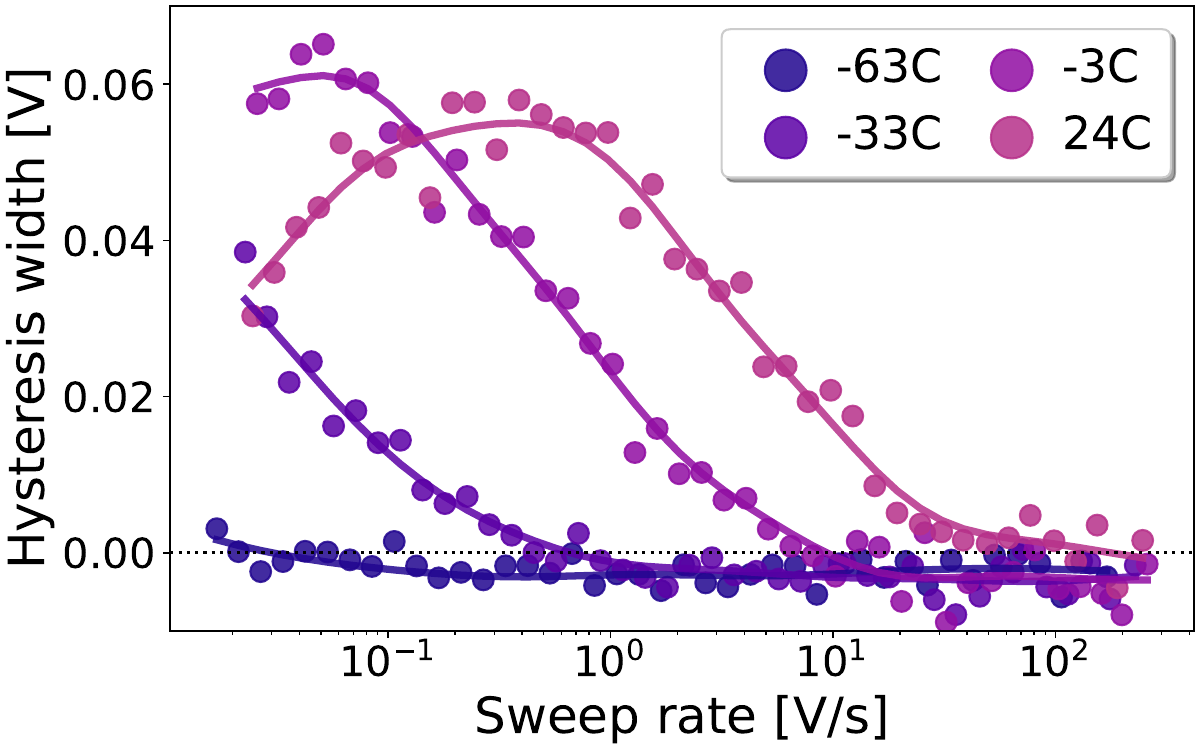}
\end{subfigure}
\begin{subfigure}[b]{.245\linewidth}
\includegraphics[width=1\linewidth]{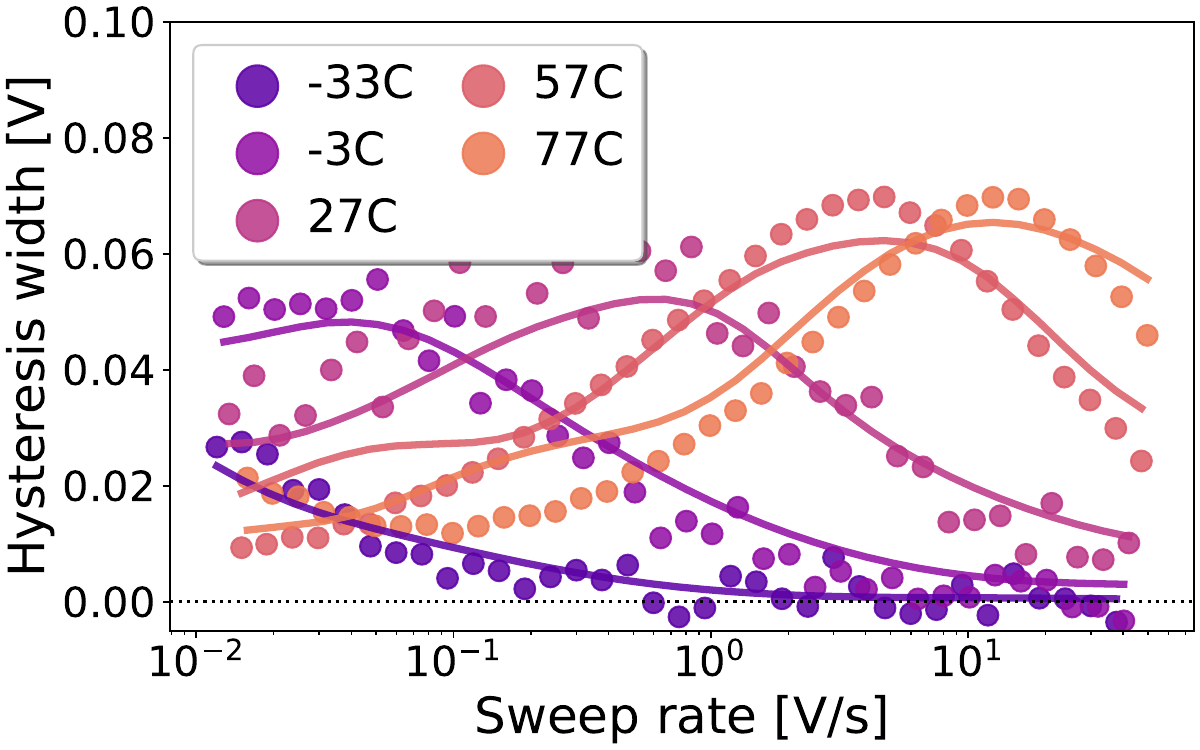}
\end{subfigure}

\begin{subfigure}[b]{.245\linewidth}
\includegraphics[width=1.0\linewidth]{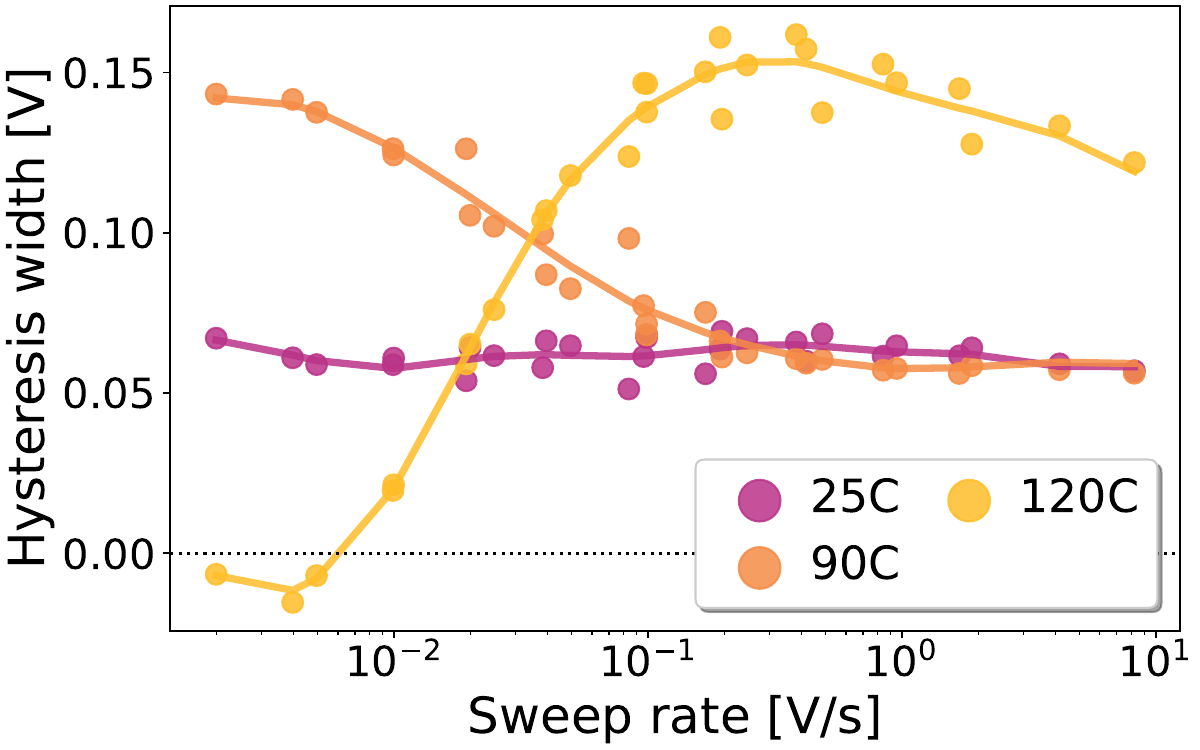}
\caption{}

\end{subfigure}
\begin{subfigure}[b]{.245\linewidth}
\includegraphics[width=1.0\linewidth]{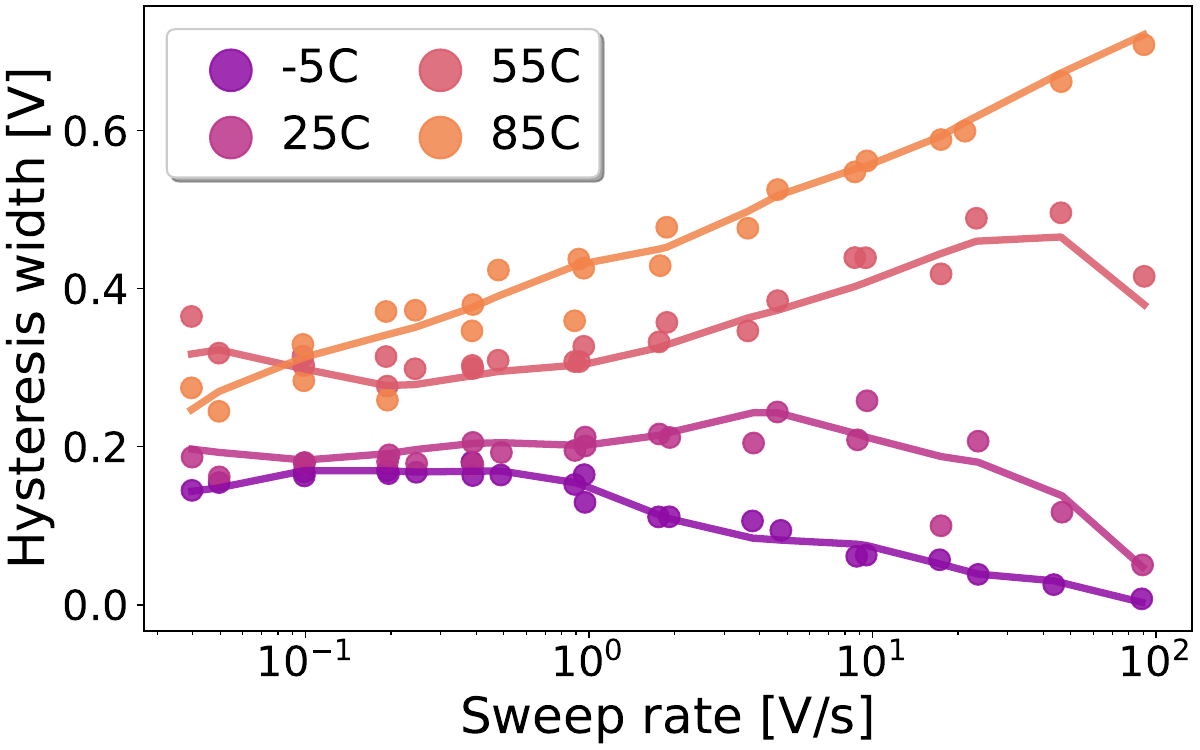}
\caption{}

\end{subfigure}
\begin{subfigure}[b]{.245\linewidth}
\includegraphics[width=1\linewidth]{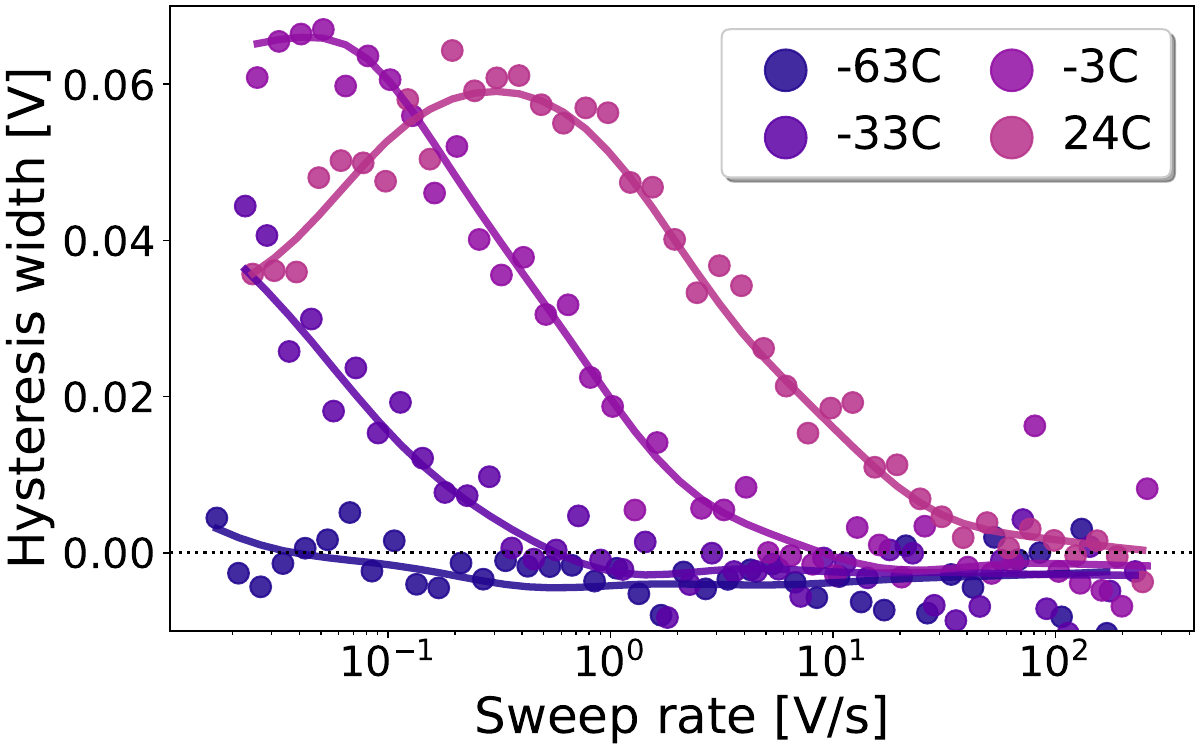}
\caption{}

\end{subfigure}
\begin{subfigure}[b]{.245\linewidth}
\includegraphics[width=1\linewidth]{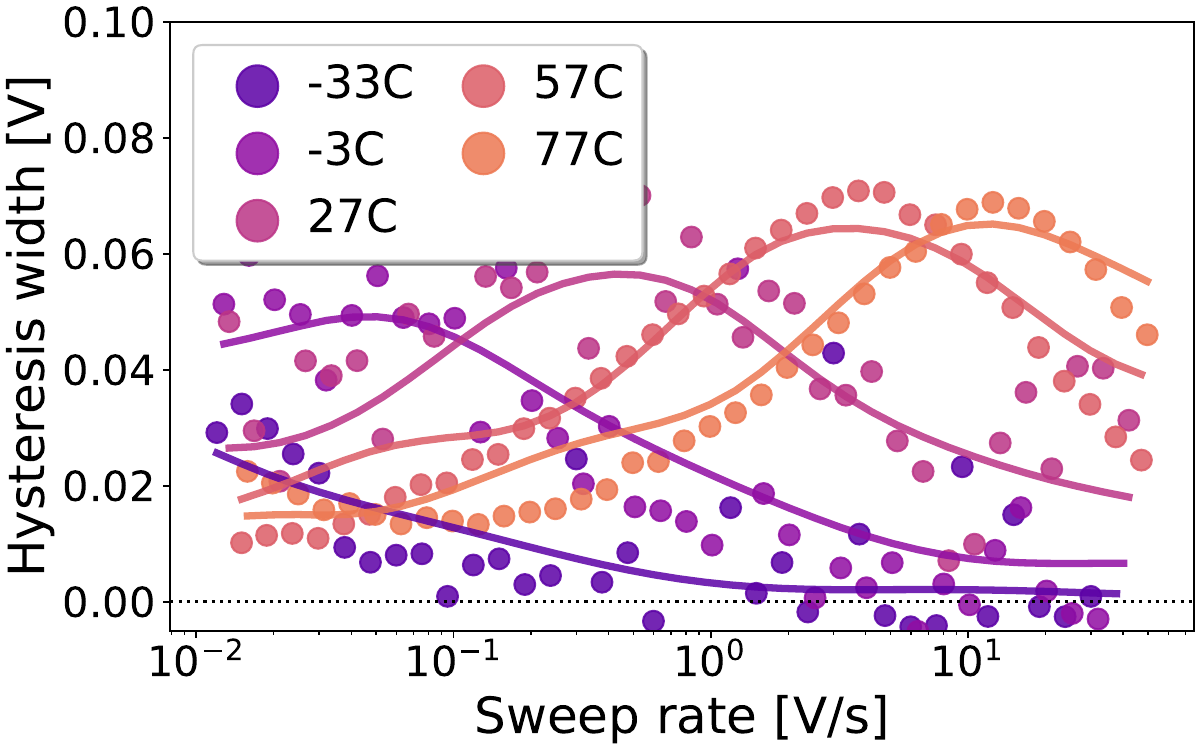}
\caption{}

\end{subfigure}
\caption{Simulated hysteresis at different readouts (top to down: the highest to the lowest \ith{}) as a function of sweep-rate at various temperatures for \textbf{(a)} the \plan{}, \textbf{(b)} the \fin{},\textbf{(c)} the \sgaa{}, and \textbf{(d)} the  \gaa{}.}
\end{figure}

\newpage
\subsection{ESiD Results for Individual Readouts}\label{supp:hys_eter}
\begin{figure}[!h]
\begin{subfigure}[b]{.245\linewidth}
\includegraphics[width=1.0\linewidth]{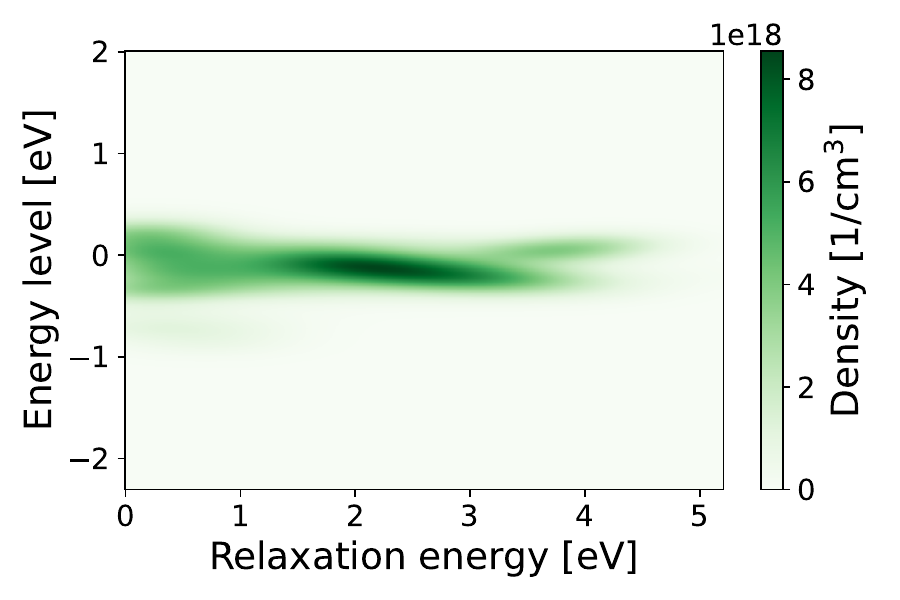}
\end{subfigure}
\begin{subfigure}[b]{.245\linewidth}
\includegraphics[width=1.0\linewidth]{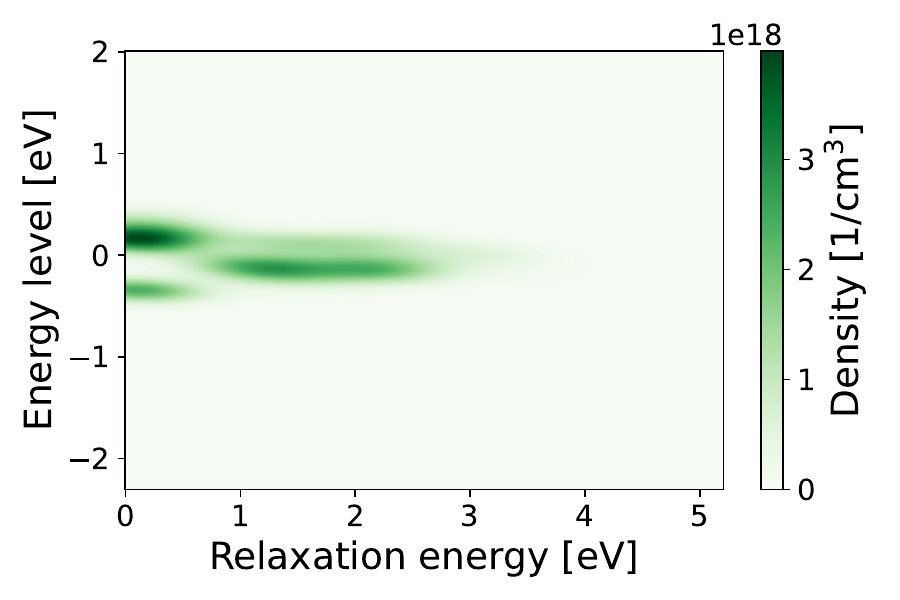}
\end{subfigure}
\begin{subfigure}[b]{.245\linewidth}
\includegraphics[width=1\linewidth]{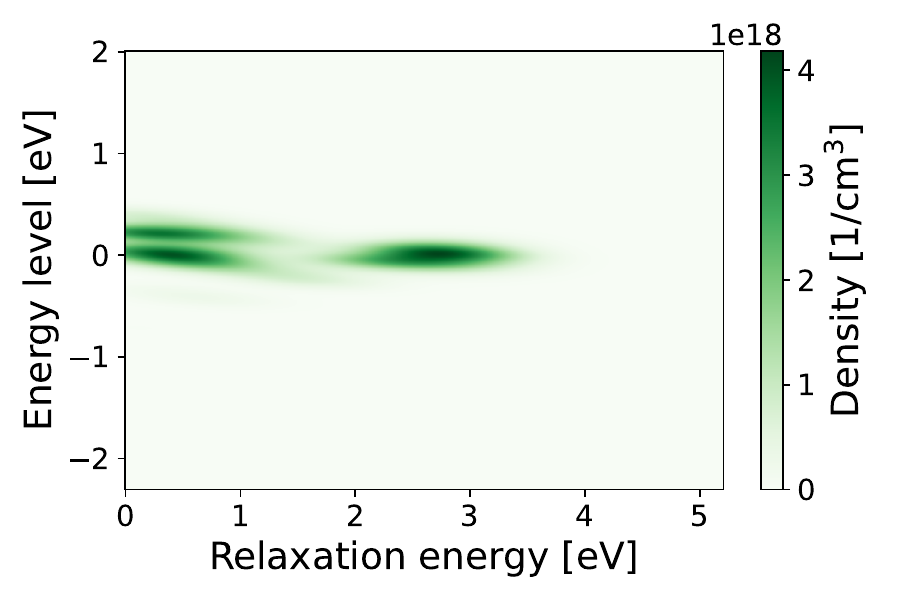}
\end{subfigure}
\begin{subfigure}[b]{.245\linewidth}
\includegraphics[width=1\linewidth]{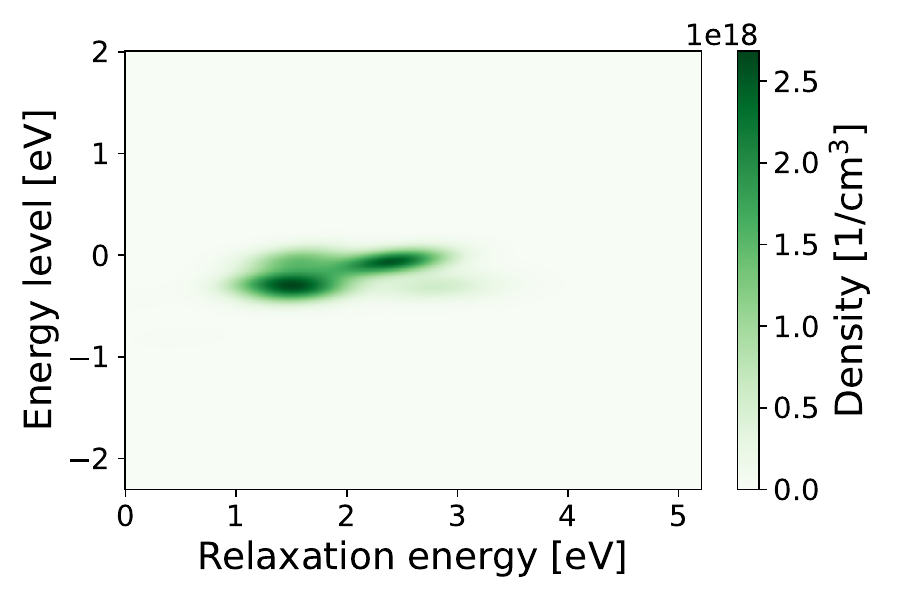}

\end{subfigure}
\begin{subfigure}[b]{.245\linewidth}
\includegraphics[width=1.0\linewidth]{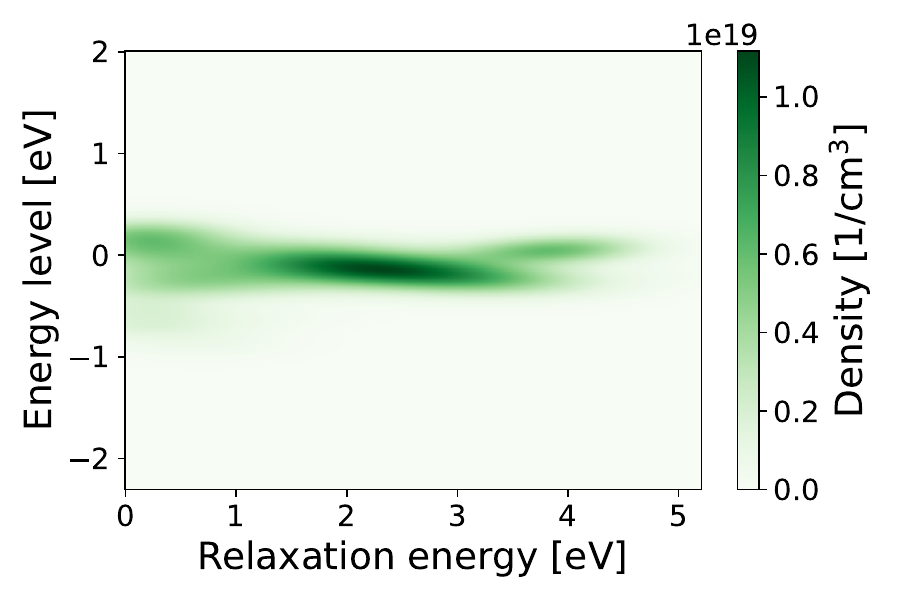}
\end{subfigure}
\begin{subfigure}[b]{.245\linewidth}
\includegraphics[width=1.0\linewidth]{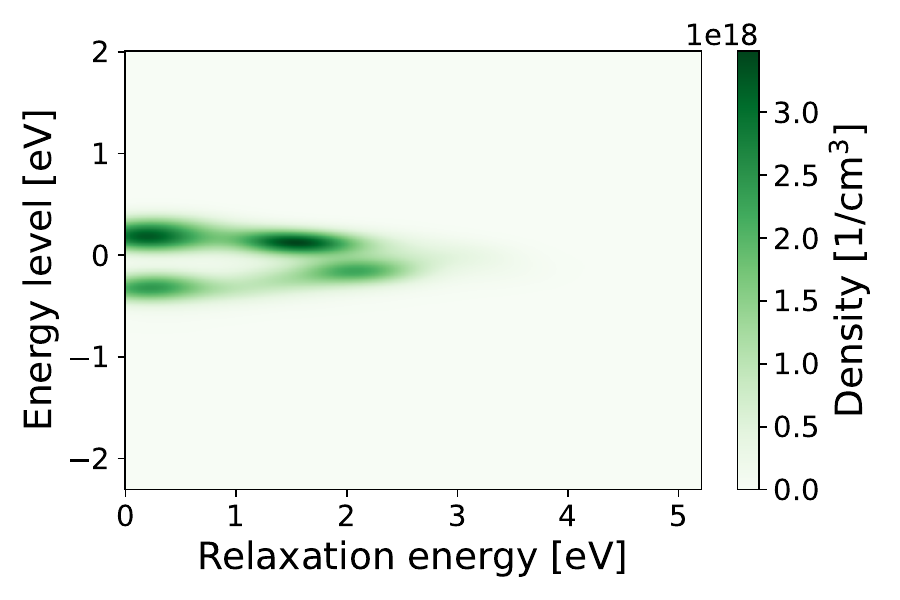}
\end{subfigure}
\begin{subfigure}[b]{.245\linewidth}
\includegraphics[width=1\linewidth]{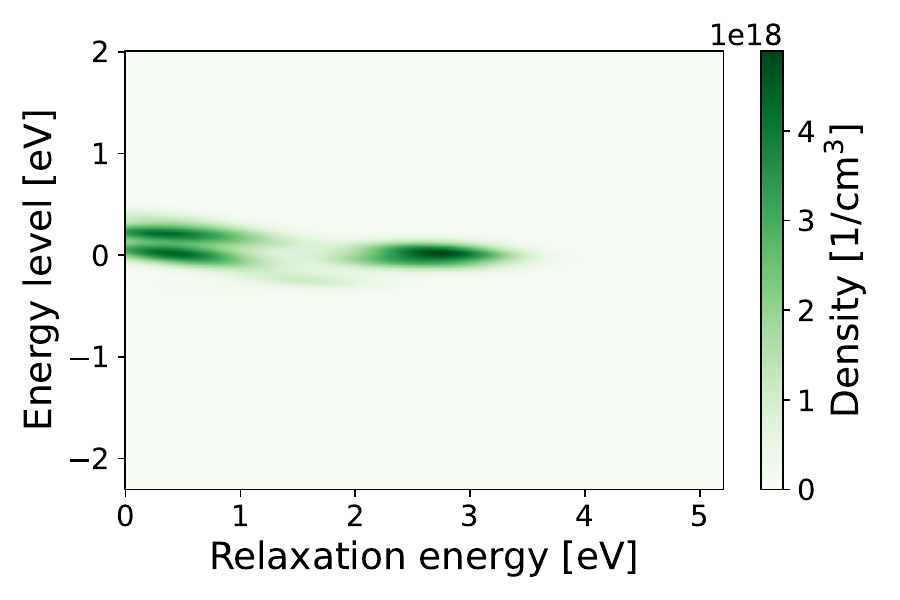}
\end{subfigure}
\begin{subfigure}[b]{.245\linewidth}
\includegraphics[width=1\linewidth]{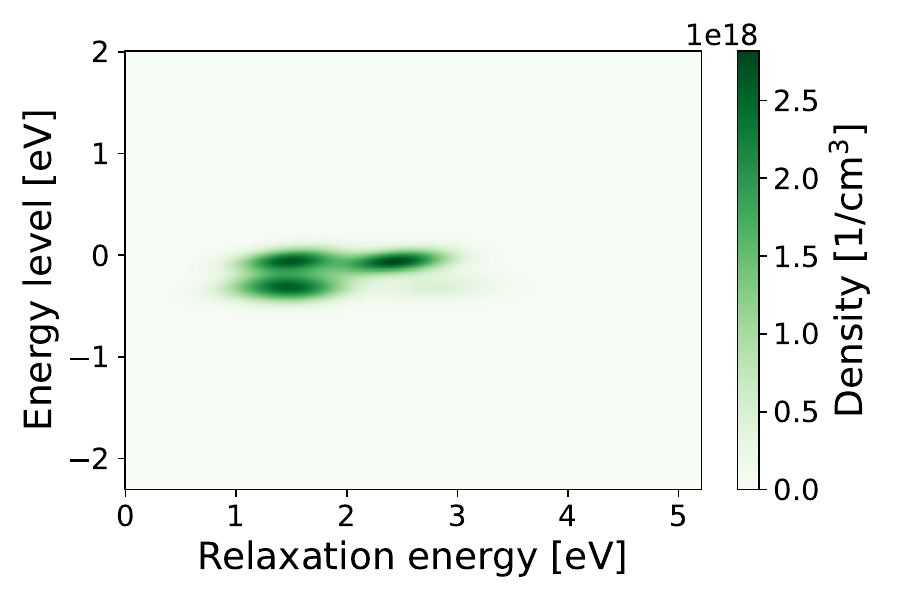}
\end{subfigure}

\begin{subfigure}[b]{.245\linewidth}
\includegraphics[width=1.0\linewidth]{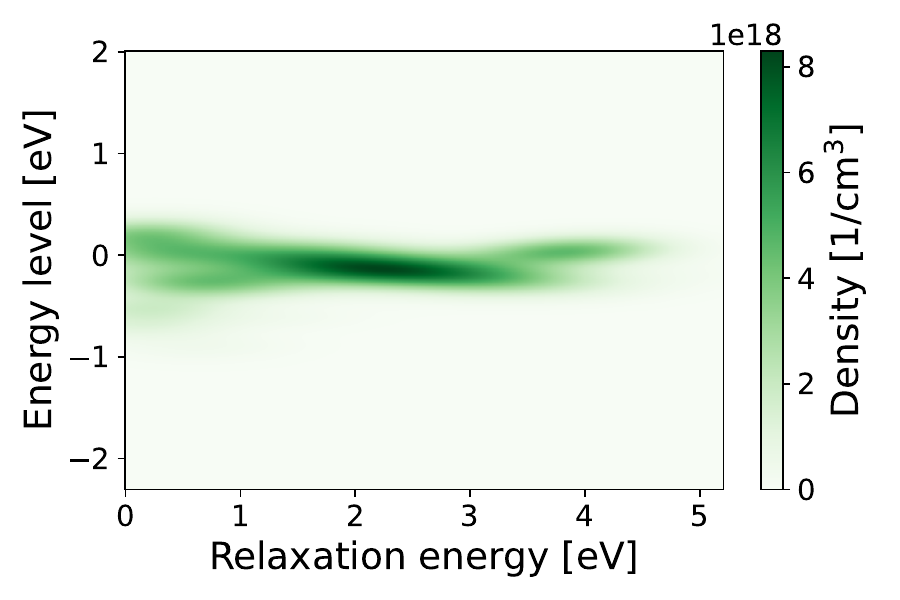}
\end{subfigure}
\begin{subfigure}[b]{.245\linewidth}
\includegraphics[width=1.0\linewidth]{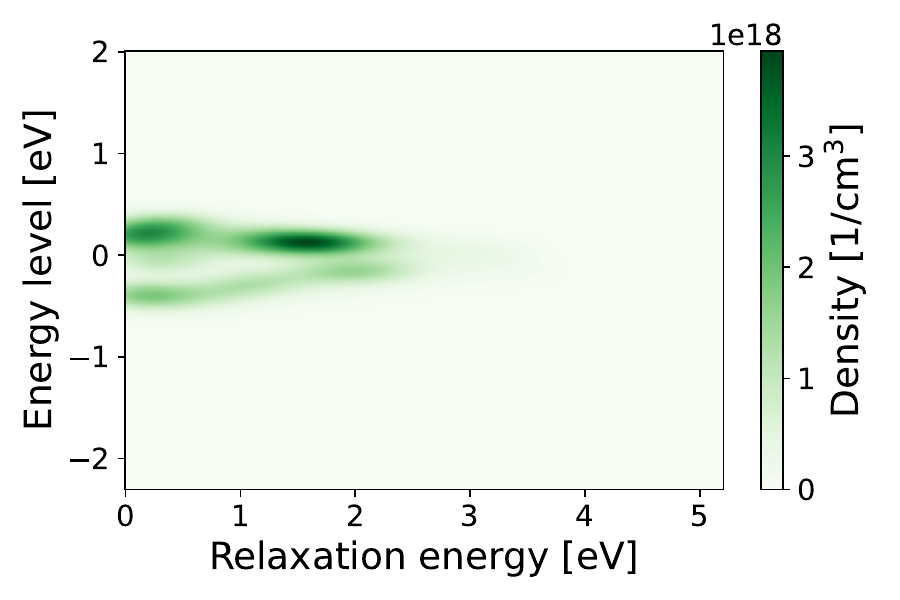}
\end{subfigure}
\begin{subfigure}[b]{.245\linewidth}
\includegraphics[width=1\linewidth]{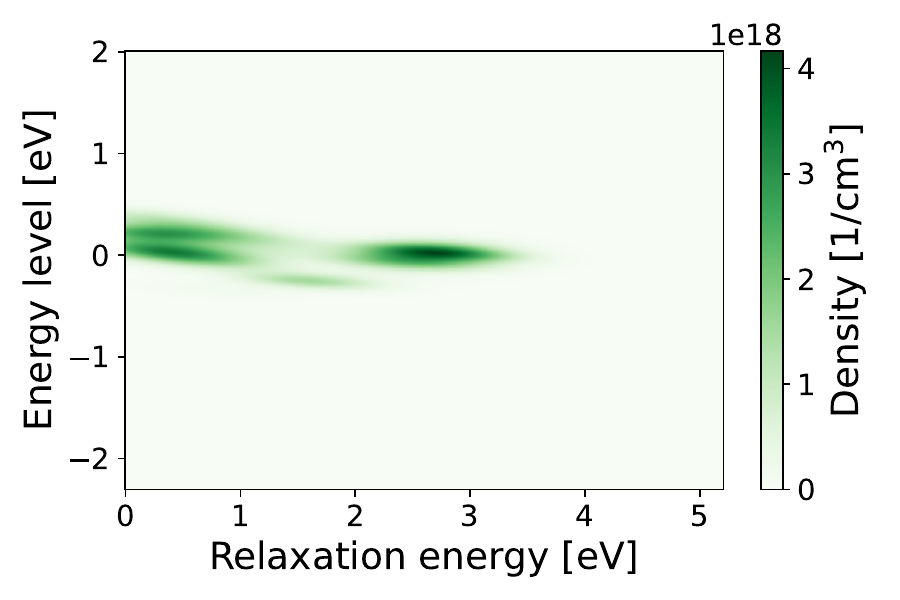}
\end{subfigure}
\begin{subfigure}[b]{.245\linewidth}
\includegraphics[width=1\linewidth]{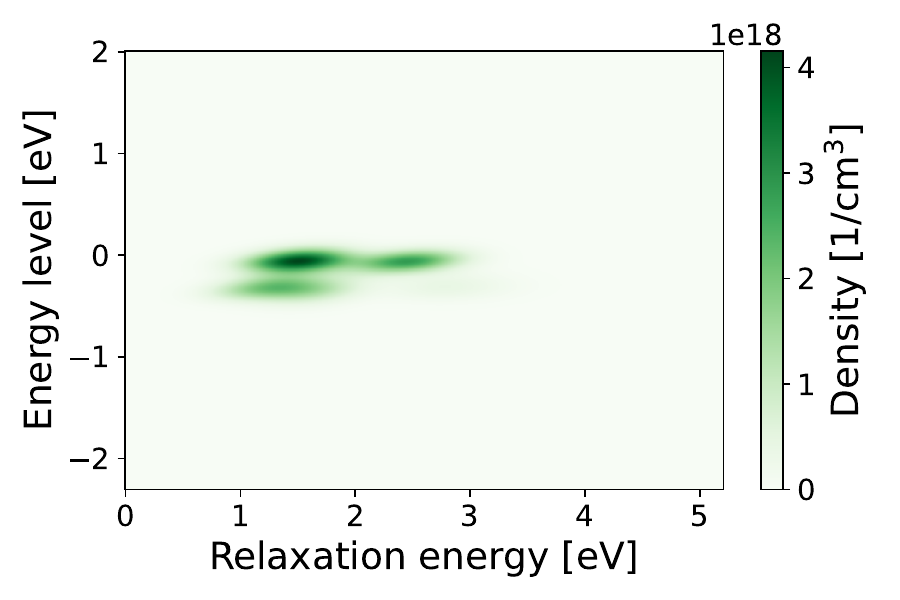}
\end{subfigure}

\begin{subfigure}[b]{.245\linewidth}
\includegraphics[width=1.0\linewidth]{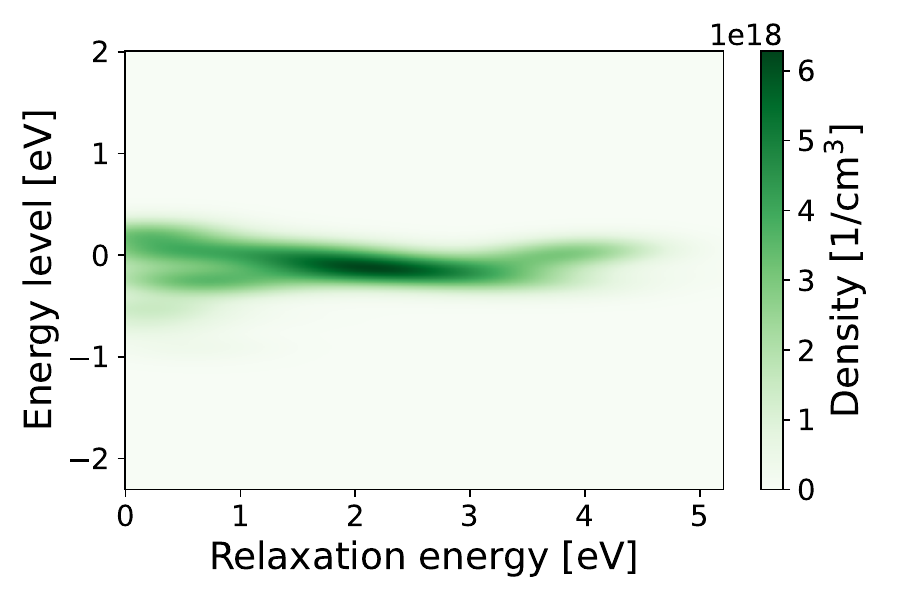}
\end{subfigure}
\begin{subfigure}[b]{.245\linewidth}
\includegraphics[width=1.0\linewidth]{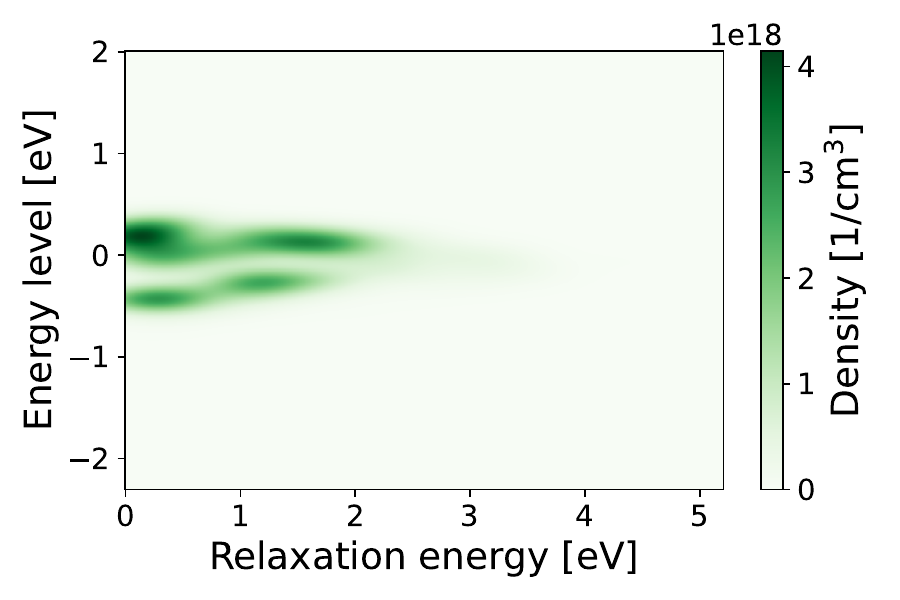}
\end{subfigure}
\begin{subfigure}[b]{.245\linewidth}
\includegraphics[width=1\linewidth]{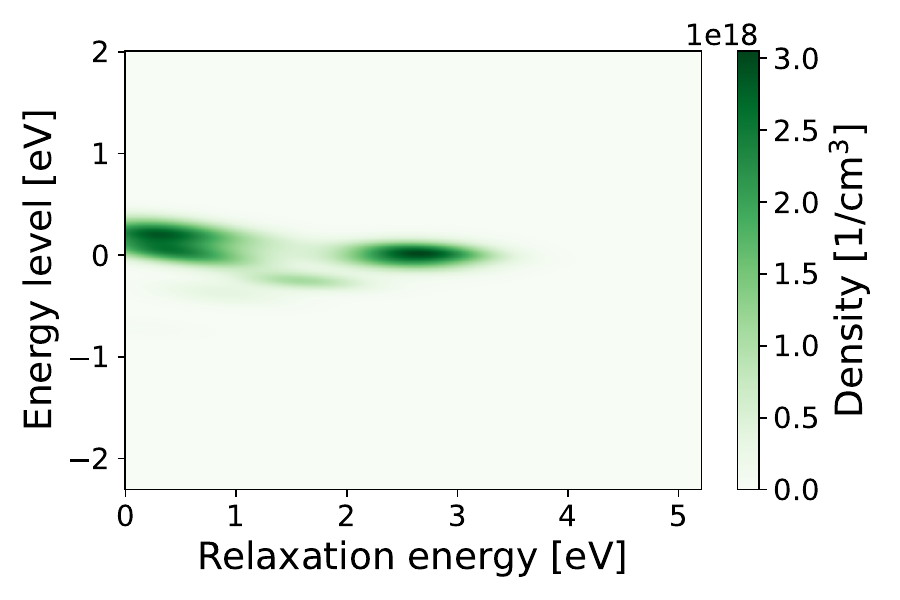}
\end{subfigure}
\begin{subfigure}[b]{.245\linewidth}
\includegraphics[width=1\linewidth]{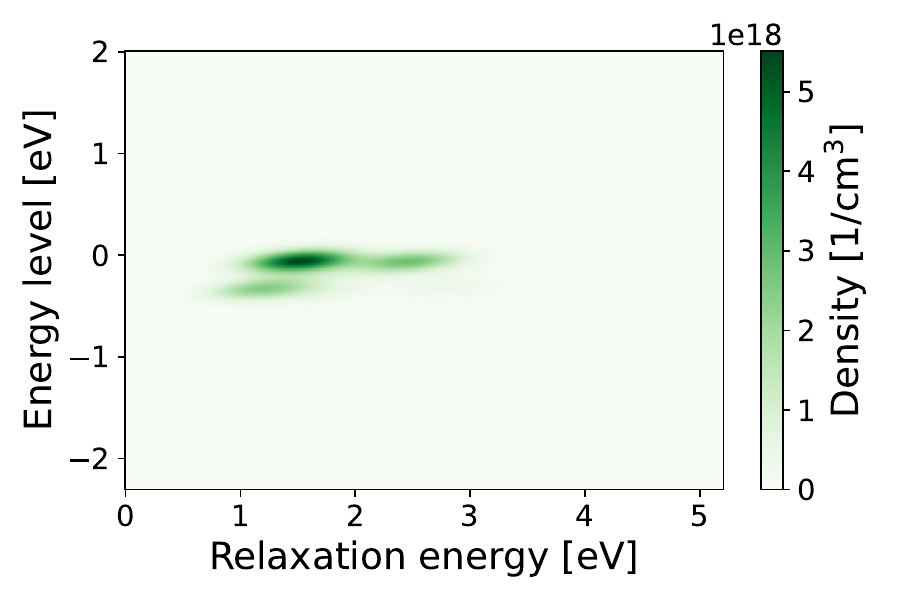}
\end{subfigure}

\begin{subfigure}[b]{.245\linewidth}
\includegraphics[width=1.0\linewidth]{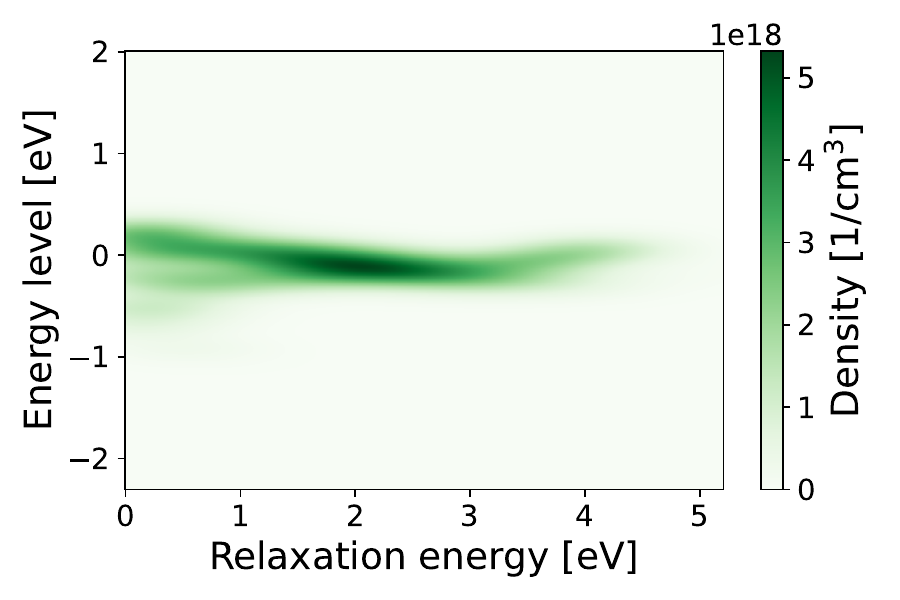}
\end{subfigure}
\begin{subfigure}[b]{.245\linewidth}
\includegraphics[width=1.0\linewidth]{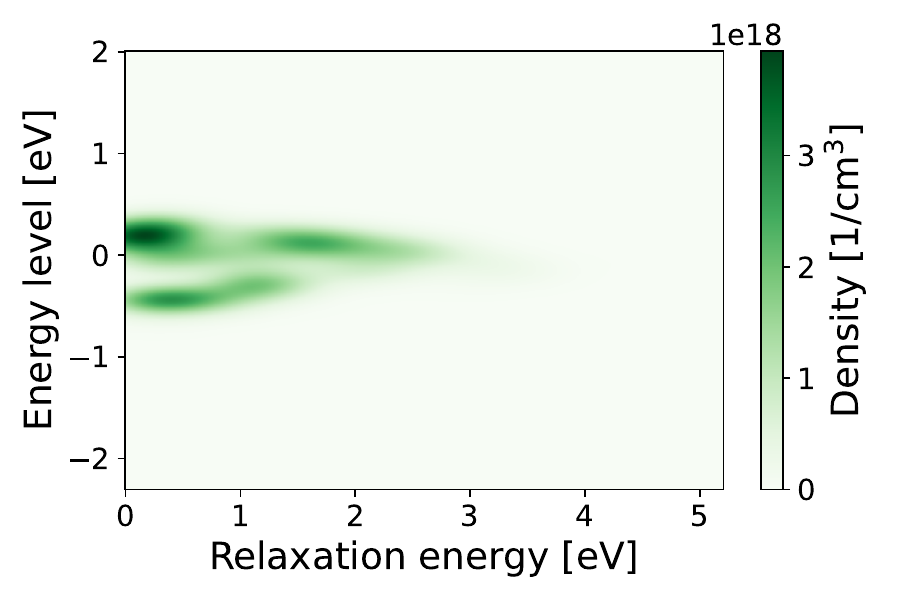}
\end{subfigure}
\begin{subfigure}[b]{.245\linewidth}
\includegraphics[width=1\linewidth]{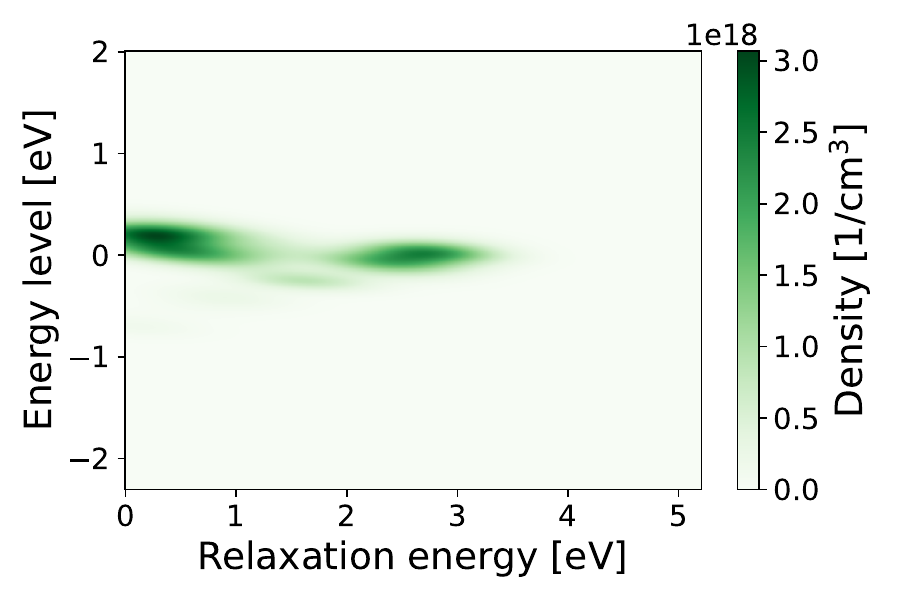}
\end{subfigure}
\begin{subfigure}[b]{.245\linewidth}
\includegraphics[width=1\linewidth]{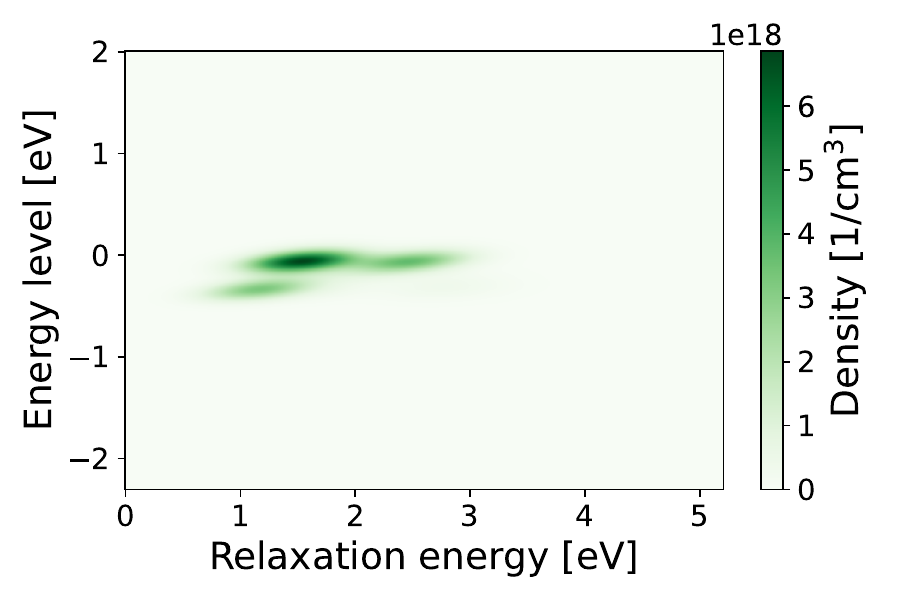}
\end{subfigure}

\begin{subfigure}[b]{.245\linewidth}
\includegraphics[width=1.0\linewidth]{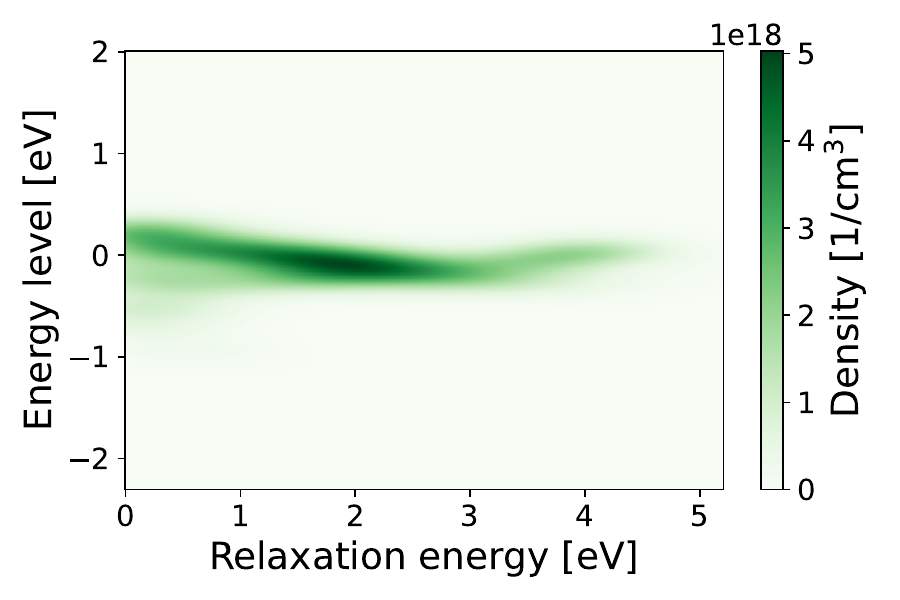}
\end{subfigure}
\begin{subfigure}[b]{.245\linewidth}
\includegraphics[width=1.0\linewidth]{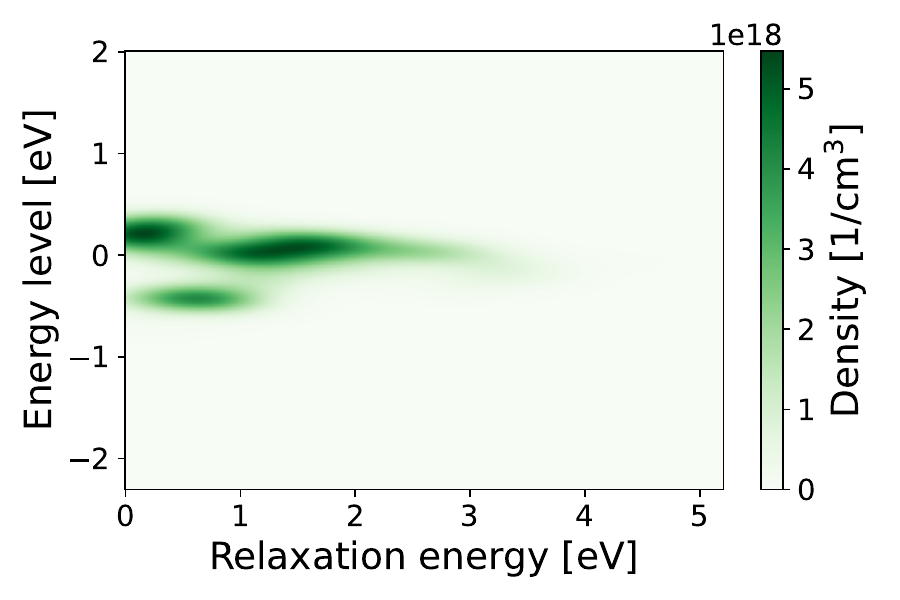}
\end{subfigure}
\begin{subfigure}[b]{.245\linewidth}
\includegraphics[width=1\linewidth]{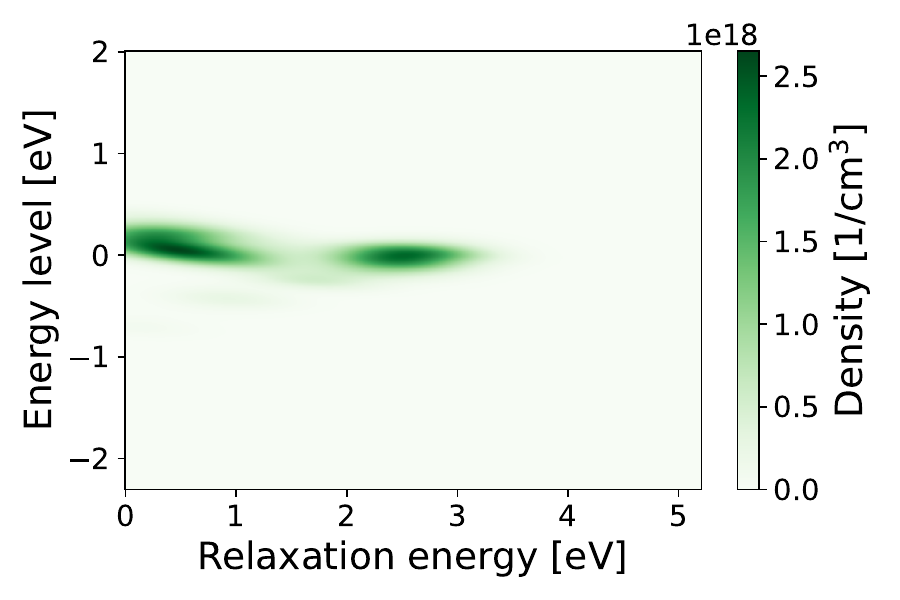}
\end{subfigure}
\begin{subfigure}[b]{.245\linewidth}
\includegraphics[width=1\linewidth]{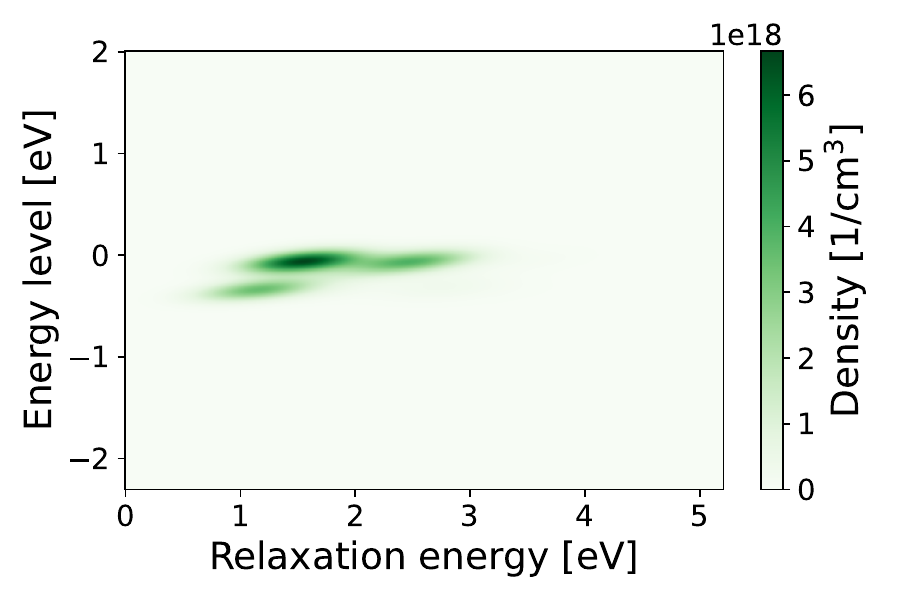}c
\end{subfigure}

\begin{subfigure}[b]{.245\linewidth}
\includegraphics[width=1.0\linewidth]{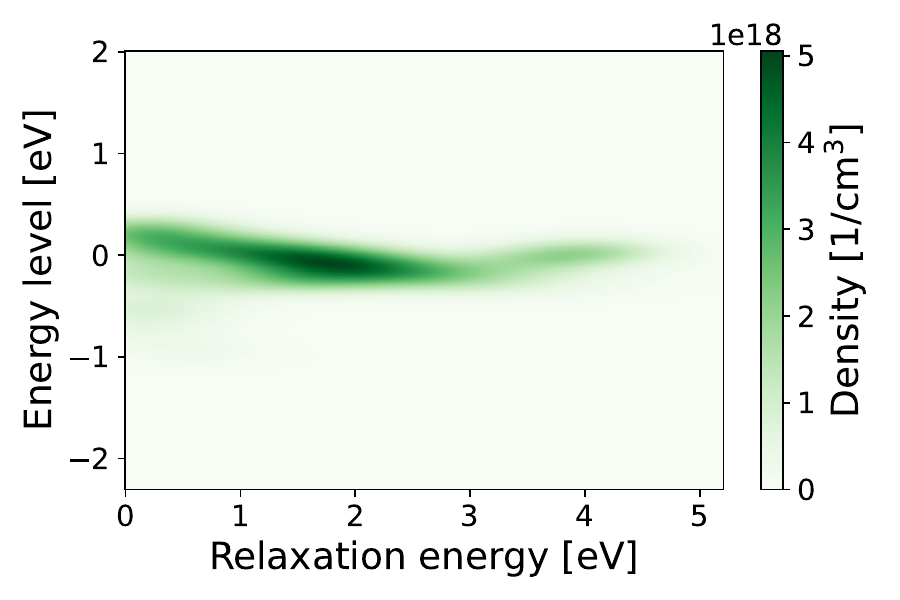}
\caption{}

\end{subfigure}
\begin{subfigure}[b]{.245\linewidth}
\includegraphics[width=1.0\linewidth]{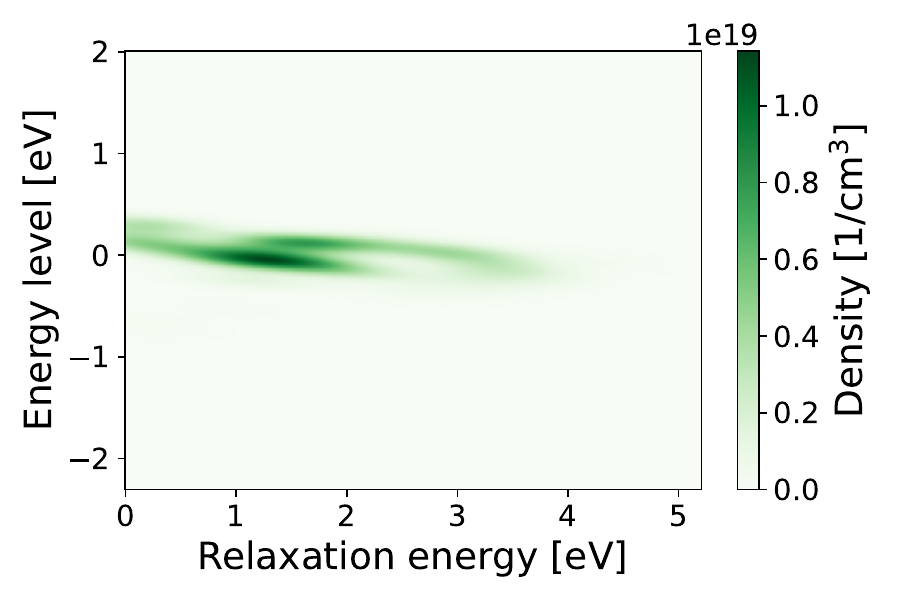}
\caption{}

\end{subfigure}
\begin{subfigure}[b]{.245\linewidth}
\includegraphics[width=1\linewidth]{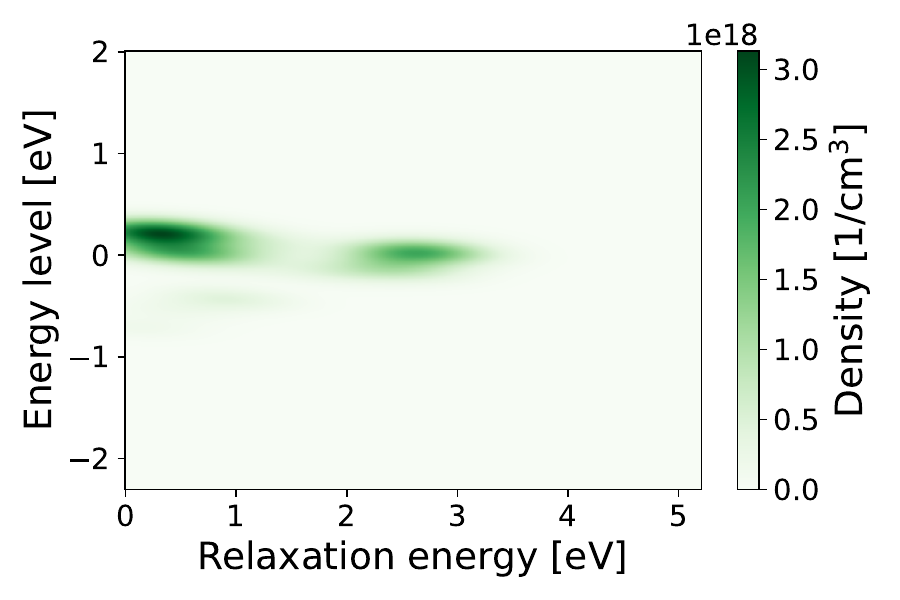}
\caption{}

\end{subfigure}
\begin{subfigure}[b]{.245\linewidth}
\includegraphics[width=1\linewidth]{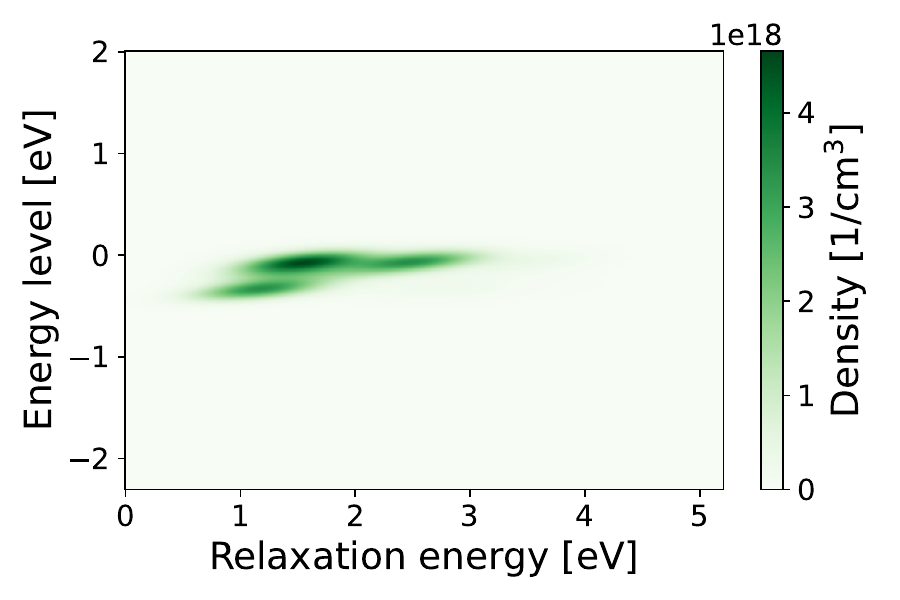}
\caption{}

\end{subfigure}
\caption{Extracted trap profiles for individual readouts.
Illustration of the energy levels and relaxation energies of the experimentally (ESiD) extracted distributions from hysteresis measurements at different readouts (top to down: the highest to the lowest \ith{}) as a function of sweep-rate at various temperatures for \textbf{(a)} the \plan{}, \textbf{(b)} the \fin{},\textbf{(c)} the \sgaa{}, and \textbf{(d)} the  \gaa{}.}
\end{figure}

\newpage
\subsection{Simulated Hysteresis Using DFT Defects}\label{supp:overal_def}
\begin{figure}[!h]
\begin{subfigure}[b]{.49\linewidth}
\includegraphics[width=1.0\linewidth]{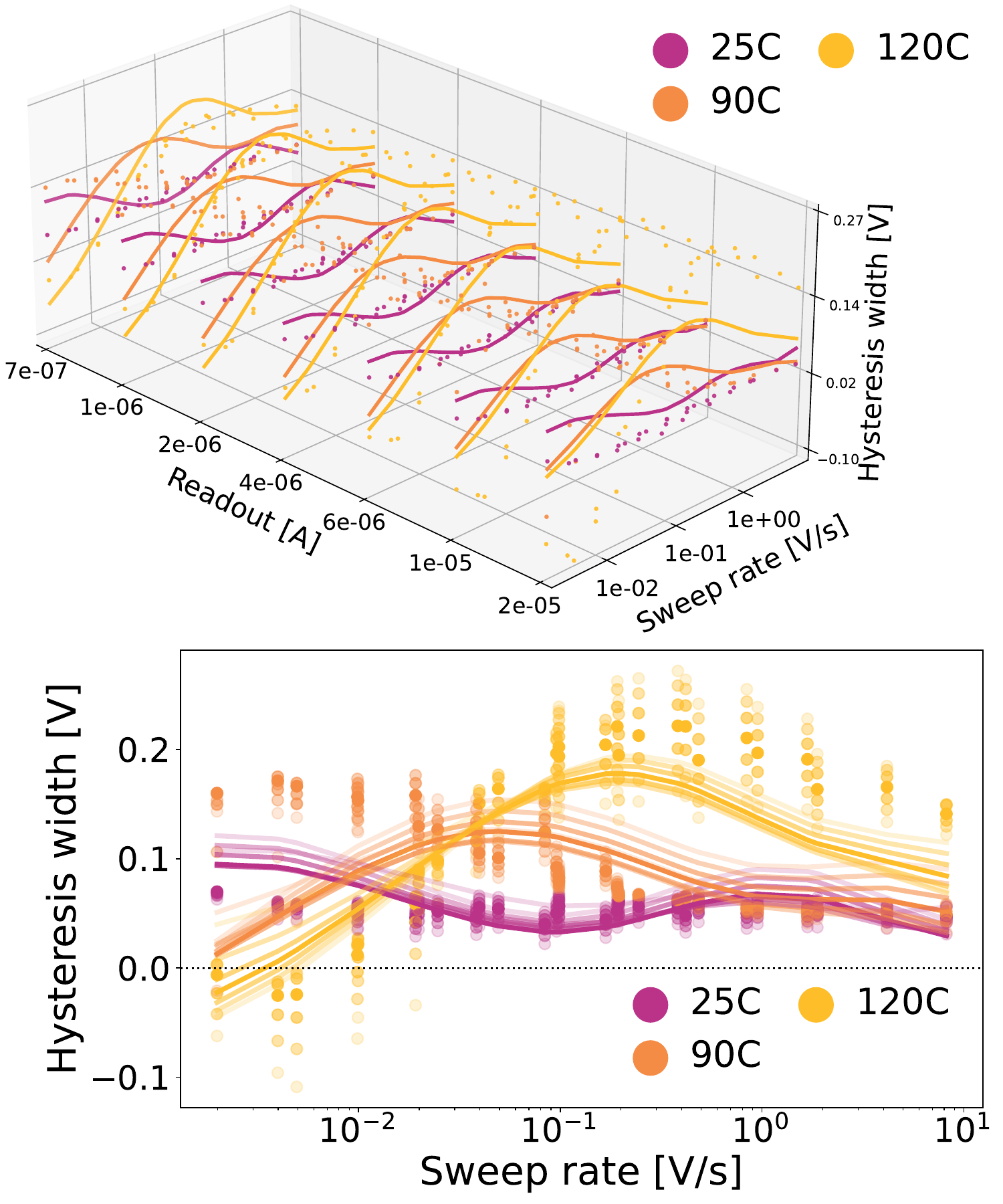}\caption{}
\end{subfigure}
\begin{subfigure}[b]{.49\linewidth}
\includegraphics[width=1.0\linewidth]{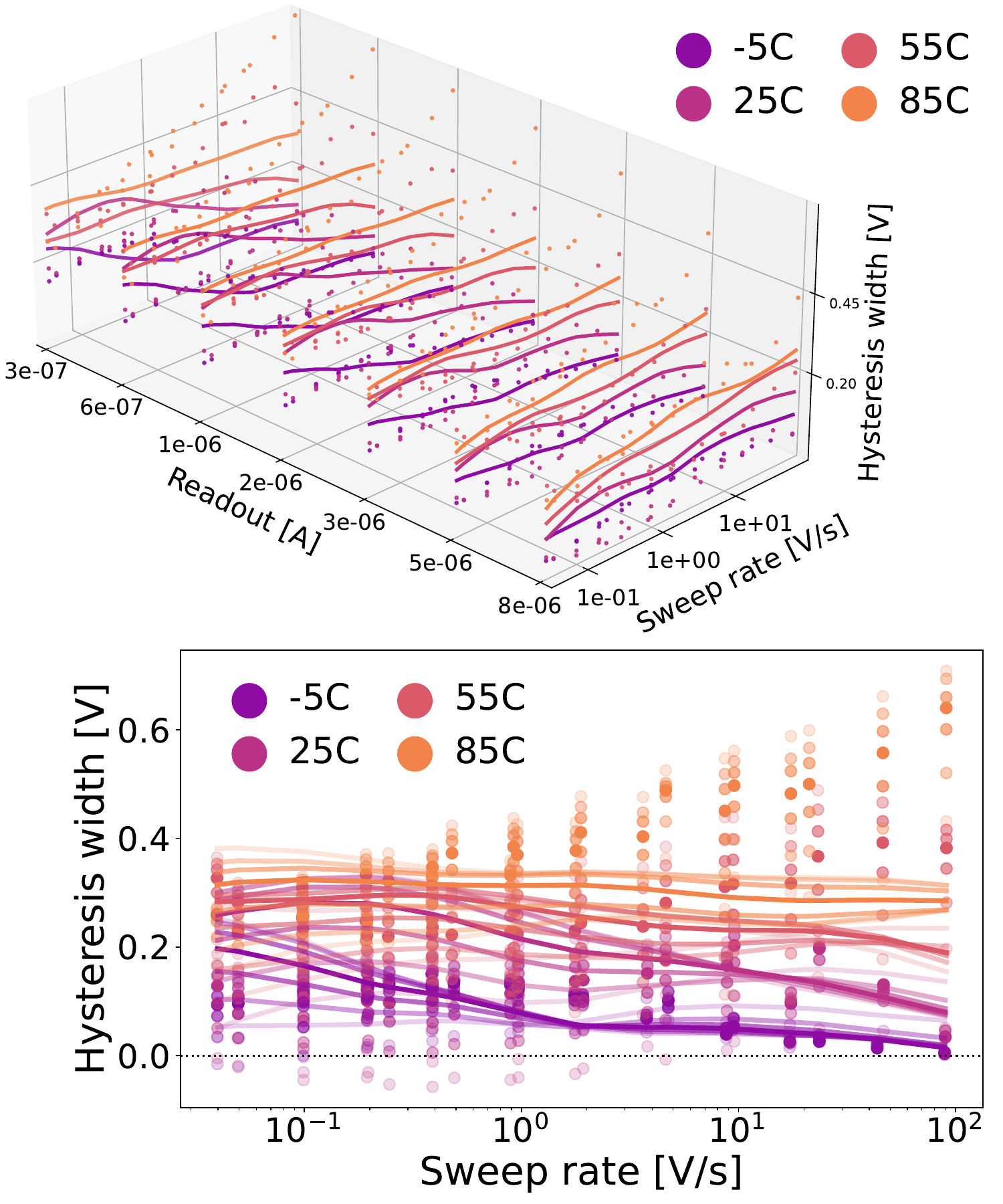}\caption{}
\end{subfigure}
\begin{subfigure}[b]{.49\linewidth}
\includegraphics[width=1\linewidth]{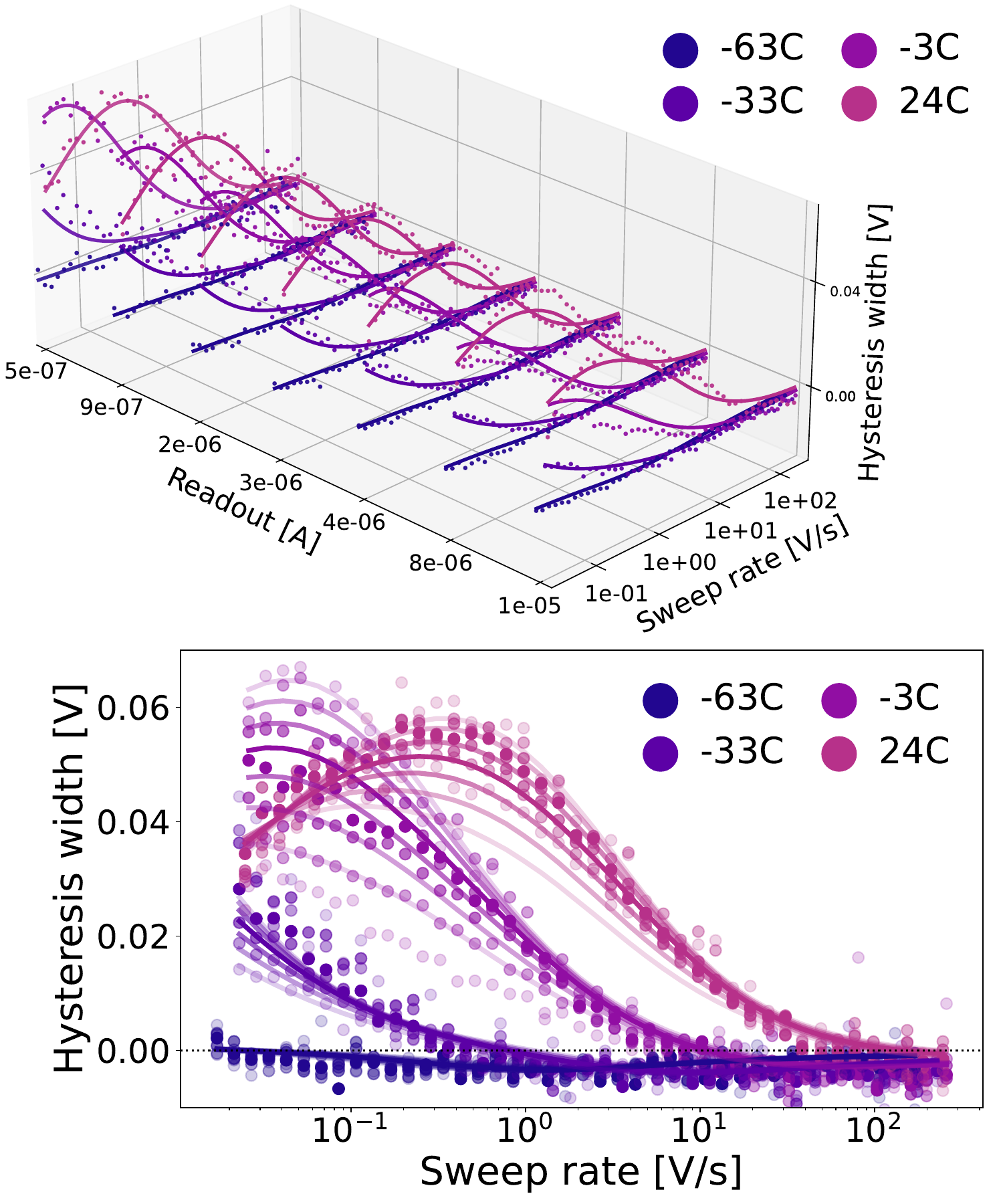}\caption{}
\end{subfigure}
\begin{subfigure}[b]{.49\linewidth}
\includegraphics[width=1\linewidth]{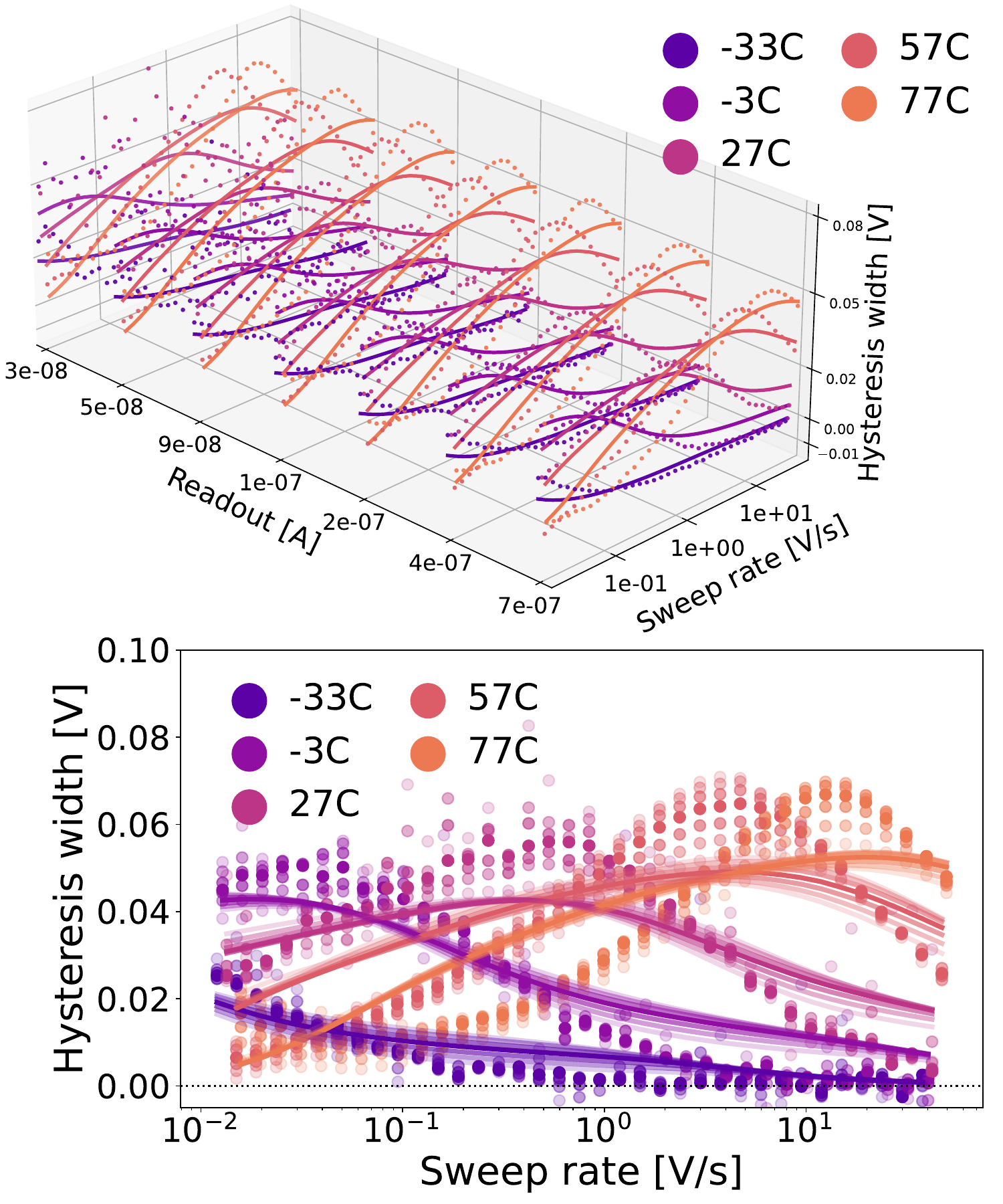}\caption{}
\end{subfigure}
\caption{Simulated (lines) hysteresis caused by predicted defects at all readouts compared to the experimentally measured values (circles) as a function of sweep-rate at various temperatures for \textbf{(a)} the \plan{}, \textbf{(b)} the \fin{},\textbf{(c)} the \sgaa{}, and \textbf{(d)} the  \gaa{}. In the upper figures the third dimension represent the current readout while in the lower figures opacity reflects proximity of the \ith{} to \itr{}.}
\end{figure}

\newpage
\subsection{Simulated Hysteresis for Individual Readouts Using DFT Defects}\label{supp:hys_def}
\begin{figure}[!h]
\begin{subfigure}[b]{.245\linewidth}
\includegraphics[width=1.0\linewidth]{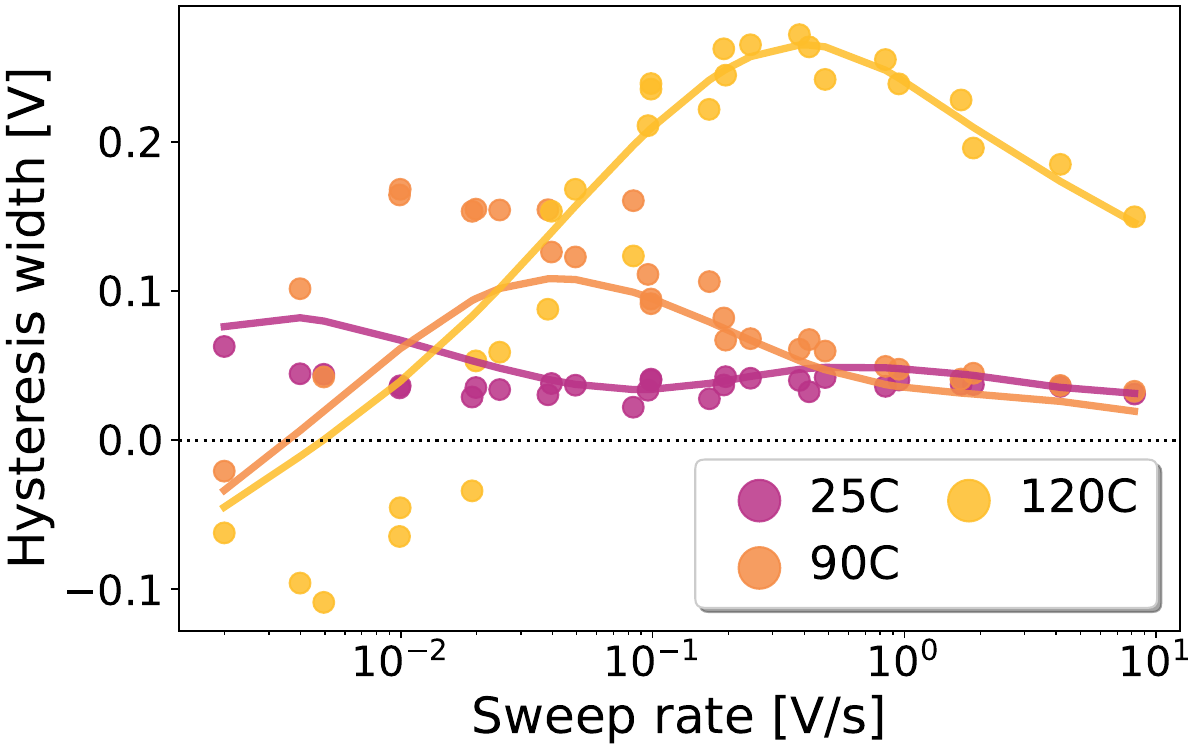}
\end{subfigure}
\begin{subfigure}[b]{.245\linewidth}
\includegraphics[width=1.0\linewidth]{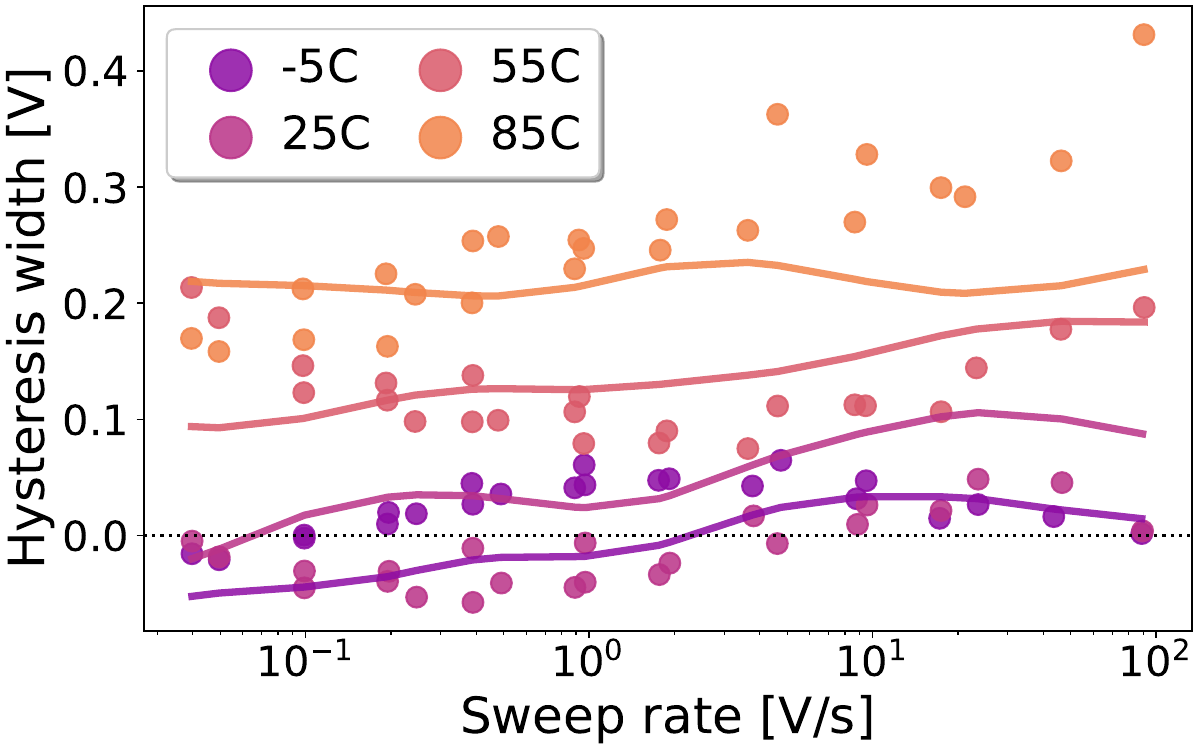}
\end{subfigure}
\begin{subfigure}[b]{.245\linewidth}
\includegraphics[width=1\linewidth]{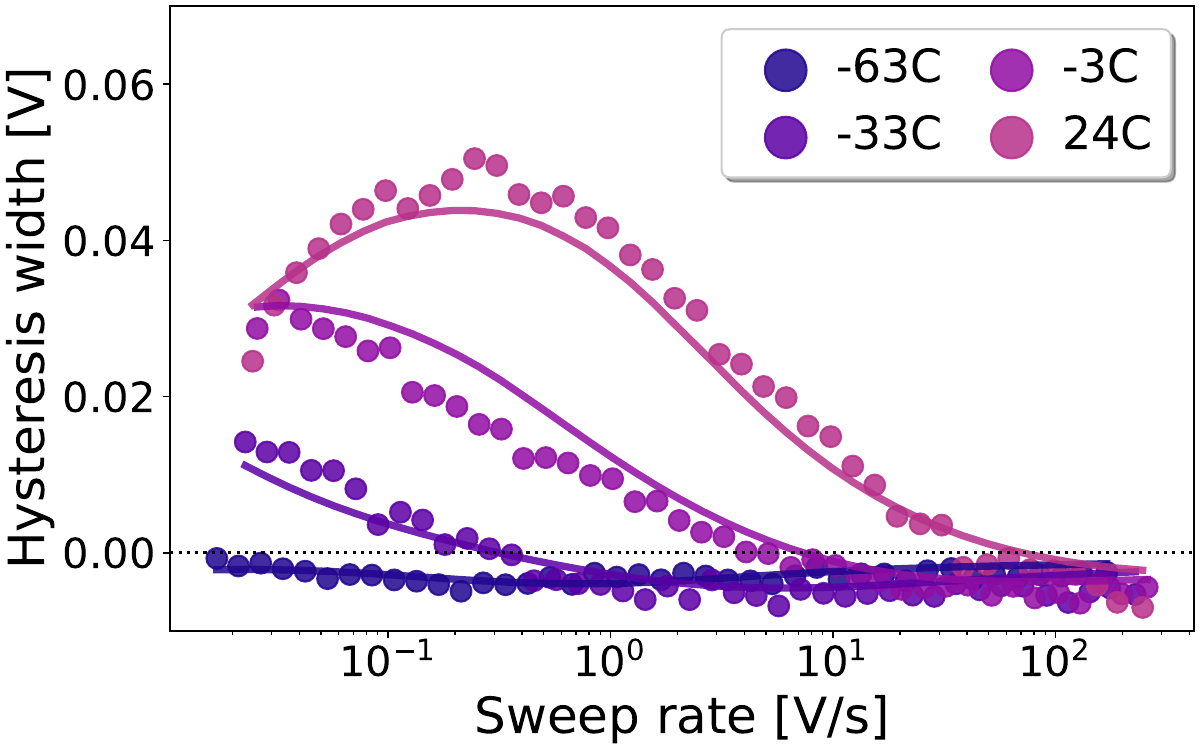}
\end{subfigure}
\begin{subfigure}[b]{.245\linewidth}
\includegraphics[width=1\linewidth]{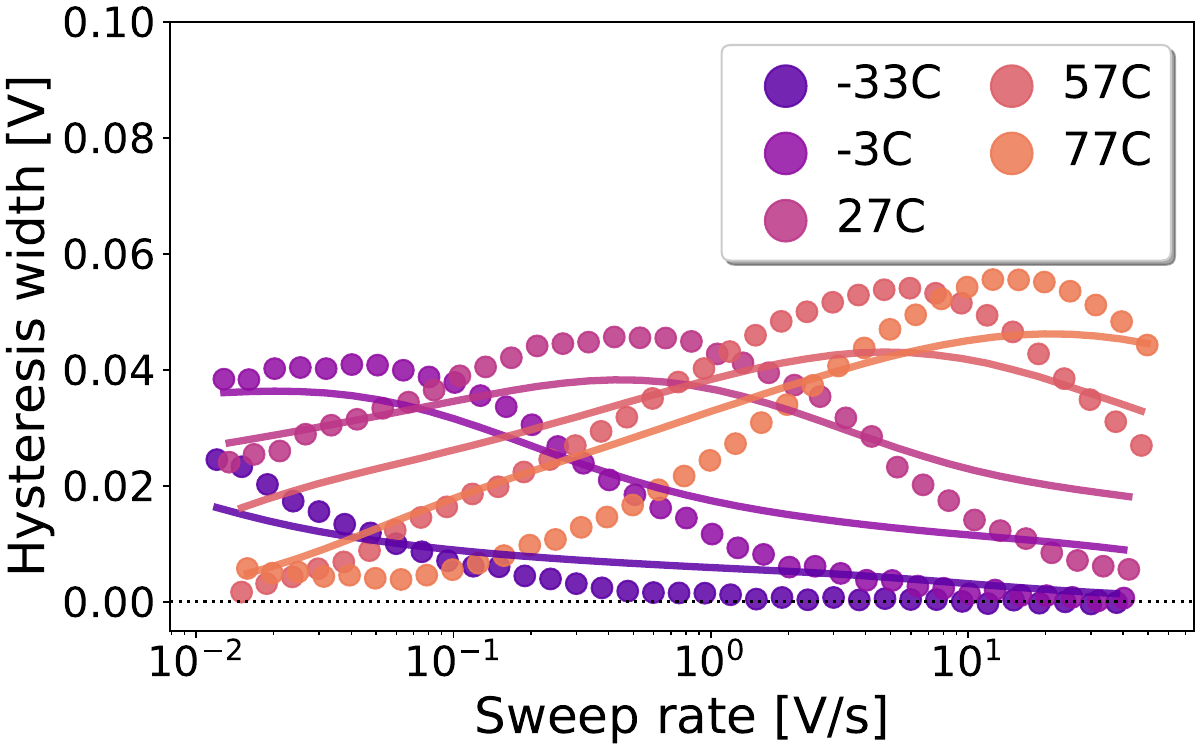}

\end{subfigure}
\begin{subfigure}[b]{.245\linewidth}
\includegraphics[width=1.0\linewidth]{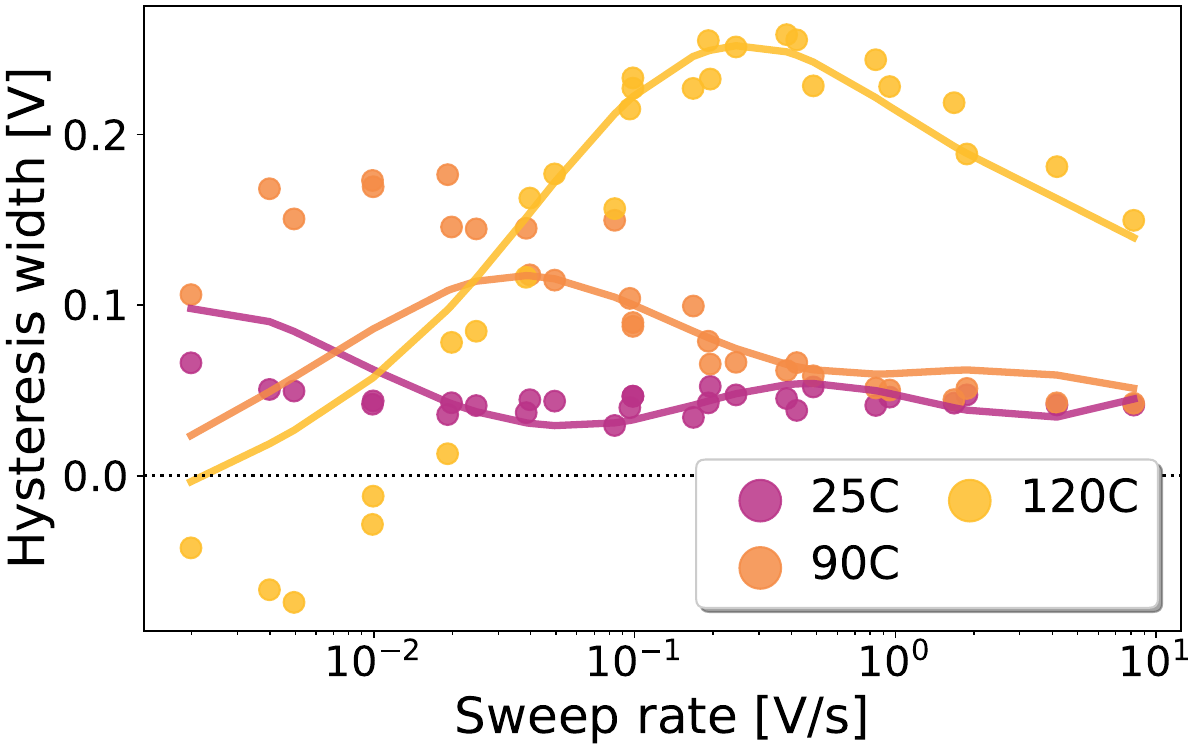}
\end{subfigure}
\begin{subfigure}[b]{.245\linewidth}
\includegraphics[width=1.0\linewidth]{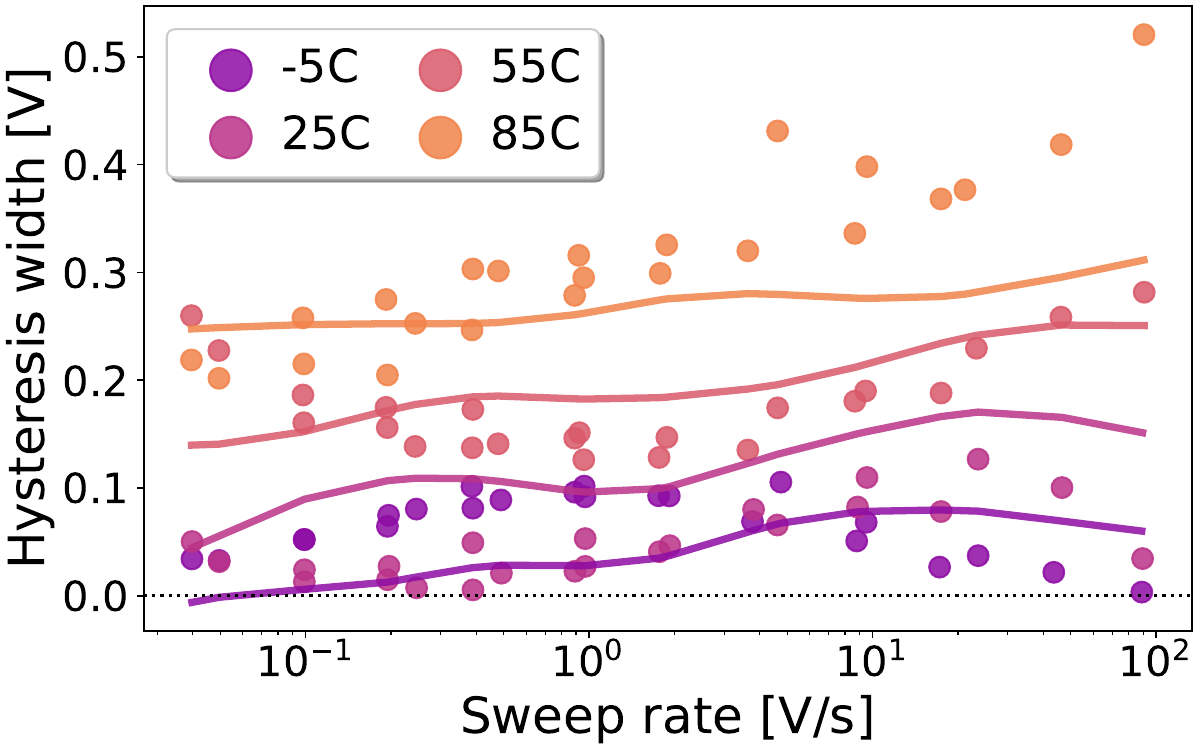}
\end{subfigure}
\begin{subfigure}[b]{.245\linewidth}
\includegraphics[width=1\linewidth]{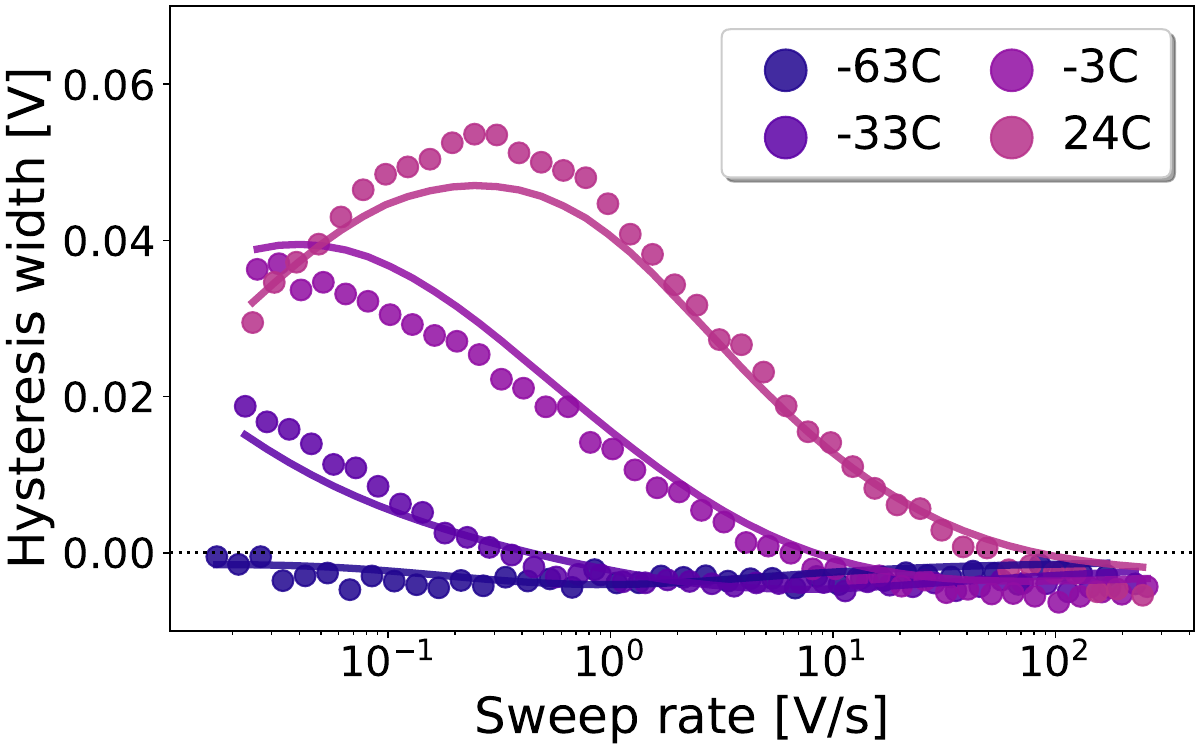}
\end{subfigure}
\begin{subfigure}[b]{.245\linewidth}
\includegraphics[width=1\linewidth]{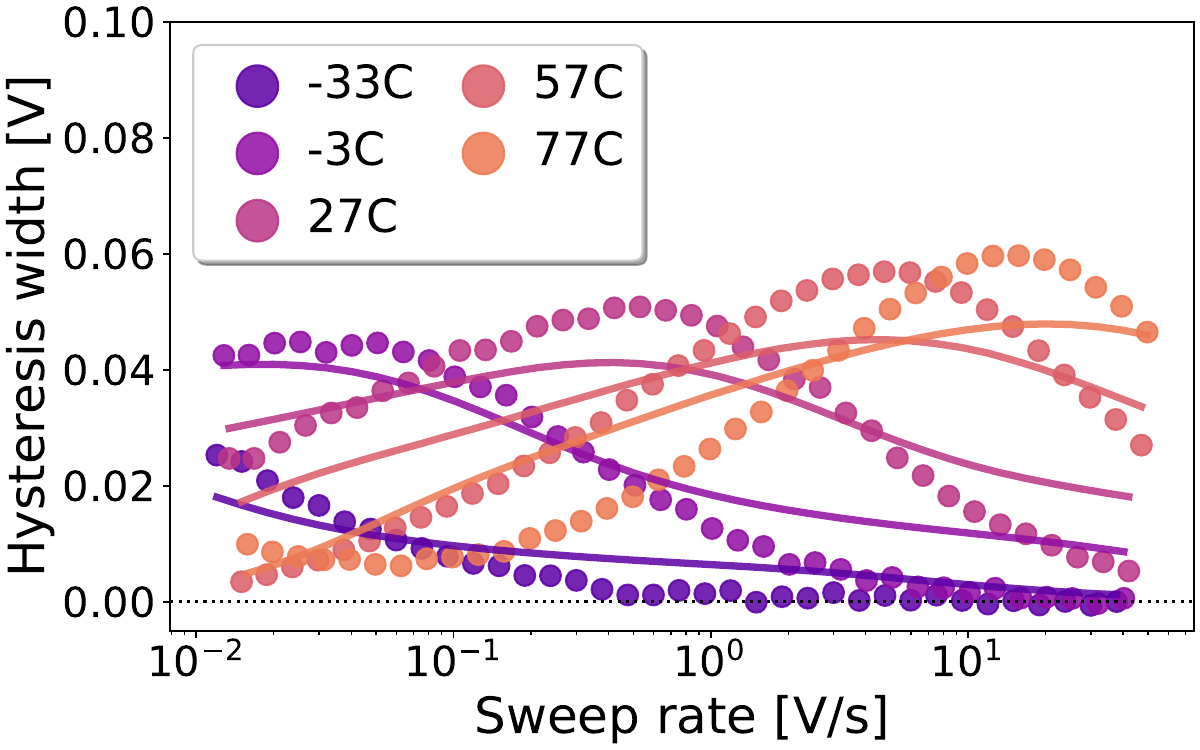}
\end{subfigure}

\begin{subfigure}[b]{.245\linewidth}
\includegraphics[width=1.0\linewidth]{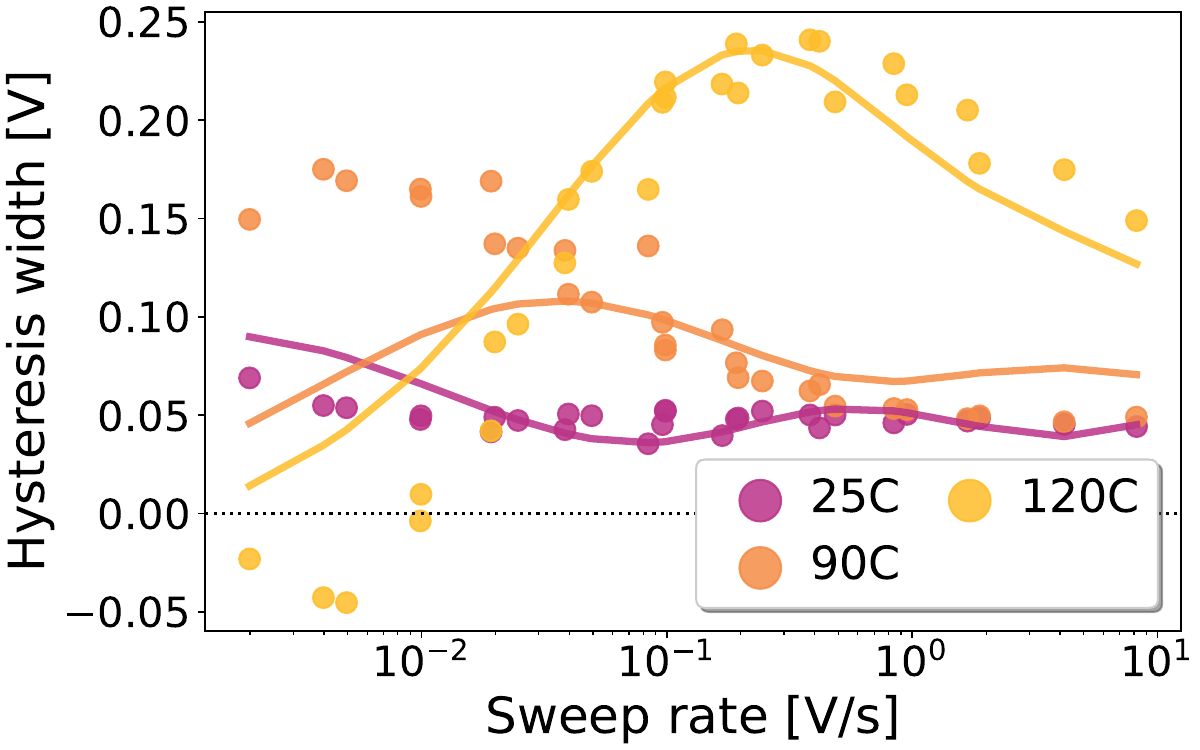}
\end{subfigure}
\begin{subfigure}[b]{.245\linewidth}
\includegraphics[width=1.0\linewidth]{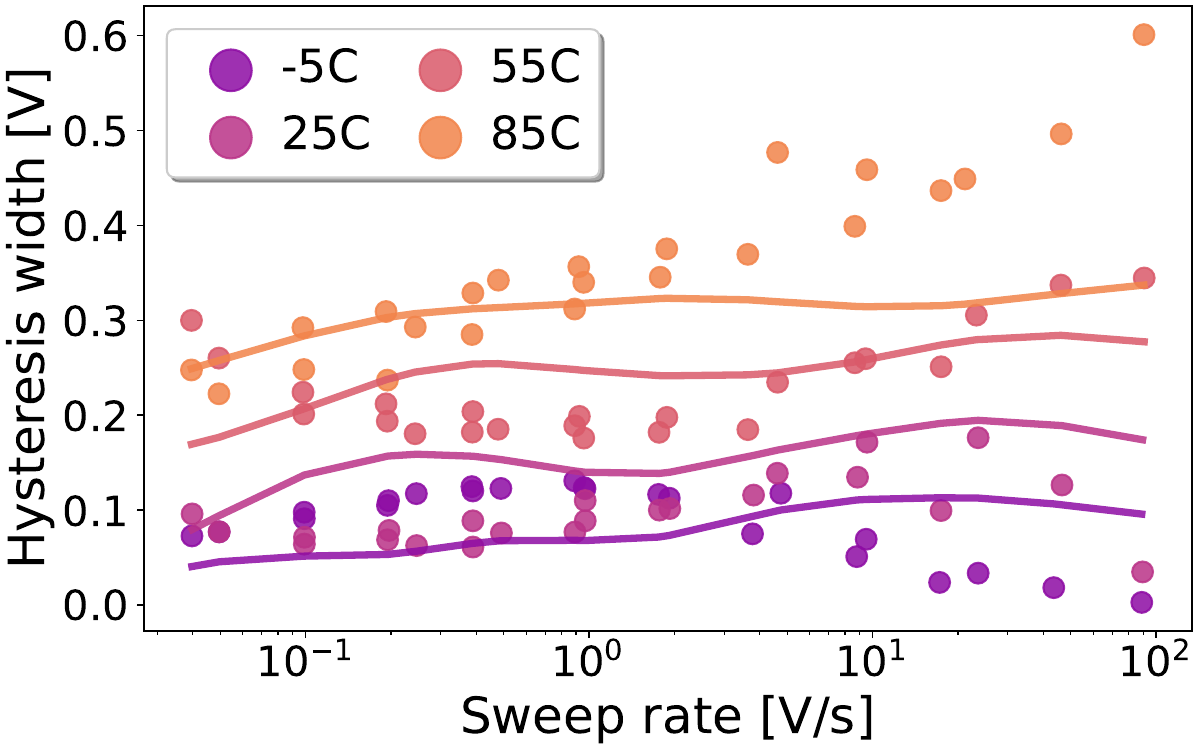}
\end{subfigure}
\begin{subfigure}[b]{.245\linewidth}
\includegraphics[width=1\linewidth]{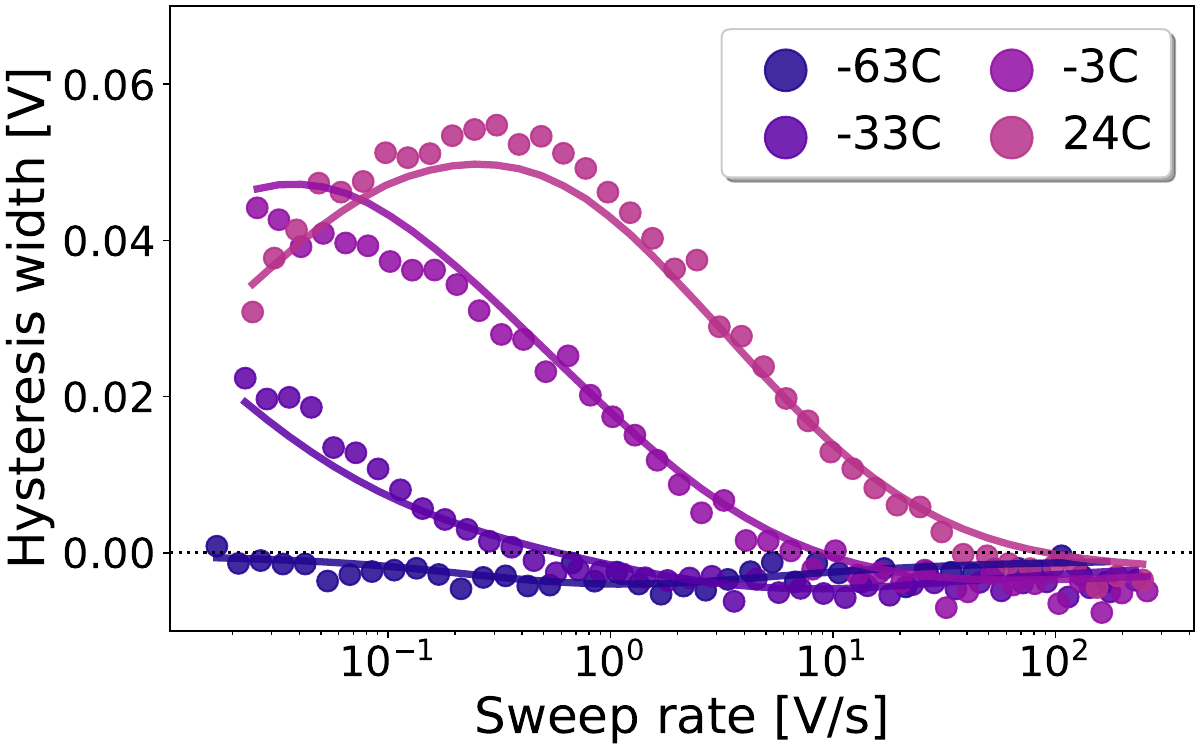}
\end{subfigure}
\begin{subfigure}[b]{.245\linewidth}
\includegraphics[width=1\linewidth]{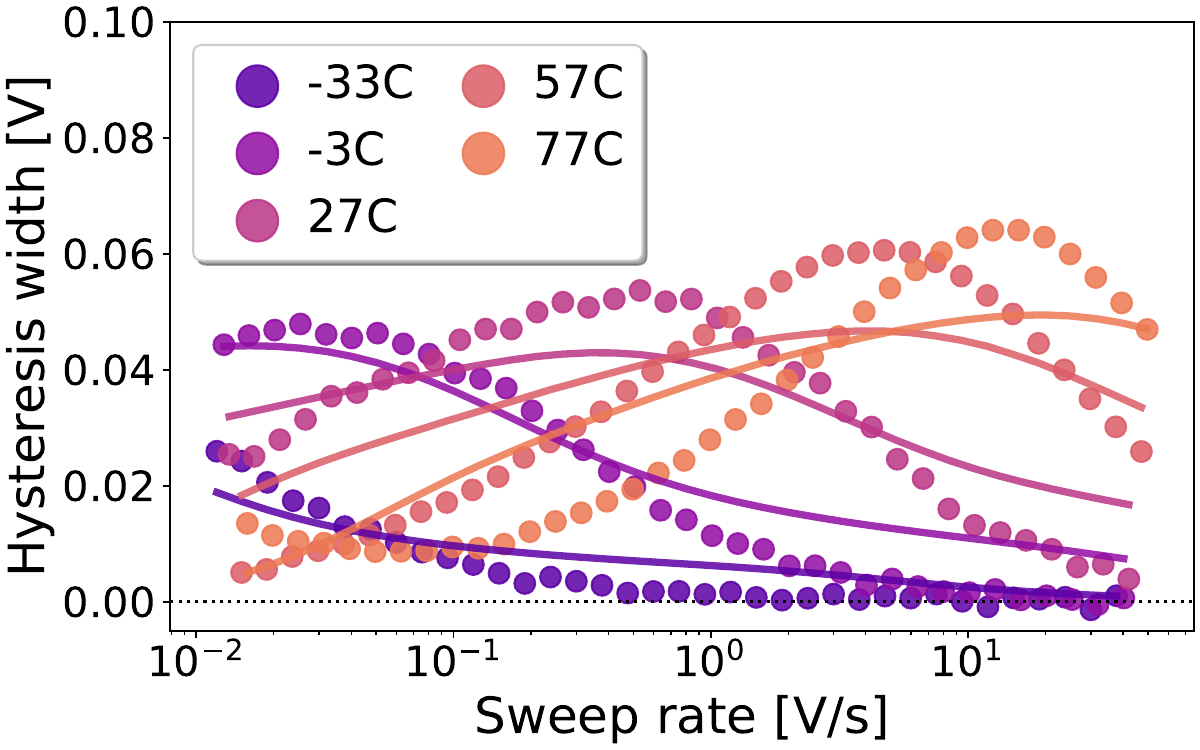}
\end{subfigure}

\begin{subfigure}[b]{.245\linewidth}
\includegraphics[width=1.0\linewidth]{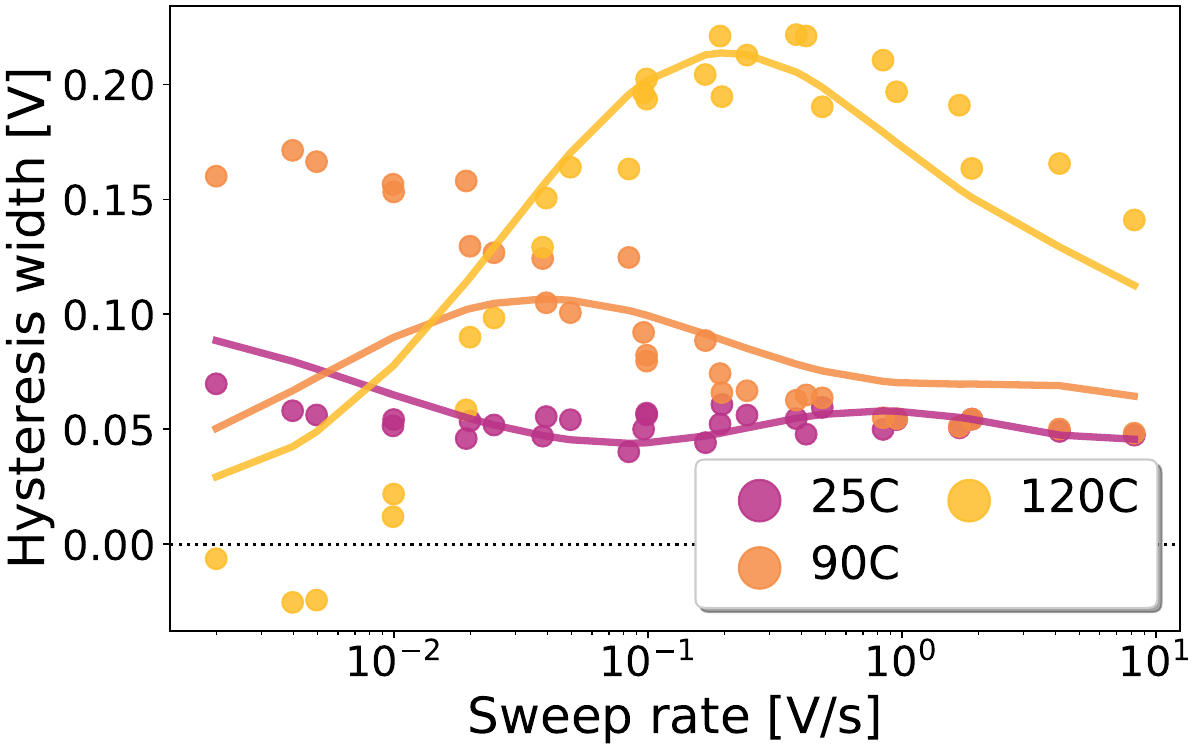}
\end{subfigure}
\begin{subfigure}[b]{.245\linewidth}
\includegraphics[width=1.0\linewidth]{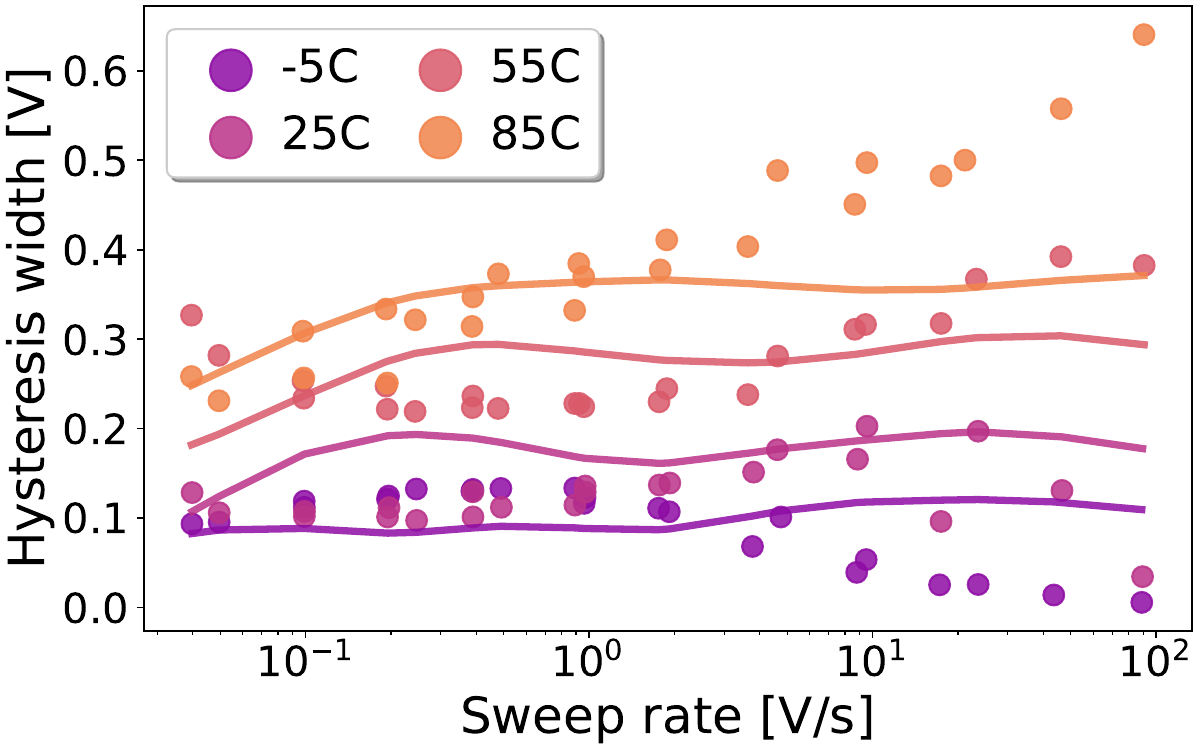}
\end{subfigure}
\begin{subfigure}[b]{.245\linewidth}
\includegraphics[width=1\linewidth]{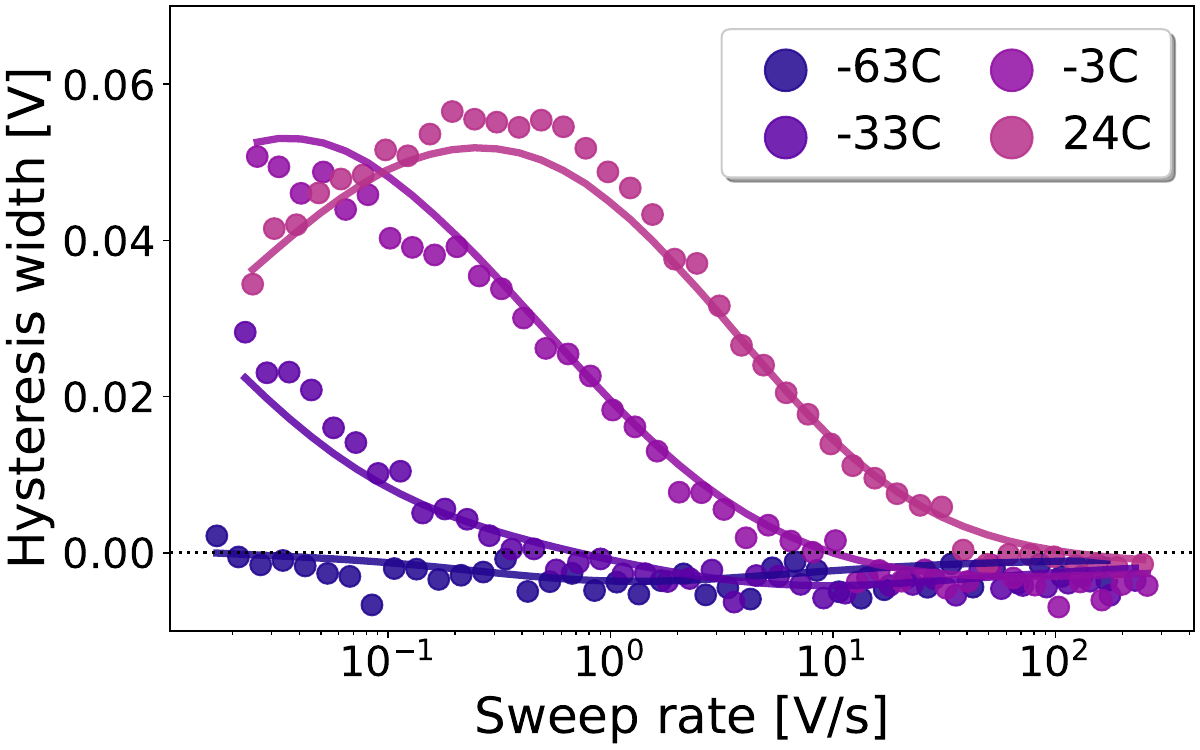}
\end{subfigure}
\begin{subfigure}[b]{.245\linewidth}
\includegraphics[width=1\linewidth]{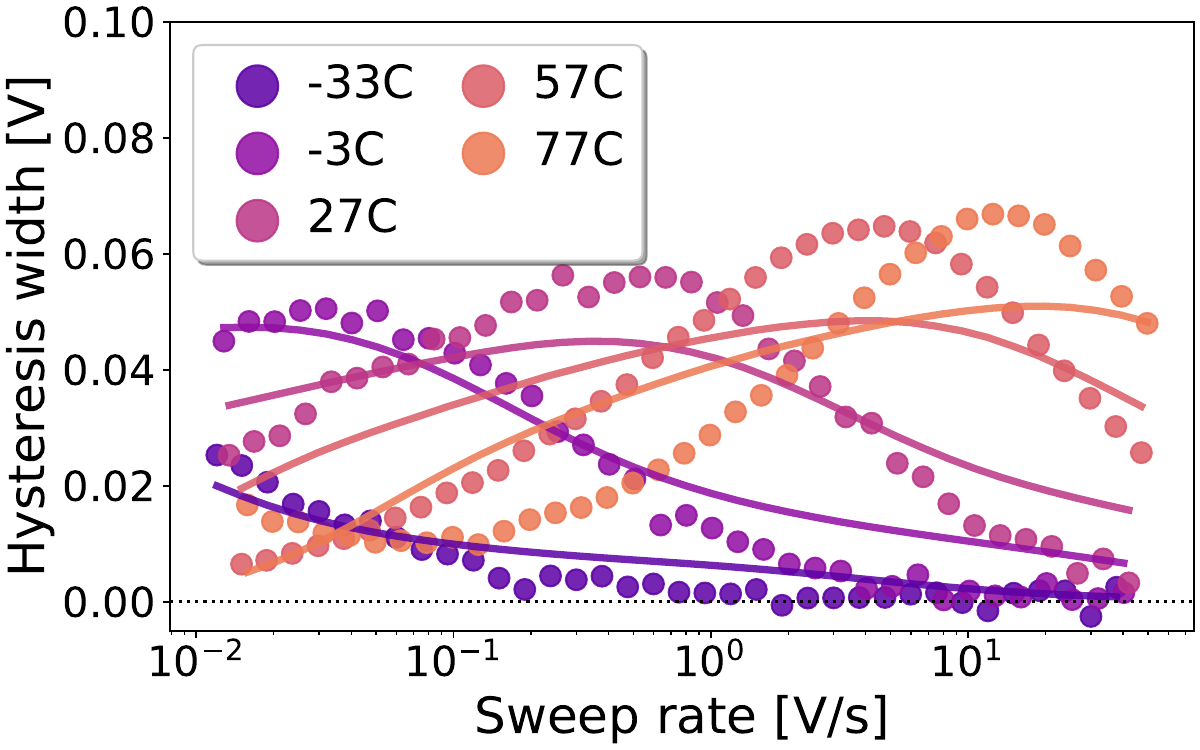}
\end{subfigure}

\begin{subfigure}[b]{.245\linewidth}
\includegraphics[width=1.0\linewidth]{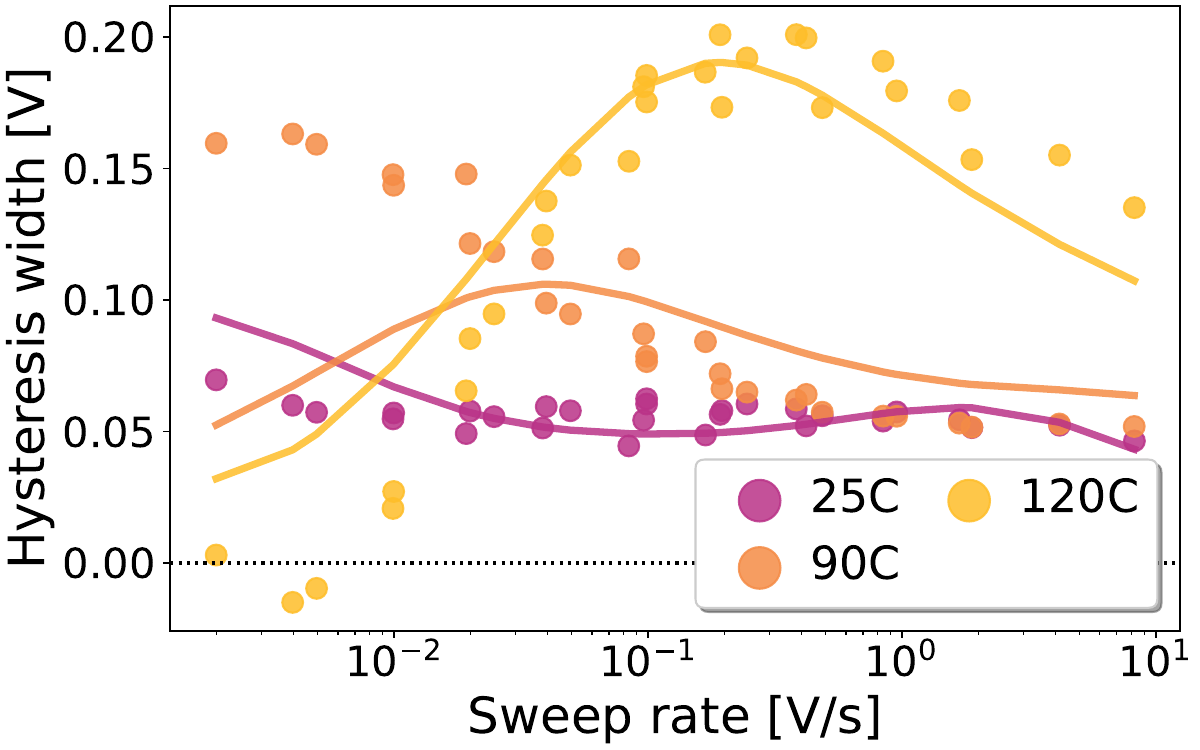}
\end{subfigure}
\begin{subfigure}[b]{.245\linewidth}
\includegraphics[width=1.0\linewidth]{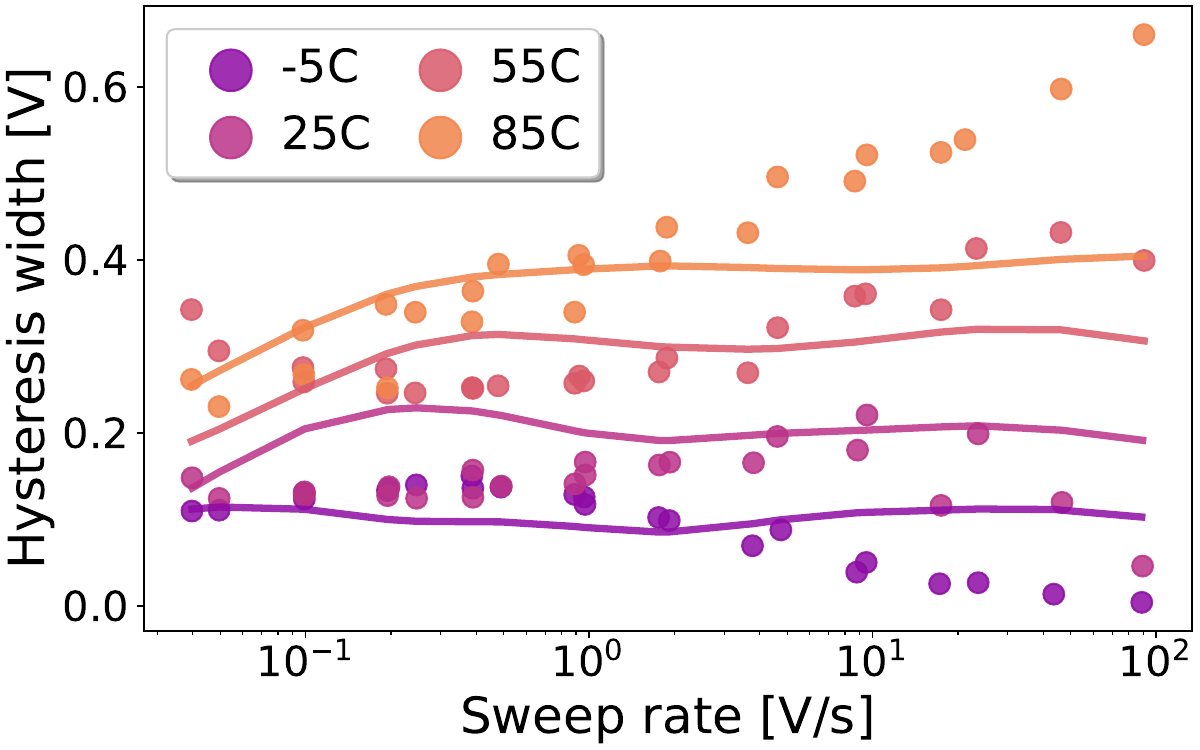}
\end{subfigure}
\begin{subfigure}[b]{.245\linewidth}
\includegraphics[width=1\linewidth]{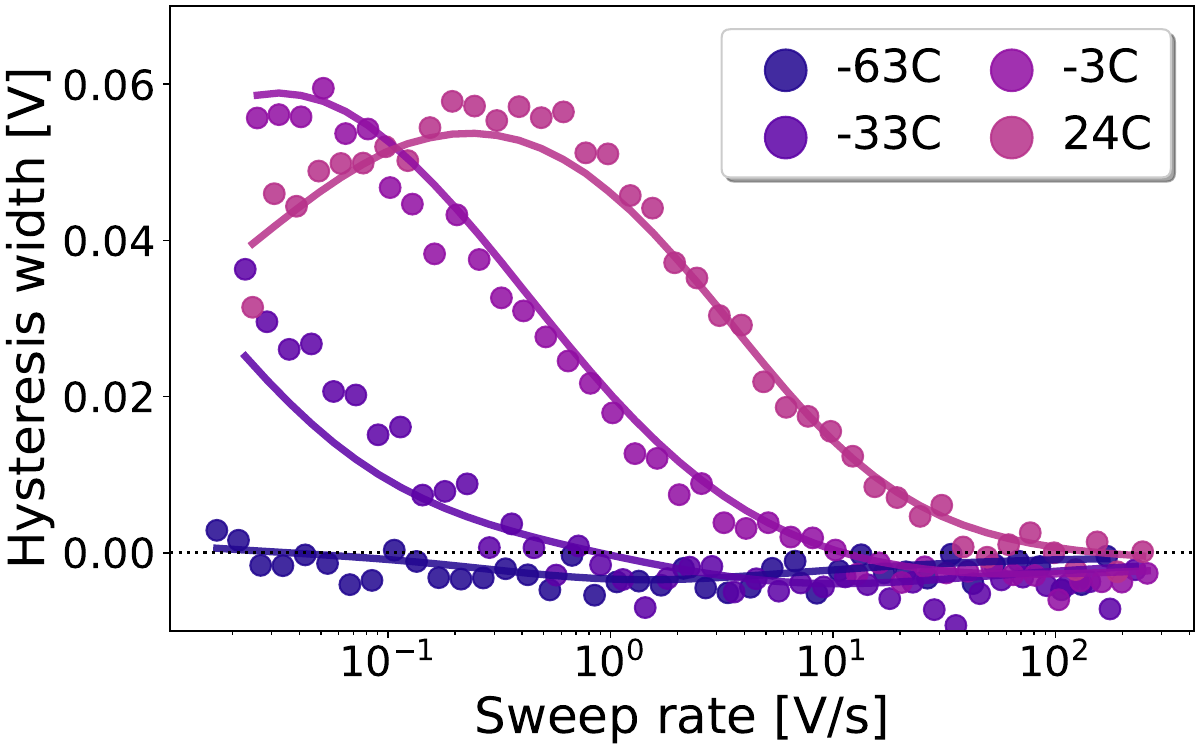}
\end{subfigure}
\begin{subfigure}[b]{.245\linewidth}
\includegraphics[width=1\linewidth]{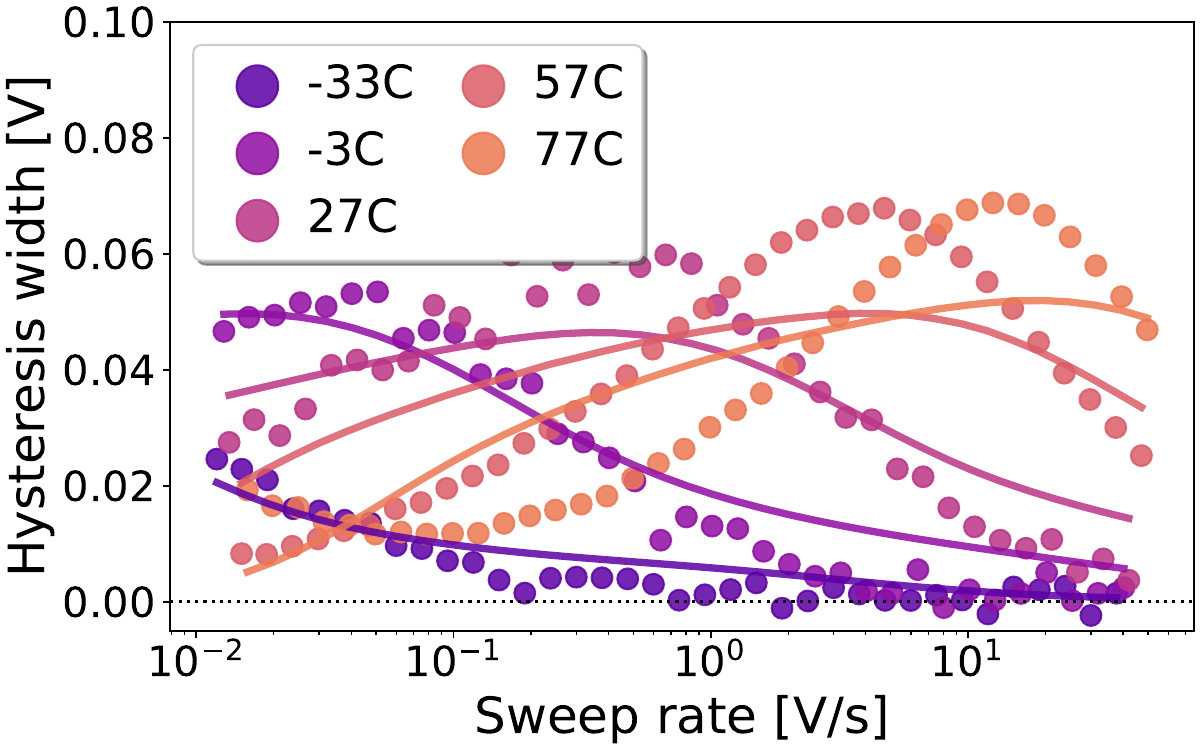}
\end{subfigure}

\begin{subfigure}[b]{.245\linewidth}
\includegraphics[width=1.0\linewidth]{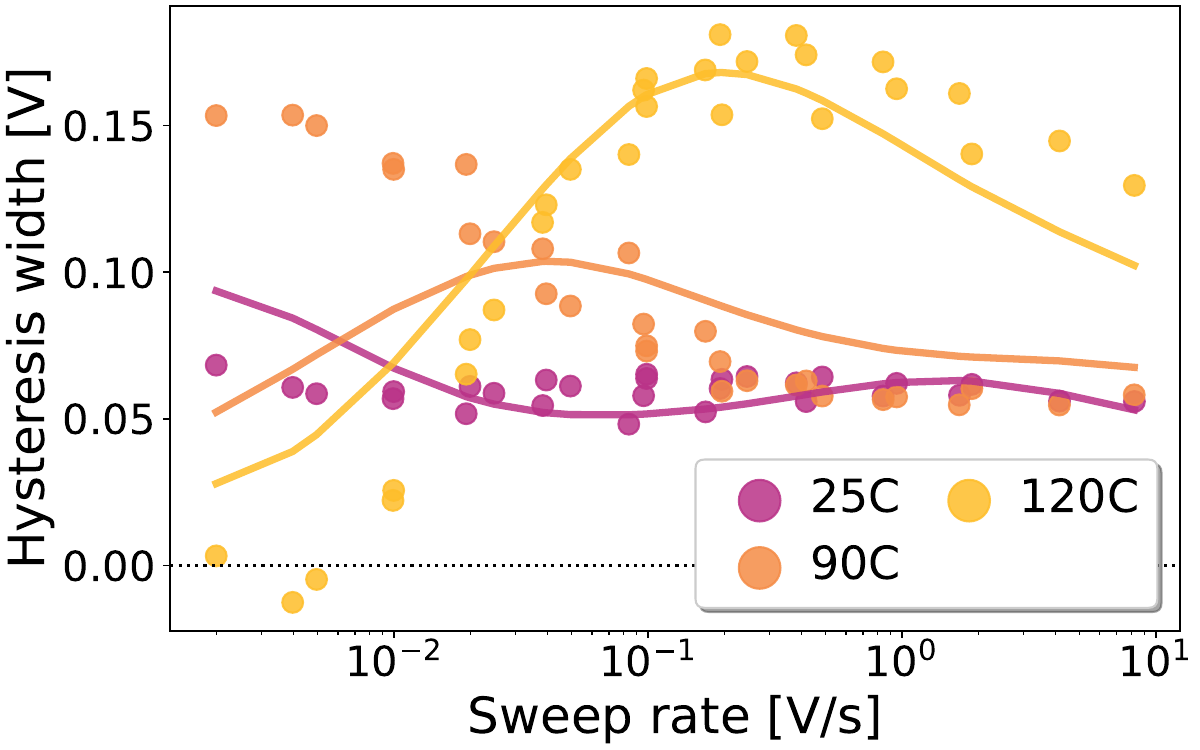}
\end{subfigure}
\begin{subfigure}[b]{.245\linewidth}
\includegraphics[width=1.0\linewidth]{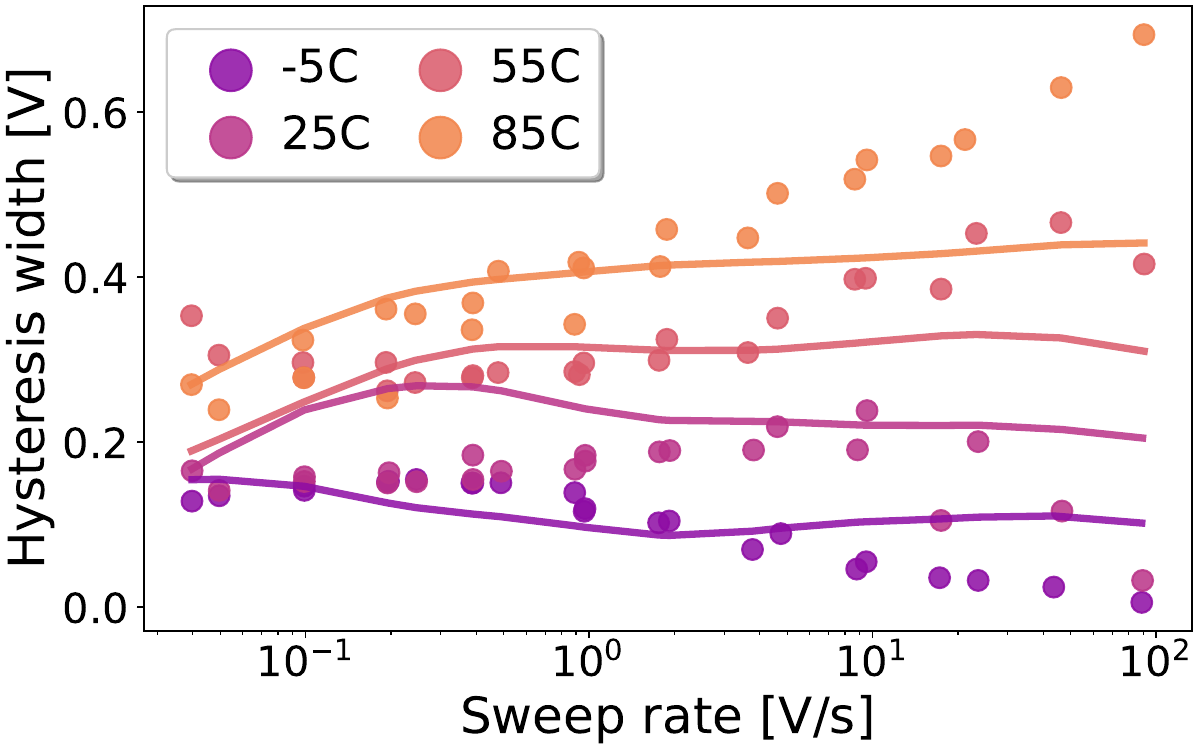}
\end{subfigure}
\begin{subfigure}[b]{.245\linewidth}
\includegraphics[width=1\linewidth]{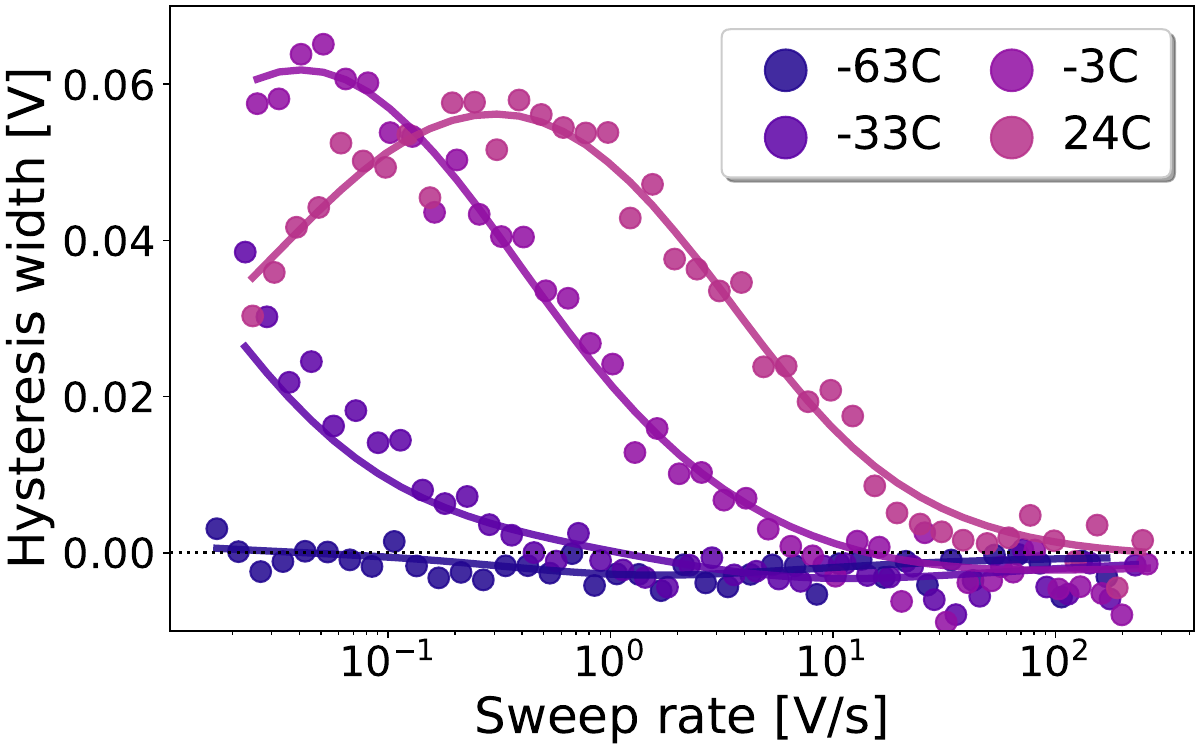}
\end{subfigure}
\begin{subfigure}[b]{.245\linewidth}
\includegraphics[width=1\linewidth]{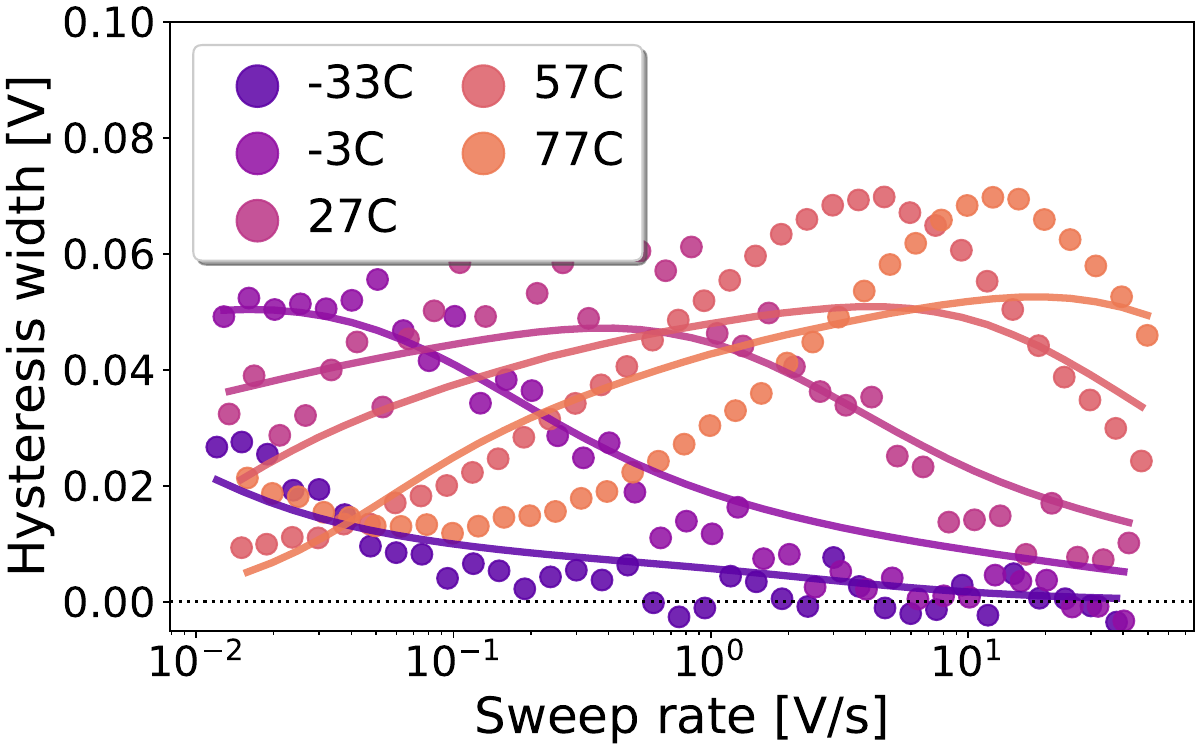}
\end{subfigure}

\begin{subfigure}[b]{.245\linewidth}
\includegraphics[width=1.0\linewidth]{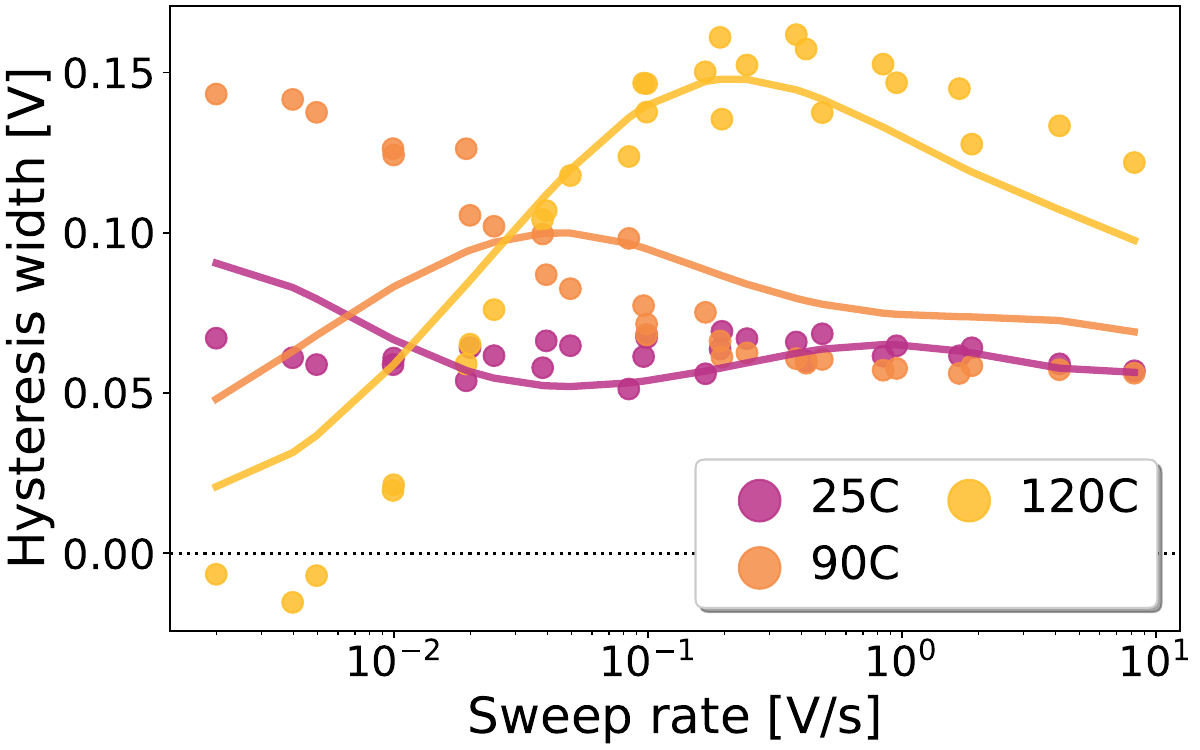}
\caption{}

\end{subfigure}
\begin{subfigure}[b]{.245\linewidth}
\includegraphics[width=1.0\linewidth]{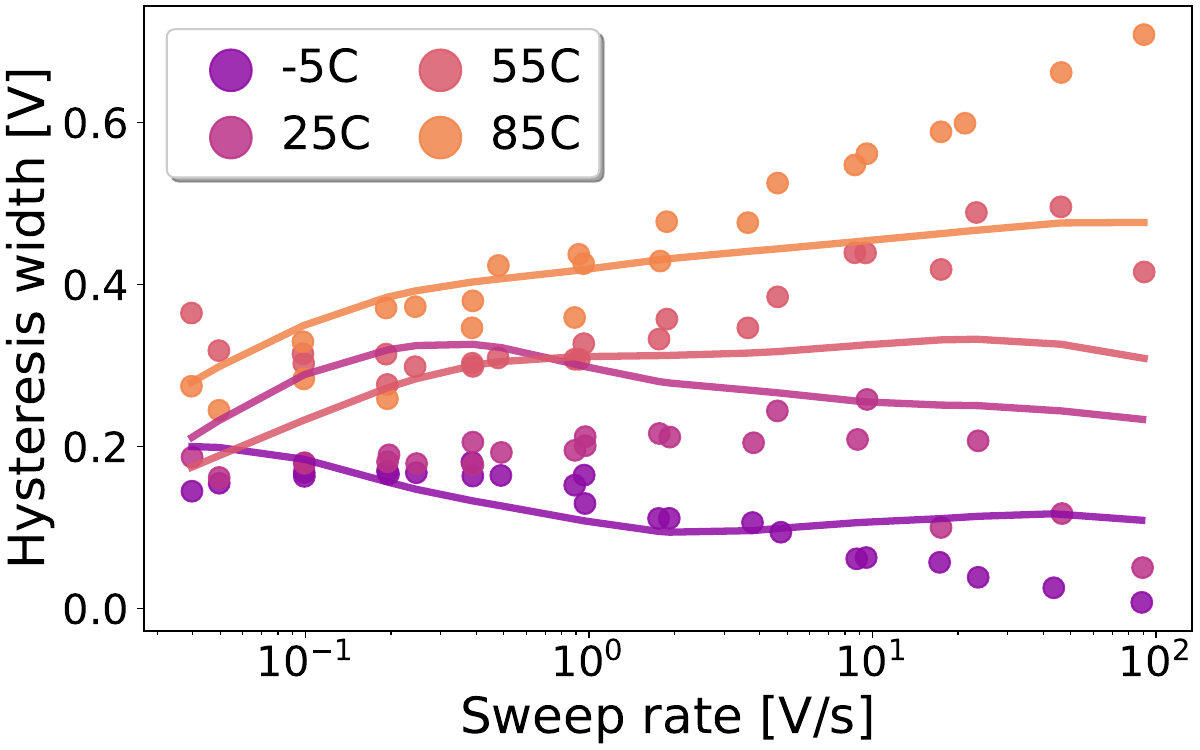}
\caption{}

\end{subfigure}
\begin{subfigure}[b]{.245\linewidth}
\includegraphics[width=1\linewidth]{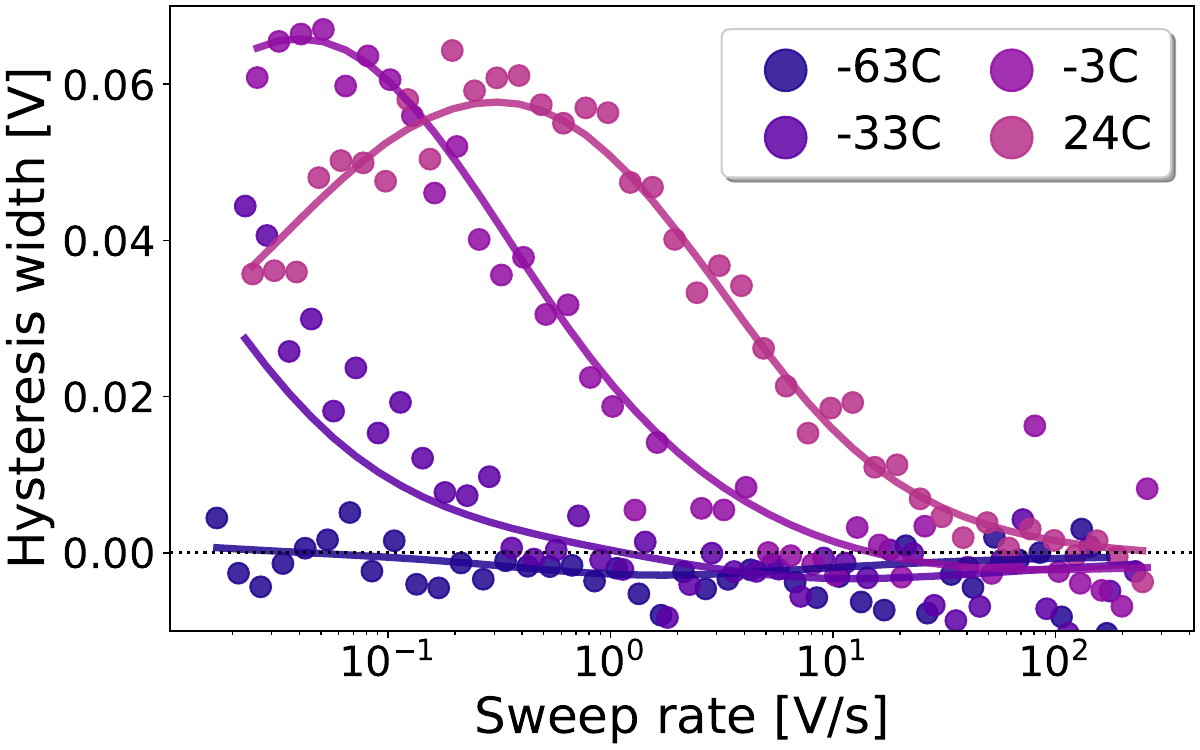}
\caption{}

\end{subfigure}
\begin{subfigure}[b]{.245\linewidth}
\includegraphics[width=1\linewidth]{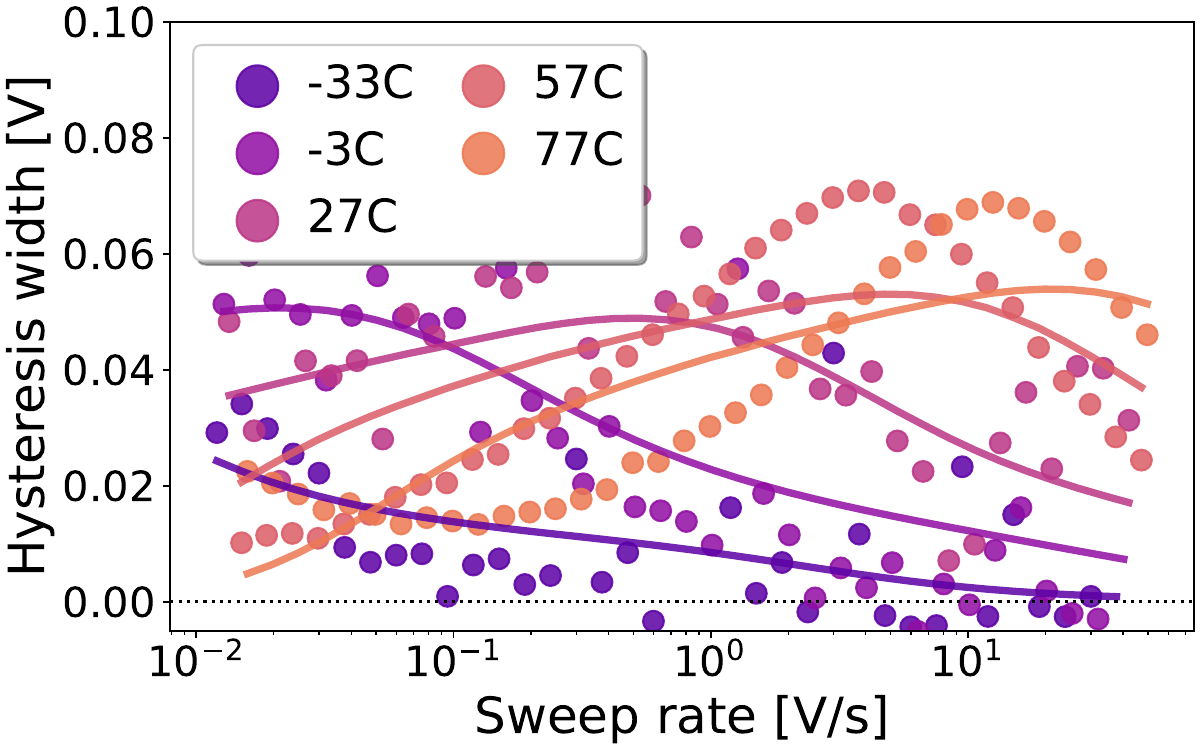}
\caption{}

\end{subfigure}
\caption{Hysteresis caused by predicted defects at individual readouts.
Measured (circles) and simulated (lines) hysteresis when only using suitably weighted DFT defects as traps, at different readouts (top to down: the highest to the lowest \ith{}) as a function of sweep-rate at various temperatures for \textbf{(a)} the \plan{}, \textbf{(b)} the \fin{},\textbf{(c)} the \sgaa{}, and \textbf{(d)} the \gaa{}.}
\end{figure}

\end{appendices}
\newpage
\bibliography{refs}
\clearpage
\end{document}